\documentclass[11pt,letterpaper,twoside]{report}

\usepackage{blindtext}

\usepackage{apptools}

\usepackage{geometry}
\usepackage{setspace}
\usepackage{titlesec}

\usepackage{tocloft}
\usepackage{indentfirst}
\usepackage[hang,flushmargin]{footmisc}

\usepackage[square,comma,sort&compress,numbers]{natbib}

\usepackage{calc}

\usepackage{bibentry}
\nobibliography*

\usepackage[labelformat=parens,list=on]{subcaption}
\usepackage{graphicx}
\graphicspath{{figs/}}
\usepackage{booktabs}
\usepackage{multicol}
\usepackage{listings}

\usepackage{multirow}

\usepackage{algorithm} 
\usepackage[noend]{algpseudocode} 
\usepackage{verbatim}

\usepackage{amsthm}
\usepackage{amsmath}
\usepackage{amssymb}
\usepackage{mathtools}
\usepackage{thmtools}
\usepackage{bm}
\usepackage{thm-restate}

\usepackage{tikz}
\usepackage{fontawesome5}
\usetikzlibrary{arrows, arrows.meta, decorations.pathreplacing, patterns, automata, positioning, fit, shapes, backgrounds, calc, patterns.meta}

\usepackage{ifthen} 

\usepackage{algorithm}
\usepackage{algpseudocode}

\usepackage{mathptmx}
\DeclareMathAlphabet{\mathcal}{OMS}{cmsy}{m}{n}
\usepackage{microtype}
\usepackage{textcomp}

\usepackage[hidelinks,pdfauthor={Andrew C. Freeman}, pdfpagemode={UseOutlines}]{hyperref} 

\usepackage[capitalize]{cleveref}
\usepackage{caption}
\usepackage{subcaption}
\usepackage{float}
\usepackage[htt]{hyphenat} 
\usepackage{array}
\usepackage{arydshln}

\usepackage{svg}
\usepackage{algorithm}
\usepackage{algpseudocode}
\usepackage{pifont}
\newcommand{\cmark}{\ding{51}}%

\usepackage{multirow}

\crefname{section}{Sec.}{Secs.}
\Crefname{section}{Section}{Sections}
\Crefname{table}{Table}{Tables}
\crefname{table}{Tab.}{Tabs.}

\usepackage{xspace}
\xspaceaddexceptions{]\}$)$}

\usepackage{comment}

\makeatletter
\algrenewcommand\ALG@beginalgorithmic{\ttfamily}
\makeatother

\usepackage{makerobust}

\usepackage{url}
\usepackage{breakurl}

\usepackage{multirow}

\usepackage{enumitem}

\usepackage[colorinlistoftodos]{todonotes}

\usepackage{pdfcomment}
\usepackage{titlecaps}

\usepackage{pifont}
\usepackage{lipsum}  
\usepackage{diagbox}
\usepackage{multirow}
\usepackage{adjustbox}

\titleformat{\chapter}[block]{\singlespacing\fillast\bfseries}{\centering \MakeUppercase{\chaptertitlename} \thechapter:~}{0em}{\uppercase}

\titleformat*{\section}{\bfseries}
\titleformat*{\subsection}{\bfseries}
\titleformat*{\subsubsection}{\bfseries}

\titlespacing{\chapter}{0in}{.62in}{11pt}

\titlespacing{\paragraph}{0in}{0.08in}{0.07in}

\allowdisplaybreaks

\newlength\schedgraphwidth
\newcommand{\Ab}[2]{\noindent  #1 \> #2 \\}
\newcommand{\Abi}[2]{\noindent #1 \hspace{1.5cm} \= #2 \\}
\newcommand{\adder}[0]{AD{$\Delta$}ER}
\newcommand{\dt}[0]{{$\Delta t$}}
\newcommand{\dtm}[0]{{$\Delta t_{max}$}}
\DeclareMathOperator*{\argmax}{\arg\!\max}

\newtheorem*{definition}{Definition}

\begin{document}

\newgeometry{left=1in,top=2in,right=1in,bottom=1in,nohead,footskip=0.5in}
\pagenumbering{roman}

\begin{titlepage}
\begin{center}


\vspace{2in}
\begin{singlespace}
    \MakeUppercase{Rethinking Video with a Universal Event-Based Representation}\\ 
\end{singlespace}

\vspace{61pt} 
Andrew C. Freeman
\end{center}


\vspace{39pt}
\begin{singlespace}
\noindent
\begin{center}
A dissertation submitted to the faculty at the University of North Carolina at Chapel Hill
in partial fulfillment of the requirements for the degree of Doctor of Philosophy in
the Department of Computer Science.
\end{center}
\end{singlespace}

\vspace{39pt}
\begin{center}
\begin{singlespace}
Chapel Hill\\
2024
\end{singlespace}
\end{center}


\vspace{39pt}
\begin{flushright}
\begin{minipage}[t]{1.8333 in}
Approved by:\\
Ketan Mayer-Patel\\
Montek Singh\\
Henry Fuchs\\
Colin Raffel\\
Guorong Wu\\
\end{minipage}
\end{flushright}

\end{titlepage}

\newgeometry{left=1in,top=8.47in,right=1in,bottom=1in,nohead,footskip=0.5in}

\begin{center}
\begin{singlespace}
\copyright 2024\\
Andrew C. Freeman\\
ALL RIGHTS RESERVED
\end{singlespace}
\end{center}

\clearpage

\newgeometry{left=1in,top=2in,right=1in,bottom=1in,nohead,footskip=0.5in}
\restoregeometry

\begin{center}
\vspace*{52pt}
{\textbf{ABSTRACT}}
\vspace{11pt}

\begin{singlespace}
    Andrew C. Freeman: Rethinking Video with a Universal Event-Based Representation\\
(Under the direction of Ketan Mayer-Patel and Montek Singh)
\end{singlespace}
\end{center}

Traditionally, video is structured as a sequence of discrete image frames. Recently, however, a novel video sensing paradigm has emerged which eschews video frames entirely. These ``event'' sensors aim to mimic the human vision system with asynchronous sensing, where each pixel has an independent, sparse data stream. While these cameras enable high-speed and high-dynamic-range sensing, researchers often revert to a framed representation of the event data for existing applications, or build bespoke applications for a particular camera’s event data type. At the same time, classical video systems have significant computational redundancy at the application layer, since pixel samples are repeated across frames in the uncompressed domain.

To address the shortcomings of existing systems, I introduce \textbf{A}ddress, \textbf{D}ecimation, $\Delta t$ \textbf{E}vent \textbf{R}epresentation (\textbf{\adder{}}, pronounced “adder''), a novel intermediate video representation and system framework. The framework transcodes a variety of framed and event camera sources into a single event-based representation, which supports source-modeled lossy compression and backward compatibility with traditional frame-based applications. I demonstrate that \adder{} achieves state-of-the-art application speed and compression performance for scenes with high temporal redundancy. Crucially, I describe how \adder{} unlocks an entirely new control mechanism for computer vision: application speed can correlate with both the scene content and the level of lossy compression. Finally, I discuss the implications for event-based video on large-scale video surveillance and resource-constrained sensing.

\clearpage


\begin{center}
\vspace*{52pt}
\textit{To my loving wife Lexie, and to God for whom we live}
\end{center}

\pagebreak


\begin{center}
\vspace*{52pt}
{\textbf{ACKNOWLEDGEMENTS}}
\end{center}

I owe the deepest gratitude to my advisors, Ketan Mayer-Patel and Montek Singh, for their unrivaled support and guidance through the years. I could not have imagined a more perfect research environment for my personality, nor a more interesting research topic to explore. I also express thanks to my committee members and research collaborators outside the department: Henry Fuchs from the University of North Carolina Department of Computer Science; Colin Raffel from the University of Toronto Department of Computer Science;  Guorong Wu and Mark Zylka from the University of North Carolina Neuroscience Center; Stephen Williams and John Slankas from the Laboratory for Analytic Sciences at North Carolina State University; and Bin Hwang, Wenjing Shi, and Ding Li from Amazon Web Services.

\clearpage

\begin{center}
\vspace*{52pt}
{\textbf{PREFACE}}
\end{center}

\noindent\textit{On the Use of the First-Person}

To be clear about my own contributions, I use the first-person singular pronoun ``I'' when referring to my actions, experiments, and observations. I use the first-person plural pronoun ``we'' when referring to myself and the reader, or when presenting shared observations of the research community at large.


\clearpage

\renewcommand{\contentsname}{\hfill TABLE OF CONTENTS \hfill}
\renewcommand{\cfttoctitlefont}{\bfseries}
\renewcommand{\cftaftertoctitle}{\hfill}
\renewcommand{\cftdotsep}{1.5}
\cftsetrmarg{1.0in}

\setlength{\cftbeforetoctitleskip}{61pt}
\setlength{\cftaftertoctitleskip}{11pt}

\renewcommand{\cftchapfont}{\normalfont}
\renewcommand{\cftchappagefont}{\normalfont}
\renewcommand{\cftchapleader}{\cftdotfill{\cftdotsep}}

\renewcommand{\cftchappresnum}{\chaptertitlename~}
\renewcommand{\cftchapaftersnum}{:}
\makeatletter
\newcommand*\updatechaptername{%
        \addtocontents{toc}{
        \protect\renewcommand*\protect\cftchappresnum{\@chapapp\ }
        \IfAppendix{\protect\renewcommand*\protect\cftchapnumwidth{5.5em}}
        {\protect\renewcommand*\protect\cftchapnumwidth{4.5em}}
        }
}
\makeatother

\setlength{\cftbeforechapskip}{15pt}
\setlength{\cftbeforesecskip}{10pt}
\setlength{\cftbeforesubsecskip}{10pt}
\setlength{\cftbeforesubsubsecskip}{10pt}

\begin{singlespace}
\tableofcontents
\end{singlespace}

\clearpage

\renewcommand{\listtablename}{LIST OF TABLES}
\phantomsection
\addcontentsline{toc}{chapter}{LIST OF TABLES}

\setlength{\cftbeforelottitleskip}{-11pt}
\setlength{\cftafterlottitleskip}{11pt} 
\renewcommand{\cftlottitlefont}{\hfill\bfseries}
\renewcommand{\cftafterlottitle}{\hfill}

\setlength{\cftbeforetabskip}{10pt}

\begin{singlespace}
\listoftables
\end{singlespace}

\clearpage

\renewcommand{\listfigurename}{LIST OF FIGURES}
\phantomsection
\addcontentsline{toc}{chapter}{LIST OF FIGURES}

\setlength{\cftbeforeloftitleskip}{-11pt}
\setlength{\cftafterloftitleskip}{11pt} 
\renewcommand{\cftloftitlefont}{\hfill\bfseries}
\renewcommand{\cftafterloftitle}{\hfill}

\setlength{\cftbeforefigskip}{10pt}
\cftsetrmarg{1.0in}

\begin{singlespace}
\listoffigures
\end{singlespace}
\clearpage

\phantomsection{}
\addcontentsline{toc}{chapter}{LIST OF ABBREVIATIONS}

\begin{center}
\textbf{LIST OF ABBREVIATIONS}
\end{center}

\begin{tabbing} 
\Abi{\texttt{ABR}}{Adaptive Bit Rate}
\Ab{\texttt{ADC}}{Analogue to Digital Converter}
\Ab{\texttt{\adder}}{Address, Decimation, $\Delta t$ Event Representation}
\Ab{\texttt{ASINT}}{Asynchronous Integration sensor}
\Ab{\texttt{APS}}{Active Pixel Sensor}
\Ab{\texttt{ATIS}}{Asynchronous Time-based Image Sensor}
\Ab{\texttt{BIQI}}{Blind Image Quality Index}
\Ab{\texttt{CABAC}}{Context-Adaptive Binary Arithmetic Coding}
\Ab{\texttt{CCD}}{Charge Coupled Device}
\Ab{\texttt{CMOS}}{Complementary Metal Oxide Semiconductor}
\Ab{\texttt{CBR}}{Constant Bit Rate}
\Ab{\texttt{DAVIS}}{Dynamic and Active Vision System}
\Ab{\texttt{dB}}{Decibels}
\Ab{\texttt{DCT}}{Discrete Cosine Transform}
\Ab{\texttt{DBSCAN}}{Density-Based Spatial Clustering of Applications with Noise}
\Ab{\texttt{DVS}}{Dynamic Vision System}
\Ab{\texttt{EOF}}{End of File}
\Ab{\texttt{FPGA}}{Field Programmable Gate Array}
\Ab{\texttt{FPS}}{Frames Per Second}
\Ab{\texttt{GB}}{Gigabytes}
\Ab{\texttt{Gb}}{Gigabits}
\Ab{\texttt{GOP}}{Group Of Pictures}
\Ab{\texttt{IID}}{Independent and Identically Distributed (random variables)}
\Ab{\texttt{MB}}{Megabytes}
\Ab{\texttt{Mb}}{Megabits}
\Ab{\texttt{MIL}}{Multiple Instance Learning}
\Ab{\texttt{ms}}{Milliseconds}
\Ab{\texttt{NIQE}}{Natural Image Quality Evaluator}
\Ab{\texttt{PSNR}}{Peak Signal-to-Noise Ratio}
\Ab{\texttt{QE}}{Quantum Efficiency}
\Ab{\texttt{QP}}{Quantization Parameter}
\Ab{\texttt{ROI}}{Region of Interest}
\Ab{\texttt{s}}{Seconds}
\Ab{\texttt{SIMD}}{Single Instruction/Multiple Data}
\Ab{\texttt{SSIM}}{Structural Similarity Index Measure}
\Ab{\texttt{VMAF}}{Video Multimethod Assessment Fusion}
\Ab{\texttt{VOD}}{Video on Demand}

\end{tabbing}

\clearpage

\phantomsection{}
\addcontentsline{toc}{chapter}{LIST OF SYMBOLS}

\begin{center}
\textbf{LIST OF SYMBOLS}
\end{center}

\begin{tabbing} 
\Abi{$\Delta$}{Change (in time, for example)}
\Ab{$D$}{Decimation, expressing intensity units accumulated by $2^D$}
\Ab{$t$}{Time}
\Ab{\dtm}{The maximum time interval that an \adder{} event may span, or (in \cref{ch:adder_compression} onwards)\\ \hspace{18.5mm} the maximum time interval for only the \textit{first} \adder{} event at a new intensity level.}
\Ab{$\Delta t_{frame}$}{Ticks per input frame, when the source is a framed video}
\Ab{$\Delta t_{ref}$}{Ticks per reference interval}
\Ab{$\Delta t_s$}{Ticks per second}
\Ab{exp}{Exponential function}
\Ab{$I$}{Intensity per unit time, found by $\frac{2^D}{\Delta t}$}
\Ab{$I_{max}$}{Maximum intensity representable by the data source over the baseline time interval}
\Ab{$I_{min}$}{Minimum intensity representable by the data source over the baseline time interval}
\Ab{ln}{Natural logarithm function}
\Ab{$M$}{\adder{} intensity contrast threshold}
\Ab{$p$}{DVS event polarity (+/-1)}
\Ab{\hypertarget{$\sigma$}{$\Sigma$}}{Set of events (inputs from the user or the environment) that can cause mode transitions}
\Ab{$\theta$}{DVS log-intensity contrast threshold}
\end{tabbing}

\clearpage

\pagenumbering{arabic}
 \chapter[~~~~~~~~~~~~Introduction]{Introduction\footnotemark}\label{ch:intro}\footnotetext{Portions of this chapter previously appeared in the proceedings of 2023 ACM Multimedia Systems \cite{freeman_mmsys23}.}

\graphicspath {{ch01_intro/images}}

Digital video is ubiquitous in modern life. We each carry a high-resolution video camera in our pocket, ready to capture any interesting or exciting moment we encounter.  We can send any video we record to a friend with the press of a button, automatically uploading it to a cloud service such as Google Photos. We have quick access to movies and television shows through Netflix, Prime Video, Hulu, and Disney+. We can access tailor-made entertainment and educational programming through YouTube. 

Besides these obvious consumer-focused services, video is pervasive in industrial, corporate, and governmental deployments. Retail stores are riddled with surveillance cameras to aid efforts against shoplifting and violence. Robots and autonomous vehicles depend on video cameras to perform obstacle avoidance and path planning. Governments leverage covert camera deployments and large video databases to spy on threats domestic and abroad. 

These varying video systems fall into two primary categories: human applications and computer vision applications. The former category has a human viewer, while the latter category is predominantly interpreted by a computer to perform an automated task. Arguably, a third category exists at the intersection of human and computer vision. Such pipelines have processing steps to emphasize certain types of information before a video is viewed by a human. An example is computer-assisted video triage for surveillance by intelligence agencies: a large video corpus may be filtered for relevant time segments before a human analyst examines the segments for activities of interest.

Despite the multiplicity of video \textit{applications}, the quality metrics, compression schemes, and underlying video representations have persisted through the decades with the same fundamental assumptions. Standard frame-by-frame quality metrics aim to capture the perceived perceptual video quality for a human viewer. Quality of experience metrics attempt to quantify temporal artifacts such as buffering and rate thrashing, which may be distracting or frustrating to a human viewer. Based on these metrics, expert groups developed compression standards to remove details of little notice to human viewers and have a predictable rate-distortion tradeoff. These compression techniques all utilize synchronous image inputs and assume high temporal redundancy between consecutive frames.

In the early days of digital video, human viewership was the main focus of attention for video codec researchers. The compression standard H.261 was introduced in the late 1980s, bringing with it many of the techniques that are now ubiquitous and offering the first practical means of video coding and playback \cite{h261}. Thirteen years later, the state-of-the-art face detection algorithm could process low-resolution images at a rate of only 15 frames per second \cite{viola}. The computational and methodological challenges of early computer vision work relegated researchers to the low-rate processing of individual decoded images. Meanwhile, video compression standards continued to push the boundaries of resolution, frame rate, and bit rate for human viewership with new and refined techniques. Regardless, both application sectors relied on image frames.

Once the needs of computer vision became evident, some attempts were made to push vision application semantics into video codecs. Examples include object-based video coding (OBVC) and region of interest (ROI) coding in H.264 \cite{obvc}. These mechanisms did not gain much traction and were removed in the later H.265 standard. However, one may achieve similar effects with a highly-customized encoder through the manual setting of regional quality parameters. This fine tuning, in fact, became the standard route for application-driven video compression in subsequent years. Still, image frames were the decompressed representation used for visual display and vision applications.

One can argue that discrete images are in use because that is the modality used by video cameras since their analog inception: a camera captures a sequence of images in rapid succession, and if the rate at which they are displayed to a human is high enough, it will appear as a continuous motion to the human visual system. We should not take this imaging modality as a given, however. Recent years have seen the rise of novel video cameras which do \textit{not} capture image frames. Instead, these ``event cameras" record independent, asynchronous pixel streams \cite{survey}. The advantages to this type of sensing include extremely high dynamic range, high speed, and low power \cite{survey}.

The research conducted heretofore on event cameras, however, has continued the trend of a disconnect between the goals of computer vision and traditional video coding. The event camera community largely consists of vision and robotics researchers, who are trying to develop novel applications which can leverage this unique data type. Attempts at lossy compression of the data from these cameras either fall back on traditional frame-based encoders or merely discard data in the decompressed representation. Even the efforts of vision researchers seem at odds with developments in the hardware space: they are developing applications for particular camera sensing modalities and data types, while the hardware advances towards entirely different, more informative event-based modalities.

I approach this scattered landscape with a number of questions. Why are we converting asynchronous data into frames? Why are we building applications and compression schemes for particular cameras, given the rapid rate of hardware development? Can we gain some advantage by converting traditional framed video into an event representation? If we remove compatibility with a human-viewer application, can we build an entirely new codec specifically for vision applications? Can we link compression quality not just with application-level accuracy, but also with application speed?

In this dissertation, I address the above questions with a novel framework: \adder{}. In its decompressed form, \adder{} is a general representation consisting of a stream of timestamped events, each of which represents an update to a single pixel's absolute intensity value. This reepresentation encompasses the data of both framed \textit{and} existing event-based video. Leveraging a common transcoder mechanism, one can easily implement a mechanism to transform any spatio-temporal data stream to \adder. Upending the common wisdom of video coding, I give priority to vision applications (rather than human viewers) in all aspects of the transcoder and compressor design. I investigate the practical implications of this representation as a natively-captured sensing modality and as a transcoded representation for existing framed and event-based camera sources, various data prediction and compression schemes, backwards compatibility with traditional frame-based applications, and bespoke asynchronous applications.

\section{Definitions}
In the literature, authors often conflate the term “event'' with a particular type of event camera. I provide the following definitions to remove ambiguity.

\textbf{Spatio-temporal data:} The term used when the smallest addressable unit of a given data stream has both a spatial and temporal component. 

\textbf{Pixel model:} A means of transforming input intensity data into some output. A pixel model can be either implemented in hardware (e.g., CCD, CMOS, DVS pixels) or in software.

\textbf{Intensity unit:} The base unit of luminosity data which is ingested by a pixel model. For realizations of pixel models in sensing hardware, the intensity unit is a photon. For software pixel models, the intensity unit is typically an accumulation of photons scaled to some range (e.g., 0-255).

\textbf{Event:} Any discrete data point expressing that some intensity-based condition has been met at a specific time.

\textbf{Contrast Event:} An event which conveys that intensity has \textit{changed} by a certain amount at a given point in time, in comparison to some prior state (e.g., DVS \cite{dvs}).

\textbf{Intensity Event:} An event which conveys the amount of light accumulated over some period of time (e.g., \adder{}).

\textbf{Event video:} Any spatio-temporal representation of visual information whose primary data type is a time-based event. In this dissertation, event video data representations will be referred to by their specific data source where necessary.

\section{Motivation}

Traditional synchronous framed video representations have a fixed and uniform sample rate. Consequently, a spatial region of a video which is dynamic (changing greatly) has the same number of intensity samples as a spatial region which is relatively static (unchanging). Thus, there is much data redundancy in the raw representation, which is currently only addressed at the level of the compressor. Furthermore, the precision of a framed video is bounded by the maximum representable value in a single image frame, typically 8-12 bits per color channel.

In addition, there are fundamental representational weaknesses in the most common form of event sensors used today: \textbf{temporal contrast sensors} (e.g, DVS \cite{dvs}). These sensors do not record traditional image frames; rather, each pixel independently generates a sequence of \textit{change events}, where a bespoke time-based event triggers when the log intensity of the pixel increases or decreases by a certain amount \cite{survey}. This technique allows for extreme temporal precision (on the order of microseconds) and high dynamic range, but prevents the sensor from capturing data where the intensity is unchanging or changing slowly. Thus, many users of these cameras augment event sensors with frame-based image capture to fill in the informational gaps (e.g., DAVIS cameras \cite{survey,davis_dataset}). However, these framed images tend to exhibit motion blur, and they are typically processed separately from the event data due to the slow computational performance of fusing the information across modalities. Additionally, each event expresses the intensity change \textit{relative to} a previous event, so this raw representation does \textit{not} support lossy compression or expose a rate-distortion tradeoff. These sensors are rapidly gaining traction for vision tasks and high-speed robotics, but their extraordinary data rates (up to millions of events per second) and ever-increasing spatial resolutions (now reaching 1080p) are not adequately addressed from a systems perspective. Instead, most research applications merely convert event data back to a number of frame-based representations. These representations severely reduce the temporal resolution of the data, diminishing a major benefit to event sensing.

I present the following analogy to motivate my work. Imagine that traditional cameras from different companies all had their own proprietary raw data formats, with no common representation between them. Now imagine that many video applications could only operate on raw data from the cameras of a particular company, and that all other video applications could operate only on PNG image sequences derived from cameras' raw representations. That is effectively how the event camera ecosystem looks today: there is no codec equivalent to H.26X \cite{h266} for event video, meaning that event camera systems are not employing any significant lossy compression and are hand-built for particular cameras' raw data formats. I argue that to build the most effective real-time event video systems, the application-level video representation should \textit{also} be asynchronous, and it should enable rate-driven lossy compression. Furthermore, I emphasize that framed video produced either with a stationary camera (i.e., surveillance video) or with a sufficiently high frame rate relative to the captured motion, can equally be thought of under an event-based camera model.

\section{Thesis Statement}
\begin{quote}
\textit{
    Arbitrary spatio-temporal video data can be represented as a single asynchronous, compressible data type. Applications can operate on this unified data representation, rather than targeting a number of source-specific representations. This universal representation unlocks advantages in compression, rate adaptation, and application speed and accuracy.
}
\end{quote}

\section{Proposed System and Contributions}

I propose the \textbf{A}ddress, \textbf{D}ecimation, $\Delta t$ \textbf{E}vent \textbf{R}epresentation (\textbf{\adder{}}, pronounced “adder'') format as the “narrow waist'' representation for asynchronous video data. Its fundamental data type is an \textbf{intensity event}, meaning that data points are asynchronous and directly convey the cumulative intensity upon a pixel over some variable length of time. These events are fundamentally opposed to those of contrast-based event sensors, which represent only intensity \textit{change}. 

\begin{figure}
    \centering
    \includesvg[width=\linewidth]{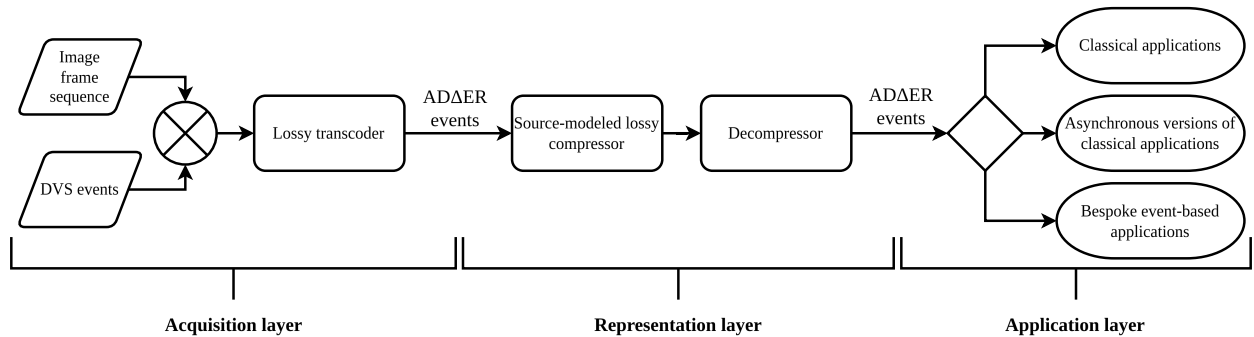}
    \caption[Abstract overview of the \adder{} framework]{Abstract overview of the \adder{} framework.}
    \label{fig:system_diagram_intro}
\end{figure}

\begin{figure}
    \centering
    \includesvg[width=0.81\linewidth]{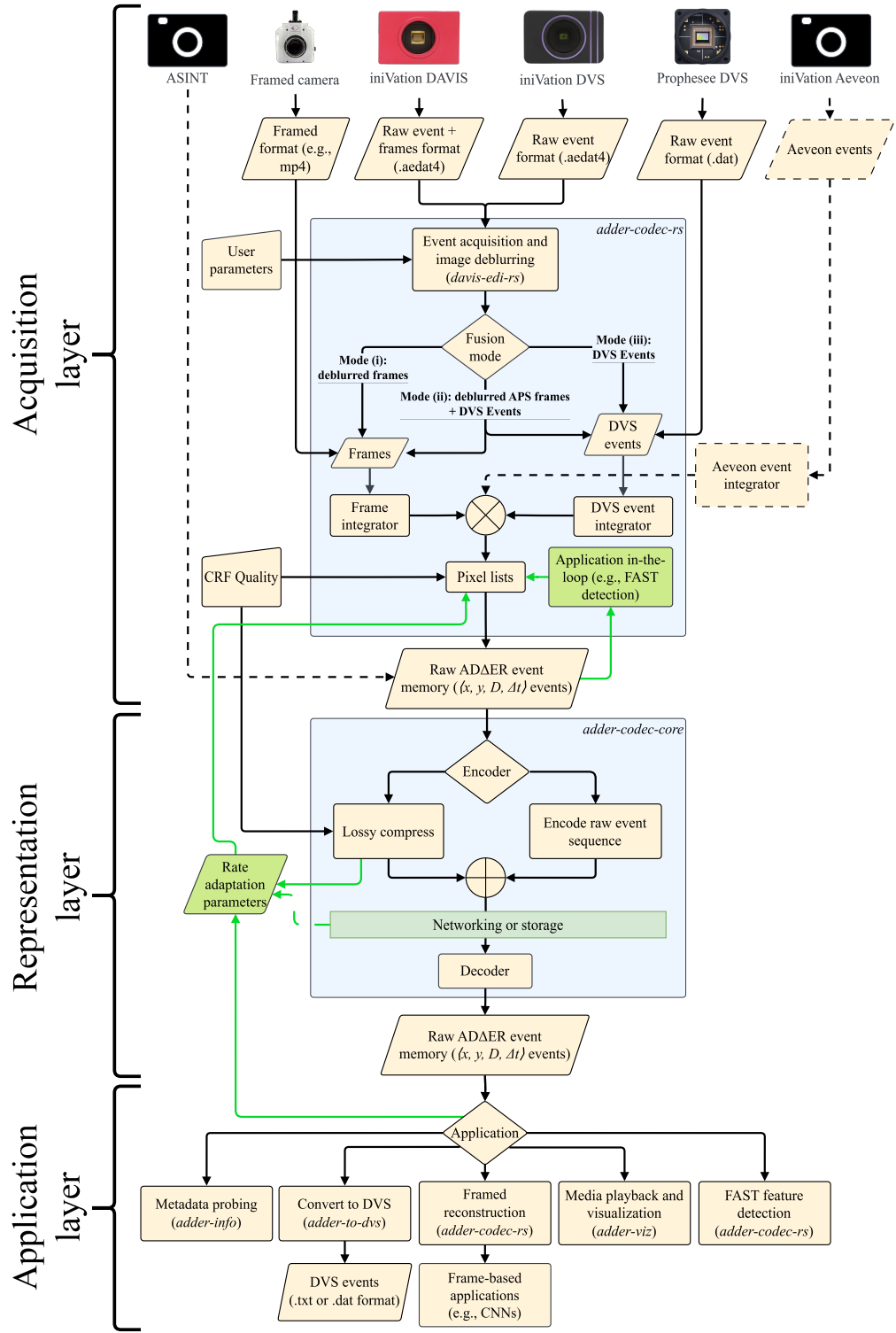}
    \caption[Detailed diagram of the \adder{} framework]{Detailed diagram of the three-layer \adder{} framework. Italicized names reflect the names of software packages in the Rust Package Registry. Dashed lines indicate future work. With \adder{}, framed and event-based video sources can be transcoded to a common representation. Since there is a single raw representation, we can have a simple source-modeled compression scheme. The representation supports bespoke event-based applications, while being backwards compatible with classical applications.}
    \label{fig:system_diagram_intro_full}
\end{figure}

\begin{figure}
    \centering
    \includegraphics[width=\linewidth]{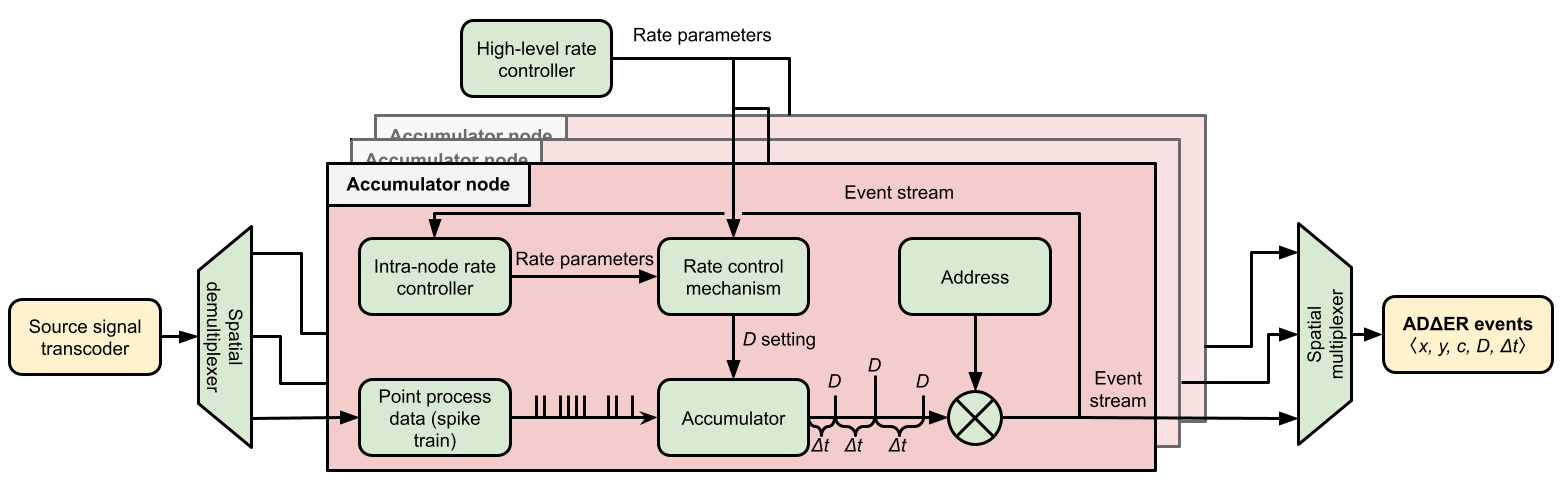}
    \caption[Abstract overview of the \adder{} transcoder]{Abstract overview of the \adder{} transcoder.}
    \label{fig:transcoder_diagram_intro}
\end{figure}

We can imagine \adder{} as an abstract pixel model which receives intensity and generates an output. The pixel model accumulates intensity until reaching a certain threshold $2^D$, determined by a software parameter $D$. This threshold, $2^D$, is effectively the per-pixel temporal light sensitivity. When the threshold is reached, the pixel outputs an event containing its spatial coordinates, its $D$ parameter, and the time elapsed since the pixel last fired an event, $\Delta t$. The combination of $D$ and $\Delta t$ conveys the incident brightness of the pixel measurement. Crucially, the $D$ parameter can be different for each pixel, and it can change overtime. This dynamic threshold property means that we can adjust the rate of data produced by each pixel according to some higher-level concerns, such as compression performance, application accuracy, and application throughput. I illustrate the \adder{} pixel model in \cref{fig:transcoder_diagram_intro}.

This seemingly simple pixel model undergirds the whole of this dissertation. I argue that the time variability and rate adaptation potential of raw \adder{} events makes it well suited as a target \textit{transcoded} representation for arbitrary video sources. It is general enough to represent within it arbitrary intensity-based video modalities. These modalities include frame-based video, contrast-based event video, multi-modal video combining the prior data types, and possibly intensity event video data of the future.

This dissertation proposes and evaluates a framework which unifies disparate sources of video under the \adder{} representation, illustrated broadly in \cref{fig:system_diagram_intro} and in greater detail in \cref{fig:system_diagram_intro_full}. The framework is divided into three primary layers.

\subsection{Acquisition Layer}

The acquisition layer encompasses the generation of \adder{} data from various data sources. The framework accepts particular input data formats for traditional framed video (e.g., mp4 files), and outputs from iniVation and Prophesee cameras which capture constrast-based events. In this dissertation, I do not research hardware architecture or real-time robotics systems for event cameras; rather, my evaluation on real-world data is conducted only on pre-recorded videos. However, I often discuss the speed characteristics of my implementations and the implications for potential live-streamed camera deployments. Furthermore, I explore a software simulation of an unrealized camera design (ASINT \cite{montek}) which natively records \adder{} events, and I propose methods for dynamically adjusting camera parameters for rate optimization. 

The other cameras shown in \cref{fig:system_diagram_intro_full}, existing and upcoming commercial cameras \cite{cmos_review,dvs,DAVIS,aeveon,survey}, do not natively capture \adder{}-style events. Instead, I have a modular interface for reading different video data types. Each sensing modality has a driver for transcoding its data to the \adder{} representation. All transcode processes leverage a generic \adder{} pixel model which accumulates intensity information over time, generating events when an intensity integration threshold is met. The source drivers, then, must only transform an incoming data point to an absolute intensity measurement and an integration time. For example, the framed video driver integrates intensities in the range $[0,255]$ over a time interval matching the inverse of the video frame rate.

Whereas traditional video systems have a single compressor which induces loss, the \adder{} transcoder serves as the first stage of a two-stage compressor. The transcode process itself is where most loss occurs, according to a quality parameter defined by a user or an application. A lower quality target reduces the temporal sensitivity of the \adder{} pixel models, thus averaging out the temporal variations in incoming intensities into fewer \adder{} events.

\subsection{Representation Layer}

Once a source video is transformed into the \adder{} representation, it can optionally be processed by the second stage of the compressor. This stage mirrors some techniques of traditional video coding, but I adapt them for the sparse, asynchronous video paradigm. The compressor organizes \adder{} events into spatiotemporal groupings and encodes event prediction residuals based on nearby events in both space and time. Importantly, the source model is ignorant to the original video modality: it does not require knowledge of whether the original input came from a framed camera or an event camera. Rather, I tune the compressor for generic \adder{} inputs. This source agnosticism makes the common \adder{} representation the ``narrow waist'' of the framework.

\subsection{Application Layer}

Finally, I introduce several interfaces for applications. Classical frame-based applications may be used by reconstructing intensity frames from \adder{} events. Alternatively, one may easily adapt classical algorithms for the asynchronous paradigm by executing them asynchronously on small spatial patches surrounding new \adder{} events, rather than uniformly processing an entire frame. I also propose bespoke event-based applications and discuss how \adder{} may serve as an input for spiking neural networks. 

Uniquely, the \adder{} system correlates information loss with the bit rate of the decompressed representation. Compressing for lower qualities actually reduces the number of raw \adder{} events, rather than simply reducing the bit rate of the compressed representation. With an online application, then, one can strike a balance between the available video bandwidth, application accuracy, and application speed. Lowering the quality of an \adder{} transcode will reduce the bandwidth requirement and application accuracy, but can make the application \textit{faster}.

\section{Dissertation Overview}

The rest of this dissertation is organized as follows. 

\begin{itemize}
    \item \textbf{\cref{ch:video_representations}} offers an introduction to the hardware sensing mechanisms, representations, and applications for framed cameras and event-based cameras.
    \item \textbf{\cref{ch:adder_transcoding}} introduces \adder{} as a universal representation for various source video data types. I discuss mechanisms for transcoding framed, DVS, and DAVIS camera data to \adder{} and evaluate the transcoded quality and speed with various parameters.
    \item \textbf{\cref{ch:adder_compression}} describes the \adder{} representation layer and application layer. I introduce an adapted feature detection algorithm for \adder{} which can be executed during the transcode loop for novel protosaliency-driven compression. I evaluate the feature detector and a lossy compression scheme on a frame-based surveillance video dataset, showing that the compression performance correlates to the speed of event-based applications by reducing the decompressed data rate. 
    \item \textbf{\cref{ch:software_description}} describes the architecture of my open-source \adder{} software framework, including various command-line utilities, a visual transcoder interface, and an interactive video player.
    \item \textbf{\cref{ch:conclusion}} concludes the dissertation with a review of the proposed system and a discussion of future work.
\end{itemize}

\chapter[~~~~~~~~~~~~Background: Data Compression and Video Systems]{Background: Data Compression and Video Systems}\label{ch:video_representations}

\graphicspath {{ch02_background/images}}

\section{Introduction}

Given the ubiquity of digital video in our modern lives, it is easy to discount the ongoing challenges of video systems. Video cameras continue to push the boundaries of resolution and data rate. Larger video files require greater storage capacity and network bandwidth. Larger data sets and faster GPUs lead to more complex deep learning models for applications. There exists an ongoing push and pull between advances in video hardware (cameras and GPUs) and software (compression and applications).

At the same time, the conventional notion of ``video'' is now in flux. Novel ``neuromorphic'' camera sensors, aiming to mimic the human retina and visual system \cite{survey}, have introduced an entirely new mode of video sensing and representation by eschewing image frames at the sensor level. Up to this point, however, these \textit{event} sensors have lacked many of the systems-level conveniences we enjoy with traditional video. 

This chapter begins with a discussion of general data compression, which is the backbone of practical video systems. Then, I introduce a simplified, 3-layer system model for generic video. Finally, I compare and contrast the existing classical and event-based video systems in the context of this model.

\section{Data Compression}

To fully understand the benefits and drawbacks inherent to various video systems, we must first understand the general principles of data compression.

 An \textbf{alphabet}, $A$, is a set of \textbf{symbols}, discrete pieces of information which one may want to express. $X$ is a random variable with a discrete number of possible outcomes which are symbols within $A$. The probability of any individual outcome $x\in X$ is notated as $P(x)$. For example, a coin flip has two outcomes (heads and tails) each with probability $0.5$. In the context of digital media, we can define \textbf{data} as the actual 1s and 0s (bits) used to represent something we are trying to convey (be it text, images, or video).

The \textbf{Shannon information content} (also known as \textbf{self-information} or simply \textbf{information}) is the measure of ``surprise'' that an outcome represents \cite{information_theory_textbook}, found as\footnotemark

\footnotetext{Throughout this dissertation, I always specify the logarithmic base in equations or use shorthand notation such as ``ln.'' The subsequent meaning of ``log'' in discussions in the text is inferrable from the context.}

\begin{equation}
    h(x) = \log_2\frac{1}{P(x)}
\end{equation}
bits. The \textbf{entropy} of $X$ is the average self-information of the possible outcomes of $X$ \cite{information_theory_textbook}. It is found as

\begin{equation}
    H(X) = \sum_{x\in A_X} P(x)\log_2\frac{1}{P(x)}
\end{equation}
bits.

\textbf{Compression} is the act of transforming an input source with an \textbf{encoder} such that its \textit{information} is represented with less \textit{data}. Compression can be either \textbf{lossless}, where the source can be perfectly reconstructed by a \textbf{decoder}, or \textbf{lossy}, where the reconstructed output is different in some way compared to the source.

Shannon's \textbf{source coding theorem} establishes a bound for the compressibility of a sequence of independent and identically distributed random variables (IID; those which have the same probability distribution and are independent outcomes). Specifically, for $N$ such variables each with entropy $H(X)$, a sequence cannot be compressed into $NH(X)$ or fewer bits without information loss \cite{information_theory_textbook}. Furthermore, we can compress to $NH(X)$ bits so long as we tolerate a small probability of error \cite{information_theory_textbook}.

The \textbf{expected length} of a binary symbol code is found as
\begin{equation}
 L(C,X)=\sum_{x\in A_X} P(x)l(x),
\end{equation}
where $l(x)$ is the bit length of the symbol $x$ \cite{information_theory_textbook}. The source coding theorem for symbol codes states that the expected code length of a lossless compression scheme cannot be less than the entropy of the source \cite{information_theory_textbook}.

Let us walk through a simple example of compression to motivate its utility.  Suppose that we want to store the following text sequence:

\centerline{\texttt{AABAABAF}}

\noindent If we use the ASCII encoding scheme, then each character in our sequence occupies a single byte (8 bits) on disk. Then, we can represent our sequence as the following binary values:

\centerline{\texttt{01000001 01000001 01000010 01000001 01000001 01000010 01000001 01000011}}

\noindent We see that under ASCII encoding, our text sequence is 8 bytes long. In other words, we have 8 bytes of \textit{data}.

\begin{figure}
\centering        \adjustbox{trim=0 13.95cm 0 0,clip}{\includesvg[width=1.0\linewidth]{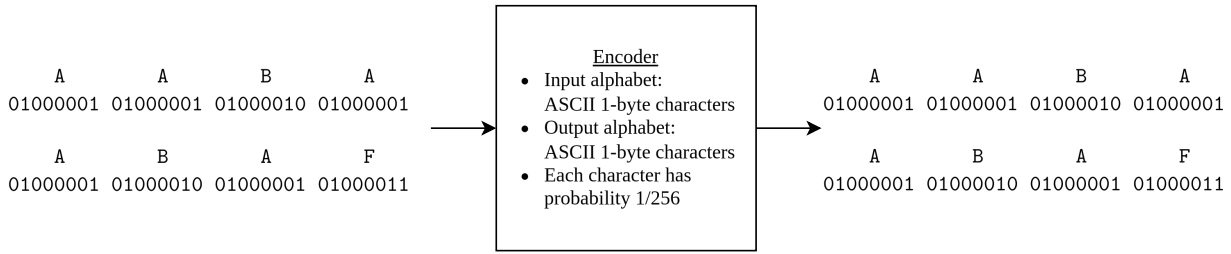}}
        \caption[A naive ASCII encoder]{A naive ASCII encoder. The output data is unchanged from the input.}
        \label{fig:encoder_1}
\end{figure}

However, not all of this data is necessarily \textit{useful}. If we run our source text sequence through a naive ASCII text encoder, as in \cref{fig:encoder_1}, the encoder may assign an equal probability to all 256 ASCII characters.  In this case, our output binary is unchanged from the input. 

\begin{figure}
\centering        \adjustbox{trim=0 10.5cm 0 3.4cm,clip}{\includesvg[width=1.0\linewidth]{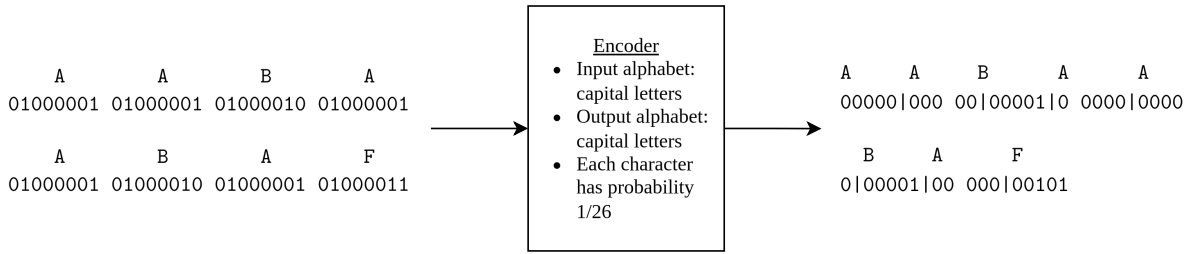}}
        \caption[An encoder for only capital letters with equal probability]{An encoder for only capital letters with equal probability.}
        \label{fig:encoder_2}
\end{figure}

Now, consider that our naive encoder has prior knowledge that the input text sequence will only contain capital letters. There are 26 such letters, so we can assign a probability of $\frac{1}{26}$ to each character. The number of bits required to represent 26 unique symbols is $\lceil \log_2 26 \rceil = 5$ bits. Therefore, we can construct a new mapping for these 26 characters to numbers in the decimal range $[0,25]$, or binary $[\texttt{00000}, \texttt{11001}]$. Our encoded output is now only 5 bytes, as shown in \cref{fig:encoder_2}.

\begin{figure}
\centering        \adjustbox{trim=0 6.9cm 0 6.9cm,clip}{\includesvg[width=1.0\linewidth]{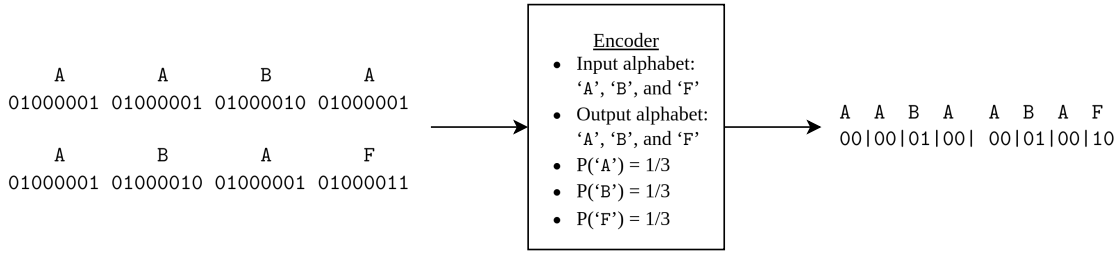}}
        \caption[An encoder for only the characters \texttt{A}, \texttt{B}, and \texttt{F}, with equal probability]{An encoder for only the characters \texttt{A}, \texttt{B}, and \texttt{F}, with equal probability.}
        \label{fig:encoder_3}
\end{figure}

Now suppose that our input sequence is known to only contain the characters \texttt{A}, \texttt{B}, and \texttt{F}, with equal probability for each. Since we only have three characters, we now need only $\lceil \log_2 3\rceil = 2$ bits to represent any individual character. We can map our three characters to the binary range $[\texttt{00},\texttt{10}]$. As shown in \cref{fig:encoder_3}, this scheme can compress our input to to only 2 bytes, a 4:1 compression ratio from the original.

\begin{figure}
\centering        \adjustbox{trim=0 3.4cm 0 10.5cm,clip}{\includesvg[width=1.0\linewidth]{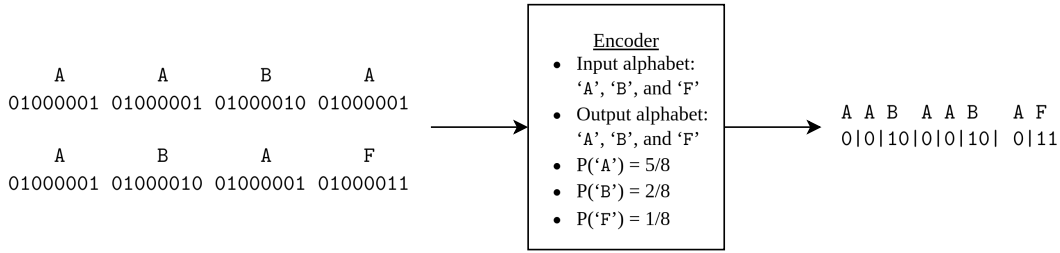}}
        \caption[An encoder for only the characters \texttt{A}, \texttt{B}, and \texttt{F}, with unequal probabilities]{An encoder for only the characters \texttt{A}, \texttt{B}, and \texttt{F}, with unequal probabilities.}
        \label{fig:encoder_4}
\end{figure}

If we examine the original source again, we see that the distribution of characters is not even. There are more \texttt{A}s than \texttt{B}s, and more \texttt{B}s than \texttt{F}s. If our encoder has a more realistic notion of the probability of each symbol, we can achieve even greater compression. \cref{fig:encoder_4} shows that, since \texttt{A} occurs with high probability, we can choose to represent \texttt{A} with a \textit{single } bit, \texttt{0}. Since we have three symbols, however, we must use 2 bits to represent at least one code word, making this a \textit{variable-length code}. No other code word may start with \texttt{0} in this case, since the decoder will not know the length of the word to be decoded (1 or 2 bits) ahead of time, making this a \textbf{prefix code}. Therefore, our code words for \texttt{B} and \texttt{F} must start with the bit \texttt{1}. With this scheme, we can encode our source with only 11 bits, a compression ratio of 5.8:1 (\cref{fig:encoder_4}).

\begin{figure}
\centering        \adjustbox{trim=0 0 0 14.0cm,clip}{\includesvg[width=1.0\linewidth]{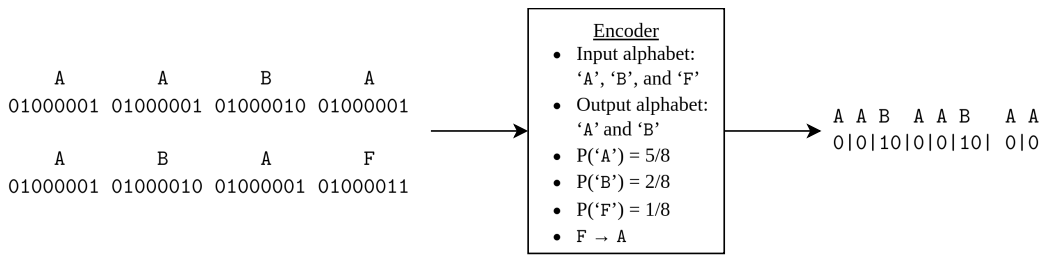}}
        \caption[An encoder for only the characters \texttt{A}, \texttt{B}, and \texttt{F}, with unequal probabilities]{An encoder for only the characters \texttt{A}, \texttt{B}, and \texttt{F}, with unequal probabilities. The \texttt{F} symbol is transformed to an \texttt{A} before encoding.}
        \label{fig:encoder_5}
\end{figure}

Lossless compression techniques can only get us so far, however. If we allow some potential for loss, we can further compress our data. For example, if we decide that the \texttt{F} character is not very important for our purposes, we can instruct our encoder to transform any \texttt{F} into an \texttt{A} before encoding, as in \cref{fig:encoder_5}. This allows us to further reduce the compressed size to just 10 bits, for a compression ratio of 6.4:1.

In the case of the source coding theorem, error is introduced by the encoder with some non-zero \textit{probability}. In contrast, when I discuss loss in this dissertation, I refer to \textit{deliberate} (non-probabilistic) decisions made by an encoder to reduce the information content. In our example, the \texttt{F} character \textit{always} is transformed to be an \texttt{A}, and all other characters have 0 probability of incurring loss. 

While lossy compression is likely undesirable for text data (a garbled mess of letters is not exactly readable), it is enormously beneficial for images and video. The human visual system is not very sensitive to slight variations in color, for example, allowing us to reduce such variations in a lossy manner with minimal impact on the perceived quality.

\subsection{Arithmetic Coding}
The goal of \textbf{entropy coding} is to lossless-compress data to a size close to the entropy of the source. Today, a common and highly efficient entropy coding technique for video is \textbf{arithmetic coding} \cite{arithmetic_coding}. 

An arithmetic coder has a \textit{probability model} for the symbols in the alphabet, as we had in \cref{fig:encoder_4}. The encoder divides the decimal range $[0,1)$ between these probabilities. Let us construct a new probability model. We add an end-of-file (EOF) symbol to indicate when the entire message has been encoded. I will elucidate this symbol after walking through the encoding process. Suppose our probability intervals are as follows:
\begin{equation}
\begin{aligned}
    \texttt{A}: &  [0.0, 0.5)\\
    \texttt{B}: &  [0.5, 0.75) \\
    \texttt{F}: &  [0.75, 0.9) \\
    \texttt{EOF}: & [0.9, 1.0)
    \end{aligned}
\end{equation}

And suppose we want to encode a very simple message:

\centerline{\texttt{ABF}}

The coder subsequently divides the sub-intervals according to the probability model and the next symbol encountered, constructing a floating point number (with arbitrary precision) lying within an interval. By design, any particular encoded number can correspond only to one possible sequence of symbols. If we begin encoding our message, we see that our first symbol is \texttt{A}. Therefore, we subdivide the interval $[0.0, 0.5)$ based on our probability model. Then we have that the intervals for the second symbol are:
\begin{equation}
\begin{aligned}
    \texttt{A}: &  [0.0, 0.25)\\
    \texttt{B}: &  [0.25, 0.375) \\
    \texttt{F}: &  [0.375, 0.45) \\
    \texttt{EOF}: & [0.45, 0.5)
    \end{aligned}
\end{equation}

Since our next symbol to encode is \texttt{B}, we take the \textit{second} sub-interval and subdivide it further, obtaining:
\begin{equation}
\begin{aligned}
    \texttt{A}: &  [0.25, 0.3125)\\
    \texttt{B}: &  [0.3125, 0.34375) \\
    \texttt{F}: &  [0.34375, 0.3625) \\
    \texttt{EOF}: & [0.3625, 0.375)
    \end{aligned}
\end{equation}

The final symbol in our message is \texttt{F}. We take the \textit{third} interval and subdivide it in like manner, obtaining:
\begin{equation}
\begin{aligned}
    \texttt{A}: &  [0.34375, 0.353125)\\
    \texttt{B}: &  [0.353125, 0.3578125) \\
    \texttt{F}: &  [0.3578125, 0.360625) \\
    \texttt{EOF}: & [0.360625, 0.3625)
    \end{aligned}
\end{equation}

Since we have encoded our entire message, we must somehow indicate that we are finished. This is where we leverage the \texttt{EOF} symbol. We have the range $[0.360625, 0.3625)$ for \texttt{EOF} at this stage, so if the number we encode lies within this range, we can perfectly decode our original message. For example, we may choose to encode the number $0.362$ to represent our message. 

The decoder goes through the same interval subdivision process, finding the exact sequence of symbols that produces an interval containing our number. The number $0.362$ lies with in the first sub-interval for the first symbol, indicating this must be an \texttt{A}, the second sub-interval for the second interval (\texttt{B}), and so on until we find that we are within an \texttt{EOF} sub-interval. In practice, this decoding process is iterative, reading one bit at a time. Whenever enough bits are read (extending the floating point number) to sufficiently narrow the interval, it has decoded a new symbol.

Crucially, the encoder and decoder must have the same probability model for valid decoding. Either both ends of the codec must have the same fixed model, or the sent message must communicate its probability model to the decoder in some form. Often, the latter method has a high communication overhead

\subsection{Adaptive Arithmetic Coding}

While entropy is a strong principle undergirding information theory, we almost always violate the assumptions behind an entropy measurement. Specifically, real-world data is virtually never IID. In the above examples, our variables did not share the same probability distribution (i.e., not identically distributed). One can also imagine many instances where variables are not independent. If we are encoding letters of English text, for example, the letter `q' is extremely likely to be followed by a `u' \cite{information_theory_textbook}. With a robust \textit{source model}, we can often surpass the supposed entropy of a source. Furthermore, if our probability model can \textit{adapt} to the inputs it receives, then we can do even better.

With \textit{adaptive} arithmetic coding, we define some mechanism for changing our probability model based on the symbols we encode or decode. This method is well-suited for stream codes, where we may not have the luxury of examining the entire message at once to compute (and communicate) an optimal probability model.

Let us motivate this adaptation idea with a large-scale text example. With a simple grep search, we can find that the English Standard Version of the Bible (including extratextual headings) contains the capital letter `J' 7076 times. If we look at multi-letter sequences, we find that it contains the sequence `Je' 3562 times, `Jes' 1292 times, `Jesu' 1189 times, and `Jesus' 1189 times. A well-tuned adaptive model may, over time when encoding the text, start to weight the symbol `e' with probability near to 0.5 when the previously-encoded symbol is `J.' Specifically, we can interpret this as the \textbf{conditional probability } $P(\text{`e'}|\text{`J'}) = 0.5$. Likewise, the model may adapt to the distributions of multiple preceding characters in a word: `Jesu' is always followed by `s.' The encoder can spend virtually zero bits to encode the `s.'

\subsection{Context-Adaptive Binary Arithmetic Coding}

Suppose that we want to be able to encode sequences of English text interchangeably with sequences of French text. Since all the words are different, we want different probability models depending on the language. With \textbf{context-adaptive binary arithmetic coding} (CABAC) we can leverage multiple source models with a single encoder, swapping them out as needed according to the \textit{context} of the data we are encoding. In practice, the contexts are managed in software as separate probability tables. Each context is adaptive according to its own source model.

To manage video data efficiently, one must leverage many compression strategies in harmony. One must be able to induce loss according to patterns and assumptions in the data. This process establishes the contexts for a CABAC. The range of encodable symbols in each context must be sufficiently small so that the coding speed is performant. The source model for each context must also be well-suited to the content, but not too slow to compute. I view each of these steps as a bit of an art form, requiring some level of creativity to distinguish underlying patterns efficiently.


\section{Generic Video System Model}

Rather than addressing a single problem in isolation, this dissertation covers end-to-end video systems in their entirety. In colloquial terms, I describe ``end-to-end'' in this context as encompassing the following steps:

\begin{enumerate}
    \item Where the original video data comes from (acquisition)
    \item What happens to the video data to get it from place A to B (representation)
    \item What purpose the video is used for (application)
\end{enumerate}

\begin{figure}
    \centering
    \includegraphics[width=1\linewidth]{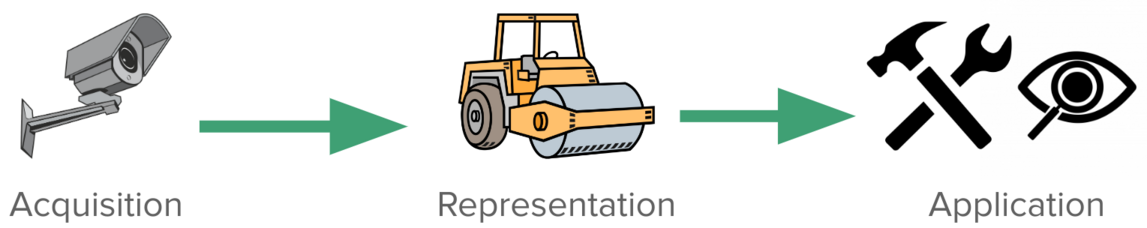}
    \caption{Generic model of the major components of a video system.}
    \label{fig:generic_system_model} 
\end{figure}

\noindent I diagram these three main phases in \cref{fig:generic_system_model} and describe them below.

\subsection{Acquisition Layer}

Video acquisition describes the production of a raw video representation. A \textit{raw} representation is the uncompressed data produced by a video source and processed by an application. A \textit{video source} may be a camera capturing natural data from the real world. Alternatively, the source may be computer-rendered data, such as animation, gameplay, screen recordings, visual effects, or other computer graphics. At this stage of our model, we can imagine that our video data is stored as some raw representation in computer memory, ready to be processed by the next stage.

\subsection{Representation Layer}

Once we have video data, we must do something with it. We can either send it somewhere, or immediately process it locally in its raw form. An example of the latter option is an interface display on a smartphone. Sixty or more times per second, a phone device renders a graphical representation of a user interface and immediately displays it to the user. After an image is displayed, it is discarded. Such local applications may be said to bypass the representation layer. However, they require a direct implementation on the video acquisition hardware. This dissertation does not explore such systems, but focuses on systems where processing and applications occur in some context other than the video generation hardware, which brings with it many problems with practical resource constraints.

We can divide the category of \textit{sending} video into two lanes: storage and transmission. One may \textit{store} a video in order to view or process it later. In this case, the raw video data is transformed from its raw representation into some other representation for storage. Typically, this transformation involves \textit{lossy compression} to reduce the size of the video data.

Alternatively, one may \textit{transmit} a video from one device (the \textbf{server}) to another device entirely (the \textbf{client}). For example, live streamers on YouTube and Twitch transmit their videos over the web in real time to thousands of client devices through a content distribution system. Importantly, video storage typically precedes video transmission, even in the case of livestreaming. The server maintains a  local buffer for the raw video data, which it then compresses and prepares for transmission. This buffering process introduces latency, which is compounded by delays in the\ network transmission and decoding at the client.

\subsection{Application Layer}

In this video system context, the application layer is the ultimate \textit{purpose} of the video, where we use it for some particular task. We often think of video in terms of entertainment, where the ``application" is a human viewing the video. However, human vision may constitute many other purposes, such as security monitoring, education, and telecommunication. On the other hand, many applications are built for computer vision. These applications leverage visual processing or deep learning to ``see" information in video such as object detection and activity recognition. Lastly, some applications fall under the realm of human-machine teaming, where the video is processed by both human and computer vision. These applications include video triage for surveillance systems \cite{triage_article}.


\section{Classical (Framed) Video Systems}\label{sec:framed_systems}

With this generic system in mind, we can explore  \textit{classical} video systems as a specific example of this model. The classical model is the standard used by the industry and academics the last several decades since the introduction of digital video. Its defining characteristic is the underlying raw representation: image frames.

An image frame is a 2D representation of visual information. \textit{Pixels} arranged in a 2D matrix may be comprised of several \textit{channels} for color and transparency information, and each channel of each pixel in the matrix holds a value of a fixed size (e.g., 8-14 bit integers or floating point numbers).

\subsection{Acquisition Layer}

Under the classical paradigm, a video is fundamentally a sequence of discrete image frames. This representation arose from the very first ``motion pictures'' in the late 1800s, such as The Horse in Motion \cite{horse_in_motion}, which were recorded with many cameras capturing a single image on a photochemical plate in rapid succession. Gradually, movie cameras developed to run spools of film across the imaging plane, exposing each segment of film stock for a preset amount of time. With the advent of digital video cameras, moving film was replaced with a stationary image sensor. As with film, each pixel in a classical image sensor is exposed synchronously with the other pixels. This ensures that, generally speaking, any individual frame is temporally coherent.

Likewise, graphics and display hardware developers found that discrete image frames were well-suited to the computational and display technologies available at the time. \cref{fig:framed_model} illustrates the production of image sequences from framed cameras in the acquisition layer. Below, I describe the hardware sensing mechanisms of various types of framed cameras.

\subsubsection{CCD Sensors}

The pixels in Charge Coupled Device (CCD) sensors obtain an absolute count of the photons reaching them during the exposure of the entire sensor. The pixels ``integrate" light over the exposure time, then the charges of all pixels are flushed and measured one row at a time in a readout. There is a single amplifier and analog-to-digital converter (ADC) for the entire sensor, located off-grid from the sensor array. This off-grid location allows for a high-quality implementation free from the physical constraints of on-sensor ADCs. Integrated charges are transported from pixels across the sensor (one column at a time) to this ADC, yielding low measurement noise and high sensitivity \cite{blanc2001ccd}. In a traditional CCD, when a pixel becomes completely saturated before the integration time is complete, the pixel is subject to bleeding voltage, or ``blooming." Blooming occurs when a pixel receives a charge greater than it can hold during the exposure, causing nearby pixels (usually in the same column of the pixel array) to receive an increased charge from the excess \cite{blooming_overview,blooming_and_smear}. The effects of blooming can be reduced by accumulating many shorter exposures, though this requires significant post-processing and reduces the accuracy of the final image. While there are CCD cameras with built-in anti-blooming, the technology greatly hurts its quantum efficiency (QE) \cite{antiblooming}, a measure of the sensitivity of a sensor. Put simply, a sensor with a QE of 95\%  will measure 95 photons for every 100 photons it is exposed to, meaning it is highly accurate. However, a CCD with anti-blooming gates (which collect the excess charge that builds up in overexposed pixels) may have a lower QE \cite{antiblooming}.

CCD sensors are predominantly used for scientific photography applications where high sensitivity is crucial, such as astronomy or medical imaging. Due to their high power consumption and propensity for unwelcome artifacts such as blooming, CCDs are not commonly employed in consumer devices.

\subsubsection{CMOS Sensors}
In contrast to CCDs, Complementary Metal Oxide Semiconductor (CMOS) sensors have a signal amplifier in each pixel, and a separate ADC for each column of the array \cite{blanc2001ccd}. The digital conversion is performed in parallel by the multiple ADCs, allowing for faster image capture and reset \cite{cmos_review}. These sensors suffer from higher noise than CCDs; however, they consume much less power, are cheaper to manufacture, and do not suffer from blooming artifacts \cite{cmos_review}. Today, CMOS sensors are ubiquitous in consumer photography and video recording devices.

\subsubsection{ISO}
The \textit{ISO} (pronounced ``I-S-O'' or ``eye-so'') of a framed sensor denotes its sensitivity to light. The name derives from the International Organization for Standardization, although it is not an acronym. The ISO of a digital camera is user- or automatically-adjustable, and it follows the ISO system used for image film. Specifically, the ISO determines the strength of the signal amplifier(s) for the pixels in the sensor. Increasing the ISO in isolation makes the pixels more sensitive, yielding brighter images, but increases the signal noise.

\subsubsection{Shutter Speed}\label{sec:shutter}
The \textit{shutter speed} (or exposure time) determines how long the sensor plane will be exposed to light to capture an image. The name derives from the ``shutter'' commonly found in professional cameras. The shutter is a physical barrier blocking the sensor from incoming light, which moves out of the way during image exposure. For single-image photography, the shutter offers precise timing of image exposure and protects the sensor from dust and debris when not in use. When recording video, however, modern cameras keep the shutter open, and instead control the exposure times of the sensor electronically.

A slow shutter speed means that the sensor is exposed for a relatively long amount of time, whereas a fast shutter speed indicates that the sensor is exposed only briefly. With all other settings being equal, slowing the shutter speed will produce a brighter image, but it will also increase the blur of motion which occurs during the exposure time. 

In CMOS sensors, the sensor array is typically read one row at a time. If there is fast motion, this sequential readout can result in the ``rolling shutter'' effect, whereby objects appear skewed due to non-synchronous capture. In contrast, CCD sensors and some modern CMOS \cite{Butler_Baskin} sensors have \textit{global shutter}. Here, the entire image frame is captured at once and buffered for readout, eliminating the rolling shutter effect. Global shutter tends to be cost-prohibitive for consumer devices.

One typically adjusts shutter speed and ISO together in an attempt to capture a video with low motion blur, low noise, and high dynamic range.

\subsubsection{Frame Rate and High-Speed Cameras}\label{sec:high_speed_cameras}
Since traditional videos are merely sequences of synchronous image frames, it follows that the speed of motion which can be observed by a framed camera is also directly tied to its imaging rate. The \textit{frame rate} of a video is usually measured in frames per second (FPS). Consumer video is typically recorded at 24-60 FPS, while higher frame rates are considered ``high-speed'' or ``slow motion.'' Importantly, the frame rate is not merely the inverse of the shutter speed. For example, a 30 FPS video must have a shutter speed faster than $\frac{1}{30}\textsuperscript{th}$ of a second. After each image frame is captured, it takes some amount of time to reset the sensor and prepare to record the next image. Additionally, a $\frac{1}{30}\textsuperscript{th}$ second shutter speed is too slow for moving video subjects. Thus, we may imagine that our 30 FPS video is captured with a $\frac{1}{500}\textsuperscript{th}$ second shutter speed. Our first image exposure concludes at $0.002$ seconds, but our second image exposure does not begin until $0.0\overline{33}$ seconds. Therefore, between any two consecutive images in our video, the camera fails to capture $0.031\overline{33}$ seconds of information. If we want to capture any of this data while keeping moving objects from becoming blurry, we must increase the frame rate.

Although a higher frame rate increases the amount of time for which one captures information, it brings with it several major drawbacks. For one, any increase in the recorded frame rate accordingly increases the amount of data which must be stored. While I describe framed video compression standards in \cref{ch:video_representations},  I simply note here that any increase in the data output of a sensor also increases the computational time spent compressing a video. Past a certain frame rate and resolution, even the most powerful video encoding hardware will fail to keep up with the raw data rate produced by a camera. As such, professional high-speed cameras record raw (uncompressed) image frames to a Random Access Memory (RAM) buffer. After the recording is over, this buffer is cleared and the raw frames are encoded in a standard format. Therefore, the recording time is limited by the amount of RAM available, and the compression can only be performed offline. For example, the Phantom T4040 camera can capture video at $9,350$ FPS at a resolution of $2560\times 1664$. With 256 GB of internal RAM, however, the camera in this configuration can only record for \textit{4.3 seconds} \cite{phantom}. Higher frame rate captures up to $444,440$ FPS are possible, but only when the resolution is greatly reduced \cite{phantom}.

Secondly, the faster shutter speeds necessary for high-frame-rate capture make the resulting images far darker. Each halving of the exposure time also halves the overall image brightness. Mere ISO adjustments cannot overcome very dark images without undue noise. Therefore, a videographer must illuminate the recorded scene with additional lights, often limiting the camera to a fixed position and view.

Finally, high-speed capture is exceedingly expensive. The aforementioned Phantom T4040 has an estimated cost upwards of $\$80,000$ for the camera body alone. A high-quality lens with a wide aperture (allowing more light through and faster shutter speeds) can easily cost more than $\$5,000$. External lights can cost $\$1,000$ or more each. High-capacity SSD storage (for the enormous video files) can cost thousands more. Even for a well-funded research lab or corporation, these expenses can be prohibitive.

\subsubsection{Raw Representations}

    The raw, decompressed representation for traditional video is a sequence of discrete images. Each image can be interpreted as an $m \times n \times c$ array of values, for $m$ rows, $n$ columns, and $c$ color channels. Traditionally, a monochrome image is represented with $c=1$, whereas a color image has $c=3$, typically with separate channels for the blue, green, and red color components. An image is typically stored in contiguous memory, in row-major order. Each color component is typically represented as an unsigned integer of size 8-16 bits, although these integers may be normalized to a floating point representation in the range $[0,1]$  for certain image processing tasks.

Dynamic range is inseparable from the capture sensitivity of the source, providing a practical upper limit to the intensity range of the representation \cite{dynamic_range}. Often, one records video at a high bit depth (up to 12 bits), then masters the video for viewing on both SDR and HDR displays \cite{tone_mapping}. The mastering process often involves adjusting brightness levels to show desired levels of detail. In this case, a higher-precision representation allows a filmmaker to have more flexibility in adjusting the dark and bright regions of the image during post-processing. Additionally, high precision helps to diminish the apparent edges within regions of intensity gradient.

The temporal component of a video frame is usually implicit, based on the index of the image in a sequence and the known frame rate of the video. For example, the fifth image in a video at 24 frames per second (FPS) can be said to span the time interval $[0.167, 0.208)$ seconds, or simply associated with the timestamp $0.167$ seconds. The temporal \textit{placement} of a video frame, however, is divorced from any sense of the video sensing parameters, assuming the video source is a traditional camera. The exposure time of the video frame is not conveyed in the raw representation without additional metadata. 
    
   Since pixel samples have a fixed frame rate, there is no inherent pixel averaging of stable video regions across multiple frames, causing high-contrast and high-speed scenes to suffer greatly from noise in the dark regions of the video. Even the advent of extremely high frame rate cameras did not bring with it a new paradigm of video representations \cite{high_fps_hevc}. Rather, these cameras produce large, spatio-temporally redundant framed video files using standard codecs. While some video container formats allow for variable frame rate video, the pixel samples remain temporally organized across frames \cite{vvc}, and software support for such videos is limited.

   Discrete images are an effective representation for fast processing on modern hardware. Since the data is temporally synchronous, one may leverage a GPU for parallelized convolution operations, with applications to image processing and deep learning. High frame rates can be a severe limitation, however, as each doubling of the frame rate requires a doubling of the decompressed data to be processed. Furthermore, one must design applications to be frame rate agnostic. For example, a small spatial movement between two consecutive images at a low frame rate implies a lower real-world speed than that same movement between consecutive images at a high frame rate. The frame rate challenge is especially difficult when training machine learning models for video with recurrent neural networks or convolutional neural networks, which assume a fixed input sample rate \cite{temporal_film,large_video_classification,deep_multiplicative}.

\subsection{Representation Layer}

\begin{figure}
    \centering
    \includesvg[width=0.6\linewidth]{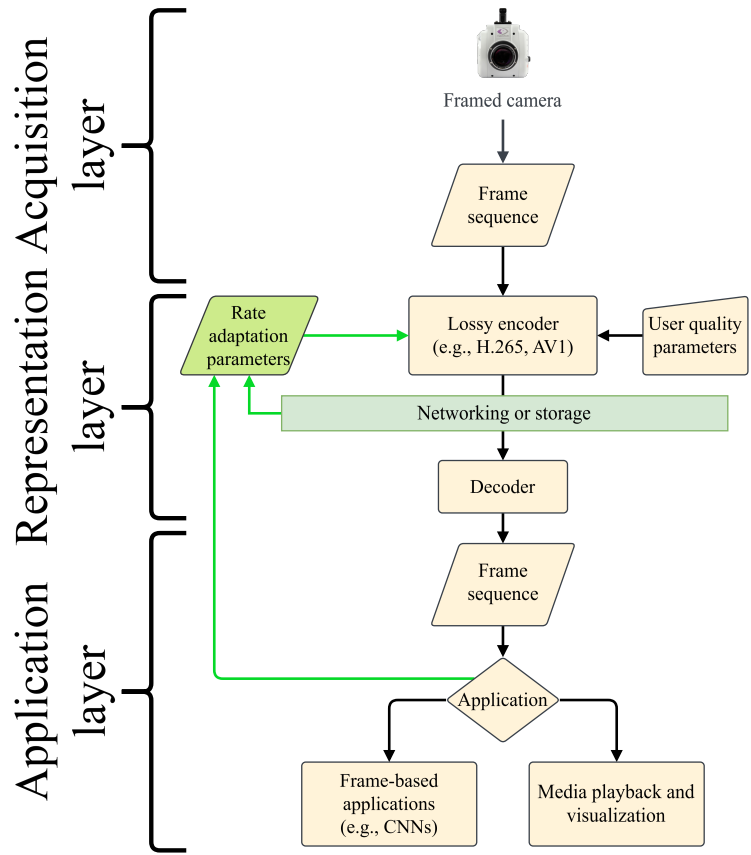}
    \caption[Framed system model]{Framed system model. Any information on temporal redundancy is lost during decompression. Computer vision and human vision applications typically receive the same frame-based input representation.}
    \label{fig:framed_model} 
\end{figure}

We see that a raw framed video representation has high temporal and spatial redundancy. This redundancy can produce enormous data rates, especially at high frame rates and resolutions. Since the early days of video systems, researchers have sought to mitigate these rate concerns through compression, so that the bandwidth and storage capabilities of consumer devices can receive and process video with reasonable speed. The compression, transmission, and decompression of the video data occurs in the representation layer of our system model (\cref{fig:framed_model}). Here, I describe some of the frame-based compression techniques and the quality metrics used to evaluate them.

\subsubsection{Quality Metrics}\label{sec:framed_quality_metrics}

The traditional video paradigm was born out of an effort to make multimedia consumption accessible to people on consumer devices. The standard goal of a video codec, then, is to balance compression ratios with quality. We can measure the image quality, based on visual appearance of the decoded frames, or quality of experience (QoE), based on the subjective experience of a human viewer. The latter focuses on the interplay of changes in resolution, bit rate, and latency \cite{qoe}. In this dissertation, I focus more on computer vision applications than human viewing, so here I discuss visual metrics over QoE metrics.

The purpose of a visual quality metric is to convey the amount of distortion introduced by lossy compression. \textit{Distortion} in this context is any change from the original source. Common signs of distortion in framed video include intensity banding, block artifacts, and aliasing.

Typical image quality metrics include the Peak Signal-to-Noise Ratio (PSNR) \cite{psnrssim}, the Structural Similarity Index Measure (SSIM) \cite{psnrssim}, and more recently the Video Multimethod Assessment Fusion (VMAF) \cite{vmaf_reproducibility,vmaf}. These metrics aim to represent \textit{perceptual} quality; that is, how a human viewer would perceive the video. In each of these metrics, distorted image frames in the decompressed domain are compared to ``signal'' input frames to a compressor. These metrics are known as full-reference metrics, since they compare pixel-level details between a signal image and a distorted image. PSNR is the most common, yet arguable least informative, framed quality metric. PSNR correlates poorly with human perception, which SSIM improves upon with windowed pixel comparisons \cite{ssim}. VMAF leverages a machine learning model trained to capture human perceptual quality, and it can account for temporal features and consistency between frames \cite{vmaf}.

In contrast, some reference-free video quality metrics exist to evaluate the perceptual quality without the need for the uncorrupted signal. These metrics include Natural Image Quality Evaluator (NIQE) \cite{niqe} and Blind Image Quality Index (BIQI) \cite{biqi}. These techniques capture the statistical patterns within natural images and penalize deviations from such patterns in arbitrary images.

\subsubsection{Color Models}

Framed sensors typically record images in the RGB color space, where each pixel is composed of three subpixel values for Red, Green, and Blue color channels. A Bayer pattern on the sensor filters incoming light into the RGB components, and this color arrangement is replicated with color subpixels in device displays \cite{10114931}. The intermixing of these subpixel intensities gives human viewers the illusion of full-spectrumcolor.

For video representation, however, we typically transform the raw images into the YCbCr color spaces. Here, the Y channel contains the luminance (brightness) of a pixel. The Cb and Cr channels convey the color. Video codecs often use this color space for \textit{chroma subsampling}, conveying the color components at a lower frequency than the luma component. For example, the 4:4:4 sampling strategy preserves the full sample rate for all the components, while the 4:2:0 sampling strategy preserves only one sample of the Cb and Cr components for every four samples of the Y component, reducing the raw data rate by half \cite{7991046}. Upon decoding, a single CbCr sample is copied for its three neighboring Y samples.  Subsampling takes advantage of the fact that the human visual system is more sensitive to variations in intensity (brightness) than to variations in color. Thus, video systems are designed to tolerate higher loss in the color domain.

\subsubsection{Compressed Representations}\label{sec:framed_compression}

To achieve high compression, frame-based video codecs seek to reduce \textit{spatial} and \textit{temporal} redundancy. An individual frame may be compressed as an I-frame (intra-frame), P-frame (predictive frame) or B-frame (bidirectional predictive frame). I-frames occur at regular intervals, such as once a second. Encoding an I-frame does not require reference to any other frame, so they serve as reset points for entropy coding. On the decoding side, this enables video scrubbing, where a user may jump to a specific time in a video; the first frame decoded after the time jump is a new I-frame. 

As the name implies, I-frames leverage only intra-frame spatial compression, similar to compressing a standalone JPEG image. A common technique is the Discrete Cosine Transform (DCT), which converts a spatial block of pixels into its frequency components \cite{richardson}. The component coefficients are ordered from low to high frequencies. The resulting coefficients are quantized, reducing their magnitude (and thus entropy). A quantization parameter (QP) determines the severity of quantization according to pre-determined lookup tables. Higher quantization results in lower quality, but a higher compression ratio.

P-frames and B-frames achieve inter-frame temporal compression. P-frames encode the difference residuals between the current frame and some previous reference frame. B-frames operate in like manner, but have a reference frame in the future as well. As such, they may be encoded in an order different from the display order. Both inter-frame types use motion compensation to predict the movement of spatial blocks of pixels between the reference frame and the current frame, given by a motion vector. The codec encodes the prediction residuals, the differences between the current block and the reference. Modern codecs (H.265, H.266, AV1, etc.) may also perform frequency transforms and quantization on these temporal prediction residuals \cite{h265,h266,av1}.

Compressed data may be stored in one of several container formats, such as MP4 \cite{mp4_container}, AVI \cite{avi}, and MKV \cite{mkv}. These formats wrap underlying video and audio streams in time segments, which enable playback scrubbing and stream switching. They also encode metadata information about the underlying data, but do not contain a decoder in and of themselves.

\subsubsection{Rate Adaptation}

In a simple video encoding task, one often uses a constant bit rate (CBR) parameter. In this case, the encoder varies the level of quantization throughout encoding such that the bit rate throughout the video remains near a given target. For example, suppose we instruct an encoder to use a CBR of 1 Mb/s. If a one-second packet of encoded video has a data size of 1.1 Mb, then the encoder will increase the QP offset (i.e., increase the amount of loss) to bring the data rate of the next packet below 1 Mb/s. Likewise, if the packet has an encoded size of 0.9 Mb, the encoder may lower the QP offset to take advantage of the additional headroom. While this scheme makes playback and bandwidth requirements consistent and predictable, it can result in undesirable variations in visual quality. That is, more complex or dynamic scenes will exhibit more compression artifacts than less information-dense scenes.

In contrast, streaming settings and video on demand (VOD) systems may have varying computational resources and server-client-bandwidth. In these cases, the system continuously monitors the bandwidth of the server-client connection, adjusting the target encoding bit rate to maximize the available bandwidth while maintaining low latency. In practice, we often assume a single producer, multiple consumer system (i.e., multiple clients receiving a given video stream). Here, it is impractical to encode individualized streams for each client. Rather, the server encodes several streams at different bit rate levels, often with varying resolutions. Then, each client requests packets at the bit rate appropriate for their current connection strength, processing power, and buffer size. This mechanism is known as \textit{adaptive bit rate} (ABR) streaming. Common examples of ABR standards include MPEG DASH \cite{mpeg_dash} and Apple HLS \cite{apple_hls}.

\subsection{Application Layer}

Video is increasingly utilized for computer vision tasks such as object detection, action recognition, 3D reconstruction, and scene segmentation \cite{cv_survey}. These systems typically rely on convolutional neural networks (CNNs) to learn feature embeddings from decoded image frames. In some cases, researchers employ recurrent neural networks (RNNs) to reduce the temporal redundancy across a temporal window of input images. Yi et al. presented a framework for task-driven rate directives in video compression \cite{task_driven_compression}, to bridge the gap between human and computer vision.  Most computer vision algorithms depend on a framed interpretation of the world, digesting video as a sequence of frames \cite{cv_survey}. \cref{fig:framed_model} illustrates how these computer and human vision applications have the same input representation. Despite great progress in the efficacy of such systems, it is disingenuous to suggest that their structures emulate the human eye's continuous and dynamic sensing \cite{survey}. 

Since modern video codecs can efficiently compress temporal redundancies, vision applications may save computational resources by processing compressed representations directly. CNNs have shown computational benefits when directly processing DCT embeddings \cite{learning_freq}, codec prediction frames and motion vectors \cite{compressed_recognition,c3d}, and custom neural representations \cite{compressed_vision,chen2021nerv}. As Wiles et al. note, existing frame-based applications and networks cannot operate on such compressed representations \cite{compressed_vision}. The compression and application layers are largely divorced in classical video systems, and compression performance does not correlate directly with application speed. This quality is most severe in systems where the video \textit{does} achieve high temporal compression, such as surveillance and high-speed video. An application may incorporate a differencing mechanism to remove temporal redundancy, but such a scheme does not scale with changes to the input frame rate. 

In a similar vein, recent work has explored \textit{sparse convolution}, computing convolution results only on localized patches of saliency \cite{visapp19,3DSemanticSegmentationWithSubmanifoldSparseConvNet,focalsparse,deltacnn,jointsparse}. However, in all of these cases, the underlying representation is temporally synchronous. Therefore, doubling the video frame rate necessitates a doubling of the frames processed by the application, slowing the computation speed even if the motion content is similar. In contrast, if the compressed bitrate directly were to correlate to the \textit{decompressed} bitrate (i.e., the amount of data the application must process), then researchers could develop applications which are more rate-adaptive and predictable.



\section{Event Video Systems}\label{sec:event_video_systems}

Whereas traditional video systems have image frames as their fundamental raw representation, recent years have seen the rise of a new video paradigm organized around cameras which capture data asynchronously. These \textit{event-based video systems} exceed traditional systems in dynamic range and sensing speed. However, they introduce new challenges with compression, rate adaptation and applications. Here, I describe the components of the event video system model and compare them to the framed system model described above (\cref{sec:framed_systems}).

\subsection{Acquisition Layer}

\begin{figure}
    \centering
    \includesvg[width=0.6\linewidth]{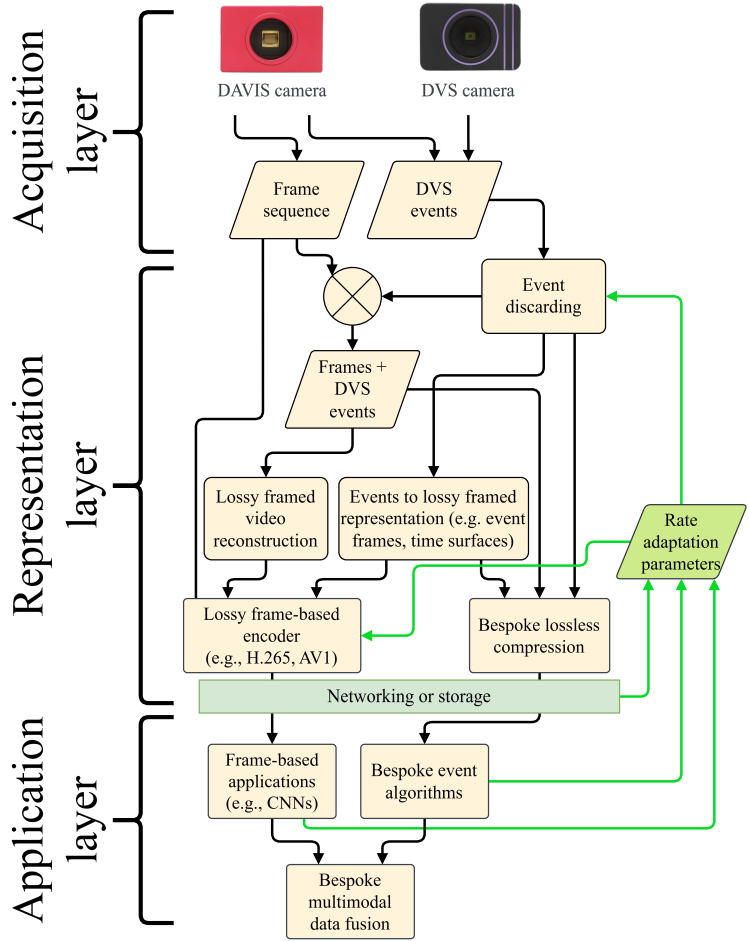}
    \caption[Existing event-based system model]{Existing event-based system model. Frequently, an event system may leverage both framed and event sensing. The choice of intermediate representation(s) and compression techniques depends largely on the target application. If framed data is to be processed alongside raw event data, the results of two or more application modalities must be fused by some higher-level application. }
    \label{fig:event_model} 
\end{figure}

``Event'' cameras are neuromorphic in nature, which is a radical departure from the classical sensing paradigm. Rather than record data synchronously across all pixels, each pixel instead captures information asynchronously from the others. The predominant sensing setups are illustrated in \cref{fig:event_model}.

\subsubsection{DVS}

Dynamic Vision Sensors (DVS) prioritize localized capture rate over global accuracy. Each pixel continuously measures the log voltage excited by the incident intensity \cite{dvs}. Each pixel measures the change in this voltage from a stored reference voltage, and outputs a corresponding signal if the \textit{change} exceeds a given threshold \cite{dvs}. The pixel array communicates these signal change ``events'' across each row, where they are appended with the pixel's spatial address \cite{dvs}. An arbiter for the array reads only one row of pixels at a time, tagging all events for a given row with the same timestamp at a microsecond resolution \cite{dvs}.

We can formulate the above description with the following notation. DVS cameras continuously measure the log intensity, $\widetilde{L} = \ln(I)$, incident upon a pixel $\langle x,y\rangle$, until the brightness changes  such that $\widetilde{L}(x, y, t) - \widetilde{L}(x, y, t- \Delta t) = p\theta$, where $p \in \{-1,+1\}$ is the polarity of the brightness change, $\theta > 0$ is the threshold, and $\Delta t$ is the time elapsed since the pixel last recorded such a brightness change (\cref{fig:dvs}) \cite{survey}. When the brightness changes across the threshold $p\theta$, the pixel outputs an asynchronous \textit{event}, represented by the tuple $\langle x, y, p, t\rangle$. The parameter $\theta$ is determined by a variety of camera gain settings, together with the effects of scene illumination. It is thus not easily determinable by the camera user, and may change drastically during the course of a video recording.

        The DVS approach has significant advantages. First, change events are quickly fired by pixels without incurring the frame latency of a synchronous sensor.  Second, a high dynamic range is achieved due to logarithmic compression of the intensity.  Third, when the intensity at a pixel is relatively stable, few events are fired, thereby conserving bandwidth.
        
        The key weaknesses of DVS are poor pixel noise characteristics and the difficulty of reconstructing actual intensity values.  The pixel values suffer from noise because log conversion is performed by a subthreshold MOS transistor, and the sensed value is an instantaneous voltage sample as opposed to an accumulation over an integrating interval \cite{dvs}.  Due to the differencing nature of the sensing, the high-quality reconstruction of actual intensity values becomes difficult as noise accumulation leads to drift.
        
        Because of these shortcomings, DVS sensors are typically used when capturing the edges of moving objects is of primary importance.  This primarily includes robotics applications where fast response times are necessary and human visual quality is not important, such as object detection and tracking, gesture recognition, and SLAM (simultaneous localization and mapping). On the other hand, numerous learning-based \cite{Rebecq19pami,Rebecq19cvpr,Scheerlinck20wacv} and direct \cite{Scheerlinck,reinbacher,async_kalman_filter} methods for framed reconstructions of DVS data are proposed in the literature, but these are limited to reconstructing 100-1000 frames per second even on state-of-the-art hardware and low-resolution ($240\times 180$ pixels) sensors \cite{Rebecq19pami}.

    Today, DVS cameras are commercially available from iniVation, Samsung, and Prophesee. iniVation is a start-up born out of a collaboration with ETH Zurich, where the pixel technology was originally developed. Prophesee is a subsidiary of Sony. The commercial offerings of DVS cameras are still researcher-oriented. For example, the cameras from iniVation require a separate computer connected over USB to drive the camera and record its data. Recent camera developments focus on increasing the sensor resolution and reducing the size. Compared to the cost of high-speed framed cameras as described in \cref{sec:high_speed_cameras}, DVS-based sensors are extremely affordable. During the course of my research, my lab acquired one such sensor for $\sim\$4,500$, an order of magnitude cheaper than the Phantom T4040 \cite{phantom_article}, while providing more than double the effective temporal resolution at a similar spatial resolution.

\begin{figure}
     \centering
     \begin{subfigure}[t]{0.32\textwidth}
         \centering
         \includegraphics[width=\textwidth]{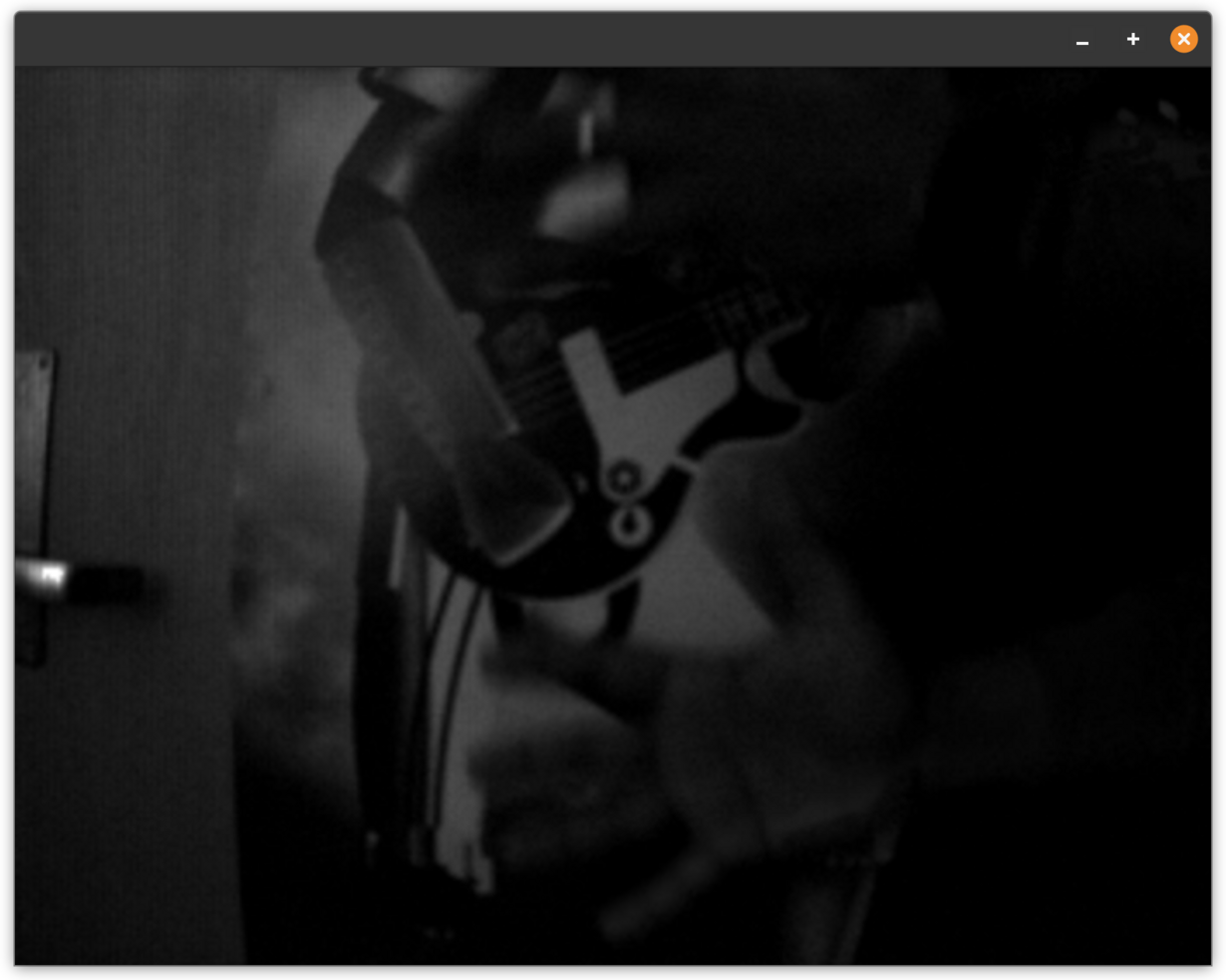}
         \caption{APS frame}
         \label{fig:aps}
     \end{subfigure}
     \hfill
     \begin{subfigure}[t]{0.32\textwidth}
         \centering
         \includegraphics[width=\textwidth]{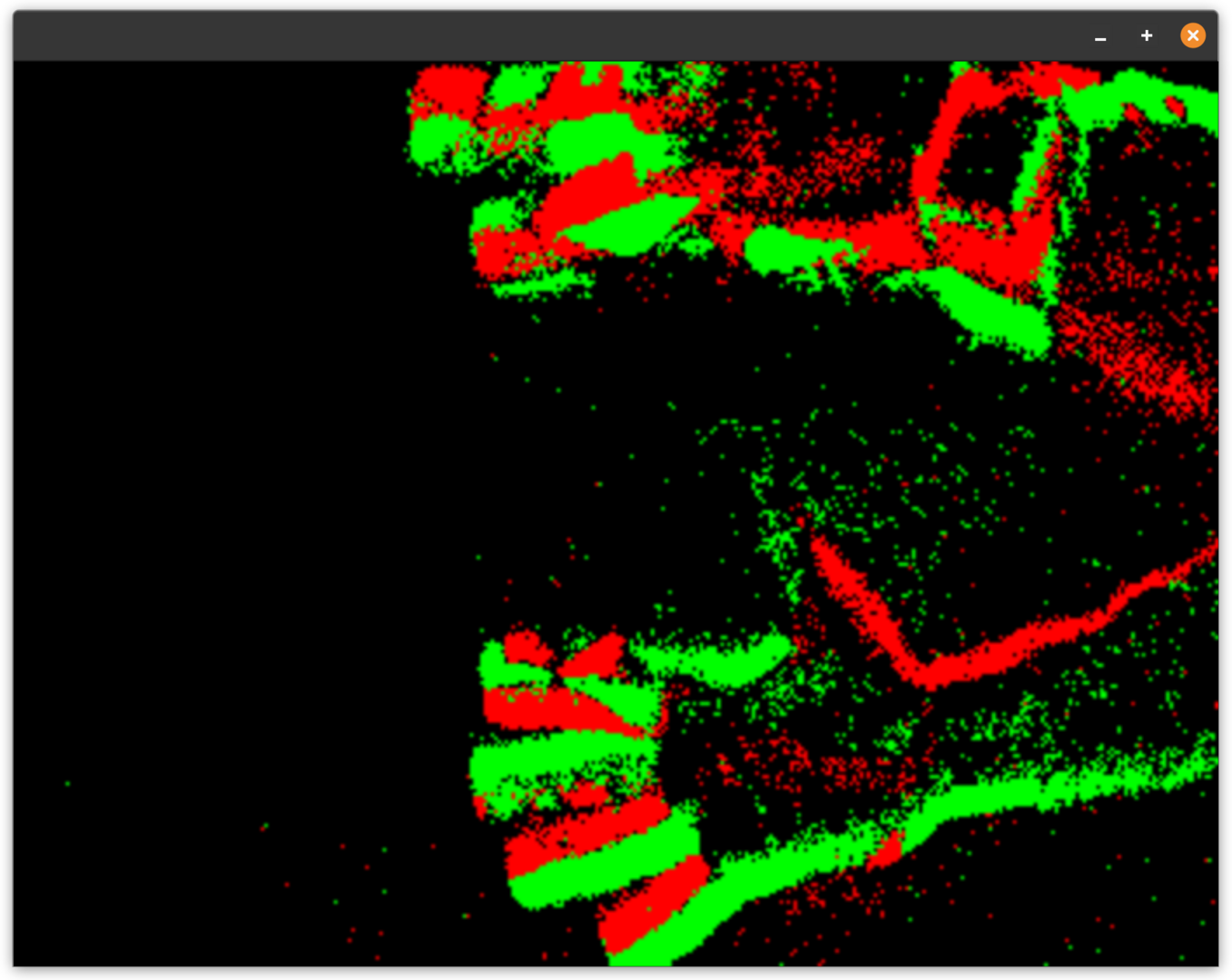}
         \caption{Event frame}
         \label{fig:events}
     \end{subfigure}
     \hfill
     \begin{subfigure}[t]{0.32\textwidth}
         \centering
         \includegraphics[width=\textwidth]{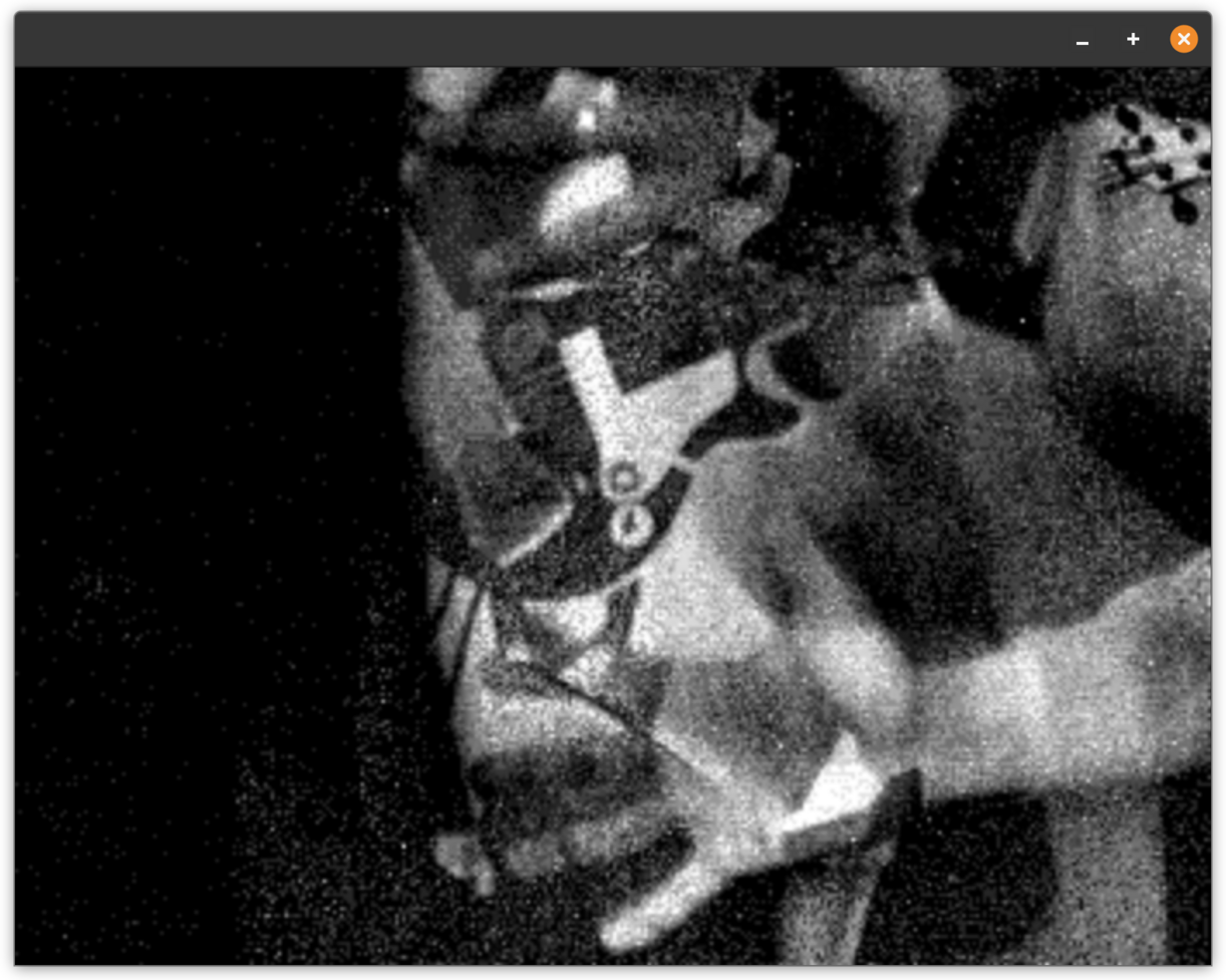}
         \caption{Accumulated event intensities}
         \label{fig:accumulated}
     \end{subfigure}
     \caption[Various displays of DAVIS camera data in the iniVation DV software.]{Various displays of DAVIS camera data in the iniVation DV software. The video shown is from the DVSMOTION20 dataset \cite{dayton_dataset}. (a) The APS image shows intensities for the whole frame, but has blurred motion and low dynamic range. (b) 3-state mapping of events occurring over a 1/60th second interval. A red pixel indicates that a negative polarity event occurred (pixel got darker), a green pixel indicates a positive polarity event (pixel got brighter), and a black pixel indicates that no event occurred. (c) The frame accumulates events and converts the log representations of the instantaneous intensity values to a linear intensity for display. Exponential decay reduces the influence of older events. The accumulated events show higher dynamic range and sharpness than APS, but much higher noise due to the DVS subpixels.}
     \label{fig:davis_demo}
\end{figure}
    
    \subsubsection{DAVIS and ATIS}
    
        Dynamic and Active Pixel Vision Sensors (DAVIS) aim to improve the usability of DVS technology by co-locating active (APS) pixels alongside the DVS pixels on the sensor chip \cite{survey,DAVIS}. These active pixels capture discrete frames at a fixed rate, just as a traditional camera. The APS and DVS data outputs have separate readout lines on opposite sides of the sensor \cite{davis_a}. The APS images are captured with a global shutter (\cref{sec:shutter}) \cite{davis_a}. While capturing APS images may increase DVS noise \cite{davis_a}, APS data can improve the quality of intensity reconstructions by preventing drift in DVS accuracy \cite{Scheerlinck,Pan_EDI,high-framerate_recon,inherent_compression}. The resulting intensity reconstructions can exhibit very high dynamic range, and are useful in vision applications where broader scene understanding is important. Visual examples are shown in \cref{fig:davis_demo}.
        
        Asynchronous Time-based Image Sensors (ATIS) take a different approach than DAVIS for capturing intensities alongside DVS events. Integrator subpixels on these sensors express absolute intensities by outputting two AER-style events, where a short time between the events corresponds to a bright pixel while a long time between events corresponds to a dark pixel \cite{survey}. The pixels do not integrate for a fixed time period, but rather integrate until reaching a given global (and fixed) exposure threshold \cite{atis2}. Although these intensity measurements are accurate, they trigger only when the DVS subpixel records an intensity change event \cite{atis}. Thus, ATIS does not provide absolute intensities for every pixel at a given point in time, but only at high-contrast moving edges. Additionally, ATIS pixels are larger than those of DAVIS \cite{davis_a}.
        
        ATIS and DAVIS aid vision tasks, such as object detection, by augmenting contrast-based events with absolute intensity measurement. However, they produce two distinct streams of data, and require the development of a sophisticated supporting software infrastructure to fuse the two streams to fully take advantage of their capabilities. For DAVIS in particular, the intensity frames help one to focus the camera lens, use classical vision algorithms alongside event-based algorithms for comparison or augmentation, and achieve better performance in intensity reconstruction through the fusion of event and framed data. DAVIS has emerged as a prominent event-based sensing technology for research, while ATIS has not seen widespread commercial adoption.

\subsubsection{Vidar/Spiking Camera} \label{sec:vidar}
The Vidar camera aims to rectify the lack of events for static regions of a scene with time-based intensity events \cite{vidar1, vidar2}. Each pixel has an accumulator which integrates light until reaching a global threshold. When a given accumulator reaches the threshold, that pixel fires an event denoting that time. Thus, Vidar can capture absolute intensities of an entire scene. However, the underlying event representation is a binary event frame produced at fixed time intervals \cite{spiking_camera}, so the sensor suffers from enormous data redundancy in the static regions and the temporal granularity is much lower than that of a DVS camera.

\subsection{Representation Layer}\label{sec:dvs_representation_layer}

\subsubsection{Intermediate Representations}\label{sec:existing_event_intermediate_representations}

      While some applications process DVS events directly with spiking neural networks (SNNs) \cite{Duwek_2021_CVPR,Barbier_2021_CVPR,spiking1}, many resort to grid-based representations of the events, which are more amenable to classical vision pipelines and do not require specialized neuromorphic network hardware \cite{Gehrig_2019_ICCV,eventframe,eventcar,EV-FlowNet,eventflow,TORE_volumes}. \cref{tab:representations} (in the next chapter) outlines several of these grid-based representations and highlights the tradeoffs researchers make between temporal precision and ease of processing.

      Let us focus our attention on the ubiquitous DVS events. An SNN \cite{dvs_snn} or asynchronous filter \cite{Scheerlinck} may process individual events sequentially. Event \textit{frames}, on the other hand, are discrete 2D images which accumulated the event counts or polarities of all pixels over a uniform period of time \cite{survey}. Similarly, a voxel grid may provide a more dense representation with interpolation between event frames \cite{survey}. Time surfaces, in contrast, are 2D images which map the timestamp of the last event received for each pixel \cite{time_surface}. The synchronous nature of event frames, voxel grids, and time surfaces provides straightforward compatibility with standard neural network architectures and GPU architectures \cite{survey}. 
      
      Finally, much work in the literature aims to reconstruct natural-looking intensity images from event data. This work is driven by the goal of greater performance with classical vision applications, such as object detection. E2VID uses a convolutional recurrent network with groups of event tensors which are variable in time \cite{Rebecq19pami}. The framed video reconstruction improves the performance of object classification and visual-inertial odometry tasks, however the reconstruction time is slow and dependent on the output frame rate \cite{Rebecq19pami}. The event-based double integral (EDI, described in greater detail in \cref{sec:edi}) is a direct method for fusing data from DVS events and low-rate APS images from a DAVIS camera \cite{Pan_EDI}. This work addresses the issue of the APS frames being blurry, relative to the DVS events, and proposes a method for ``deblurring'' the APS images based on the events fired over the images' exposure intervals. Additional work has explored the fusing of the complementary data from Vidar and DVS sensors for image reconstruction. Kang et al. proposed a unified imaging system which combines a DVS sensor and Vidar sensor with a beam splitter to capture the same view with both cameras simultaneously \cite{retinomorphic_sensing}. The authors used a recurrent neural network (RNN) to dynamically switch between the two data streams for each pixel according to whether or not the pixel is static or changing, and achieved higher visual quality in video reconstructions at a given data rate with their joint system compared to Vidar alone. However, their proposed representation maintains the separate Vidar and DVS event representations within it, has no mechanism for content-directed dynamic rate control, and relies on framed video reconstruction to represent high-rate visual information from the DVS event stream for use in vision applications. A more recent work offers some improvements for Vidar-DVS fusion \cite{neuspikenet}, but its speed characteristics were not disclosed.

\subsubsection{Compression}

Although researchers have explored rate control schemes for DVS streams, the proposed systems perform rate adaptation only at the application level \cite{glover}, or at the camera source by adjusting biases for $\theta$ \cite{dvs_feedback_control,dvs_rate_patent}, rather than exploring a rate-distortion control scheme for event representation and transmission. Since DVS events express intensity \textit{change}, to incur loss on one event has a compounding effect on later events for a given pixel. Therefore, robust compression is an extremely difficult and, to this point, underexplored task. \cref{fig:event_model} illustrates some of the general compression systems used and their corresponding input representations.

Khan et al. proposed a method for ``lossless'' DVS compression which used polarity event frames in conjunction with a frame-based video encoder (H.265) \cite{khan}. This method is actually \textit{lossy} in the generation of the event frames, due to temporal quantization. Additionally, it has poor performance at high frame rates. Similar lossy compression methods for DVS either quantize the temporal components \cite{towards_dvs_lossy,Schiopu_1} or discard some events entirely. Recent work by Schiopu et al. has explored lossless DVS compression optimized for an on-camera hardware implementation \cite{Schiopu_2,Schiopu_3}.

For DAVIS cameras, researchers may fuse the active and dynamic information as discussed above, but often still fall back to framed representations to incorporate the disparate streams. As before, this approach temporally quantizes the events, removing much of the high-rate advantage of DVS capture. Banerjee et al. uniquely proposed a system for application-driven DAVIS rate control on a networked client \cite{quadtree_compression,banerjee2021joint}, selectively discarding DVS events of low saliency based on the APS images and bandwidth constraints. However, their scheme does not unify the event and the framed information under a single representation, meaning that the streams still require separate application-level logic.

     Additionally, the application ecosystem for event vision is fragmented between numerous raw and lossless-compressed formats. Data sets are variously distributed in a human-readable text format \cite{davis_dataset,v2e}, a \texttt{.bag} format for the Robot Operating System \cite{davis_dataset,Rebecq19pami,lee2022vivid,v2e}, an \texttt{.hdf5} format for learning-based Python tasks \cite{ddd20,v2e}, an \texttt{.aedat2} format designed for operability with the jAER software \cite{ddd20,v2e}, or a \texttt{.mat} format for use in MATLAB applications \cite{Pan_EDI,async_kalman_filter}. Event camera manufacturers iniVation and Prophesee each have their own software suites providing applications, but each operates only on their proprietary camera data formats. While some tools exist to transform raw camera data to one of the above formats, none of them fundamentally changes the underlying data types (i.e., temporal contrast events).

\subsection{Application Layer}

Due to the difficulty, time, and expense of acquiring large-scale datasets with an event camera, there have been a number of works related to simulating DVS and DAVIS sensors based on framed video inputs \cite{pmlr-v87-rebecq18a,10.3389/fnins.2021.702765,v2e}. Existing frameworks for event camera data focus on generalizing learning-based application interfaces and evaluation mechanisms for particular event cameras; that is, the input event data representation is unchanged, and only the applications are modular \cite{8784777,8460541,7862386}. 

Some vision applications have been built from the ground up with these contrast-based event sensors in mind, using spiking neural networks (SNNs) \cite{Duwek_2021_CVPR,Barbier_2021_CVPR,spiking1}. Many applications, however, convert the sparse event data to a frame-based representation for use with traditional processing schemes and CNNs \cite{Gehrig_2019_ICCV,eventframe,eventcar,EV-FlowNet,eventflow,TORE_volumes}. Cannici et al. and Messikommer et al. recognized that standard CNNs use many redundant computations when processing event data, and proposed sparse convolutional network architectures for DVS with applications in object detection and recognition \cite{asyncconv1,Messikommer20eccv}. \cref{fig:event_model} outlines the representations used for various application modalities.


\section{Conclusion}
This chapter discussed requisite background material in data compression, classical video systems, and novel event-based video systems. I discussed how classical systems were designed primarily for human vision, meaning that computer vision applications process much temporally redundant data. On the other hand, newer event-based video systems achieve spatiotemporal sparsity, but they often fall back to inefficient frame-based representations for video applications. Even if they preserve the original event representation, these applications are custom-built for a particular sensing modality, making them not applicable for future event sensors.

     \chapter[~~~~~~~~~~~~\texorpdfstring{\adder}{ADDER}: A Universal Video Representation]{\texorpdfstring{\adder}{ADDER}: A Universal Video Representation\footnotemark}\label{ch:adder_transcoding}\footnotetext{Significant portions of this chapter previously appeared in the proceedings of 2020 ACM Multimedia \cite{freeman_emu} and 2023 ACM Multimedia Systems \cite{freeman_mmsys23}.}

\graphicspath {{ch03_representation/images/}}

\section{Introduction}

We have explored the landscape of video systems and found a chasm between two opposite modes of thinking. On one side lies framed video, intimately tied to the sensing mechanism of the first video cameras. Temporal redundancy is reduced only in the compressed representation, compressors are built around human perceptual quality metrics by design, and applications struggle to process high frame rate decompressed data with speed. On the other side lies existing event-based video, intimately tied to the sensing mechanisms of contrast-based event sensors. Temporal redundancy is here reduced efficiently in the decompressed representation. However, such data is generated only by event cameras or camera simulators, robust lossy compression is nigh impossible, and many systems cast contrast events into a number of framed representations for application-level processing. 

As with DVS systems, I argue that intermediate representations can be helpful from a compression and application standpoint. The enormous data rates of a DVS camera, for example, make downstream processing difficult. However, the complexity and fragmentation of the existing event-based ecosystem hurts the generalizability of various compression techniques and applications. Firstly, applications are designed for individual DVS intermediate representations, such as polarity frames or time surfaces. Therefore, any compression of the input events must occur during the transformation of the events into an intermediate representation, or else bespoke compression techniques must be employed for each intermediate representation. By casting events into a framed representation, these systems heavily quantize the temporal information. A framed representation without temporal quantization would produce a video at one million FPS, which is impossible for frame-based computation hardware to process in real time. For example, the modern NVIDIA A100 GPU was benchmarked at inferencing ResNet50 with low-resolution image frames at a rate of only 1714 FPS \cite{Qblocks}. Secondly, applications and compression techniques are designed specifically for contrast-based DVS sensors. This single-camera focus means that any future advancements in event-based sensing hardware will require the reinvention of these applications. Thirdly, each event camera manufacturer uses a different proprietary data format, and online systems lack a common camera interface to unify them.

Thus, we see that classical video representations (i.e., frames) are not optimized for high-rate vision applications. Meanwhile, modern event video representations (i.e., contrast events) are not optimized for human viewing or event-based lossy compression. I argue that this gap leaves ample room for an alternate representation that can leverage the strengths of both approaches. To achieve practical utility for the existing framed and event video landscapes, this representation must be a \textit{software representation}. That is, one must be able to \textit{transcode} a relevant video type to this unified representation. In the following, I outline the specific goals of such a representation, emphasizing the main points in bold.

\subsection{Requirements for a Unified Video Representation}

The chief weakness of framed video is its lack of temporal sparsity in its decompressed representation. That is, consecutive decoded frames typically have high temporal redundancy. Spatiotemporal vision applications must then process this redundant data, even if it does not contribute to the application-level results. Therefore, I seek to \textbf{increase data sparsity in the decompressed domain for framed video sources}. To achieve this, the decoded representation should be \textbf{event-based}; pixels must have asynchronous values.

Existing contrast-based event video makes lossy compression difficult. Since events express intensity change relative to some prior baseline and prior events, it is difficult to predict the application-level impact of quantizing an event timestamp or discarding an event altogether. As such, most of the source-modeled compression work for event video has focused on lossless compression \cite{Schiopu_2,Schiopu_3}. I seek to simplify lossy compression and expose a \textbf{predictable rate-distortion tradeoff}. The unified event representation must therefore express temporal independence. For a given pixel, incurring loss in one event should not impact the meaning of all subsequent events. Therefore, \textbf{events should express absolute intensity}, rather than intensity change.

Frame-based applications are ubiquitous in computer vision, and they can provide crucial baselines against which we can evaluate a new representation. Therefore, a unified event representation must be trivially \textbf{backwards compatible with framed applications}. That is, we should be able to quickly transform a temporal window of events into an image frame for classical processing.

Event video also currently struggles to incorporate multimodal data from contrast events and intensity frames (e.g., DAVIS cameras). Event-based vision applications can often achieve stronger performance when augmenting events with intensity frames \cite{survey}, but the methods are representationally fragmented. One may generate a high frame rate video sequence by ``deblurring" the intensity images with DVS events, which have a high temporal resolution \cite{Pan_EDI}. In this case, the system inherits all the weaknesses of high frame rate classical video, outlined above. Alternatively, one may process the intensity images and DVS events as separate input representations within an application. In the latter case, one must leverage bespoke methods for fusing the disparate data within the application \cite{hmnet}. I argue that multimodal \textbf{data fusion should occur before the application}, during the generation of the intermediate representation.

Currently, event-based video applications are designed around a single sensing modality, such as DVS or DAVIS. Meanwhile, advances in asynchronous sensing have enabled the development of novel designs which address the weakness of the mainstay contrast-based cameras \cite{montek,aeveon,vidar1}. So that event-based applications are not nullified by sensor improvements, a unified representation should be completely \textbf{source agnostic}. That is, one should have a simple interface to transcode the data from existing and future sensing modalities into this representation.


\begin{table}[t]
      \caption[Comparison of event video representations]{Comparison of event video representations. The \adder{} representation in transcode mode (ii) achieves the most flexibility for event-based applications, while it can also be trivially transformed to a frame-based intensity representation for use in classical applications.}
      \label{tab:representations}
      \begin{tabular}{p{2.0cm}|p{2.3cm}p{1.8cm}p{1.8cm}p{1.8cm}p{1.8cm}p{1.5cm}}
        \toprule
        Name &Representation & Continuous? & Full-sensor intensities? &Transformable to DVS? &Compatible with CNNs? &Compatible with SNNs?\\
        \midrule
        DVS & $\langle x,y,p,t \rangle$  & \cmark &   & — &   & \cmark\\
        DAVIS & $\langle x,y,p,t \rangle,$ $[w, h, I]$  & DVS only & APS only & — & APS only & DVS only \\
        DAVIS image fusion & $[w, h, I]$  &   & \cmark &   & \cmark &   \\
        DVS event frames & $[w, h, \Delta L]$  &   &   &   & \cmark &  \\
        DVS time surfaces & $[w, h, t]$ &   &   &   & \cmark &  \\
        DVS voxel grid & $[w,h,n,$ $\sum p\text{ or }\frac{\partial L}{\partial t}]$ &   &   &   & \cmark &   \\
        ATIS & $\langle x,y,p,t \rangle,$ $\langle x,y,I,t \rangle$ & \cmark &   & — &   & \cmark \\ \hdashline
        \adder{} event frames & $[w, h, \frac{2^D}{\Delta t}]$ &   & \cmark &   & \cmark &   \\
        DAVIS (i) $\rightarrow$ \adder& $\langle x, y, D, \Delta t\rangle$ & \cmark & \cmark &  &   & \cmark\\
        DAVIS (ii) $\rightarrow$ \adder& $\langle x, y, D, \Delta t\rangle$ & \cmark & \cmark & \cmark   &   & \cmark\\
        DVS (iii) $\rightarrow$ \adder& $\langle x, y, D, \Delta t\rangle$ & \cmark &   & \cmark &   & \cmark\\
        \bottomrule
      \end{tabular}
    \end{table}

\section{Inspiration: Future Event Sensors}

\subsection{ASINT}

I contrast the common DVS event cameras to the sensor architecture proposed by Singh et al. \cite{montek}, and for convenience I refer to this architecture as ASINT (for `Asynchronous Integration'). Unlike DVS pixels, the pixels of an ASINT sensor record absolute luminance values, more like a traditional CMOS camera sensor. Each pixel has its own integrator and decimator. The integrator accumulates incoming photons as photoelectrons, and the decimator uses a decimation factor $2^D$ to divide the event frequency, determining the threshold of light the pixel must take in before firing an event \cite{Smith2017ASM}. Hence, such a sensor is \textit{not} invariant to scene illumination, but can directly encode scene luminosity without the need for conventional active pixels like DAVIS \cite{survey, sensitive_color_davis, new_color_davis}, estimation \cite{scheerlinck2018continuoustime}, or a complex  neural network to reconstruct video \cite{rebecq2019high}. The events output by the ASINT sensor are tuples of the form $\{x, y, D, \Delta t\}$, where $x$ and $y$ are the pixel's coordinates in the sensor, $D$ is the decimation factor, and $\Delta t$ is the time elapsed during the pixel's light accumulation. The incident light intensity, $I$, of a pixel may be computed by dividing the decimation factor by the time delta, $\Delta t$, between two consecutive events for that pixel, written as $I \propto \frac{2^D}{\Delta t}$. The data produced by ASINT, then, is a sequence of \textbf{intensity events}. The ASINT model is contrasted with DVS and framed sensors in \cref{fig:sensortypes}.

The ASINT design has a number of advantages over the DVS design. Chiefly, since an ASINT sensor records absolute luminance values, there is no need for an estimation, augmentation, or learning-based method for reconstructing image frames from events. Rather, reconstructing an image frame is as simple as averaging the intensity of the events spanning a particular length of time for each pixel. ASINT is distinct from ATIS in that the integration threshold, $D$, can vary \textit{dynamically} over time for each pixel, according to scene dynamics. The event rate can thus be optimized by internal controls or an external application. Compared to framed sensors, the sensor can achieve extremely high dynamic range. Over a given interval of time the pixels corresponding to the darkest parts of the scene can maintain a low $D$-value and obtain an accurate $\Delta t$ measurement while the pixels corresponding to the brightest parts of the scene can maintain a high $D$-value and potentially fire many events. This helps mitigate noise in the lower range of intensities and prevent overexposure in the higher range. Rather, if the $D$-adjustment scheme is set appropriately, one can always recover detail in both extremes of the scene during framed image reconstruction \textit{after} the time of data capture. 

While the sensing mechanism of ASINT unlocks a new realm of event-based sensing, a working prototype has not yet been constructed.

\subsection{Aeveon}

Recently (in mid-2023), the primary manufacturer of DVS and DAVIS sensors announced the upcoming Aeveon sensor \cite{aeveon}. Unlike DVS, this camera natively captures floating-point intensity events. One can then reconstruct natural-looking high dynamic range images from these events, while maintaining a much higher speed and lower power usage than a traditional frame-based camera. Notably, the sensor independently tunes the exposure time of each pixel according to its incident brightness and its perceived saliency. The reported design of this sensor appears extremely similar to ASINT, bolstering the notion that intensity event sensing is the up-and-coming technology.

\section{Proposed representation: \texorpdfstring{\adder{}}{ADDER}}

\begin{figure}
    \centering
    \includegraphics[width=1\linewidth]{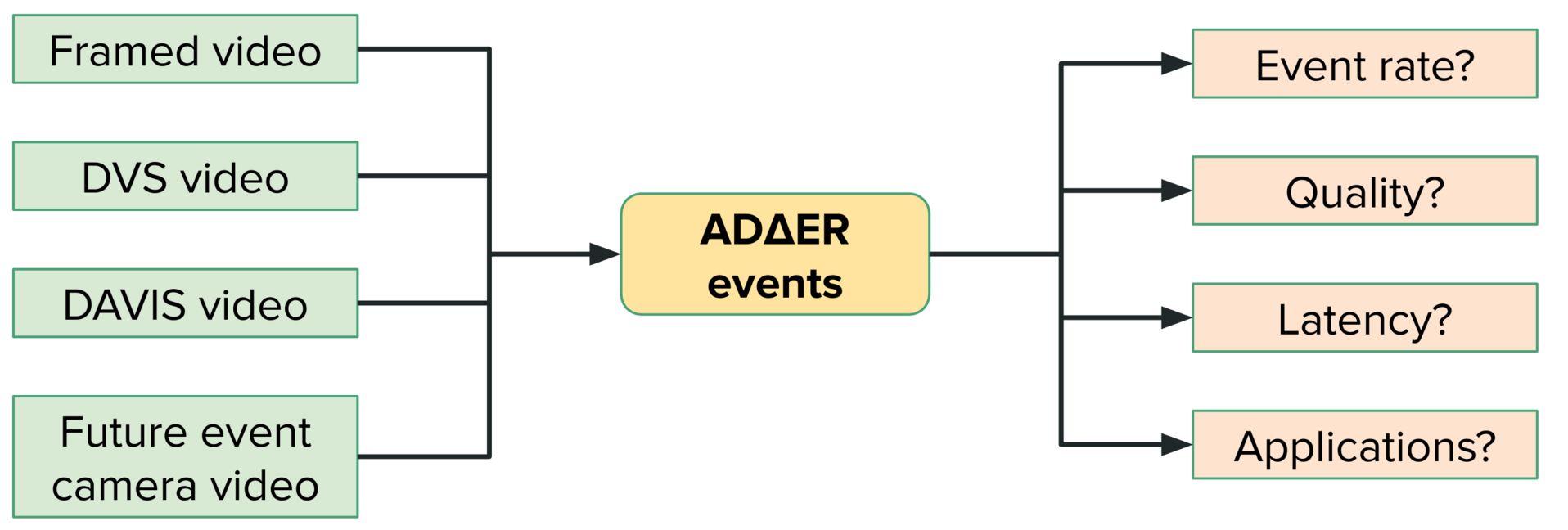}
    \caption[Coarse overview of an \adder{} system]{Coarse overview of an \adder{} system. I transcode multiple video types to a single representation and explore the implications of the representation on event rate, quality, latency, and applications.}
    \label{fig:simple_adder} 
\end{figure}

     Although I recognize the utility of a DVS-style representation for contrast-based events, we have seen that such a format is not amenable to representing absolute intensities. With a DVS-style approach, inducing loss is feasible only in terms of mitigating sensor noise and allowing an application to ignore events it deems unimportant. That is, removing a DVS event in a given pixel's stream will necessarily affect the interpretation of later events in that pixel's stream, since each event conveys intensity information \textit{relative} to a previous event. For the same reason, even the DVS-style ATIS representation, which can encode intensity information, is ill-suited as a general, adaptable, and compressible representation. Specifically, these representations do not support receiver-driven lossy compression of their events. I argue that, if one must already transform the raw data to prepare it for an application, one should instead transcode it to a fast, compressible, non-proprietary event data representation which is both application agnostic and camera agnostic. Then, applications may ingest this single representation, and support for future event sensing modalities can be implemented in the data transcoder rather than in individual applications. This model is illustrated in \cref{fig:simple_adder}.

\begin{figure}
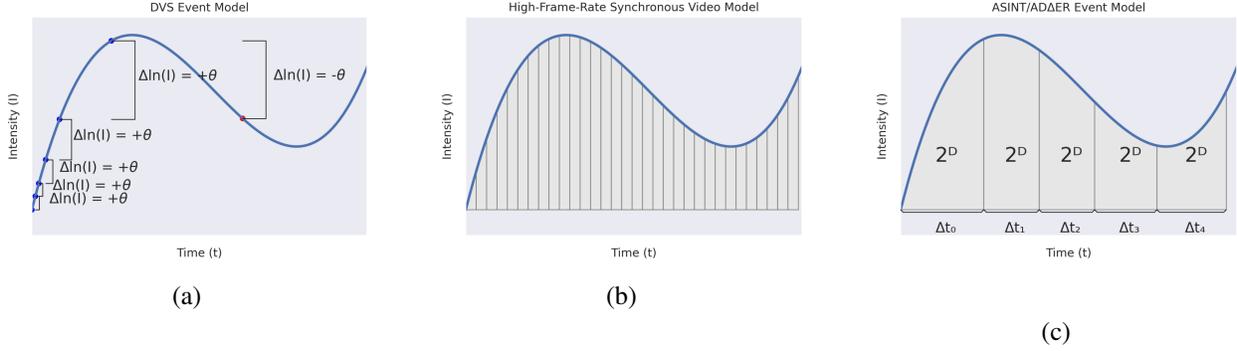

     \centering
     \begin{subfigure}[t]{0.30\textwidth}
         \centering
         \includegraphics[width=\textwidth]{data_models2/dvs.png}
         \caption{}
         \label{fig:dvs}
     \end{subfigure}
     \hfill
     \begin{subfigure}[t]{0.30\textwidth}
         \centering
         \includegraphics[width=\textwidth]{data_models2/framed.png}
         \caption{}
         \label{fig:framed_diagram}
     \end{subfigure}
     \hfill
     \begin{subfigure}[t]{0.30\textwidth}
         \centering
         \includegraphics[width=\textwidth]{data_models2/adder.png}
         \label{fig:adder_asint}
         \caption{}
     \end{subfigure}
     \caption[Comparison of video models]{Comparison of video models. (a) A new DVS event is fired whenever ln(I) changes by the threshold $\theta$.  Since differences are sensed, intensity reconstruction is challenging. (b) A framed video contains one sample for each pixel at synchronous points in time, with fixed lengths between samples. (c) An \adder{} indicates that the accumulated intensity units reach $2^D$.  $D$ controls the per-pixel sensitivity, which I show as constant here for illustrative purposes. In the \adder{} model, however, $D$ adapts to changes in the intensities being integrated, to reduce the event rate.}
     \label{fig:sensortypes}
\end{figure}

At the same time, the proposed ASINT sensor \cite{montek} (and similarly, Aeveon \cite{aeveon}) suggests a powerful underlying representation. I divest this representation from ASINT, and I propose the \textbf{A}ddress, \textbf{D}ecimation, $\Delta t$ \textbf{E}vent \textbf{R}epresentation (\textbf{\adder{}}, pronounced “adder'') format as the “narrow waist'' representation for asynchronous video data.  As illustrated in \cref{fig:adder_asint}, a pixel $\langle x,y,c\rangle$ continuously integrates light, firing an \adder{} event $\langle x,y,c,D,\Delta t\rangle$ when it accumulates $2^D$ intensity units (e.g., photons), where $D$ is a \textit{decimation threshold} and $\Delta t$ is the time elapsed since the pixel last fired an event. I measure $t$ in clock ``ticks,'' where the granularity of the clock tick length is user adjustable. Unlike ATIS events, a \textit{single} \adder{} event directly specifies an intensity, $I$, by
        $I = \frac{2^D}{\Delta t}$.
    The key insight of this model is \textbf{the dynamic, pixelwise control of $D$} during transcode. Raising $D$ for a pixel will decrease its event rate, while lowering $D$ will increase its event rate. With this multifaceted $D$ control, I can ensure that pixel sensitivities are well-tuned to scene dynamics, avoiding the rate and compression issues of the ATIS model while maintaining the strengths of intensity events.

With this representation, I have a unified format for which to build compression schemes and applications targeting a wide variety of camera technologies.  With source-specific transcoder modules and an abstract \adder{} library, I can cast video from arbitrary camera sources into the single, unified \adder{} data representation. This unified representation can effectively capture the temporal resolution of event camera data, while supporting straightforward compatibility with both convolution and spiking neural networks (\cref{tab:representations}).

To generate optimal \adder{} streams, I developed several transcoding methods and buffering techniques. In the remainder of this chapter, I discuss my work on the abstract \adder{} pixel model and how \adder{} event streams are generated with self-directed $D$ control mechanisms. Then, I describe the particular transcoder modules for framed video, DVS video, and DAVIS video. Finally, I explore techniques for intensity prediction and the discrete event simulation of a proposed sensor which natively captures \adder{}-style event data.


\section{Acquisition Layer}

In the acquisition layer, we can transcode from one or multiple different data sources into the common \adder{} representation. \cref{fig:system_diagram_intro_full} illustrates the fusion of these disparate sources.

\subsection{\texorpdfstring{\adder{}}{ADDER} Transcoder Fundamentals}\label{sec:transcoder_fundamentals}
    
    To explain how to transform data into \adder, I must first describe my novel pixel model and parameters common to all transcode types.
    
    \subsubsection{Event Pixel List Structure Definition}\label{sec:pixel_list_structure}
    We will first examine the integration of a single \adder{} pixel model. I represent an \adder{} pixel state with a linked list. Each node in the list consists of a decimation factor, $D$, an intensity integration measurement, $I$, and a time interval measurement, $\Delta t$. The connections between nodes carry an \adder{} event representing the intensity that the parent node integrated \textit{before} the child node was created.  When initializing a new list, we set $D$ to be the maximal value possible to represent the first intensity we intend to integrate. To maintain precision, the values $I$ and $\Delta t$ in the list are floating point numbers. Since \adder{} is source-agnostic, we interpret the incoming values as generic ``intensity units'', rather than tying them to a real-world measure of intensity, such as photons, and we can scale the input values of individual data sources according to our needs. The pixel lists form the generic transcoder model depicted at the heart of the acquisition layer in \cref{fig:system_diagram_full_repeat}.
    
    I define a node of the list by the following:

\begin{definition}
    A node $T$ is a tuple of the form $\langle D, I, \Delta t\rangle$. An edge $E_n$ between nodes $T_{n}$ and $T_{n+1}$ is a tuple of the form $\langle D', \Delta t' \rangle$. For a list with $N$ nodes, $N \in \mathbb{N}$, the following invariants hold:

    \begin{equation}\label{eq:iv1}
        I_i = \sum\limits_{n=i}^{N-1}2^{D_{n}'} + I_{N}
    \end{equation}

    \begin{equation}\label{eq:iv2}
        \Delta t_i = \sum\limits_{n=i}^{N-1}\Delta t_{n}' + \Delta t_{N} - \varepsilon, \text{where $-N + i \leq \varepsilon \leq 0$ represents floating point truncation error}
    \end{equation}

    \begin{equation}\label{eq:iv3}
        D_i > D_i' \geq D_{i+1}, \text{and each change to a particular $D_i$ is a monotonic increase}
    \end{equation}

    \begin{equation}\label{eq:iv4}
        D_i \geq \lceil\log_2 I_i\rceil
    \end{equation}
    
\end{definition}

    \subsubsection{Event Pixel List Structure Example}
    
\begin{figure}
        \centering
        {\includesvg[width=1.0\linewidth]{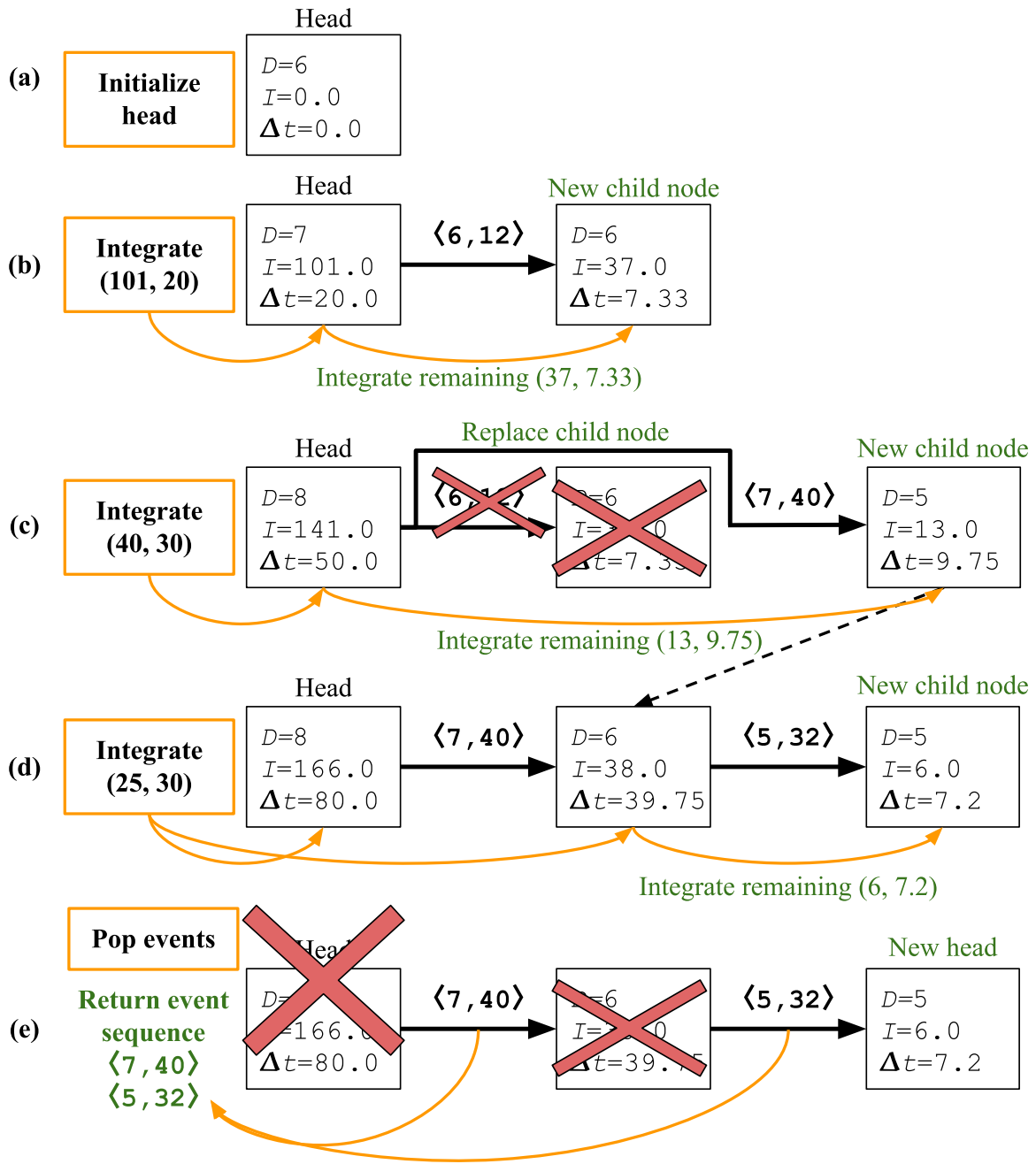}}
        \caption[Event pixel list structure]{Structure of event pixel lists under continuous integration.}
        \label{fig:pixel_list}
    \end{figure}

    Let us walk through a visual example of this list structure in \cref{fig:pixel_list}. Suppose that our first intensity to integrate is 101. We initialize our head node with $D=\lfloor\log_2(101)\rfloor = 6$ in \cref{fig:pixel_list}a. 
    
    Now, suppose we integrate the intensity 101, spanning 20 time units (ticks). The head only accumulates $64/101$ of the intensity units before saturating its $2^D=64$ integration. Thus, the time spanned for this partial integration is $(64/101)\cdot 20 = 12.67$ ticks. We can represent an \adder{} event (without its spatial coordinate), illustrated in angle brackets, as the node connection in \cref{fig:pixel_list}b. These events consist of integer numbers, so we use $\Delta t' = \lfloor\Delta t \rfloor = 12$. At this stage, we create a child node to represent the remaining integration. The child takes on the head node's $D$ (i.e., $D=6$).  We integrate the remaining $101-64=37$ intensity units for the child node, spanning $(37/101)\cdot 20 = 7.33$ ticks. Finally, we increment the head node's $D$ value to satisfy \cref{eq:iv3} and integrate the remaining $I=37, \Delta t = 7.33$ intensity. We see that \cref{eq:iv1} holds true and \cref{eq:iv2} holds true for $\varepsilon = -0.66$
    
    Let us now integrate 40 intensity units over 30 ticks, as in \cref{fig:pixel_list}c. The head node saturates its $2^D=128$ integration, so we \textit{replace} the child node with a new node, connected by the new event $\langle D'=7, \Delta t' = \lfloor 20 + (27/40)\cdot 30\rfloor = 40 \rangle$. In this way, we can minimize the number of \adder{} events required to represent an intensity sequence, maximizing the $D'$ and $\Delta t'$ values of the events. As before, we integrate the remaining intensity after spawning the child to both the head and the child nodes. \cref{eq:iv2} now holds for $\varepsilon = -0.25$.

    Finally, let us integrate 25 intensity units over 30 ticks, as in \cref{fig:pixel_list}d. We first integrate the head event. It does \textit{not} reach its $2^D=256$ integration threshold, so we do not replace its event nor replace the child node. Then we integrate the child node with the same intensity, reaching its $2^D=32$ threshold with  $19$ intensity units and an additional $(19/25)\cdot 30 = 22.8$ ticks. Thus, we increment $D$, create an event for the child, $\langle D'=5, \Delta t' = \lfloor 9.75 + (19/25)\cdot 30\rfloor = 32 \rangle$, and spawn another node in the list to integrate the remaining $6$ intensity units across $(6/25)*30=7.2$ ticks. We see that \cref{eq:iv1} holds since $I_1 = 2^{D'_1} + 2^{D'_2} + I_3 = 2^7 + 2^5 + 6.0$ and $I_2 = 2^{D'_2} + I_3 = 2^5 + 6.0$.

    While $D$ increases monotonically for a given node (\cref{eq:iv3}), it does not always increment by 1. If we were to integrate a large intensity, $D$ may increase by more than 1, but any increase in $D$ will induce a replacement of the node's children and edge events. By contrast, if we integrate a very small intensity, we may not create any new children or edge events. In this case, each existing node simply integrates the new intensity to update its $I$ and $\Delta t$ values.

    The node connections are a queue that a transcoder outputs when there is a significant change in the intensities being integrated. When such a criterion is met, we can dequeue the events from the list, and the tail node becomes the new head, as in \cref{fig:pixel_list}e. This linked list structure minimizes the number of \adder{} events required to represent a sequence of intensity integrations.
    
    \subsubsection{User-Tunable Parameters}\label{sec:user_parameters}
    A user may set a number of transcode parameters according to their needs. 
    
    $\Delta t_{s}$: Number of ticks per second. This parameter defines the temporal resolution of the \adder{} stream. 
    
    $\Delta t_{ref}$: Number of ticks for a standard length integration (e.g., an input frame exposure time, when the source is a framed video). This parameter, along with $\Delta t_{s}$, determines the accuracy and data rate.
    
    $M$: \adder{} contrast threshold. When a pixel's incoming normalized intensity exceeds its baseline intensity by this threshold (either positively or negatively), then the pixel's event queue will be output, its integration state reset, and its baseline set to the new intensity. Unlike the DVS model, which detects changes in log intensity, we look at the change in absolute intensity. In this chapter, $M$ is uniform for every pixel. \cref{ch:adder_compression} explores variable $M$ values. Below, I evaluate the effect of $M$ on transcoded quality and event rate.
    
    \dtm: The maximum $\Delta t$ that any event can span. Suppose that there is a static scene; this parameter is directly correlated with the event rate. That is, halving \dtm{} would result in a doubling of the event rate. In practice, however, the user will simply want to ensure that \dtm{} is sufficiently large enough so that the desired amount of intensity averaging will occur in temporally stable regions of the video, but small enough so that the given application will receive pixel updates at a fast enough rate.
    
    \subsubsection{Raw Binary Representation}\label{sec:binary_representation}
    A raw \adder{} file begins with a header containing metadata that describe the resolution of the video, the file specification version, the endianness, $\Delta t_{s}$, $\Delta t_{ref}$, and $\Delta t_{max}$. Immediately after this header, there is a sequence of raw \adder{} events in the following format.
    
    \textit{x}: unsigned 16-bit \textit{x} address.
    
    \textit{y}: unsigned 16-bit \textit{y} address.
    
    \textit{c}: optional unsigned 8-bit color channel address in the range $[0,2]$. If the video is monochrome, then the $c$ value is absent.
    
    $D$: unsigned 8-bit decimation value.
    
    $\Delta t$: time spanned since the last event of this pixel.
    
    I use a packed representation, so that each event is 9-10 bytes, depending on the presence of the color channel. However, I emphasize that this style of event representation is highly compressible, as demonstrated in \cite{FreemanLossyEvent,Freeman2021mmsys}. I note that the range of natural values for $D$ is $[0,127]$, since $2^{127}$ is the maximum representable unsigned integer in my language of choice, although $D$ values will typically span the range $[0,30]$. I additionally reserve $D=254$ as a special symbol that indicates that the event represents a 0-intensity, which cannot be communicated with $D\in[0,127]$ alone.
    
    \subsubsection{Implementation Details}\label{sec:implementation_details}
    
    I implemented the entire framework in the Rust language. Non-Rust dependencies include OpenCV \cite{opencv_library} for image ingest, processing, and visualization, and FFmpeg \cite{ffmpeg} for framed video coding. The framework is fast, highly parallel, and optimized to take advantage of CPU resources. For all speed evaluations in this chapter, I used an 8-core AMD Ryzen 2700x CPU, with output files written to a RAM disk.

\begin{figure}
    \centering
    \includesvg[width=0.81\linewidth]{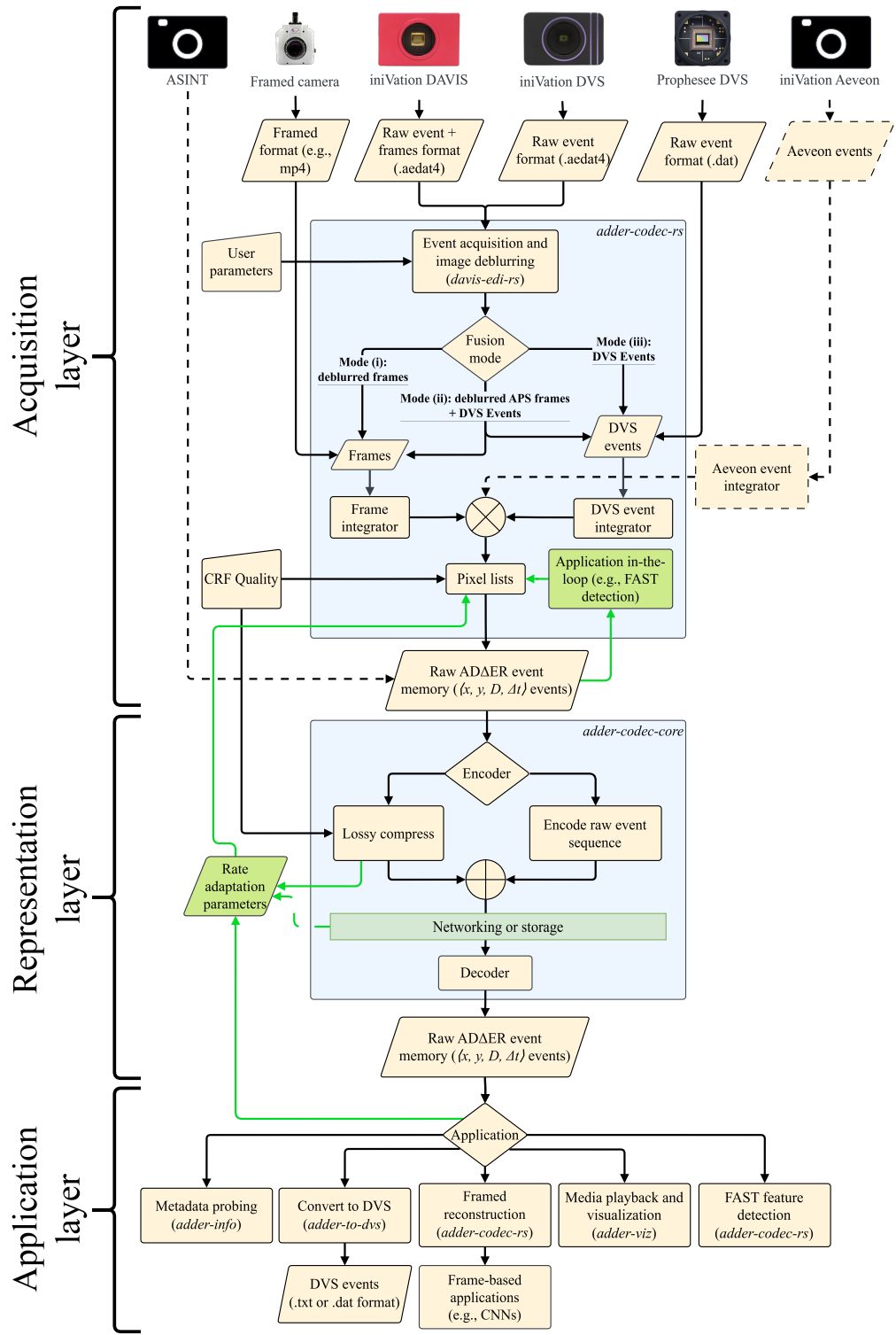}
    \caption[Detailed diagram of the \adder{} framework]{Detailed diagram of the three-layer \adder{} framework, repeated here from \cref{fig:system_diagram_intro_full} for reader convenience. Italicized names reflect the names of software packages in the Rust Package Registry. Dashed lines indicate future work. With \adder{}, framed and event-based video sources can be transcoded to a common representation. Since there is a single raw representation, we can have a simple source-modeled compression scheme. The representation supports bespoke event-based applications, while being backwards compatible with classical applications.}
    \label{fig:system_diagram_full_repeat}
\end{figure}
    
\subsection{Transcoding from Framed Video}\label{sec:framed_video_sources}
I begin by describing how I apply the \adder{} pixel model to framed video sources.

\subsubsection{Transcoder Details}\label{sec:framed_transcoder_details}
I conceptualize a video frame as a matrix of intensities integrated over a fixed time period, assuming that the intensity for each pixel may change only at instantaneous moments between frames (\cref{fig:framed_diagram}). $\Delta t_{ref}$ defines the integration time of each frame, and I set $\Delta t_{s} = \Delta t_{ref}F$ ticks for the source video frame rate, $F$. Since each \adder{} pixel integrates exactly one intensity per input frame, I can easily parallelize this process by integrating groups of pixels on many CPU threads, then collecting the resulting \adder{} events into a single vector to write out after integrating each frame. Since each pixel is processed independently in the \adder{} model, temporally interleaving the events of different pixels is unnecessary. For example, Pixel A may fire events $A_1$, then $A_2$, while Pixel B may have an event $B_1$ that falls between the times of $A_1$ and $A_2$. In this case, it is valid to encounter $A_1$ and $A_2$ before $B_1$ in the \adder{} stream, but I will never encounter $A_2$ before $A_1$.

\subsubsection{Framed Reconstruction and Preserving Temporal Coherence}\label{sec:framed_reconstruction}
To easily view and evaluate the effects of my \adder{} transcode with traditional methods, I can perform a framed reconstruction of the data. For this, I simply maintain a counter of the running timestamp, $T$ for each pixel in the \adder{} stream. When reading an event $\langle x,y,D', \Delta t'\rangle$ for a given pixel, I normalize the intensity per frame to $I_{frame} =\frac{2^{D'} \Delta t_{ref}}{\Delta t'}$, then set this as the pixel value for all frames in $[T/\Delta t_{ref}, (T+\Delta t')/\Delta t_{ref}]$.

I must, however, ensure that intensity changes only occur along frame temporal boundaries. Otherwise, there would be temporal decoherence as certain pixels fire slightly earlier than others, unless the source video is recorded at an extremely high frame rate. To this end, I ignore any intensity remaining for a given pixel list state after it generates a new \adder{} event, thus ensuring that the temporal start of each event corresponds to the beginning of a frame in the source. I can simply infer the time between events when processing the events, by rounding $T$ up to the next multiple of $\Delta t_{ref}$. My framed reconstruction program handles this case automatically, since both the source data type and $\Delta t_{ref}$ are encoded in the header of the \adder{} file.

\begin{figure}
        \centering
    \includegraphics[width=\linewidth]{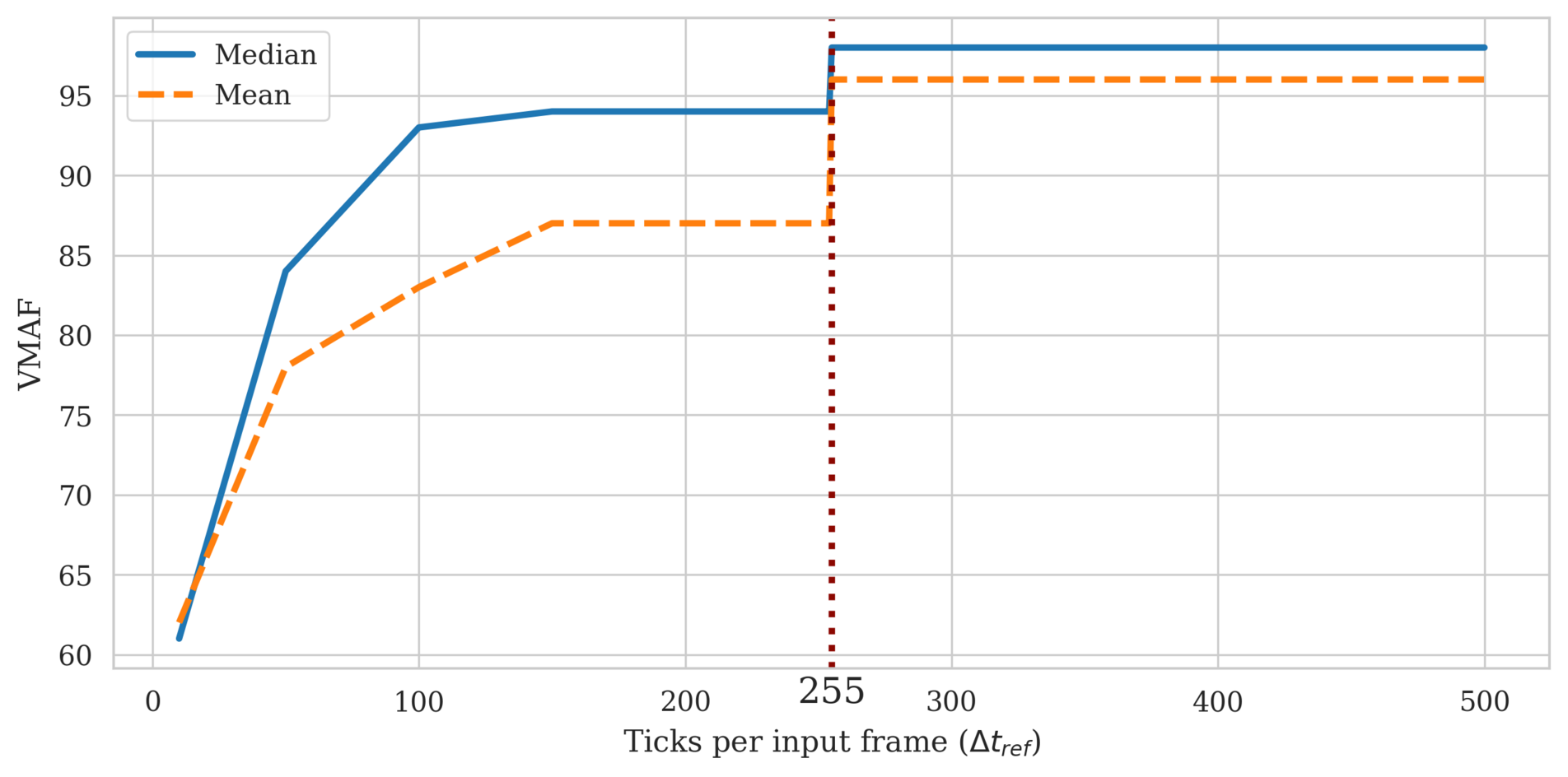}
        \caption[$\Delta t_{ref}$ vs. VMAF quality]{Effect of $\Delta t_{ref}$ on VMAF perceptual quality of \adder{} framed reconstructions. Source videos are 8 bits per color channel.}
        \label{fig:tpf_vmaf}
    \end{figure}

\begin{figure}
                \centering
                \begin{subfigure}[b]{0.38\textwidth}
                    \centering
                    \includegraphics[trim={0 0 50.5cm 0},clip,width=\linewidth]{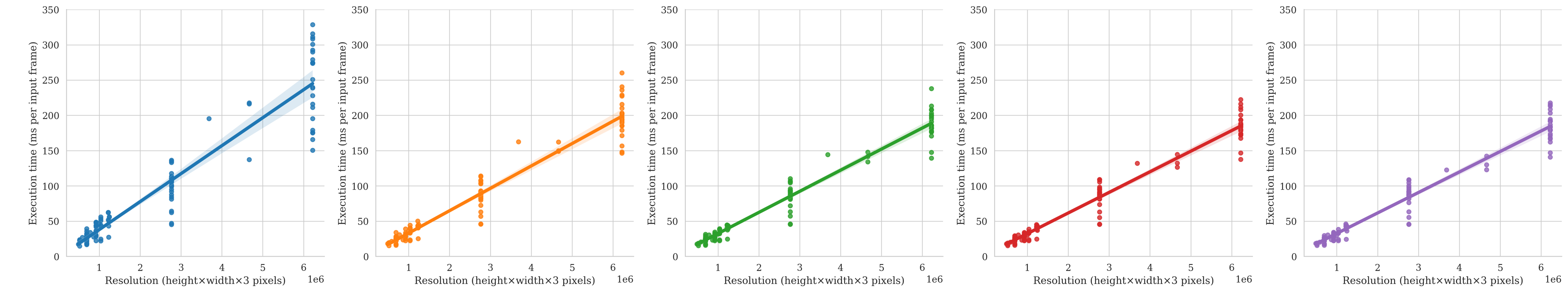}
                    \caption{$M = 0$ }
                \end{subfigure}

                \begin{subfigure}[b]{0.38\textwidth}
                    \centering
                    \includegraphics[trim={13.8cm 0 38.0cm 0},clip,width=\linewidth]{framed_eval/execution_times_white.png}
                    \caption{$M = 10$ }
                \end{subfigure}
                \hfill
                \begin{subfigure}[b]{0.38\textwidth}
                    \centering
                    \includegraphics[trim={26.4cm 0 25cm 0},clip,width=\linewidth]{framed_eval/execution_times_white.png}
                    \caption{$M = 20$ }
                \end{subfigure}
                \hfill
                \begin{subfigure}[b]{0.38\textwidth}
                    \centering
                    \includegraphics[trim={39cm 0 12.6cm 0},clip,width=\linewidth]{framed_eval/execution_times_white.png}
                    \caption{$M = 30$ }
                \end{subfigure}
                \hfill
                \begin{subfigure}[b]{0.38\textwidth}
                    \centering
                    \includegraphics[trim={51.4cm 0 0cm 0},clip,width=\linewidth]{framed_eval/execution_times_white.png}
                    \caption{$M = 40$ }
                \end{subfigure}
                \label{fig:framed_execution_times}
                \caption[Resolution vs. execution time]{Effect of framed input resolution on execution time, at various $M$ values. Performance scales linearly with resolution.}
            \end{figure}

\subsubsection{Optimizing \texorpdfstring{$\Delta t_{ref}$}{DeltaT-ref}}
Since I define $\Delta t_{s}$ relatively to $\Delta t_{ref}$, it is crucial to understand the effect of $\Delta t_{ref}$ on the transcoded representation's quality. I transcoded a diverse set of 10 framed videos to \adder{} with $M=0$ (for the most accurate representation) at various choices of $\Delta t_{ref}$. I then performed framed reconstruction of the \adder{} streams to use the VMAF \cite{vmaf} metric to calculate the perceptual quality of the reconstructions compared to the source videos. These results are illustrated in \cref{fig:tpf_vmaf}. I found that at least 255 ticks per input frame is a necessary parameter for strong reconstruction quality, due to the \textit{bit depth} of the source videos (namely, 8 bits per channel). The maximum intensity of the source frame is $\frac{255}{255} =1$ intensity units per tick in $\Delta t_{ref} = 255$, which I can represent with $\langle D=7, \Delta t =128\rangle$. The minimum intensity is 0, which I may represent with $\langle D=254, \Delta t =255\rangle$. However, if $\Delta t_{ref} = 254$, I see that the maximum intensity $\frac{255}{254} =1.004$ is not representable with an integer $\Delta t$ value. Thus, I must set $\Delta t_{ref}$ large enough that I may represent any source intensity without losing accuracy, such that $\frac{I_{max}}{\Delta t_{ref}} \leq 1.0$.

\subsubsection{Evaluation}\label{sec:framed_evaluation}
\begin{figure}
        \centering
        \includegraphics[width=\linewidth]{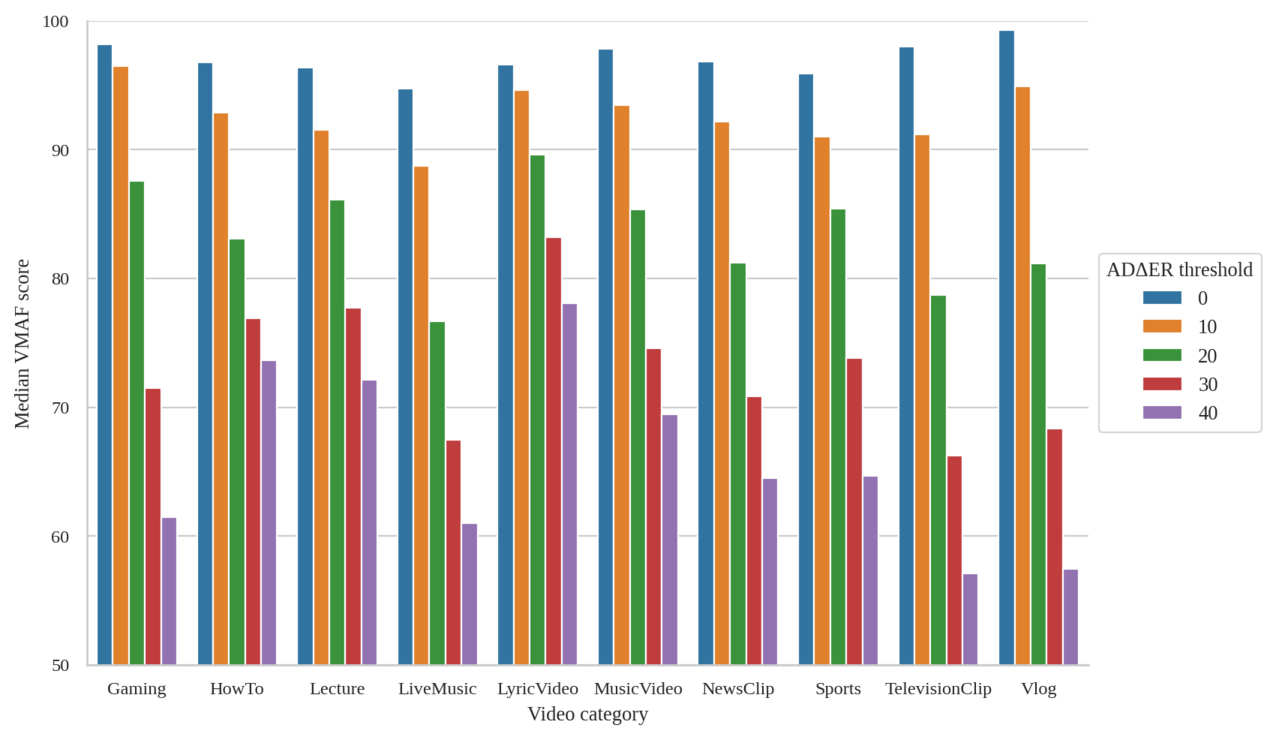}
        \caption[Categorical VMAF scores]{Median VMAF score for each framed video category after transcoding to \adder{} at various $M$ values. Quality decreases as $M$ increases for every category. Less dynamic video categories such as \texttt{LyricVideo} have little quality degradation as $M$ increases, while more dynamic categories such as \texttt{TelevisionClip} and \texttt{Vlog} degrade greatly.}
        \label{fig:median_vmaf}
    \end{figure}

To evaluate my \adder{} representation on framed video sources, I employed 112 videos from a subset of the YT-UGC video compression data set \cite{yt-ugc-dataset}. These 20-second videos span 10 categories (listed in \cref{fig:median_vmaf}) and have resolutions 360p, 480p, 720p, and 1080p. I used the variants of the videos provided with H.264 compression at the CRF-10 quality level, and each video has 24-bit color.

\begin{figure}
                \begin{subfigure}[b]{0.48\textwidth}
                    \centering
                    \includegraphics[width=\linewidth]{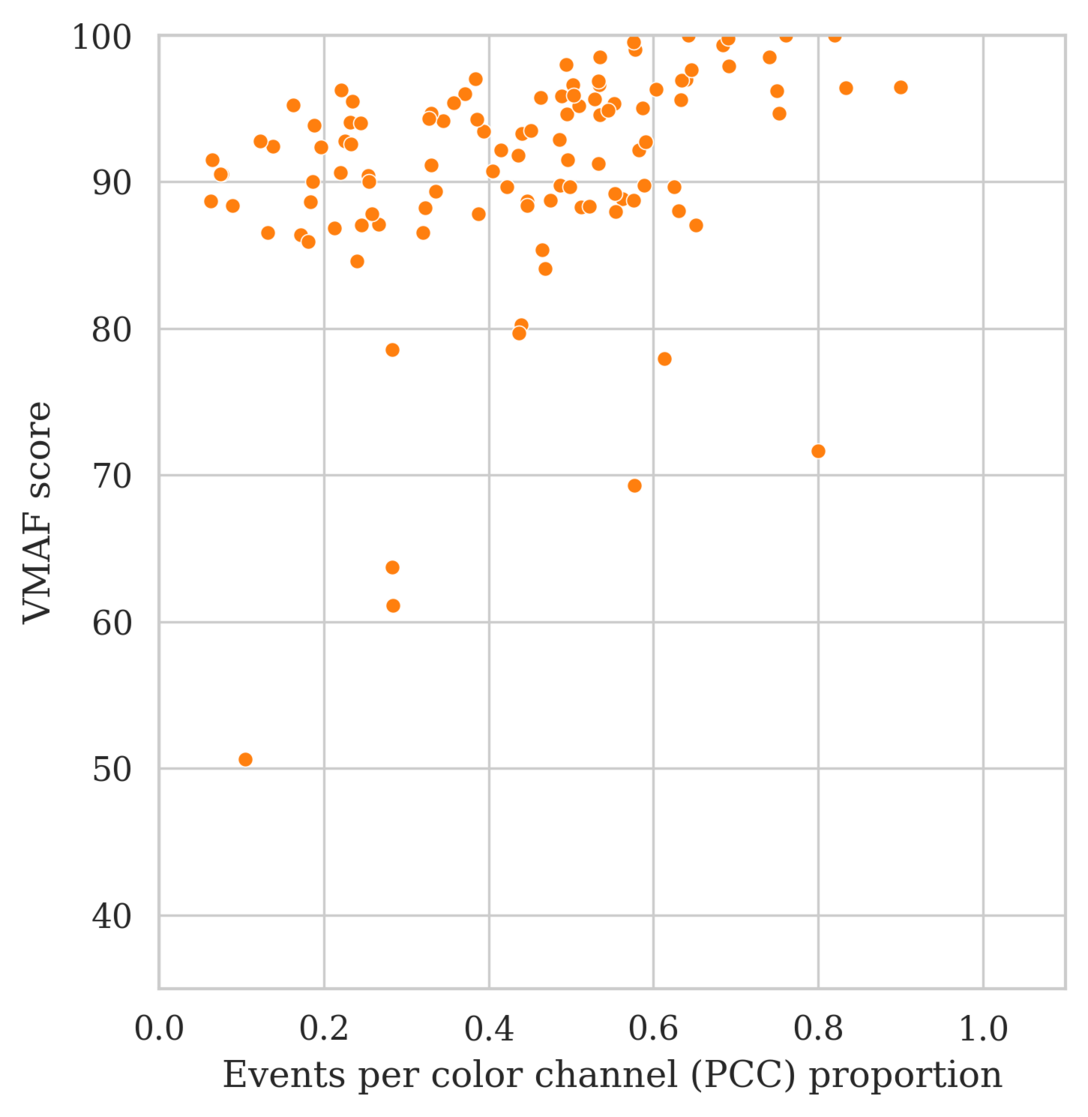}
                    \caption{$M = 10$ }
                \end{subfigure}
                \hfill
                \begin{subfigure}[b]{0.48\textwidth}
                    \centering
                    \includegraphics[width=\linewidth]{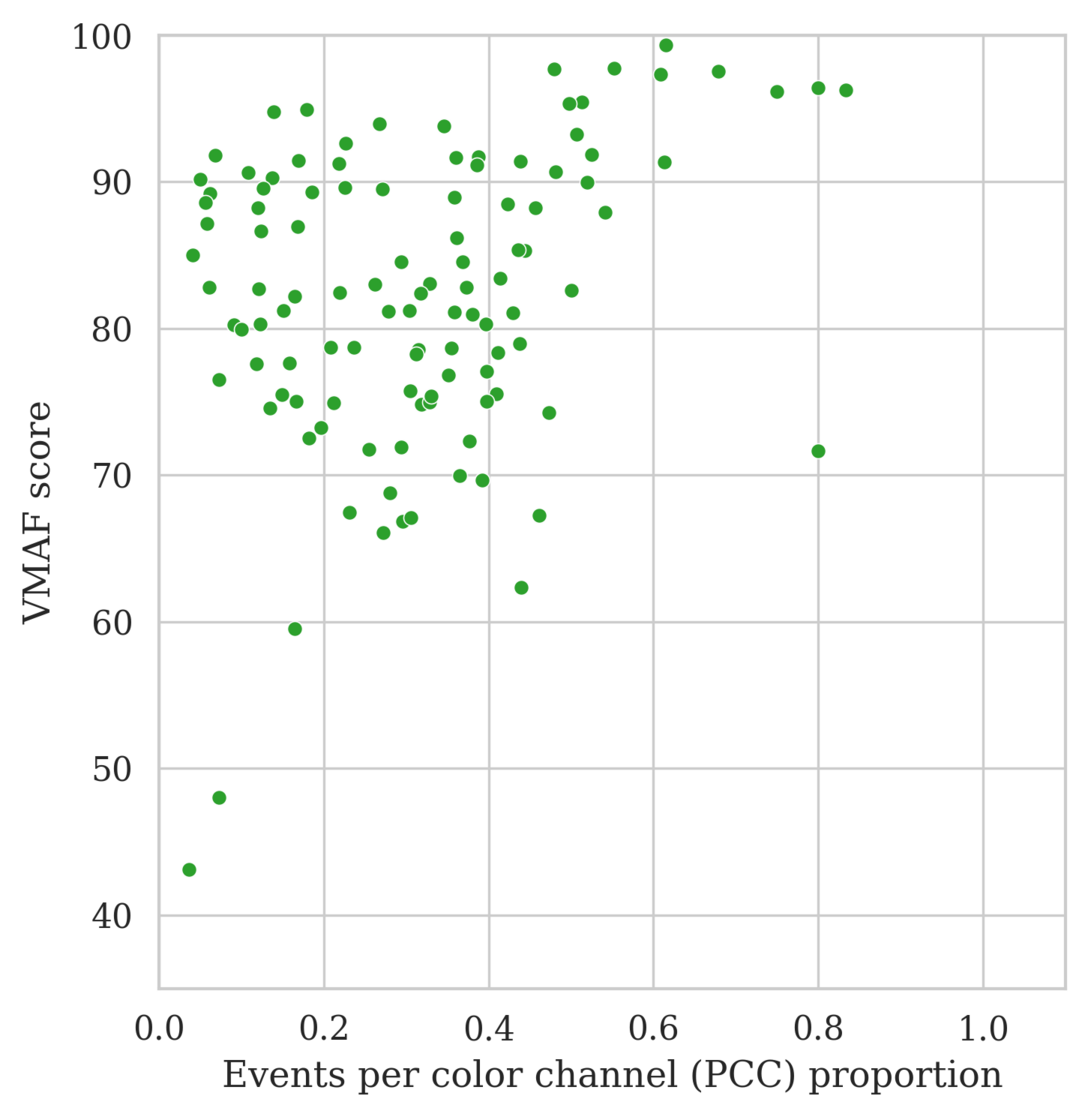}
                    \caption{$M = 20$ }
                \end{subfigure}
                \hfill
                \begin{subfigure}[b]{0.48\textwidth}
                    \centering
                    \includegraphics[width=\linewidth]{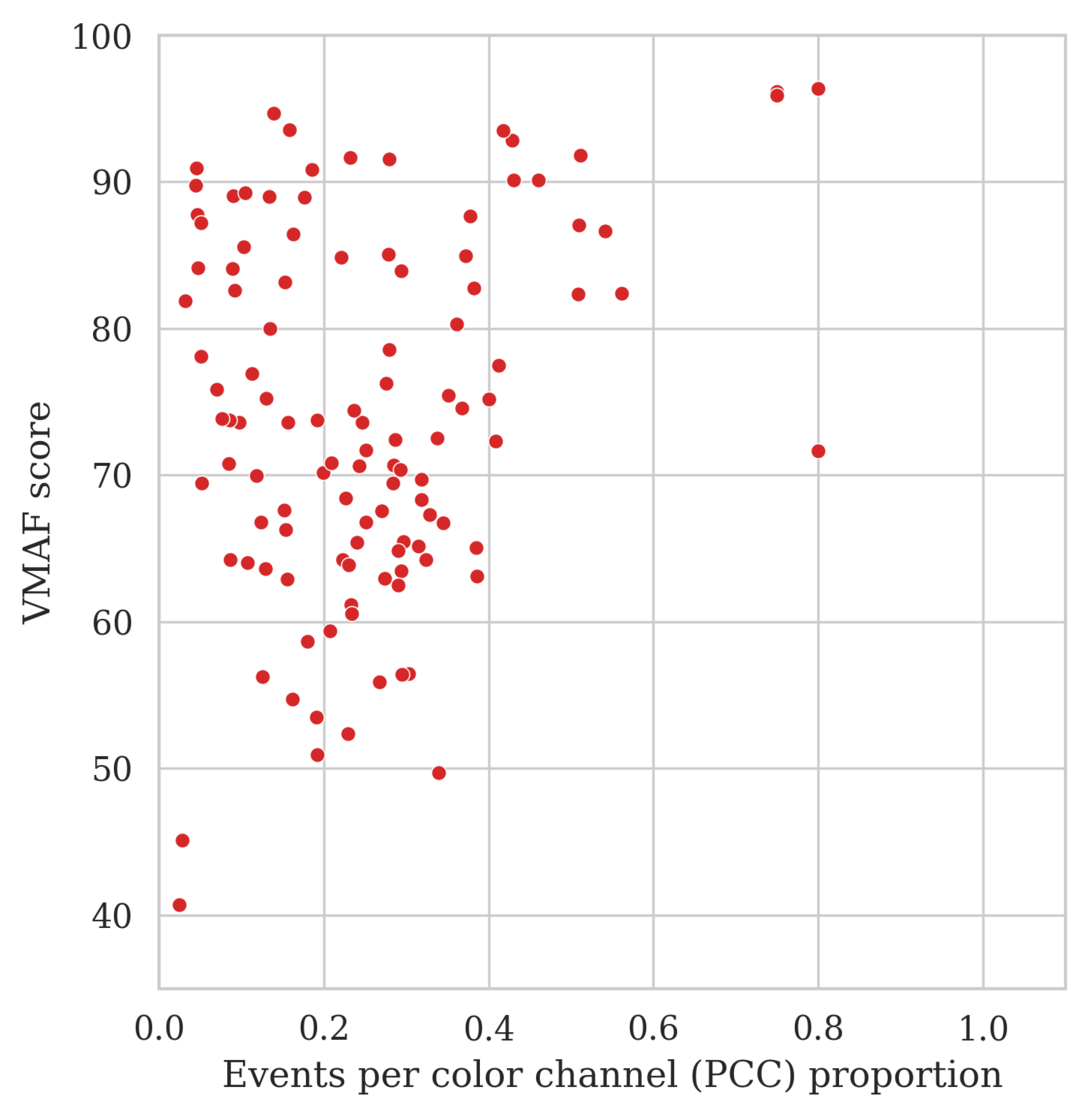}
                    \caption{$M = 30$ }
                \end{subfigure}
                \hfill
                \begin{subfigure}[b]{0.48\textwidth}
                    \centering
                    \includegraphics[width=\linewidth]{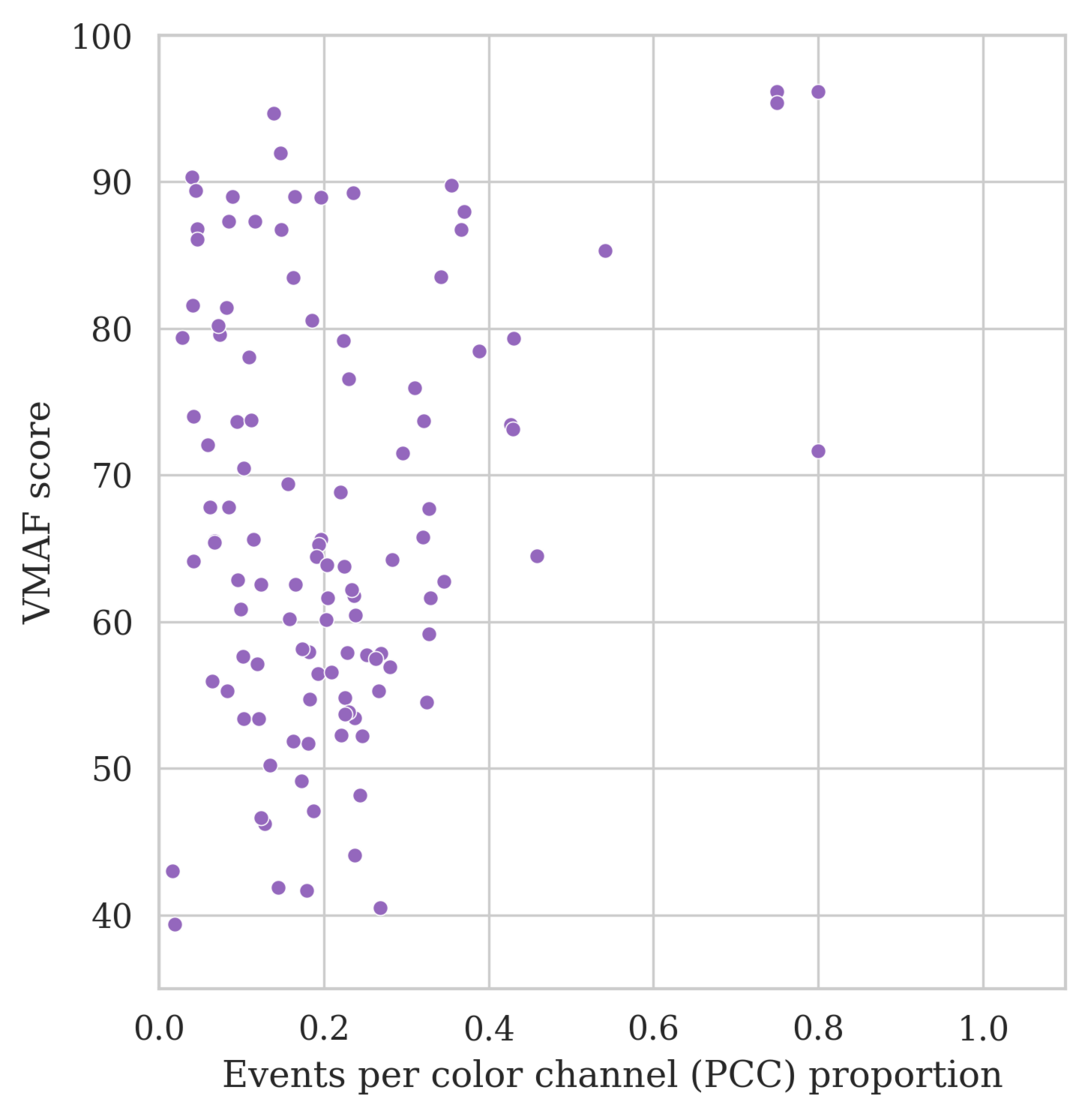}
                    \caption{$M = 40$ }
                \end{subfigure}
                \label{fig:framed_event_rate_vmaf}
                \caption[Framed source results at various $M$ values]{Scatter plots of the rate-distortion for all videos at various $M values$.}
            \end{figure}

\begin{figure}
    \centering
    \includegraphics[width=\linewidth]{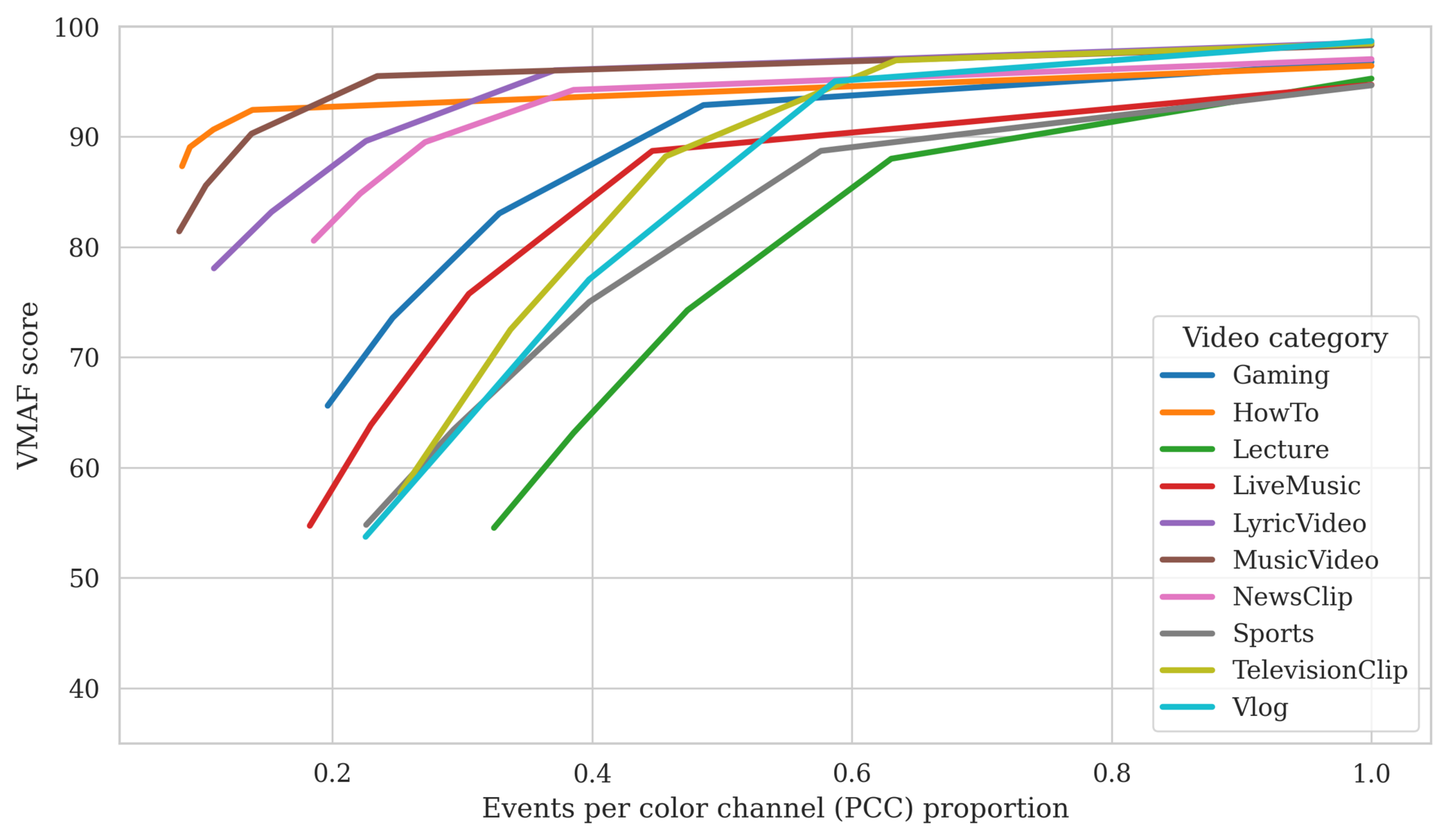}
    \caption[Rate-distortion curve for each dataset category]{Particular rate-distortion curves for one video in each category of the dataset. Increasing $M$ yields lower-quality, but lower-rate \adder{} videos.}
    \label{fig:framed_event_rate_vmaf_particular}
\end{figure}

I transcoded each video to \adder{} with parameters $\Delta t_{ref} = 255$ and $\Delta t_{max} = \Delta t_{ref} \cdot 120 = 30600$, such that the longest $\Delta t$ representable by any generated \adder{} event can span 120 input frames. As described in \cref{sec:framed_transcoder_details}, I set $\Delta t_{s} = \Delta t_{ref}F$ for each video. For example, a 24 FPS video has $\Delta t_{s} = 6120$. For each video, I varied the \adder{} threshold parameter $M \in \{0, 10, 20, 30, 40\}$. I recorded the execution time on my test machine (\cref{sec:implementation_details}) for each transcode, as illustrated in \cref{fig:framed_execution_times}. I see that larger values of $M$ yield slightly faster transcode operations, and that the time required to process an input frame scales linearly with video resolution.
    
Finally, I reconstructed framed videos from my \adder{} representations as described in \cref{sec:framed_reconstruction}, so that I can examine the quality properties of my representation. I gathered the VMAF score of each \adder{} video in the $M$ range, compared to its reference video. \cref{fig:median_vmaf} shows that the median VMAF score for each video category is extremely high, in the 95-99 range, and that perceptual quality decreases as $M$ increases.

If we examine the number of \adder{} events per pixel color channel at various levels of $M$ as a proportion compared to $M=0$  (\cref{fig:framed_event_rate_vmaf,fig:framed_event_rate_vmaf_particular}), we see a clear rate-distortion curve dependent on the choice of $M$. By merely setting $M=10$, I see a median reduction in \adder{} rate by $54\%$, and a median reduction of VMAF score by only 4.5 across my data set. In my tests, I observe a maximum \adder{} precision of 44 dB or 14 bits, demonstrating a significant increase over the 8-bit precision of the source videos. The precision tends to increase as pixels can integrate (and thus average) longer sequences of intensities with higher $M$, thus widening the range of represented values.

\begin{figure}
    \centering
    {\includegraphics[width=\linewidth]{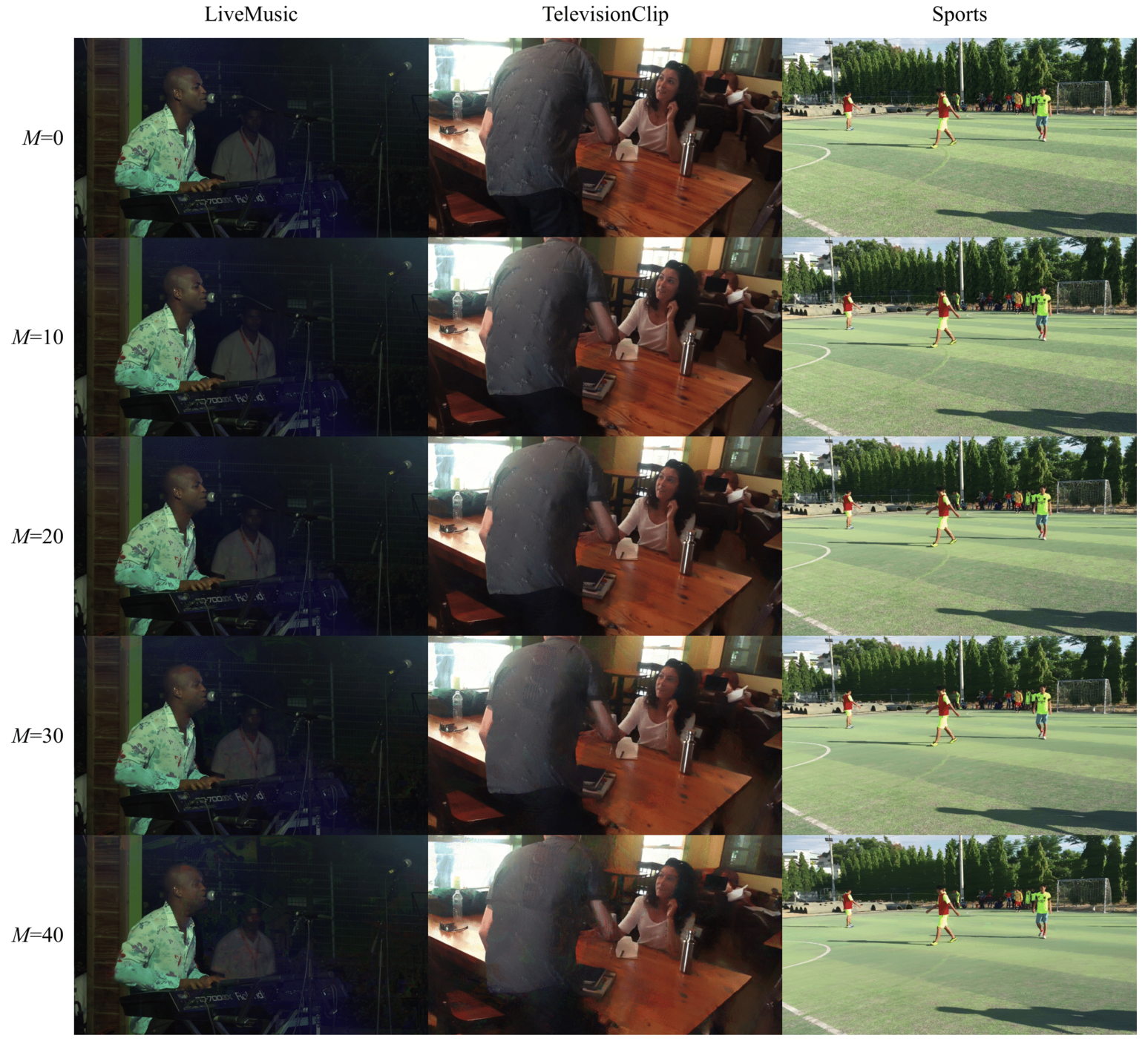}%
    \label{fig:framed_matrix}}
    \caption[Qualitative results]{Qualitative results: Instantaneous samples of framed videos transcoded to \adder{} at various $M$ values. High $M$ yields more smearing and ghosting artifacts in low-contrast regions.}
\end{figure}

As a qualitative example, \cref{fig:framed_matrix} shows instantaneous frame samples from the \adder{} transcodes with my range of $M$ values
. With higher $M$, I observe a greater ``smearing'' effect in low-contrast regions, less overall detail, and multicolored ghosting artifacts from scene transitions or fast camera motions. Surprisingly, the \texttt{LiveMusic} and \texttt{Sports} examples shown are both outliers for the VMAF score with $M=0$
, despite appearing high quality to a casual observer. This is due to the imprecision of some events with $\Delta t > \Delta t_{ref}$, near high-contrast transition points, since I round $\Delta t$ to an integer to form an \adder{} event. Furthermore, since I represent each color channel with an independent event pixel list, a small error in one color channel of a pixel will have a compound effect on the VMAF perceptual quality of that pixel. These two videos exhibit jittery camera motion, suggesting that rapid intensity changes may make such quantization errors more frequent.

\subsection{Transcoding from DVS/DAVIS Video}\label{sec:event_video_adder}
While in \cref{sec:framed_video_sources} I demonstrated that \adder{} can effectively represent framed video sources asynchronously, I here discuss \adder{}'s utility in representing video sources which are already asynchronous. Specifically, I explore the representation of the DAVIS 346 camera's DVS event and APS frame data in the \texttt{.aedat4} file format.

\subsubsection{Event-Based Double Integral}\label{sec:edi}
As I discussed in \cref{sec:existing_event_intermediate_representations}, many applications that utilize mixed-modality sources (frames and events) either reconstruct an image sequence at a fixed frame rate or process the frame data separately from the events. Furthermore, while APS frames provide absolute intensity measurements across the whole sensor, they can often be blurry in difficult exposure scenarios: low scene illumination, fast motion, or narrow aperture. Meanwhile, the higher sensitivity of DVS pixels and their fast responsiveness allow them to capture events with subframe precision. Thus, DAVIS-based reconstruction methods must address the crucial problem of \textit{deblurring} the APS frame data. Most of these efforts, however, utilize slow machine learning techniques which severely limit the practical reconstructed frame rate, obviating a primary benefit to using event cameras \cite{wang2020event,song_ecir}. One non-learned method for framed reconstruction, however, is the Event-based Double Integral (EDI) \cite{Pan_EDI}. The authors introduced the EDI model to find a sharp ``latent'' image, $L$, for a blurry APS image, by integrating the DVS events occurring during the APS image's exposure time. Crucially, this paper involves an optimization technique for determining the $\theta$ value of the DVS sensor at a given point in time; that is, the method deblurs an APS image with various choices of DVS sensitivity, $\theta$, and the sharpest latent image produced corresponds to the optimal choice of $\theta$. Both the APS deblurring and $\theta$ optimization techniques are vital steps in my \adder{} transcoder pipeline for event video sources.

\begin{figure}
        \centering
        \includegraphics[width=\linewidth]{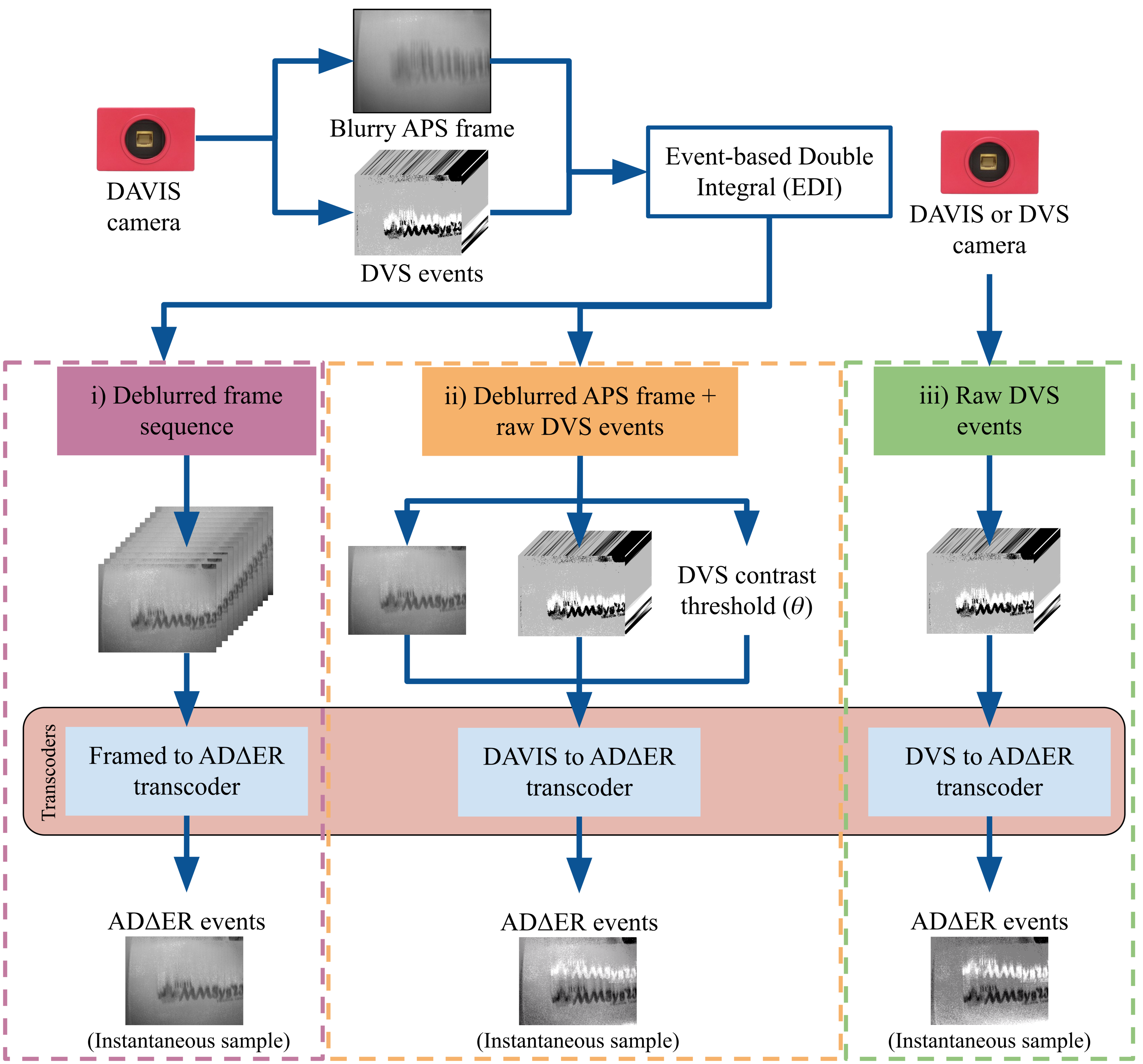}
        \caption[Pipeline for transcoding event camera data to \adder{}]{Pipeline for transcoding event camera data to \adder{}. My transcoder supports three modes. Modes (i) and (ii) incorporate deblurred APS frames from a DAVIS sensor, while mode (iii) uses DVS events alone.}
        \label{fig:davis_pipeline}
    \end{figure}

\subsubsection{My Implementation}

Unfortunately, as with much of the literature on event-based systems, the EDI model was not implemented with practical, real-time performance in mind. Rather, the authors released their implementation as an obfuscated MATLAB program, which can only reconstruct a few frames of video per second. Furthermore, their implementation uses a custom MATLAB format for the DAVIS data, underscoring my argument about the prevalence of domain-specific event data representations in \cref{ch:video_representations}. Therefore, I implemented the EDI model from the ground up in Rust, designing my program to decode data in the packet format produced directly by the DAVIS 346 camera. My implementation can reconstruct a frame sequence for this resolution at 1000+ FPS, two orders of magnitude faster than the baseline MATLAB implementation. To allow for realistic and repeatable performance evaluation, I simulate the packet latency of pre-recorded videos to reflect their real-world timing. The user can also generate one deblurred image for each APS frame, rather than a high-rate frame sequence. My implementation supports a connected DAVIS 346 camera as input, unlocking the potential for practical, real-time DAVIS applications.

\subsubsection{Transcoder Details}\label{sec:davis_transcoder_details}
I identify three ways in which I can transcode DAVIS data to \adder{}. Each mode utilizes the same per-pixel integration scheme as described in \cref{sec:transcoder_fundamentals}, but the actual intensities that I \textit{integrate} depend on the mode I choose. In each case, I receive input data from my EDI implementation running alongside my transcoder. Unlike my framed source transcoder, where I determine $\Delta t_{s}$ from my choice of $\Delta t_{ref}$ and the video frame rate, here I set $\Delta t_{s} = 1\times 10^6$ to match the temporal resolution of the DAVIS 346. Then, $\Delta t_{ref}$ defines the desired length of a deblurred frame in modes (i) and (ii) below, and the intensity scale in mode (iii) below. I illustrate my pipeline in \cref{fig:davis_pipeline}. This pipeline is included as a portion of the acquisition layer in \cref{fig:system_diagram_full_repeat}.

\paragraph{Mode (i): Deblurred frame sequence \texorpdfstring{$\rightarrow$ \adder}{to ADDER}}\label{sec:mode_1}
If I reconstruct a high-rate framed sequence with EDI, I can trivially use the framed-source transcoder technique as described in \cref{sec:framed_transcoder_details}. Qualitatively, this method produces the best-looking results, since the EDI model inherently unifies positive- and negative-polarity events in log space. That is, we can accumulate a pixel's multiple events occurring over a given frame interval to arrive at a final intensity value for the given frame. Any intensities outside the APS frame's intensity range $[0,255]$ are clamped only at this point.

\paragraph{Mode (ii): Deblurred APS frame + DVS events \texorpdfstring{$\rightarrow$ \adder}{to ADDER}}\label{sec:mode_2}

We may also choose to use EDI to simply deblur each APS frame, and input the DVS events occurring outside that frame's exposure time individually in my \adder{} transcoder. Since DVS events express log intensity relative to a previous latent value, we maintain a matrix of the most recent log intensity for each pixel, by which we calculate the intensity represented by the incoming DVS event. When we ingest my deblurred frame, we scale the frame intensities $I$ to the range $L \in [0,1]$, and set the latent log intensity for each pixel by $\widetilde{L} = \ln{(1 + L)}$. I interpret a DVS event as specifying the exact moment an intensity \textit{changes}. Suppose that, for a given pixel, we ingest a deblurred frame intensity of $I_0 = 20$ that spans time $[t'-\Delta t_{ref}, t']$, where the frame has 8-bit values and $\Delta t_{ref} = 1000$. Then, we have latent value $\widetilde{L_0} = \ln{(1+\frac{20}{255}} = 0.0755$. Suppose we have a sequence of DVS events $\langle p, t\rangle$ occurring after time $t'$, $\{E_1 = \langle 0, t' + 500\rangle, E_2 = \langle 1, t'+ 800\rangle, E_3 = \langle 1, t' + 1200\rangle\}$. When we encounter $E_1$, we first \textit{repeat} the previous integration to fill the time elapsed. That is, we integrate $\frac{L_0 \cdot 500}{\Delta t_{ref}} = 10$ intensity units over 500 ticks. 

Then, we simply set the new latent intensity as follows. Supposing $\theta = 0.15$, the latent value $L_1$ becomes $L_1 = \exp{(\widetilde{L_0} - 0.15)} - 1.0 = -0.0718$. Since we cannot integrate negative intensities, however, we must clamp $L_1$ and $\widetilde{L_1}$ to $0.0$. At this stage, we do not know how long the pixel maintains this intensity, so we do not immediately integrate an additional value. Rather, when we encounter $E_2$, we integrate $\frac{L_1 \cdot 255 \cdot 300}{\Delta t_{ref}} = 0$ intensity units over 300 ticks. As before, we set the new latent log value $\widetilde{L_2} = \widetilde{L_1} + 0.15 = 0.15$, such that $L_2 = \exp{(\widetilde{L_2})} - 1.0 = 0.1618$. When we encounter $E_3$, then, we integrate $\frac{L_2 \cdot 255 \cdot 400}{\Delta t_{ref}} = 41.27$ intensity units over 400 ticks.

Although the clamping mechanism causes a worse visual appearance than mode (i) (\cref{sec:mode_1}), as shown in the white and black pixels of moving edges in \cref{fig:davis_matrix}, this method preserves the temporal resolution of the source events. By contrast, intensity timings in mode (i) are quantized to the EDI reconstruction frame rate.

\paragraph{Mode (iii): DVS events \texorpdfstring{$\rightarrow$ \adder}{to ADDER}}\label{sec:mode_3}
Finally, I can choose to ignore the APS data, or use a DVS sensor alone, and integrate the DVS events directly. In this case, my EDI program functions solely as the camera driver or event file decoder. At sub-second intervals, I reset the latent log image $\widetilde{L}$ in the \adder{} transcoder to $\ln(0.5)$, representing a mid-level intensity. Then, I calculate the absolute intensity of the following DVS events relative to this mid-level intensity. As I do not deblur APS frames in this mode, I cannot find the optimal $\theta$ value by which to interpret these DVS events. Therefore, I use a constant $\theta =0.15$ to calculate the changes in log intensity. As shown in \cref{fig:davis_pipeline}, I only gather intensity information for pixels that change greatly enough to trigger DVS events, and unchanging pixels will carry a mid-level gray value.

\begin{figure}
        \centering
        \includegraphics[width=\linewidth]{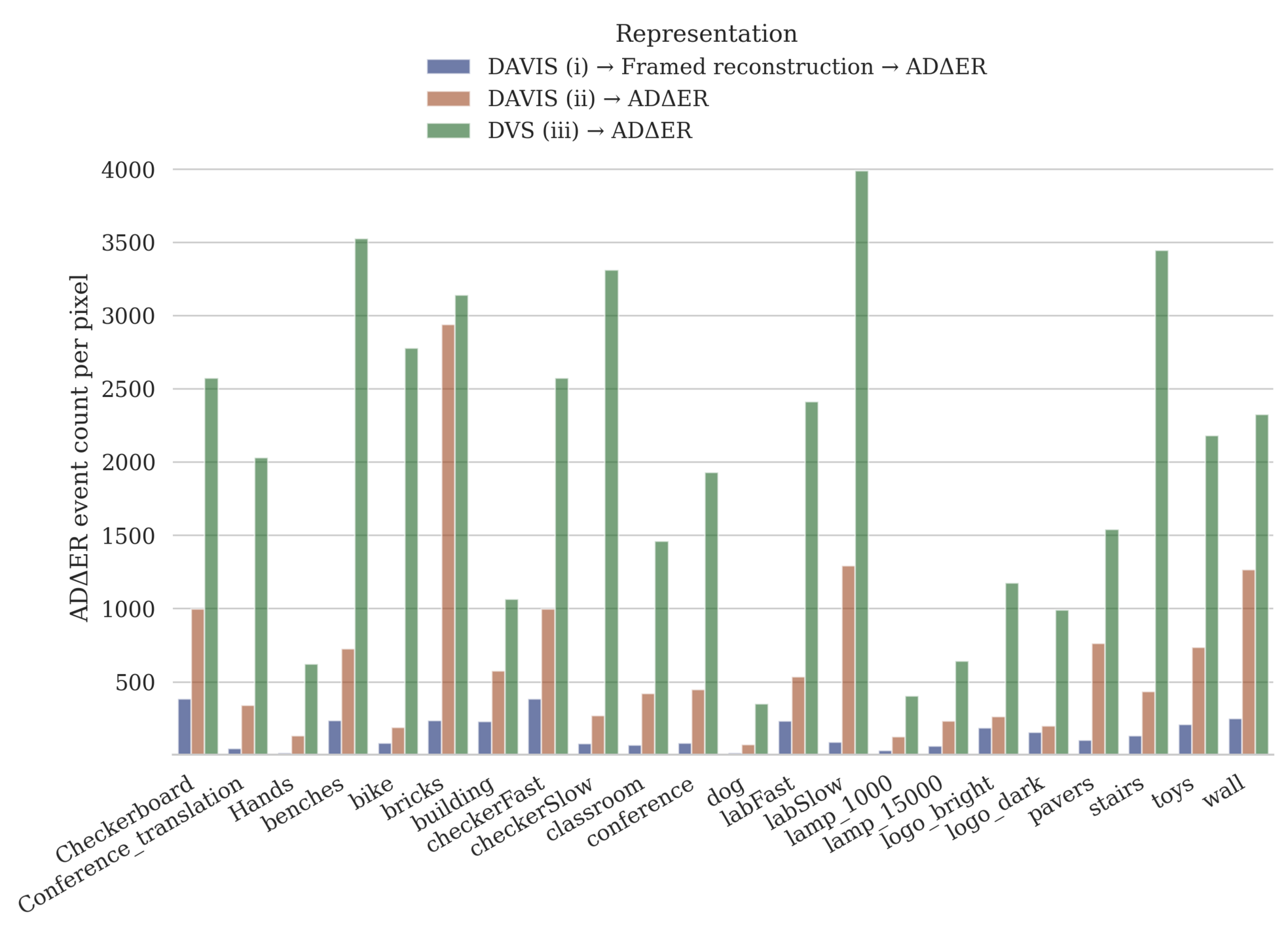}
        \caption[\adder{} event rates]{\adder{} event rates for each transcode mode and each video of my data set, where $\Delta t_{ref} = \Delta t_{s}/500$ and $M=40$. Mode (iii) yields dramatically higher \adder{} rates, despite not encompassing absolute intensities in slowly-changing or static regions of the scene.}
        \label{fig:event_source_adder_rates}
    \end{figure}

\subsubsection{Evaluation}
As my chief aim was to enable the development of fast, practical event-based systems, I needed to test my method on a real-world source representation. However, as discussed in \cref{ch:video_representations}, most event vision data sets use an intermediate representation that is slower to ingest. Thus, I employ the DVSNOISE20 data set \cite{dvsnoise_dataset}, which provides raw \texttt{.aedat4} files in the same format that the DAVIS 346 camera produces. This data set provides 3 recordings for each of 16 scenes, though I used only one recording per scene in my evaluation. I also add to the data set 6 of my own recordings with varied APS exposure lengths and more extreme lighting conditions.

I transcoded each DAVIS video to \adder{} using mode (ii) with \adder{} threshold parameter $M \in \{0, 10, 20, 30, 40\}$, $\Delta t_{ref} \in \{100, 1000\}$, $\Delta t_{s} = 1\times 10^6$ (microsecond resolution), and $\Delta t_{max} = \Delta t_{s} \cdot 4 = 4\times 10^6$. At $M=10$ for both choices of $\Delta t_{ref}$, I observed a median 49-50\% decrease in the \adder{} event rate compared to $M=0$. At $M=40$, I observed a median 66\% decrease in \adder{} event rate compared to $M=0$, demonstrating the utility of my temporal average scheme.

\begin{figure}
        \centering        \includegraphics[width=\linewidth]{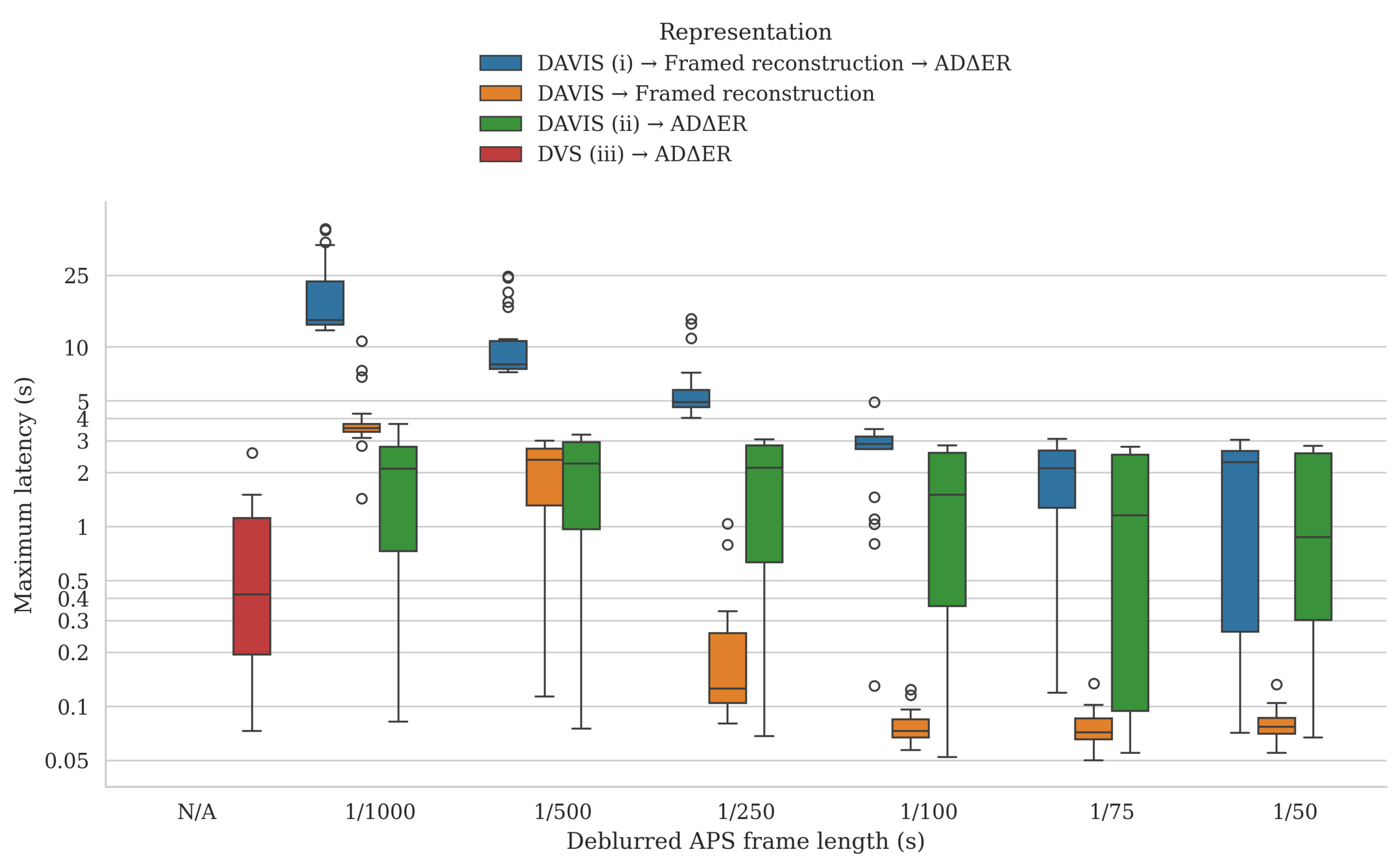}
        \caption[$\Delta t_{ref}$ vs. latency]{Effect of $\Delta t_{ref}$ on the latency of \adder{} transcoding. }
        \label{fig:davis_allmodes_latency}
    \end{figure}

Secondly, I transcoded each DAVIS video to \adder{} under modes (i) and (ii) with a fixed $M=40$, and $\Delta t_{ref} \in \{\Delta t_{s}/1000,$ $\Delta t_{s}/500,$ $ \Delta t_{s}/250,$ $\Delta t_{s}/100,$ $\Delta t_{s}/75,$ $\Delta t_{s}/50\}$ ticks. Additionally, I ran direct framed reconstructions of DAVIS in EDI and I transcoded each video once with mode (iii). I recorded the maximum latency between reading a packet from the \texttt{.aedat4} source video in my EDI program and completing the transcode of that packet's data. In \cref{fig:davis_allmodes_latency}, I observe that both framed reconstruction and transcode mode (i) incur dramatically more latency as the effective frame rate increases, which corroborates my findings in \cref{sec:framed_evaluation}. Furthermore, I observe worse latency for sequences with high DVS event rates. By contrast, mode (ii) incurs relatively constant $2.1$ second latency across all values of $\Delta t_{ref}$, yet has less variance as $\Delta t_{ref}$ increases. The speed of mode (iii) does not depend on $\Delta t_{ref}$, and I see that I can achieve a median transcode latency of only 0.4 seconds. These results support my argument that it is computationally expensive to employ a high-rate, framed intensity representation for event data. While such methods unavoidably quantize the temporal components of the event sources, modes (ii) and (iii) of my scheme can transcode the DAVIS/DVS event data to an intensity representation, \adder{}, with a temporal resolution matching that of the source.

\begin{figure}
        \centering        \includegraphics[width=0.9\linewidth]{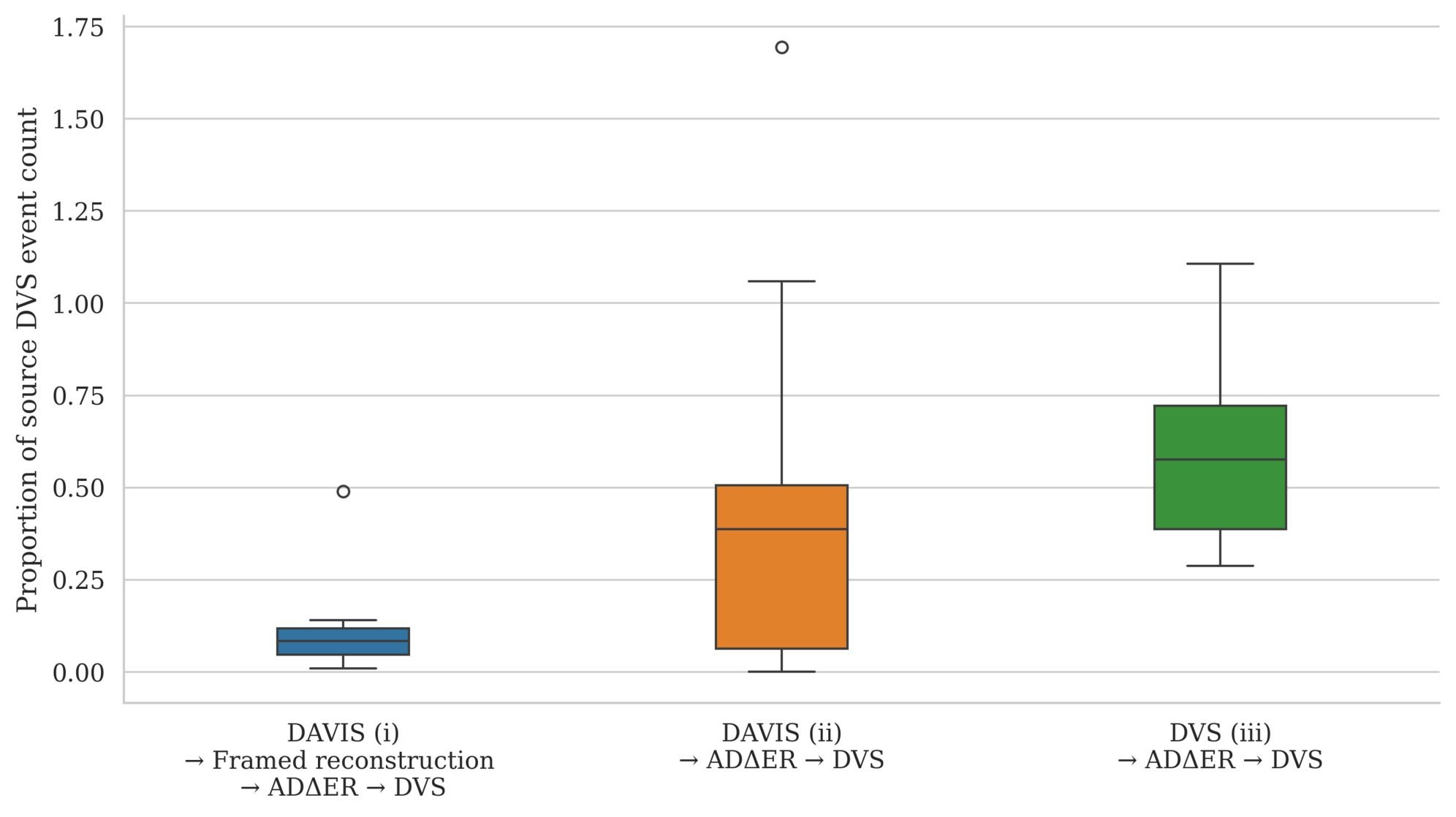}
        \caption[Proportion of recovered DVS events]{Proportion of recovered DVS events after \adder{} transcoding, as compared to the source \texttt{.aedat4} DVS event rates.}
        \label{fig:back_to_dvs_event_rates}
    \end{figure}

I see direct evidence for this claim in \cref{fig:back_to_dvs_event_rates}, which shows the proportion of DVS events reconstructed from the \adder{} representations after transcoding with $M = 40$ and $\Delta t_{ref} = \Delta t_{s}/500$ ticks, compared to the number of DVS events in the \textit{source} DAVIS videos. I can quickly reconstruct a DVS stream from \adder{}, since the intensity changes to trigger DVS events will occur only at the temporal boundaries of my sparse \adder{} events. However, I must assume a constant $\theta$, as in transcode mode (iii) (\cref{sec:mode_3}). Mode (i), which quantizes events into fixed-duration frames, can recover only a small fraction of the DVS events in the source. On the contrary, modes (ii) and (iii) have high proportions of recoverable DVS events, even with a high $M$. \cref{fig:event_source_adder_rates} illustrates that transcode mode (ii) generally necessitates far fewer \adder{} events than mode (iii) with these same settings. Thus, my \adder{} codec with mode (ii) can simultaneously utilize the APS frames to compress the DVS stream and utilize the DVS stream to deblur the APS frames, while fusing both streams into a unified, asynchronous intensity representation.

\begin{figure}
    \centering
{\includegraphics[width=\linewidth]{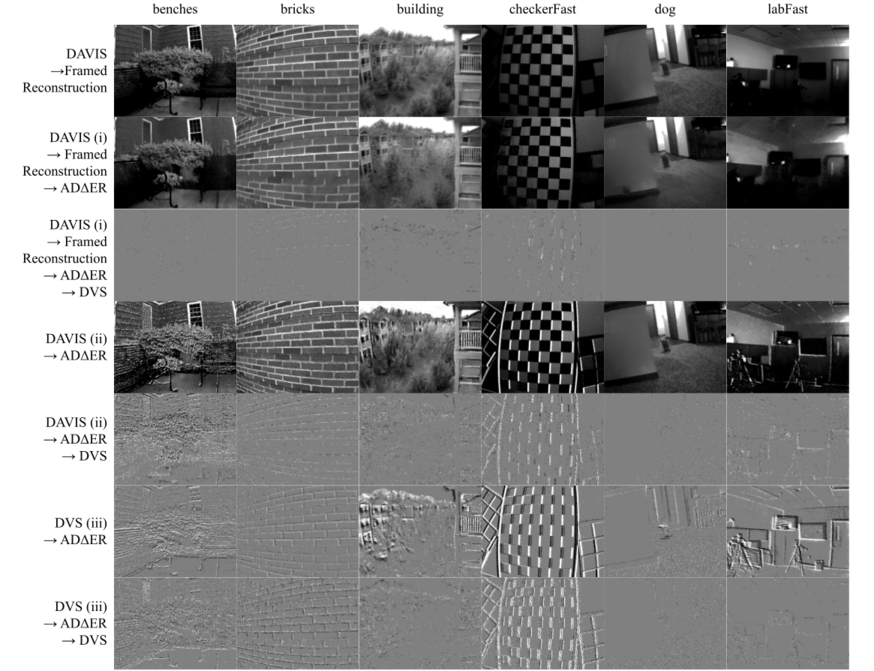}%
    \label{fig:davis_matrix}}
    \caption[Qualitative results on DAVIS sources]{Qualitative results: Event videos transcoded to framed video (first row) and to \adder{} under my three transcode modes (rows 2, 4, and 6) with $\Delta t_{ref} = \Delta t_{s}/500$ and $M = 40$. I transcode my \adder{} representations back to DVS (rows 3, 5, and 7) and see that modes (ii) and (iii) preserve much more of the temporal contrast detail of the DVS streams. Mode (ii) has a worse visual appearance than mode (i), due to DVS intensity drift.}
\end{figure}

\cref{fig:davis_matrix} shows qualitative results on six scenes from my data set. The \adder{} transcodes shown use $M=40$ and $\Delta t_{ref} = \Delta t_{s}/500$, as in my latency experiments. Since mode (i) of my event transcoder simply uses my framed source transcoder (\cref{sec:framed_transcoder_details}) after reconstructing a frame sequence with EDI, I note that the artifacts in the \adder{} representation here stem from artifacts in EDI's \textit{framed reconstruction}, or from the temporal averaging in \adder{}. In particular, I point to the \texttt{building} images, which show the limitations of EDI in deblurring APS frames under fast motion and long exposures. I see that I recover few DVS events from mode (i), due to its temporal quantization of the source events. However, mode (ii) exhibits a similar quality of recovered DVS events compared to mode (iii), despite also conveying the absolute intensities for pixels without DVS source events.

\begin{figure}
                \begin{subfigure}[b]{0.48\textwidth}
                    \centering
                    \includegraphics[width=\linewidth]{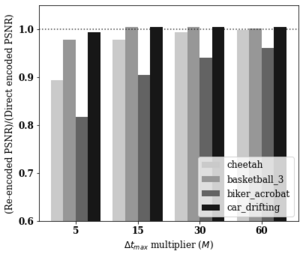}
                \end{subfigure}
                \hfill
                \begin{subfigure}[b]{0.48\textwidth}
                    \centering
                    \includegraphics[width=\linewidth]{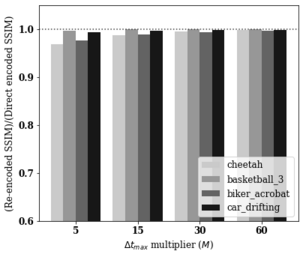}
                \end{subfigure}
                \caption[Re-encoding \adder{}]{Comparing the performance of re-encoding an \adder{} stream to a direct transcode of framed video to  \adder{}. For re-encoding, we target a larger \dtm value than the source \adder{} encoding. We evaluate the PSNR (left) and SSIM (right) quality of the reconstructions compared to the ground truth input frames, and divide the respective scores. A score less than 1.0 indicates that the re-encoded representation is of worse quality than the directly-encoded representation, while a score greater than 1.0 indicates the re-encoded representation is of better quality. }
                \label{fig:reencode}
            \end{figure}

\subsection{Re-encoding \texorpdfstring{\adder}{ADDER}}
            
        An \adder{} stream can itself serve as an input in our acquisition layer, using an \adder{} re-encoder. We read the \adder{} events as direct intensities (\cref{sec:pixel_list_structure}) into the pixel model to generate a new \adder{} stream with different parameters driving the $D$ control control. This may be useful when applications want to simplify their ingested data to ease processing or obtain a video with less spatio-temporal granularity. This transcoder also allows us to re-process \adder{} streams captured from ASINT-style sensors of the future. For example, a surveillance application may receive an \adder{} video reporting more than 1000 events per pixel, per second, but the signal may be too noisy for the application's function. The stream may then be re-encoded for each pixel to require no less than 10 events per pixel, per second, effectively reducing noise in the static regions of the scene via temporal averaging. 

        In a preliminary investigation, I implemented the \adder{} re-encoder and evaluated its performance on four videos from the Need for Speed dataset \cite{need_for_speed}. I ran the first 1000 frames of each video through our framed video transcoder. The baseline run has $\Delta t_{max} = \Delta t_{ref}$. For subsequent framed transcodes, I varied only the parameter \dtm{} by a multiplier $M \in \{5, 15, 30, 60\}$, and reconstructed their framed representations. I then re-encoded the \textit{baseline} \adder{} stream with the same $M$ multiplier values for \dtm, and likewise reconstructed their framed representations. To evaluate how well the re-encoder performs, I evaluated the PSNR and SSIM against the ground truth video for each of the eight framed reconstructions, and divided the re-encoded videos' scores by the directly-encoded videos' scores. As shown in \cref{fig:reencode}, the re-encoder performs slightly worse than the direct encoder for a low $M$, but performs roughly on par for high $M$. Overall, this a computationally fast means of reducing the bit rate of an \adder{} stream according to content dynamics. 

\subsection{Simulating an \texorpdfstring{\adder{}}{ADDER}-Style Sensor}\label{sec:asint_sim}
With the announcement of the Aeveon camera \cite{aeveon}, I recognize that specialized rate adaptation mechanisms are necessary to drive an \adder{}-style sensor in real time. On the hardware pixel level, one cannot run multiple separate integrations in parallel as in \cref{sec:pixel_list_structure} without duplicate subpixels. To this end, I propose a number of schemes to adjust pixel sensitivities based on spatiotemporal intensity predictions. To evaluate my schemes, I present a discrete event simulator for the ASINT sensor design. The ASINT sensor is depicted as a direct source of \adder{} events in \cref{fig:system_diagram_full_repeat}.

\begin{figure}
    \centering
    \begin{minipage}[t]{0.47\linewidth}
        \centering
        \includegraphics[width=.49\linewidth]{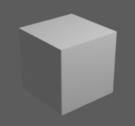}
        \includegraphics[width=.49\linewidth]{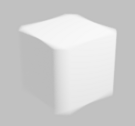}
        \caption[Blender-rendered frames]{Blender-rendered frames with shutter speeds of $\frac{1}{100}$s (left) and $\frac{1}{10}$s (right).}
        \label{img:cube_blender}
    \end{minipage}
    \hspace{0.04\linewidth}
    \begin{minipage}[t]{0.47\linewidth}
        \centering
        \includegraphics[width=.49\linewidth]{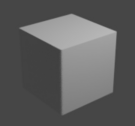}
        \includegraphics[width=.49\linewidth]{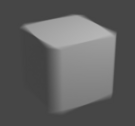}
        \caption[ASINT simulator reconstructed frames]{ASINT simulator reconstructed frames with target shutter speeds of $\frac{1}{100}$s (left) and $\frac{1}{10}$s (right).}
        \label{img:cube_emu}
    \end{minipage}
\end{figure}

\subsubsection{ASINT Discrete Event Simulation}
I employ a ray-traced Blender Cycles animation of a 3D environment to generate ground truth data for my camera simulator. Using the Photographer add-on \cite{photographer} in Blender, we can achieve accurate measures of exposure for fast shutter speed, high frame rate video. By matching Blender's virtual camera shutter speed to the reciprocal of the frame rate, we can accurately capture the scene's full light information. While this animation is rendered out as traditional frames, we can still generate accurate event data from these frames, so long as the ASINT camera reference interval we simulate is faster than or equal to the shutter speed for the rendered frames. I output the frames in monochrome 16-bit PNG format, then read those frames sequentially into my C++ ASINT camera simulator, where for simplicity I interpret the grayscale value of each pixel as the incident photon count measured on that pixel over the reference interval. I simulate photon integration for each pixel by taking a $2^D$ portion out of this photon count, calculating the time delta as a fraction of the reference interval, and firing out a 64-bit event with this data. As events fire, each pixel's $D$ value may be adjusted dynamically according to the methods described below.

\subsubsection{Event Prediction for Rate Adaptation}
To reduce the overhead of event data transmission, one can choose to dynamically update  individual pixels' $D$ values by comparing actual integrator saturation time deltas to the expected saturation time deltas. I keep a running estimate of the next firing time delta for each pixel and measure the number of consecutive, similar bits between the expected and actual firing time deltas to adjust $D$ accordingly.

Employing some simple heuristics, I typically see a firing rate of one to two events per pixel, per reference interval. For completely stable pixels, whose $D$ and $\Delta t$ values are the same for consecutive events, I do not fire subsequent repeated events but rather track in our model the number of identical events that have been measured. When the incident light intensity at the pixel changes, I write out a special event that encodes the number of times the previous event repeated before firing the new event.

Where the transition from high to low pixel intensities occurs faster than the camera simulator adjusts its $D$ values appropriately, I fire ``empty'' events with an incident intensity of 0. This is expected behavior, and when such an event is fired, I quickly throttle down the decimation factor to make the pixel much more sensitive to lower light intensities. I can also choose to throttle down the $D$ values of neighboring pixels within a user-defined radius, in an effort to avoid firing empty events for nearby pixels if their incident intensity becomes low as well.

With the foundations of the ASINT simulator in place, I began implementing additional schemes necessary to evaluate the sensor model. With the simulator, I can repeatedly emulate the camera reading the same scene with different camera settings. Since events are generated asynchronously, each event is encoded individually in a time-ordered manner. I also include a header at the beginning of the file which encodes basic metadata for the camera, including the sensor dimensions, the timestamp clock rate $C$, and $\Delta t_{max}$. The camera model ensures that there will be at least one event encoded for every pixel over each $\Delta t_{max}$ interval. With these additions in place, I devised the following $D$-control mechanisms.

\paragraph{Constant Decimation Mode}
This mode is the simple case. Each pixel maintains a constant $D$ value, set globally before data capture begins. With a low $D$ value, this ensures that the scene is captured in the highest possible fidelity, without firing any empty events. By emulating the sensor with a constant, global decimation factor low enough to fire events for the darkest intensity in a given image sequence, I can verify the accuracy of my simulator and reconstruction technique. This gives me a basis of ground truth against which to evaluate lossy event compression techniques empirically. 

\paragraph{Self-Adjustment Mode}
In software, I maintain a prediction model of each pixel individually. I track the number of stable bits in the $\Delta t$ of a pixel's event, being the number of consecutive bits, starting with the most significant, that align with our predicted $\Delta t$, which I notate $\Delta t'$. If we see that the pixel became more stable than it was for the last event, I generally (employing heuristics) increment $D$ by 1 and double the prediction for $\Delta t$. Conversely, if the pixel became less stable, I generally decrement $D$ by 1 and halve the prediction for $\Delta t$. Gradual, small changes between $D$ values helps prevent thrashing between drastically different sensitivities. However, if the pixel is not sensitive enough to the falling edge of a quickly moving object, it may fire an empty event. In this case, I throttle $D$ down to $D_{new} = \text{floor}(\log_2(D_{old}))$ and adjust our prediction $\Delta t'$ to $\Delta t'_{new} = \Delta t'_{old} / (D_{old} - D_{new})$.

\paragraph{Self-Adjustment with Radial Neighbor Adjustment Mode}\label{sec:radial_adjustment}
This mode adds a rectangle of influence within which pixels may adjust their neighbors' next $D$-values. The size of each neighborhood may shrink and grow, depending on the stability of the pixels inside, but the maximum size is set by the user when camera emulation begins. By this method, I achieve some level of \textit{motion compensation} within the data capture itself. For example, a pixel on the falling edge of a passing object may saturate at a different time than predicted. It then can adjust its neighbors' sensitivity appropriately. The goal is to capture adequate information (not have empty events) but minimize the number of events fired per pixel, per $\Delta t_{max}$ interval. I have two independently adjustable radii: the first radius defines which neighbors will have their $D$-values throttled down (for a dramatic increase in sensitivity) when a given pixel fires an empty event; the second radius defines which neighbors will have their $D$-values slightly adjusted (for a minor increase or decrease in sensitivity) according to small changes in a given pixel's sensitivity.

\paragraph{Examples}
Figs. \ref{img:cube_blender}-\ref{img:cube_emu} show reconstructed frames of a simple rotating cube animation at varying target shutter speeds. Fig. \ref{img:cube_emu} demonstrates that, due to the high bit depth of reconstructed frames, we may lower the effective shutter speed while maintaining high dynamic range and not losing highlight details in the image.

\begin{figure}
\centering
        \includegraphics[width=0.49\linewidth]{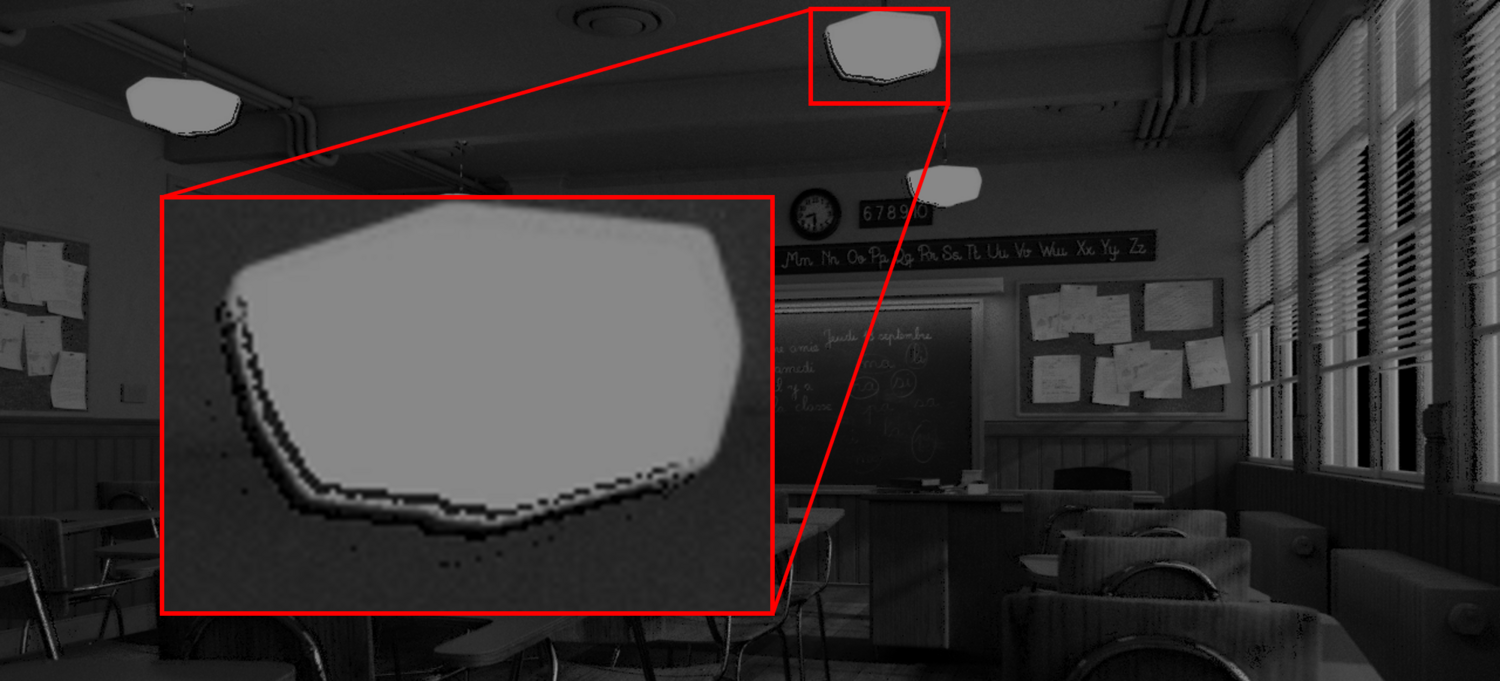}
        \includegraphics[width=0.49\linewidth]{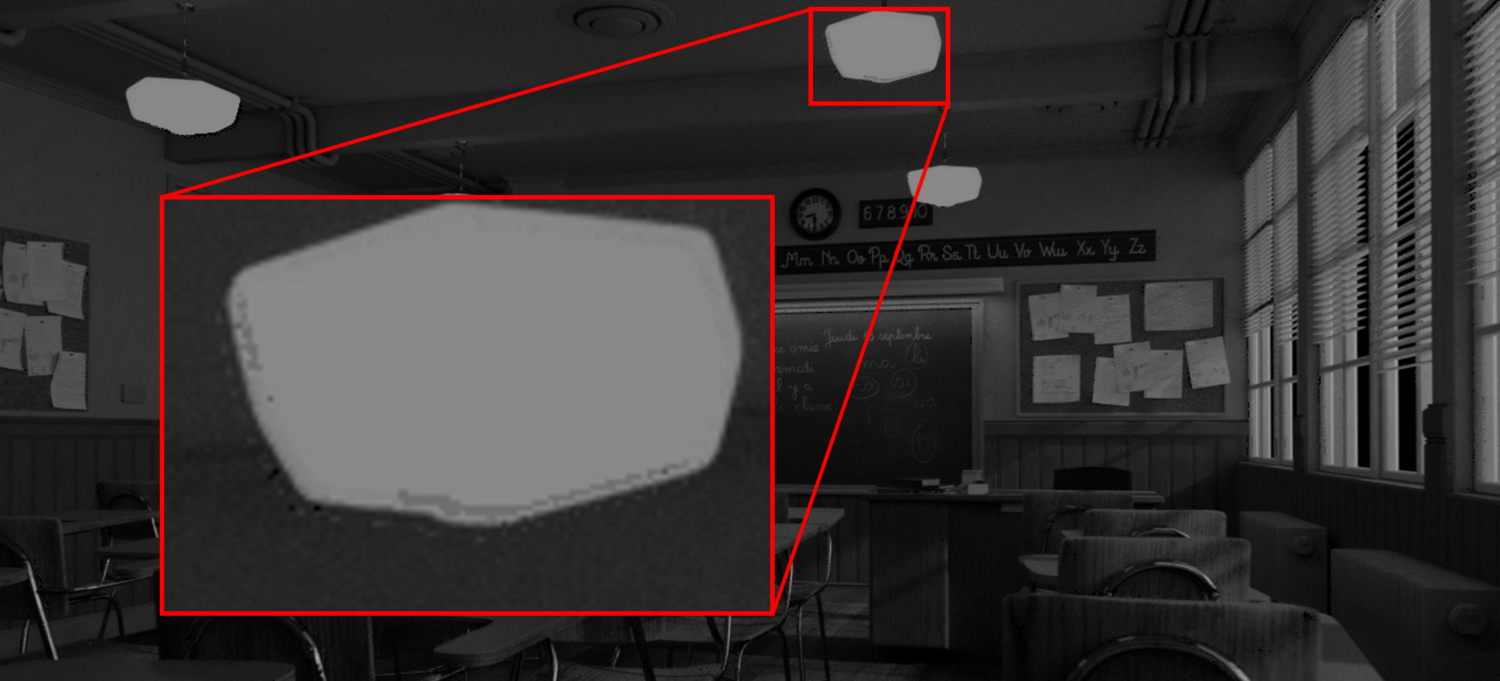}
        \caption[ASINT simulator reconstructed frames]{ASINT simulator reconstructed frames from the Blender Classroom scene with a moving virtual camera, using single-pixel decimation (left) and grouped pixel decimation with radius 20 (right).}
        \label{img:class}
\end{figure}

The black pixels on the light fixture of Fig. \ref{img:class} show the effect of empty events, and these are a direct artifact of the unique ASINT architecture. The goal of future decimation control schemes is to maintain a more detailed model of the scene and adjust $D$ values more proactively, to minimize these empty events where no light information is gathered.


\section{Transcoding as First-Stage Lossy Compression}\label{sec:first_stage_compression}

A traditional framed video system has only a single-stage compression scheme (\cref{sec:framed_compression}). The encoder induces both spatial and temporal information loss at once, then entropy codes the prediction residuals. While frequency transforms and motion estimation schemes enable robust spatiotemporal prediction, the information loss is only in the form of quantizing the prediction residuals. The accuracy of intensity samples is reduced, but the \textit{distribution} of samples in the raw representation (one sample per pixel, per channel, per frame) remains constant.

While the work described in this chapter does not include any mechanism for traditional source-modeled compression and entropy coding, I emphasize that the \adder{} acquisition layer serves as the \textit{first stage of a two-stage lossy compressor}. By setting $M > 0$, I can instruct the \adder{} model to increase its tolerance to variations in intensity over time. As \cref{fig:framed_event_rate_vmaf} clearly illustrates, higher $M$ values induce higher visual quality loss and compression. In contrast to classical video compression, however, lossy compression here actually reduces the temporal sample rate of the raw representation. A similar raw rate reduction can be achieved in DVS systems by subsampling the events. However, such a process indiscriminately \textit{discards} information, whereas the \adder{} transcode process \textit{averages} out the variations in intensity. 

This first stage of compression occurs only along the temporal domain for each pixel. The second stage, which I introduce in \cref{ch:adder_compression}, induces loss along the spatial domain.

\section{Future Work}

While the \adder{} acquisition layer indeed supports data inputs from multiple event camera manufacturers, my implementations are limited to a single data format each for iniVation (\texttt{.aedat4}) and Prophesee (\texttt{.dat}). However, other formats are in use in different camera specifications, such as \texttt{.aedat2} and \texttt{.raw}. I will work to implement data decoders for these additional formats.

Furthermore, I have implemented live camera decoding only for the iniVation cameras with the \texttt{.aedat4} packet format. I intend to replicate this functionality for connected Prophesee cameras. With this in place, I can expand the analysis of camera latency and develop mechanisms to adjust \adder{} transcode sensitivity parameters based on a parameter for tolerable latency. Furthermore, I will implement a transcoder for the Aeveon camera once it is commercially available.

I note the difficulty in empirically evaluating the quality of our transcoded representation for DAVIS/DVS sources. Since APS frames are blurred, they are not reliable references for framed quality metrics (e.g., PSNR, SSIM, and VMAF). Furthermore, existing metrics for DVS stream quality focus on noise filtering tasks and can only process a few hundred events per second \cite{spike_metric}, while real-world DVS sequences encompass several millions of events per second. Therefore, I hope to devise a practical, fast quality metric for generic \adder{} streams. To make \adder{} transcoding practical for real-time performance at high resolutions, I will work to implement Single Instruction/Multiple Data (SIMD) instructions for parallel performance and develop a hardware transcoder with a Field Programmable Gate Array (FPGA). 

\section{Conclusion}
The \adder{} format can effectively represent a variety of intensity-based video sources. Through temporal averaging of similar intensities, we can increase the intensity precision and greatly reduce the number of samples per pixel compared to grid-based representations. I demonstrate extremely low-latency DAVIS fusion, and I argue that my methods provide the computational efficiency and temporal granularity necessary for building real-time intensity-based applications for event cameras. Likewise, \adder{} exposes a simple parameter for controlling the rate-distortion tradeoff with traditional framed video sources. The acquisition layer of an \adder{} system gives one control over the data rate of the raw representation. In contrast, classical systems exhibit a rate reduction only in the entropy-coded representation.

 \chapter[~~~~~~~~~~~~\texorpdfstring{\adder}{ADDER} Compression and Applications]{\texorpdfstring{\adder}{ADDER} Compression and Applications\footnotemark}\label{ch:adder_compression}\footnotetext{Significant portions of this chapter previously appeared in the proceedings of 2021 ACM Multimedia Systems \cite{Freeman2021mmsys}, 2023 ACM Multimedia Systems \cite{freeman_mmsys23}, and 2024 ACM Multimedia Systems \cite{freeman_mmsys_2024}.}

\graphicspath {{ch04_compression/images/}}

\section{Introduction}

At this stage, I have demonstrated that \adder{} can serve as a narrow-waist representation for a variety of framed and event video sources. The raw representation uniquely allows for a flexible rate distortion tradeoff, which I showed with an evaluation of framed reconstruction quality. While the representation is \textit{novel}, I have not yet shown that it is \textit{consequential}: is there a compelling reason to transcode a framed video or event camera video to \adder?

To begin answering this question, I emphasize that my primary goal in this dissertation is to make video systems more efficient for vision tasks. Classical framed representations readily provide the compression ratios, speed,  and visual quality necessary for human applications such as entertainment and video communication. However, for computer vision systems, the driving metrics are compression ratios, speed, and \textit{application performance}. We can tolerate entirely different types of information loss, as long as the application performs within our expected level of accuracy. By designing the \adder{} video system from the ground up with vision applications in mind, my aim is to show improvements across these metrics over the existing framed and event video system models.

To this end, I present the second stage of the lossy compression scheme for \adder, in the representation layer. This compression stage takes as input raw \adder{} events from an arbitrary source. It then arranges them in a spatiotemporally organized data structure, quantizes prediction residuals for the events' time components, and performs source-modeled arithmetic coding. My naive compression scheme enables \textbf{higher compression ratios} than H.265 on scenes transcoded from low-motion video, with a negligible drop in visual quality.

In the application layer, I introduce various interfaces for running vision applications on \adder{} data. I demonstrate that classical framed applications, such as object detection, are compatible with \adder{} through framed reconstruction or frame sampling. By changing the input image representation to contain events' $D$ and $\Delta t$ components, I observe up to 4\% \textbf{higher precision} on YOLOv5 object detection. Conversely, non-learned image processing applications can be made asynchronous to take advantage of \adder's sparsity. I implement an event-based version of FAST feature detection, and find a median \textbf{speed improvement} of 43.7\% over OpenCV on a surveillance video dataset. Finally, I introduce a bespoke motion segmentation application, which uses event firing rate as an indicator of motion rather than frame differencing, demonstrating that it is feasible to build applications that work directly with event data.

As I discussed in \cref{sec:first_stage_compression}, a major advantage of a sparse video representation is the potential for dynamic rate adaptation of the \textit{raw} data. By reducing the contrast sensitivity of a pixel model, we can reduce its event firing rate. In my prior discussion, these contrast sensitivities were globally fixed. Here, I introduce the notion of \textit{application-driven rate control} at the source. An application running on either the transcode server or a client may delegate priority to certain pixels of interest. A rate controller increases the sensitivity of those pixels, or lowers the sensitivity of the other pixels. I implement rate-adaptive online systems with FAST feature detection and object tracking, demonstrating a high reduction in event rate with a minimal impact on application accuracy.

\section{Application-Oriented Codec Parameters}

When investigating lossy compression techniques and applications for \adder, I found several limitations to the software model described in \cref{ch:adder_transcoding} which hindered various aspects of practical compression. Below, I describe these limitations and the corresponding additions or modifications I made to address them.

\subsection{\texorpdfstring{$\Delta t$ vs. $t$}{DeltaT vs. t}}
 
 The earlier \adder{} work defined an event's temporal component, $\Delta t$, as the time elapsed since the pixel last fired an event. This scheme makes it straightforward to calculate the intensity expressed by the event, through simply calculating $\frac{2^D}{\Delta t}$. When reconstructing the intensities for playback or applications, the software tracks the running clock time for each pixel to ensure that events are correctly ordered.
 
 This representation lends itself poorly to lossy compression, however. Suppose that, for a given pixel, we have a sequence of $\langle D, \Delta t\rangle$ events $\{e_0 = \langle 5, 100_{\Delta} \rangle, e_1 = \langle 5, 120_{\Delta} \rangle, e_2 = \langle 5, 110_{\Delta} \rangle \}$. If we track the running time of the pixel, we see that $e_2$ fires at $t = 330$ ticks. Now, suppose without loss of generality that we incur some compression loss in the $\Delta t$ component of $e_0$, and upon reconstruction we obtain the sequence $\{e_0' = \langle 5, 70_{\Delta} \rangle, e_1' = \langle 5, 120_{\Delta} \rangle, e_2' = \langle 5, 110_{\Delta} \rangle \}$. Then, the firing time for the $e_2'$ is $t = 300$ ticks. Although we incurred loss only in $e_0$, the change-based temporal measurement has a compounding effect on all the later events for that pixel. 

To rectify this, I use an absolute $t$ representation as the temporal component of my events. In the above example, our original sequence would be $\{e_0 = \langle 5, 100_{\Sigma} \rangle, e_1 = \langle 5, 220_{\Sigma} \rangle, e_2 = \langle 5, 330_{\Sigma} \rangle \}$.  Then, if we incur the same loss on $e_0$, our reconstructed sequence would be  $\{e_0 = \langle 5, 70_{\Sigma} \rangle, e_1 = \langle 5, 220_{\Sigma} \rangle, e_2 = \langle 5, 330_{\Sigma} \rangle \}$. During playback, we simply subtract the $t$ component of the previous event of the pixel from the current event to obtain $\Delta t$ and compute the intensity as normal. In this case, the reconstructed intensity measurements would be less accurate for $e_0$ (brighter) and $e_1$ (darker), but \textit{not} $e_2$. Therefore, incurring temporal loss in one event only has a compounding effect on the reconstruction accuracy of the event immediately afterward. I utilize this absolute $t$ representation throughout the rest of this dissertation, with each event being the tuple $\langle x,y,c,D,t\rangle$. I continue using the \adder{} terminology, however, since we still compute the incident intensity based on the $\Delta t$ between two events. 

\subsection{Redefining \texorpdfstring{$\Delta t_{max}$}{DeltaT-max}}\label{sec:dtm}
Previously, I defined the parameter $\Delta t_{max}$ as the ``maximum $\Delta t$ that any event can span'' \cref{sec:user_parameters}. The reason for this definition is to ensure that a client can be guaranteed that the latest update for a pixel will be available within the \dtm{} time span. However, there is a trade-off between pixel response time and event rate. For a completely stable pixel (with an unchanging intensity value), each halving of \dtm{} will yield a doubling in its output event rate. If one made \dtm{} very high, the event rate would be lower, but the client would have to wait much longer to obtain the intensity of a newly stable pixel. In a streaming setting, this behavior leads to potentially high and unpredictable latency.

To address these streaming concerns, we redefine \dtm{} as \textit{the maximum \dt{} that the first event of a newly stable pixel can span}. Suppose we were to start integrating a pixel with 1 intensity unit per tick for 768 ticks, suppose the pixel's starting $D$ value is 8, and suppose $\Delta t_{max} = 300$. After 768 ticks, suppose that the incident intensity changes, so we must write out the pixel's event queue. Under the \textit{prior} \dtm{} scheme, our pixel would produce the event sequence $\{e_0 = \langle 8, 256_{\Sigma} \rangle, e_1 = \langle 8, 512_{\Sigma} \rangle, e_2 = \langle 8, 768_{\Sigma} \rangle \}$, where the $\Delta t$ between two consecutive events is no greater than 300 ticks.  Under my new \dtm{} scheme, the pixel would produce the sequence $\{e_0 = \langle 8, 256_{\Sigma} \rangle, e_1 = \langle 9, 768_{\Sigma} \rangle\}$, where only the $\Delta t$ of the first event at the baseline intensity level must be within the 300-tick threshold. We can now coalesce the remaining sequence of events into a single event with a higher $D$ and $\Delta t$ value. Therefore, we have the flexibility to set \dtm{} lower, ensuring low pixel latency for intensity changes, without having a high event rate for stable pixels. This helps us avoid repeatedly intra-coding the same event for stable pixels in a streaming-supported compression scheme. 

The new \dtm{} scheme mitigates the event rate, making transcoding operations and applications faster, and making the representation more amenable to compression. On the other hand, it removes any time-bound guarantee that a client dropping into an \adder{} stream will receive an event for all pixels. That is, the client will not receive events for stable pixels until their incident intensity changes.

\subsection{Adaptive Contrast Thresholds}\label{sec:contrast_thresholds}
As described in \cref{sec:user_parameters}, I previously explored the use of a constant contrast threshold, $M$, which is the same for all pixels. To now enable adaptation of content-based rate, I specify a \textit{maximum} contrast threshold $M_{max}$ and a threshold rate-of-change parameter, $M_v$. Then, a stable pixel may increase its contrast threshold by one intensity unit for every $M_v$ intervals of $\Delta t_{ref}$ time spanned, up to $M_{max}$. Applications within the transcoder loop may then forcibly \textit{lower} the $M$ of certain pixels to increase their responsiveness and accuracy as needed.

\subsection{Constant Rate Factor}\label{sec:crf}
I found that the myriad low-level parameters available within the \adder{} system are abstruse for a general user. I sought to create a simple meta-parameter and lookup table to reasonably set the underlying variables. Taking inspiration from framed codecs such as H.264 and H.265, I call my metaparameter the constant rate factor (CRF), with values ranging from 0-9. Setting CRF to 0 yields a lossless transcoded event stream, whereas setting a high CRF value will yield greater loss but a much-reduced event rate. Specifically, the CRF table determines the parameters $M$, $M_{max}$, and $M_v$ as described in \cref{sec:contrast_thresholds}, as well as the radius for feature-based rate adaptation described below in \cref{sec:feature_rate_control}. I populated the lookup table such that incrementing the CRF by 1 yields a drop in PSNR reconstruction quality of 2.5-5.0 dB on a typical framed video transcode. For this work, I evaluate CRF settings 0, 3, 6, and 9, which I will refer to as \texttt{Lossless}, \texttt{High}, \texttt{Medium}, and \texttt{Low} quality settings, respectively.

\subsection{Multifaceted \texorpdfstring{$D$}{\textit{D}} Control}
With my modifications, we see that there are several factors which influence the $D$ values of generated \adder{} events. We can loosely think of $D$ as a sum of partial components 
\begin{equation}\label{eqn:d}
D = D_{intensity} + \max (D_{stability} - D_{application}, 0).    
\end{equation}

Here, $D_{intensity}$ is the baseline $D$-value derived from the first intensity integrated for the event. For example, if we begin integrating a pixel with 223 intensity units, then we have 
\begin{equation}
D_{intensity} = \lfloor\log_2 223\rfloor = 7.    
\end{equation}

 $D_{stability}$ denotes the portion of $D$ that comes from the temporal stability of a pixel's incident intensity. For example, if $M >= 3$ and we integrate 220 intensity units from three more consecutive input frames,  we have 
 \begin{equation}
     D_{stability} = \lfloor\log_2 223 + 220*3\rfloor - D_{intensity}= 9 - 7 = 2.
 \end{equation}
The \adder{} transcoder indirectly determines $D_{stability}$ based on the contrast threshold, $M$, and the consistency of incoming intensities. That is, the longer a pixel integrates intensity, the higher its implied $D_{stability}$. Subsequently,  a higher $M$ makes a stable pixel more impervious to slight variations in incoming intensity, and it can integrate for a longer period of time.

Finally, $D_{application}$ is a lowering of $D$ according to higher-level application directives. In the example above, we might set $D_{application} = 2$ to ensure that temporal variations in intensity are not averaged out. In practice, we achieve this by manually lowering the $M$ of a pixel to make it more sensitive to variations in intensity. I note in \cref{eqn:d} that an application cannot reduce the overall $D$ beyond the baseline $D_{intensity}$, so that we do not unnecessarily increase the event rate.

My new $D$ control mechanism, in tandem with the adaptive contrast control (\cref{sec:contrast_thresholds}), gives the \adder{} framework the flexibility to allocate rate towards spatiotemporal regions of interest.


\section{Representation Layer}\label{sec:adder_reprsentation_layer}

Compared to existing DVS-based systems (\cref{sec:dvs_representation_layer}), the \adder{} representation layer has a simplified design due to its singular input representation. I outline the representation layer in \cref{fig:system_diagram_intro_full}.

\subsection{Compressed Representation}\label{sec:adder_compression}

In the landscape of video representations, \adder{} is unique in allowing for two concurrent drivers of loss. I previously introduced contrast-based loss control with $M$, which controls the rate and temporal distortion of raw events \cref{sec:user_parameters}. A single grayscale \adder{} event is 9 bytes, however, making raw files unwieldy for storage or applications. A form of entropy compression is necessary for practical systems.

We cannot simply recycle the techniques of classical video compression, however. As I noted in \cref{sec:framed_compression}, these codecs all rely on \textit{frequency transforms} such as the DCT to consolidate the information into a handful of coefficients ordered by frequency. The coefficients are then variously quantized to introduce loss. These frequency transforms rely on the assumption that the intensity data is temporally synchronous and that each intensity sample has the same level of precision (e.g., 8 bits). With \adder, we cannot make these same assumptions. Samples are asynchronous, so spatial groupings of events can represent disjoint motion between pixels. Since $\Delta t$ can vary dramatically, two events may convey drastically different intensity measurements. In my early work, I used the DCT to transform only the event $D$ components for lossy compression \cite{FreemanLossyEvent}, but I required consistent event firing rates between pixels, and the results were not very compelling.

\begin{figure*}[ht]
        \centering
        \includegraphics[width=\linewidth]{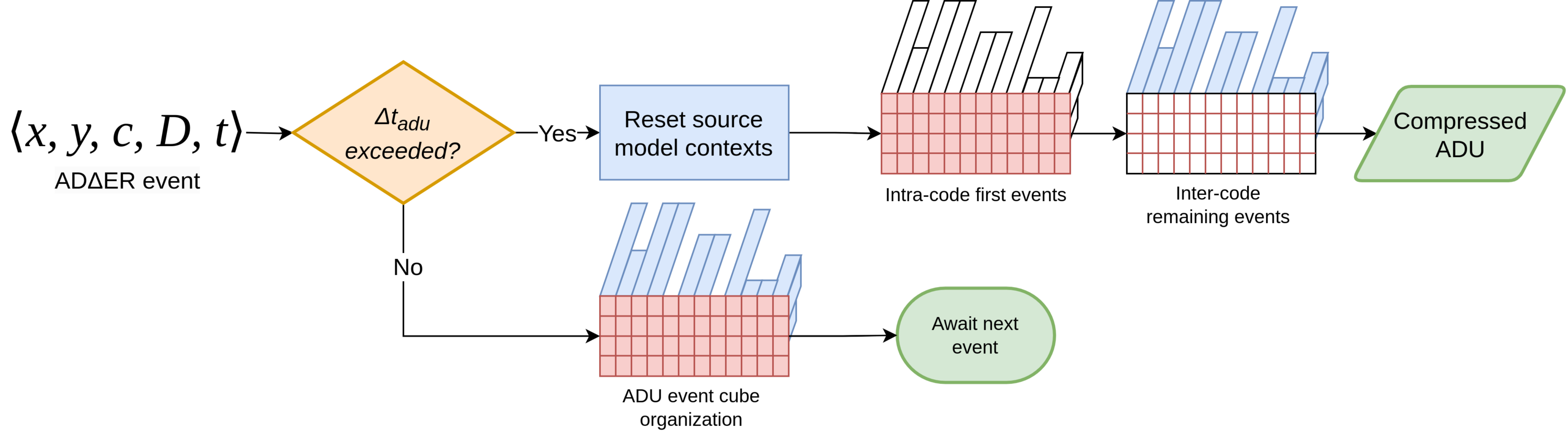}
       \caption{Simplified flowchart of the lossy compression scheme}
        \label{fig:compression_flowchart}
\end{figure*}

Instead, I argue that we must develop entirely new compression schemes from the ground up for the asynchronous paradigm. I summarize my proposed source-modeled arithmetic coding scheme for \adder{} data in \cref{fig:compression_flowchart}, and this compression step is noted in \cref{fig:system_diagram_full_repeat}.

\begin{figure*}
     \centering
     \begin{subfigure}[t]{\textwidth}
         \centering
         \includegraphics[width=\textwidth]{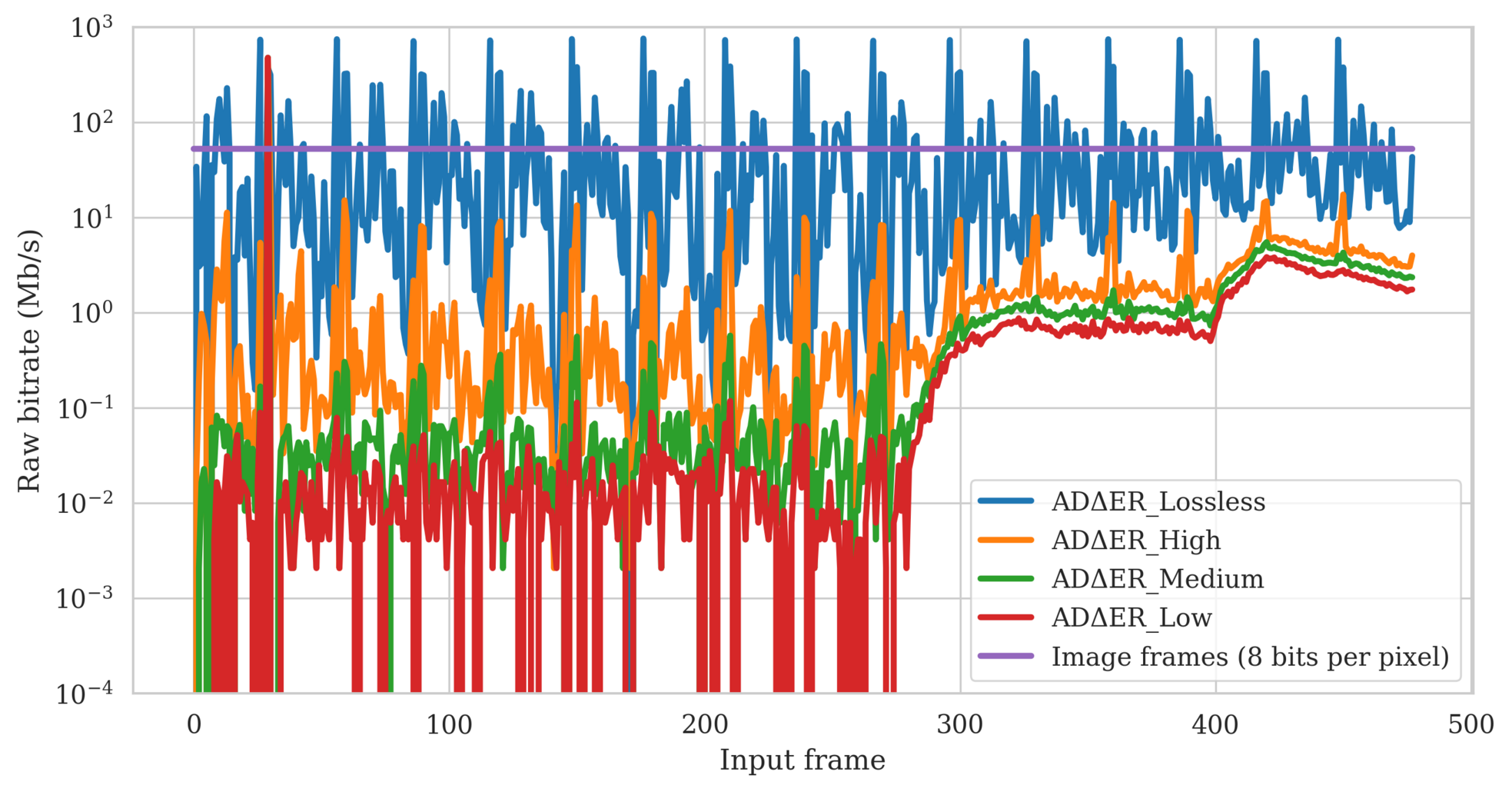}
         \caption{The bitrates of the raw \adder{} representations (before arithmetic coding) at our four quality levels. For comparison, the bitrate of a raw decoded image frame is constant.}
         \label{fig:ex_bitrate}
     \end{subfigure}
     \begin{subfigure}[t]{\textwidth}
         \centering
         \includegraphics[width=\textwidth]{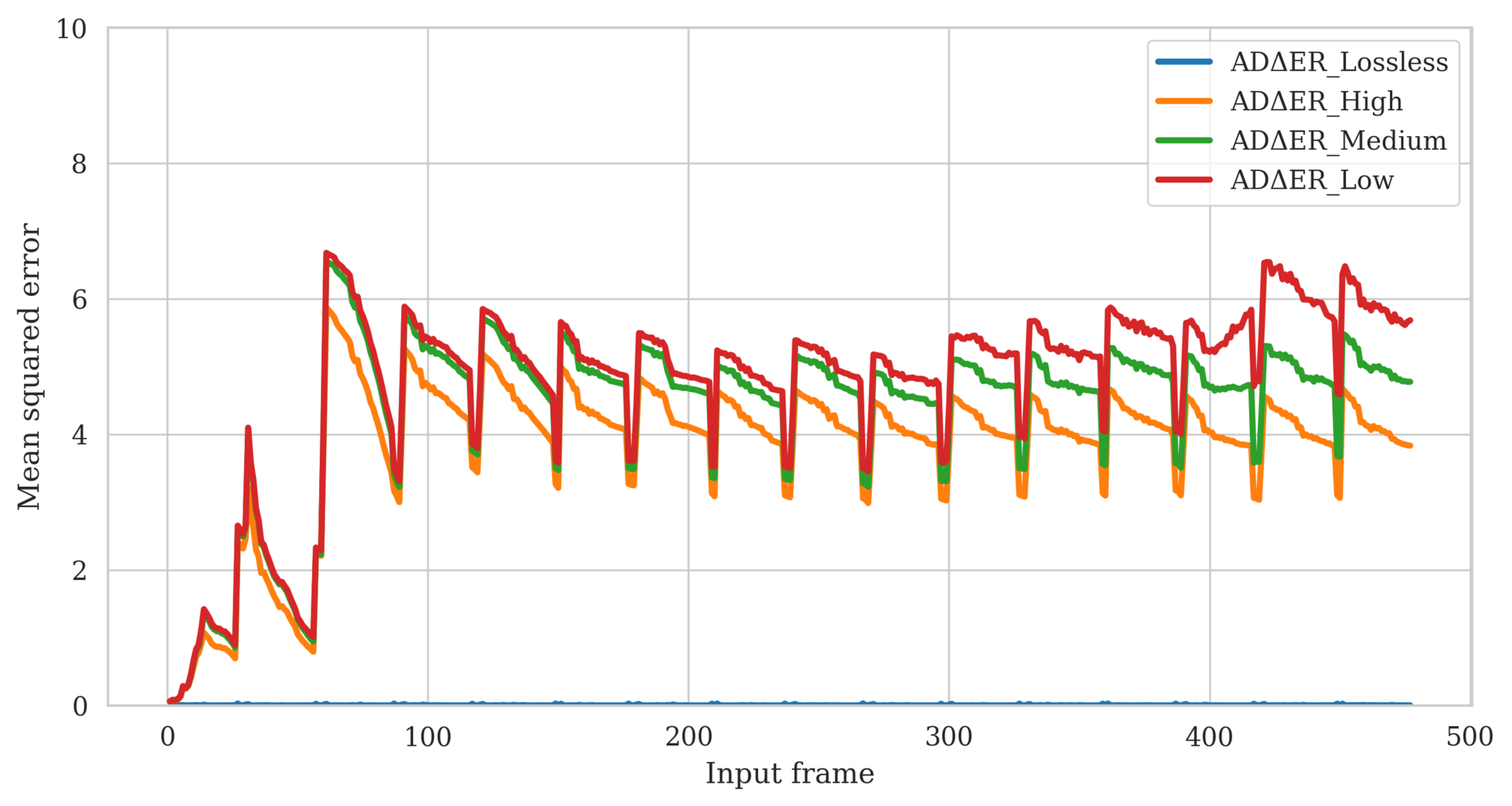}
         \caption{The mean squared error of framed reconstructions of the raw \adder{} events.}
         \label{fig:ex_mse}
     \end{subfigure}
     \caption[Key metrics on example video]{Key metrics gathered for a particular video. The \texttt{Lossless} lines are achievable with $M = 0$ as described in \cref{ch:adder_transcoding}, while the other lines result from the CRF mechanism (\cref{sec:crf}).}
     \label{fig:example}
\end{figure*}

\begin{figure*}
     \centering
     \begin{subfigure}[t]{\textwidth}
         \centering
         \includegraphics[width=\textwidth]{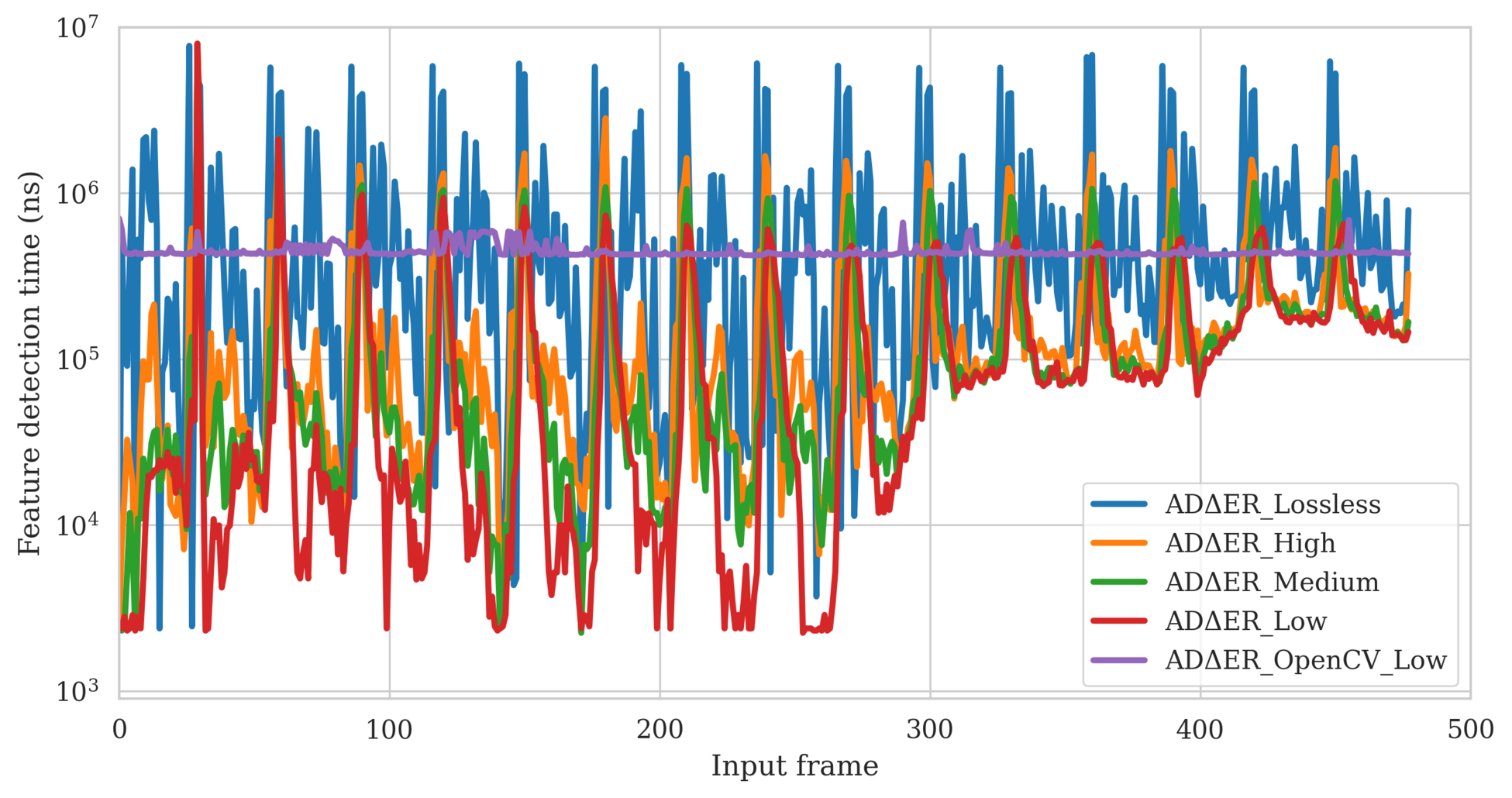}
         \caption{The execution time for FAST feature detection.}
         \label{fig:ex_feat_speed}
     \end{subfigure}
     \begin{subfigure}[t]{\textwidth}
         \centering
        \includegraphics[width=\textwidth]{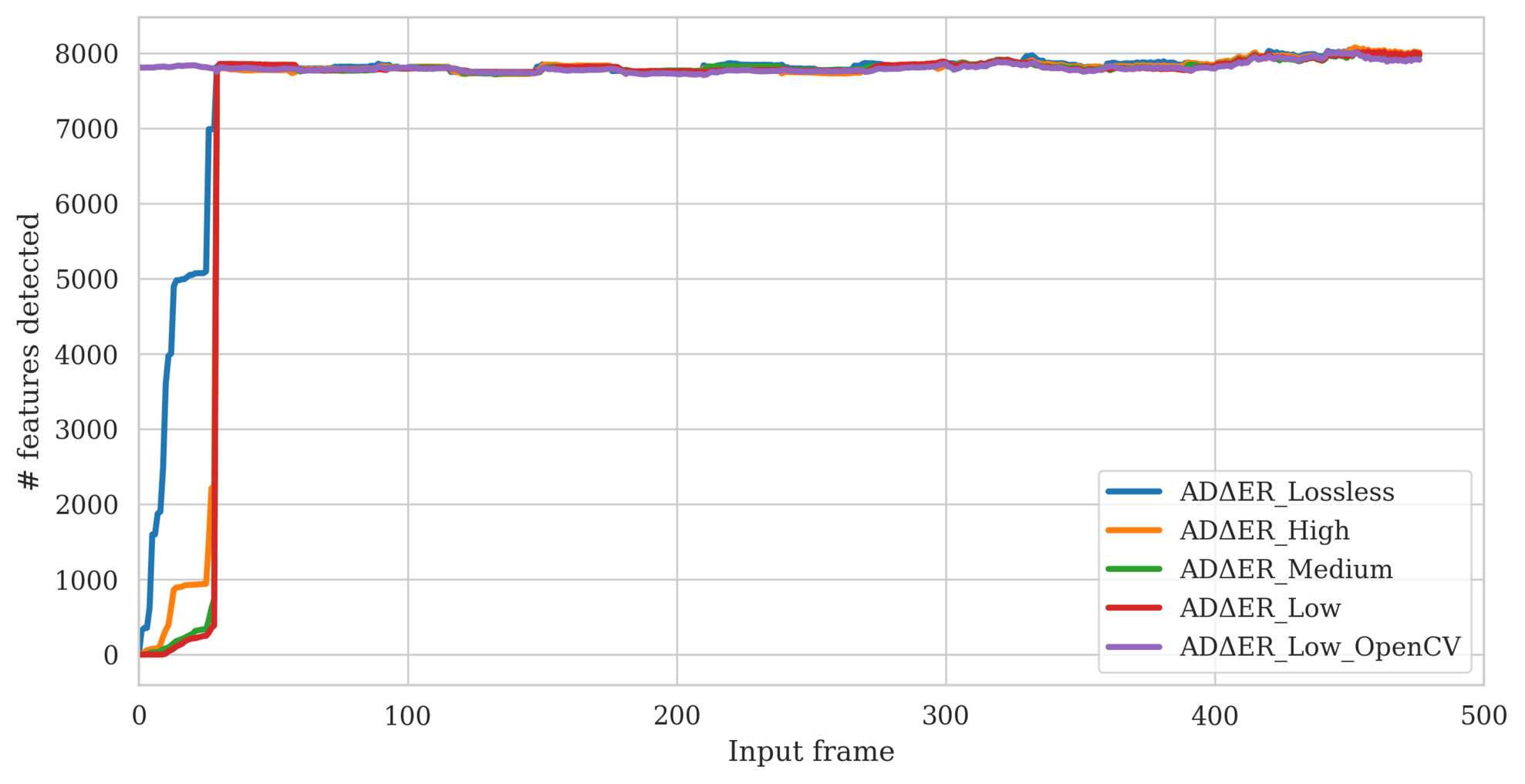}
         \caption{The total number of detected features present, over time.}
         \label{fig:ex_feat_num}
     \end{subfigure}
     \caption[Key metrics on example video, part 2]{Additional  metrics gathered for the particular video as in \cref{fig:example}. }
     \label{fig:example2}
\end{figure*}

\subsubsection{Application Data Units}
The fundamental compressed representation in my scheme is a series of Application Data Units (ADUs). I independently encode each ADU with fresh source model contexts, to support fast scrubbing and stream drop-in. The temporal span of an ADU is denoted by $\Delta t_{adu}$. In this section, I set $\Delta t_{adu} = \Delta t_{max}$, as defined in \cref{sec:dtm}.

\dtm{} here expresses a meaning similar to the I-frame interval in framed codecs. For example, suppose we are transcoding a framed video to \adder{} with $\Delta t_{ref} = 255$ ticks and $\Delta t_{adu}= 2550$ ticks. Then, each input frame spans 255 ticks, and each ADU spans 2550 ticks. Thus, our ADU contains 10 input frames of transcoded \adder{} data. If an ADU begins at time $t_0$, we encode the ADU once we encounter an event with time $t' > t_0 + \Delta t_{adu}$.

\subsubsection{Event Cubes}
Within each ADU, we organize the incoming events into \textit{event cubes}. Each event cube represents a $16 \times 16$ spatial region of pixels and a temporal range of $\Delta t_{adu}$ ticks, and we maintain independent queues of events for each pixel.

I begin encoding an ADU by intra-coding the event cubes in row major order.  Here, we encode only the \textit{first} event for each pixel in each cube. Suppose that we have spatially adjacent events $a$ and $b$, and we have already encoded $a$. I lossless-encode $D_r = D_b - D_a$ and the $t$ residual $t_r = t_b - t_a$.  If the video is in color, then our event cube contains separate pixel arrays for the red, green, and blue components, and these components are encoded in that order. I do not encode the coordinates of the events, since we have organized them spatially. 

After intra-coding all the event cubes, we inter-code the remaining events for each pixel. Here, we examine temporally adjacent events $a$ and $b$ for a single pixel. We can leverage knowledge of the prior state of the pixel to form a $t$ prediction, $p$, by 

\begin{equation}
    p_b = t_a' + \Delta t_a' \ll D_r,
\end{equation}

where $t_a'$ is the reconstructed timestamp of the pixel's last event, $\Delta t_a'$ is the reconstructed $\Delta t$ for the pixel's last event, and $D_r$ is the $D$ residual. I then determine how much loss we can apply to the $t$ prediction residual, $t_r = t_b - p_b$. For this, I iteratively right-shift the bits of the $t$ prediction residual and calculate the intensity, $I'$, that the decoder would obtain when reconstructing the $t$ given the shifted residual and the shift amount, $s$. The equation for $s$ is

\begin{equation}
    \argmax_s\bigg(I' = \frac{2^D}{(r \ll s) + p_b - t_0} : I - M_{max} < I' < I + M_{max}\bigg),
\end{equation}

where $I$ is the original intensity and $M_{max}$ is our maximum contrast threshold as described in \cref{sec:contrast_thresholds}. Without the $M_{max}$ limitation, a large prediction residual is likely to create salt-and-pepper noise when it is bit shifted. I encode $D_r$, $s$, and $t_r$ for each event remaining in a pixel's event queue before proceeding to the next pixel in row-major order.

\subsubsection{CABAC}
I use a context-adaptive binary arithmetic coder (CABAC) \cite{cabac} to perform entropy coding on my ADU data structures. I use separate contexts for the $D$ residuals, $t$ residuals, and $s$ (bit shifts).  I reserve a symbol in the $D$ residual context to denote when the decoder must move to a different spatial unit. I variously employ this symbol to indicate that an event cube does not contain any events (similar to ``skip" blocks in H.265 \cite{h265}), that an individual pixel does not contain any events, and that we have completed encoding all the events for a pixel. When all the events in an ADU have been compressed, I encode a reserved ``end of sequence'' symbol and reset the CABAC state. In this way, I support stream scrubbing and drop in, with granularity matching the ADU interval.

\subsection{Intermediate Representations}

At the intersection of the representation and application layers, we have several methods for representing decoded \adder{} events. We can reconstruct a frame sequence for compatibility with classical vision algorithms or visual playback for humans. Alternatively, we can maintain a single image frame and simply update a certain pixel when we receive a new event at those coordinates. In both cases, we can choose to update the pixel values with only the $D$ or $\Delta t$ components of the \adder{} events, rather than their represented intensities. I offer implementation details for these intermediate representations in \cref{sec:reconstruction}.

In addition, one can convert an \adder{} video to a DVS representation. This process assumes a fixed DVS contrast threshold and determines the logarithm of incoming \adder{} event intensities. When the log intensity increases or decreases beyond the fixed threshold, we simply output a DVS-style event with a timestamp matching the beginning of the time span of the \adder{} event. This process grants \adder{} backwards compatibility with existing DVS-based vision applications.


\section{Application Layer}\label{sec:adder_application_layer}

As illustrated in \cref{fig:system_diagram_full_repeat}, we have several options for interfacing with raw \adder{} data for applications. Here, I describe these techniques and introduce example applications with performance evaluations.

\subsection{Classical Object Detection}

To explore the efficacy of \adder{} for traditional deep learning models, I conducted a preliminary investigation in using \adder-based images as input for convolutional neural networks (CNNs). Specifically, I explored object detection and classification with the YOLOv5 architecture \cite{yolov5}. This CNN architecture takes as input single image frames; it does not incorporate any sense of temporal pixel stability between images (e.g., with recurrent layers), so it cannot take advantage of data redundancy in video to improve prediction accuracy. I investigate the effect of using \adder{} data as input to the model, without changing the model itself.

The YOLOv5 model uses traditional RGB image inputs. To explore color video inputs with \adder{} would require substantial modifications for YOLO to support more than three channels of input, so I focused only on grayscale video. To make my representation compatible, I find the latest \adder{} event for each pixel. I encode the derived event intensity in the R channel, the $D$ component in the G channel, and the $\Delta t$ component in the B channel of a framed image representation. The G and B channels are normalized within the range $[0,255]$. I save these images in the PNG format for easy visualization. 

\subsubsection{Evaluation}

\begin{figure*}
     \centering
     \begin{subfigure}[t]{0.49\textwidth}
         \centering
         \includegraphics[width=\textwidth]{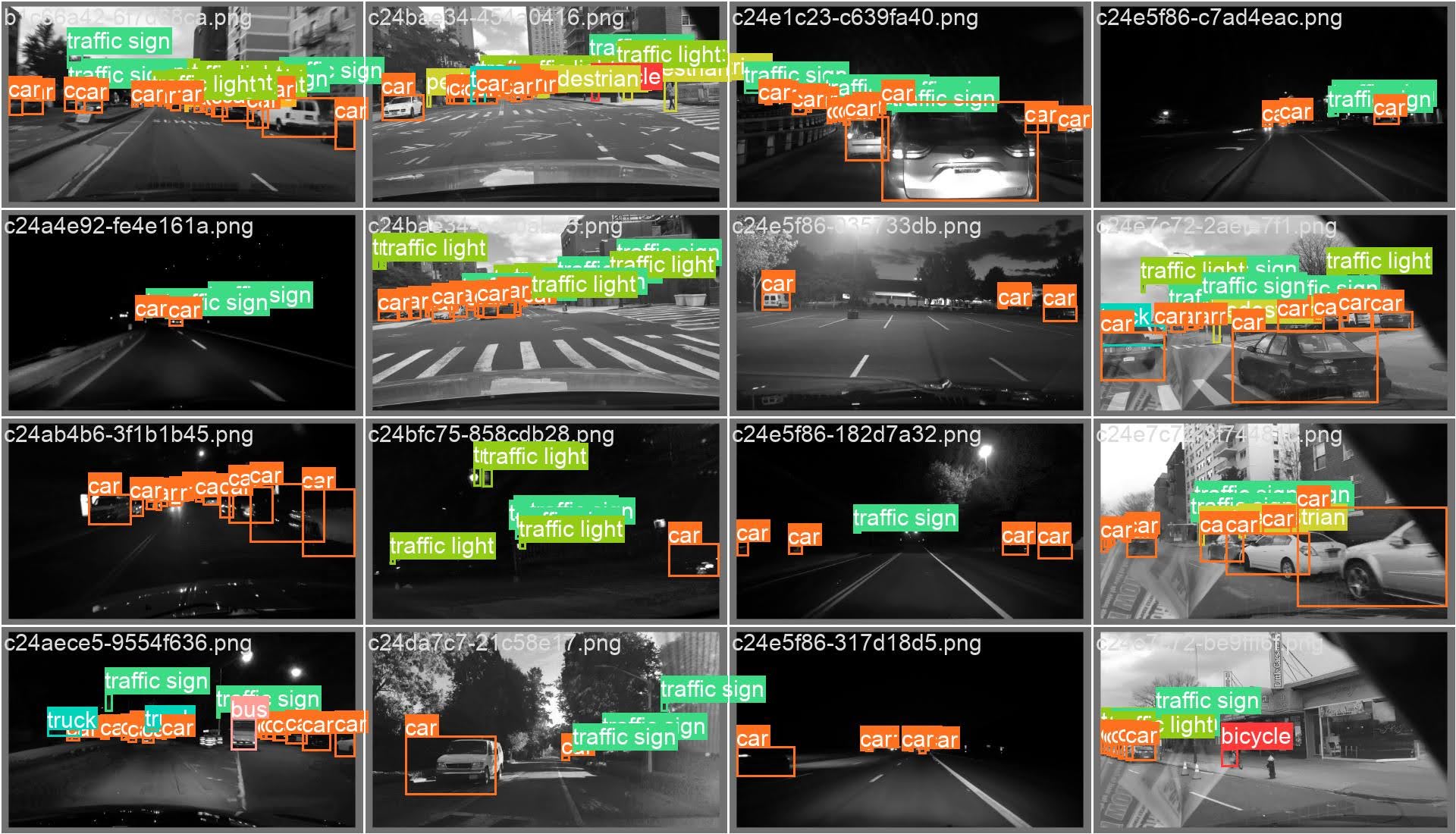}
         \caption{Ground truth}
         \label{fig:yolo_int_gt}
     \end{subfigure}
     \begin{subfigure}[t]{0.49\textwidth}
         \centering
         \includegraphics[width=\textwidth]{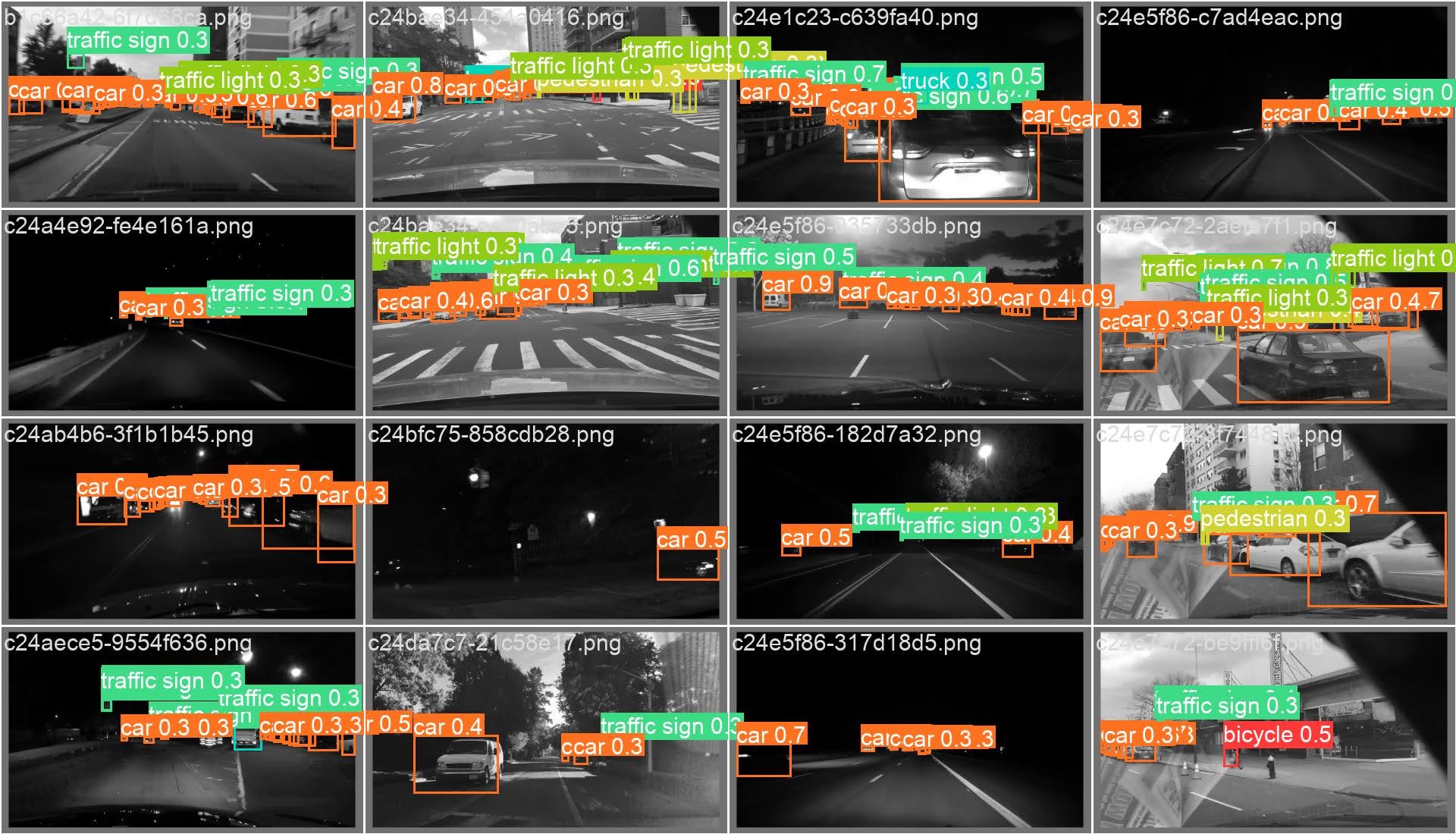}
         \caption{Predictions}
         \label{fig:yolo_int_pred}
     \end{subfigure}
     \caption[Grayscale YOLO visualizations]{Visualizations of the YOLO results on the grayscale input images.}
     \label{fig:yolo_int}
\end{figure*}

\begin{figure*}
     \centering
     \begin{subfigure}[t]{0.49\textwidth}
         \centering
         \includegraphics[width=\textwidth]{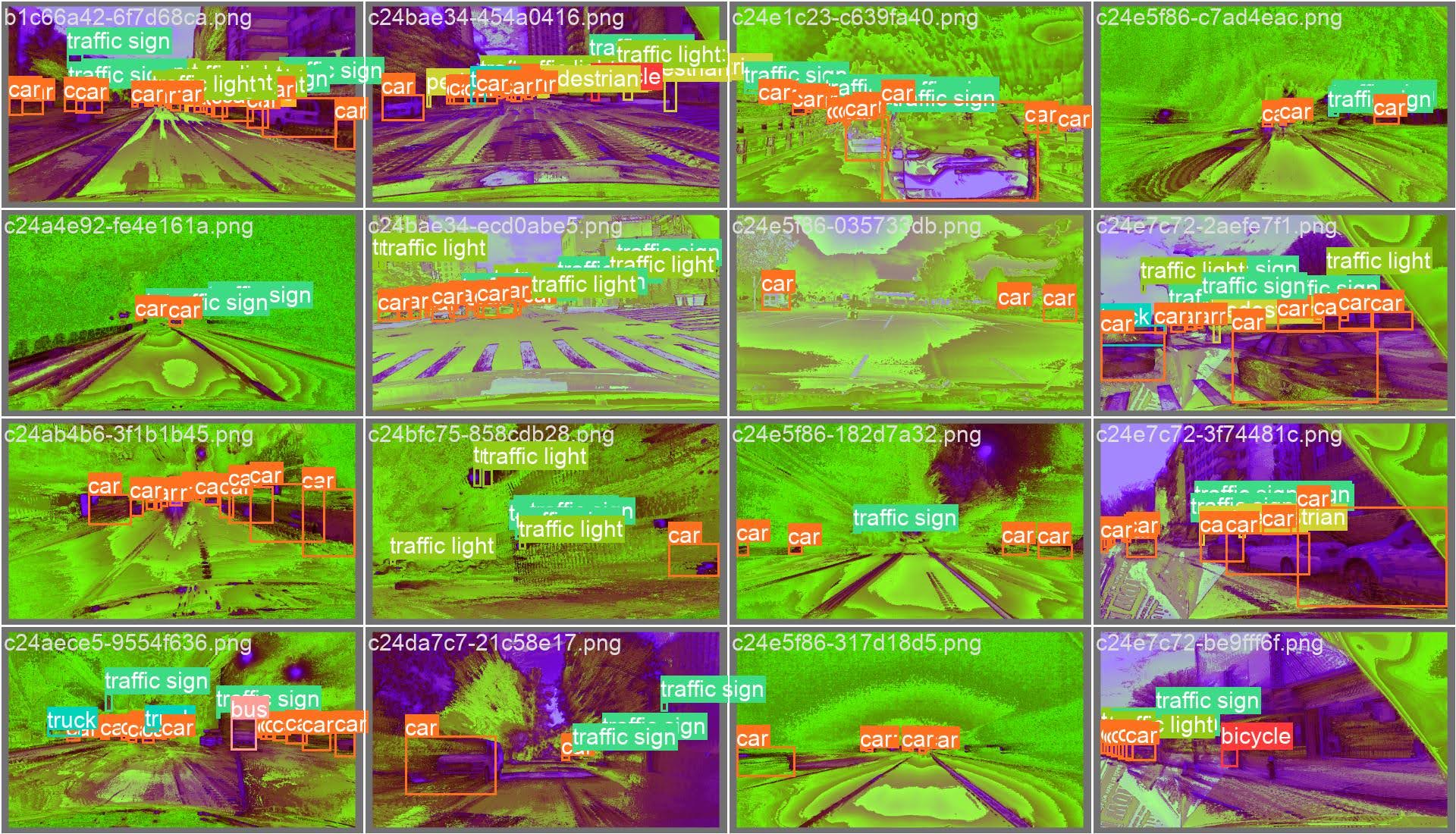}
         \caption{Ground truth}
         \label{fig:yolo_adder_gt}
     \end{subfigure}
     \begin{subfigure}[t]{0.49\textwidth}
         \centering
         \includegraphics[width=\textwidth]{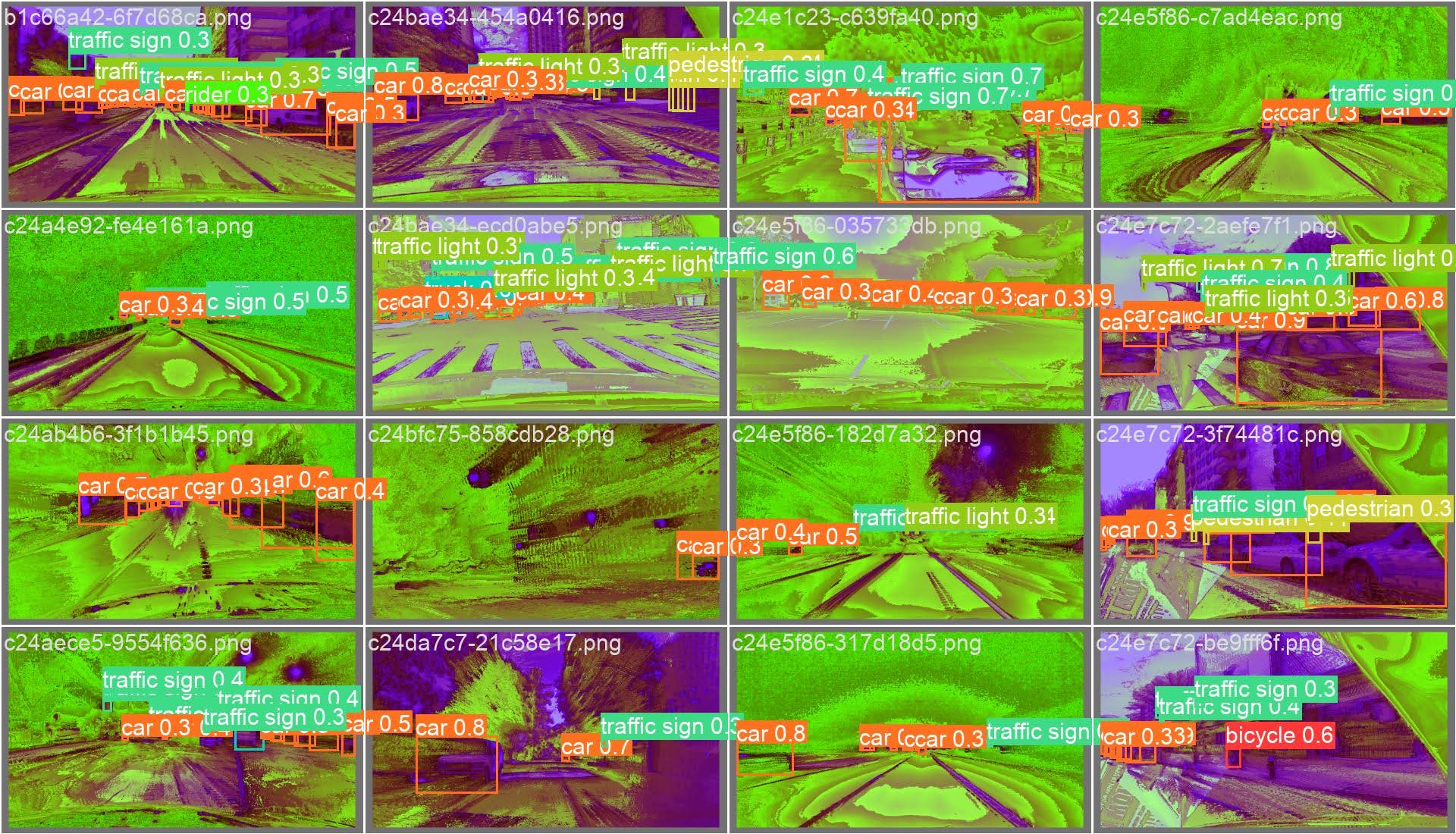}
         \caption{Predictions}
         \label{fig:yolo_adder_pred}
     \end{subfigure}
     \caption[\adder{} YOLO visualizations]{Visualizations of the YOLO results on the full \adder{} input images, with the $D$ and $\Delta t$ components encoded as separate channels.}
     \label{fig:yolo_adder}
\end{figure*}

I evaluated \adder-based object detection with YOLOv5 on the BDD100K dataset \cite{bdd100k}. This dataset consists of 100,000 videos from the forward-facing perspective of vehicles driving in the real world \cite{bdd100k}. I used the dataset's multiple-object labels and bounding boxes on one image from each video. I used a split of 70,000 videos for training, 20,000 for validation, and 10,000 for testing. For the sake of throughput with limited resources, I scaled the videos to half their original resolution (i.e., 640x360 pixels). I trained the stock (unmodified) YOLOv5 model twice: once with only the intensity component ($\frac{2^D}{\Delta t}$) of the \adder{}-transcoded data, where the R, G, and B color channels were identical (\cref{fig:yolo_int}); and once with the \adder-coded images as described above (\cref{fig:yolo_adder}). I then compared the performance of these two models on the test dataset.

\begin{figure*}
     \centering
     \begin{subfigure}[t]{0.49\textwidth}
         \centering
         \includegraphics[width=\textwidth]{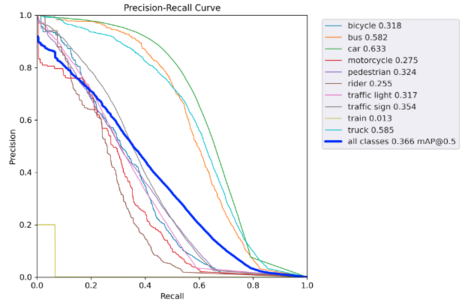}
         \caption{Grayscale images}
         \label{fig:yolo_curve_int}
     \end{subfigure}
     \begin{subfigure}[t]{0.49\textwidth}
         \centering
         \includegraphics[width=\textwidth]{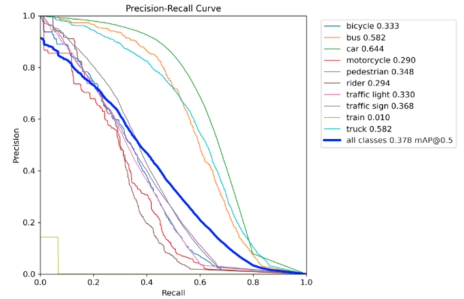}
         \caption{Full \adder{} images}
         \label{fig:yolo_curve_adder}
     \end{subfigure}
     \caption[YOLO precision-recall curves]{YOLO precision-recall curves. The numbers in the legend indicate the mAP50 scores for each class.}
     \label{fig:yolo_curves}
\end{figure*}

On the Mean Average Precision (mAP) metric at a 50\% confidence threshold (mAP50), the version of the model with separate $D$ and $\Delta t$ channels outperformed the grayscale model for all object classes in the dataset by up to 3.9\% for the \texttt{rider} class and 1.2\% overall. At a 95\% confidence threshold (mAP95), the advantage shrunk to 2.0\% for the \texttt{rider} class, but the overall improvement grew to 1.4\%. \cref{fig:yolo_curves} illustrates the precision-recall curves for both models.

These results suggest that we can improve the performance of traditional deep learning image models by simply augmenting their inputs with some notion of \textit{temporal stability}.

\subsection{Bespoke Motion Segmentation}

On the other end of the spectrum, we can develop novel methods that operate directly on \adder{} events. Here, I present a method for motion segmentation that operates on raw events.

\begin{figure*}
\centering
\subfloat[Image reconstructed from generated events]{\includegraphics[width=0.48\linewidth]{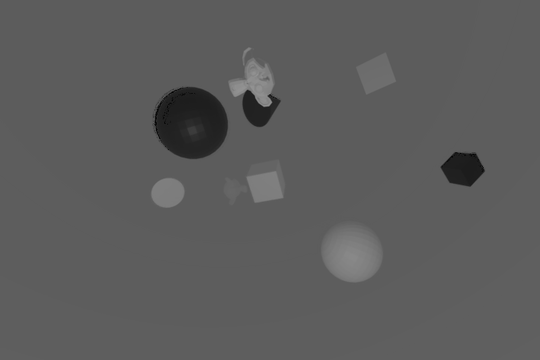}%
\label{fig:motion_a}}
\hfil
\subfloat[Binary mask thresholding event fire rates $> 2$ over this reference interval]{\includegraphics[width=0.48\linewidth]{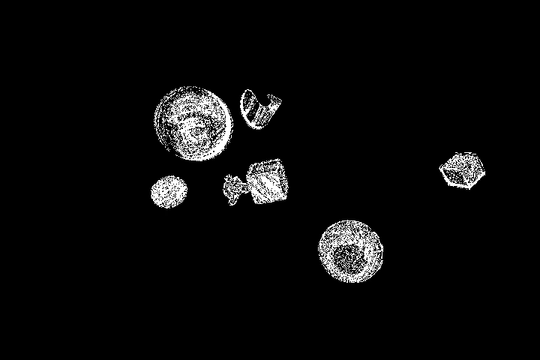}%
\label{fig:motion_b}}
\hfil
\subfloat[Closed binary mask]{\includegraphics[width=0.48\linewidth]{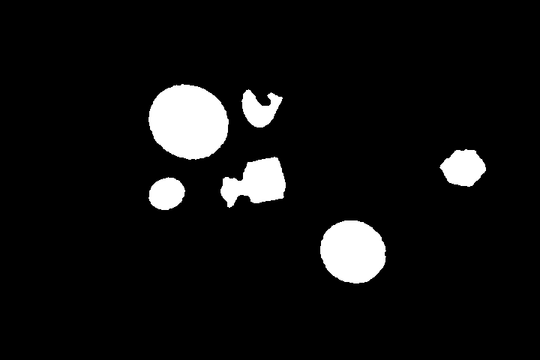}
\label{fig:motion_c}}
\hfil
\subfloat[Reconstructed frame with the mask applied]{\includegraphics[width=0.48\linewidth]{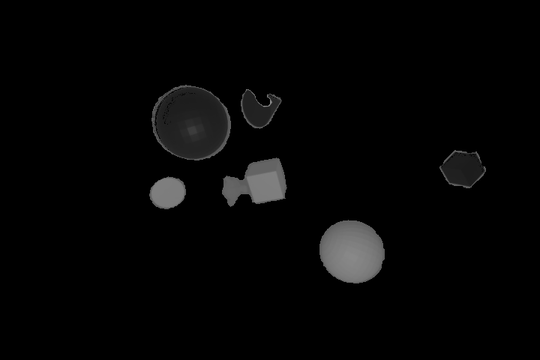}
\label{fig:motion_d}}

\caption[Motion segmentation pipeline]{Breakdown of the motion segmentation pipeline. Note that the reconstructed image (a) is not used in the calculation of (b). Rather, (b) is built from looking at the fire rate of the source events directly.}
\label{fig:motion}
\end{figure*}

A unique property of the \adder{} model combined with my decimation modes is that the sensitivity of each pixel is dependent on its previous incident intensity. This means that there is an inherent luminosity prediction for each pixel at each point in time, and the event generated by a given pixel will be determined by both this prediction \textit{and} the actual incident intensity. We can exploit this property to find areas of the image with incident intensities far from their predicted values, which is an indication of \textit{motion}. For example, if a given pixel predicts a low luminosity, it will have a low $D$-value. But if the actual incident intensity on that pixel is high, the pixel will fire multiple events as it adjusts its $D$-value to more accurately predict the new luminosity. This transition from low to high intensity indicates that there was some motion across that pixel. The same is true for a high to low intensity transition. Thus, we can merely look at the \textit{event firing rate} for each pixel over a given period of time to determine which pixels indicate movement. 

I demonstrate this phenomenon by examining one $\Delta t_{ref}$ interval of events at a time for each pixel of a simulated \adder{}-style sensor (\cref{sec:asint_sim}). This is equivalent to only examining a single still input image, except that I have crude motion data encoded in the events themselves. I build a binary matrix for the image where a pixel value is 1 if it fired more than two events over that reference interval, and its value is 0 otherwise (Fig. \ref{fig:motion_b}). This provides a noisy segmentation mask of moving objects in the scene. I clean this image with simple morphological closing (Fig. \ref{fig:motion_c}). Performing this process on a large sequence of events, I produce a video depicting binary segmentation masks for moving objects in the scene (Fig. \ref{fig:motion_d}). I note that this approach requires setting the camera's $\Delta t_{max}$ parameter appropriately to ensure that motion is captured at the desired temporal granularity. This yields accurate segmentations without the need for complex, slow methods based on multi-frame image analysis, motion vectors, and image gradients on reconstructed images. Existing approaches to event-driven motion segmentation rely on applying existing computer vision techniques to reconstructed video \cite{event_cv}, or on complex iterative optimization functions \cite{compensation}. The simplicity of my approach is due to the dynamic, per-pixel sensitivity adjustment that the \adder{} model provides, which the Dynamic Vision System cameras fundamentally lack. The result is a novel motion segmentation method that operates directly on camera events in a fast, non-iterative fashion.

\subsection{Rate Control with Online Applications}

\subsubsection{Object Tracking Rate Control for Simulated Sensor}

\begin{figure}[ht]
    \centering
    \includegraphics[width=\linewidth]{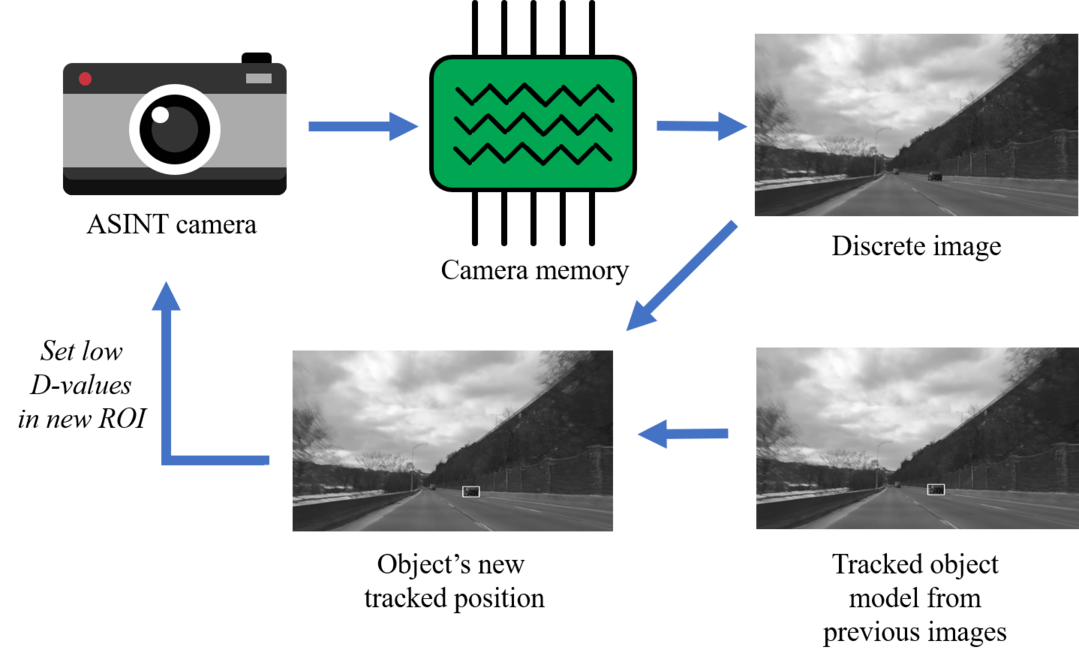}
    \caption[Diagram of the object tracking pipeline]{Diagram of the object tracking pipeline with classical vision algorithms.}
    \label{fig:tracking_diagram}
\end{figure}

I first focus on the case of a simulated \adder-style sensor (ASINT), as I introduced in \cref{sec:asint_sim}. As described by Singh et al. \cite{montek}, the most recent event for each pixel can be stored in a specially designed memory. Therefore, we can take advantage of the data format for events to maintain a live image directly from the most recent events without having to perform framed reconstruction. This image will be less temporally accurate than a framed reconstruction, but is faster to process since I only need one event per pixel, per sample. This also allows us to have drastically varying event fire rates across the scene. That is, we can have fine detail and dynamic range on the parts of the scene we are most interested in, while sacrificing some of that detail in the rest of the scene. While it is trivial to define a fixed subregion of the sensor array for which to keep $D$-values low, it is difficult to dynamically change the region of interest according to the content in the scene. The following introduces a system for adjusting the pixel sensitivites of an \adder-style sensor based on tracked object positions. I outline the system in \cref{fig:tracking_diagram}.

\paragraph{Aggressive Self-Adjustment Decimation Mode}\label{sec:aggressive}
To drastically reduce the event fire rate for certain parts of the sensor, I devised a new pixel decimation adjustment scheme. In this $D$-value adjustment mode, each pixel compares $\Delta t$ of the last event it fired to the global $\Delta t_{max}$. If $2\Delta t < \Delta t_{max}$, I increase $D$ by 1. If the pixel fires an empty event ($\Delta t \geq \Delta t_{max}$), I simply decrease $D$ by 1. This mode is useful for minimizing the number of events fired by a pixel in my object tracking application, where overall reconstructed image quality is not a primary goal.

\paragraph{Camera Parameters and Pixel Decimation}

Since we only need high-speed information on the object we are tracking, I can increase the $D$-values of the other pixels to greatly reduce the number of events fired. However, I want to keep the $D$-values of pixels within the region of interest (ROI) low enough that I can obtain accurate measurements of the object with a fine sample rate. I set the $\Delta t_{max}$ camera parameter higher than the target sample rate for the ROI. For pixels outside the ROI, I use the aggressive self-adjustment mode of decimation control as described in Sec. \ref{sec:aggressive}, so that those events will tend to be closer to $\Delta t_{max}$ and fire at a lower rate than the pixels in the ROI. By contrast, I target a higher fire rate relative to the user-specified reference interval, $\Delta t_{ref}$, for the pixels in the ROI.

\paragraph{Discrete Image Sampling}
We can periodically sample the camera memory to obtain the most recently fired event for each pixel. We can calculate the incident intensity from each of these events to obtain a discrete image. If we set an ROI, then pixels outside the ROI may fire slower than the sample frequency. These pixels change more slowly across the discrete image sequence, compared to the pixels in the ROI which will typically exhibit a new event for each sample.

\begin{figure}
    \centering
    \includegraphics[width=\linewidth]{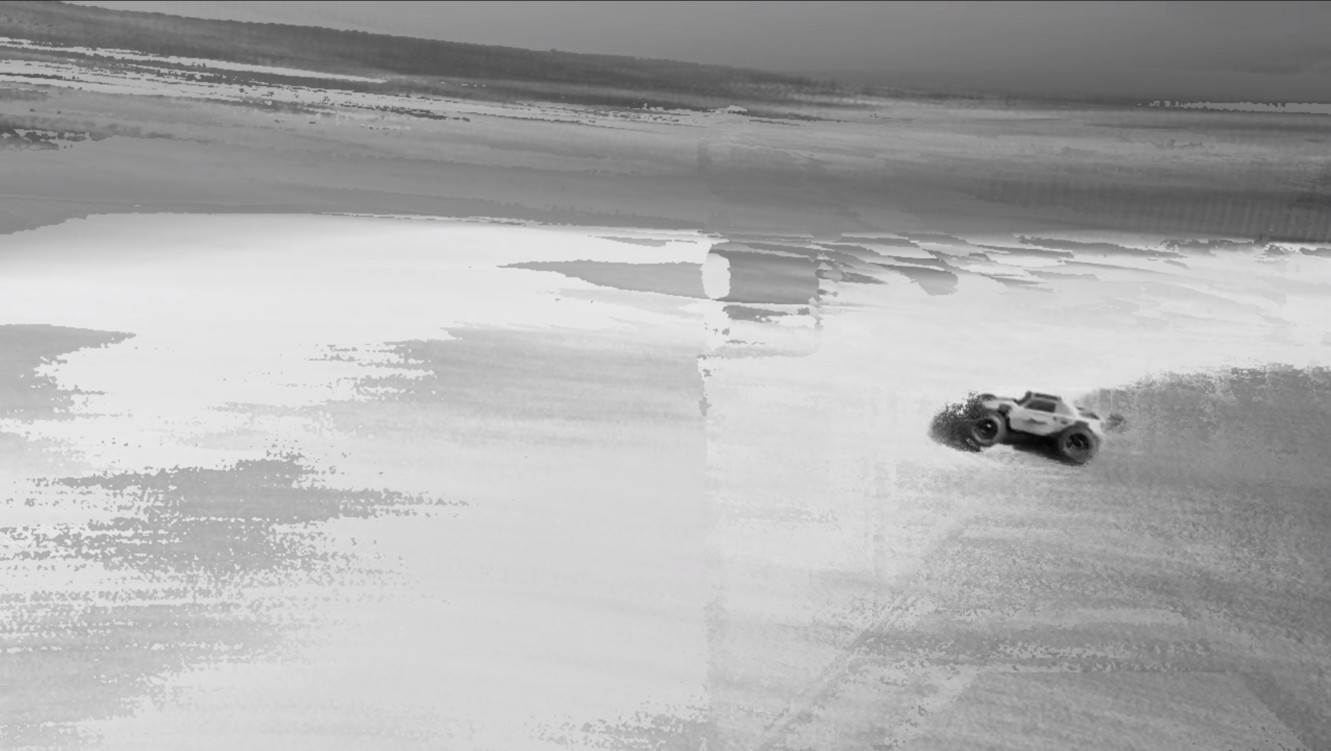}
    \caption[Temporal foveation]{Temporal foveation demonstrated on the NFS \textit{rc\_car\_rolling} scene. The pixels closest to the tracked car have a higher temporal resolution than the pixels further away, creating the smearing effect on this image reconstructed from the emulated events.}
    \label{fig:rc_car_rolling}
\end{figure}

In practice, I augment the aggressive $D$-control mechanism with an ``ROI factor," $r$, that adjusts the level of aggression on a per-pixel basis dependant on that pixel's proximity to the ROI. $r$ is an integer wherein a larger value indicates the pixel is closer to the ROI. Then for pixels outside the ROI, the maximum $\Delta t$ value by which I determine how aggressively to adjust their $D$-value is calculated as $\frac{\Delta t_{max}}{r}$. This results in the temporal resolution of the scene decreasing progressively as the pixel's distance from the ROI increases. I term the resulting effect \textbf{temporal foveation}, although the appearance of reconstructed image frames looks similar to that of traditional spatial foveation. Fig. \ref{fig:rc_car_rolling} shows an example of this effect from my experiment.

If we sample the memory at a frequency of $\frac{C}{\Delta t_{ref}}$ samples per second, we obtain a sequence of images with that frame rate, wherein the effective shutter speed for pixels \textit{inside} the ROI is approximately $\frac{\Delta t_{ref}}{C}$ seconds and the effective shutter speed for pixels \textit{outside} the ROI is approximately $\frac{\Delta t_{max}}{C\cdot r}$ seconds.

\paragraph{Classical Object Tracking}
With a sequence of discrete images, we can perform object tracking using classical computer vision techniques, which operate on traditional images. We can keep pixel sensitivities high until the object to be tracked has been detected, then set the ROI to contain this detected object. Following object detection, the only part of the scene updating at the high input frame rate is the part captured by the ROI. An object tracking algorithm can detect the change in position of the object between discrete image samples, and correspondingly update the ROI.

\paragraph{Pixel Sensitivity Adjustment}
When I update the ROI, I transmit this information back to the ASINT camera to affect the next $D$-values of pixels. In this way, I have a feedback loop where the object tracking directly informs the camera which pixels to prioritize in data capture, which then helps maintain the object tracking accuracy at high frame rates. If the object tracking is performed at a low frame rate, or if the tracked object is moving very quickly, I can set the ROI to be larger than the size of the object to account for larger object displacements between samples.

\paragraph{Experimental Evaluation}
I implemented my object tracking framework with the ASINT emulator \cref{sec:asint_sim}, using multiple instance learning (MIL) tracking in OpenCV for C++ \cite{opencv}. I employed the Need for Speed dataset \cite{need_for_speed}, which provides 240 FPS videos with ground truth bounding box data for object tracking. To isolate the object \textit{tracking} piece of the framework, I took object detection as prior knowledge by initializing my tracker with the first bounding box from the dataset ground truth.

\begin{table*}[!htp]\centering
\caption[AUC results]{AUC results. A positive AUC score difference indicates that the ground truth MIL tracking was more accurate, whereas a negative score difference indicates that tracking on temporally foveated ASINT samples had higher accuracy.}\label{tab:results_table}
\scriptsize
\begin{tabular}{lp{1.7cm}p{1.7cm}p{1.7cm}p{2.6cm}p{2.6cm}}\toprule
\textbf{Name} &\textbf{Baseline AUC} &\textbf{ASINT tracked AUC} &\textbf{AUC Score Difference}  \\\midrule
Gymnastics &83.39 &74.97 &8.43 \\
bunny &96.39 &95.97 &0.41 \\
car &89.92 &88.76 &1.16 \\
car\_camaro &30.50 &31.49 &-0.98 \\
car\_drifting &22.84 &18.09 &4.75 \\
car\_jumping &3.04 &3.04 &-0.01 \\
car\_rc\_rolling &53.68 &55.87 &-2.19 \\
car\_rc\_rotating &5.95 &35.98 &\textbf{-30.03} \\
car\_side &65.24 &63.70 &1.54 \\
cheetah &91.34 &90.78 &0.56 \\
cup &78.00 &76.92 &1.07 \\
cup\_2 &90.23 &89.61 &0.62 \\
dog &40.01 &43.57 &-3.56 \\
dog\_1 &57.60 &57.84 &-0.24 \\
dog\_2 &65.09 &62.11 &2.97 \\
dog\_3 &70.17 &71.80 &-1.63 \\
dogs &72.30 &63.24 &9.06 \\
dollar &94.93 &94.69 &0.25 \\
drone &5.97 &11.78 &-5.80 \\
ducks\_lake &57.28 &51.76 &5.52 \\
exit &64.72 &78.54 &\textbf{-13.82} \\
first &92.59 &91.31 &1.28 \\
flower &86.79 &88.24 &-1.46 \\
footbal\_skill &55.34 &25.63 &29.71 \\
helicopter &73.71 &65.56 &8.15 \\
horse\_jumping &46.23 &38.03 &8.21 \\
\bottomrule
\end{tabular}
\end{table*}

I ran my tracking program on the first 200 frames ($\approx0.833$ seconds) of video for 26 scenes from the dataset. I used the aggressive decimation adjustment mode described in Sec. \ref{sec:aggressive} with a variable $r$ to keep pixels near the ROI relatively sensitive. This allowed me to account for future motion and maintain my tracking. I discretely sampled the emulated camera memory at a rate of 240 Hz (once per input frame) and performed tracking on these discrete samples. My emulation parameters, with time units given in ticks, were:

\begin{itemize}
    \item $\Delta t_s$ (ticks per second): 12000
    \item $\Delta t_{frame}$: $12000 / 240 = 50$ 
    \item $\Delta t_{ref}$: 50
    \item $\Delta t_{max}$: 2500
    \item Decimation mode: aggressive
\end{itemize}

I calculated the bounding box overlap score compared to ground truth, and found the area under the curve (AUC) score from the overlap scores \cite{object_tracking}. To isolate the effects of my camera in the tracking framework, I examine the AUC score \textit{difference} between my scheme and the same MIL tracking algorithm run on the ground truth image frames. These results are displayed in Table \ref{tab:results_table}.

The average AUC score difference was 0.92. Since the worst possible AUC score difference would be $100$, and the best possible score difference would be $-100$, the average score difference result suggests that tracking objects in image samples that are temporally foveated under my scheme performs nearly identically to tracking objects on in traditional framed video. This aligns with my expectations, since my scheme was designed to account for the fact that the only visual information necessary to track an object is the part of the scene that actually contains the object. My method allows us to throw away unnecessary temporal information for pixels far from the tracked object, while being able to recover greater temporal measuring precision for those pixels when the object moves closer to them spatially. This allows us to perform much better than the ground truth in the {\fontfamily{qcr}\selectfont car\_rc\_rotating} and {\fontfamily{qcr}\selectfont exit} scenes, which have fast camera movement and low contrast, respectively. By contrast, my method performs very poorly in the {\fontfamily{qcr}\selectfont footbal\_skill} scene, due to its fast movement of a small object with large displacement. In this case, my $D$-adjustment scheme raised the $D$-values of the pixels too high for them to adjust quickly when the object moved across them, and I lost my tracking accuracy quickly.

\subsubsection{Asynchronous Feature Detection and Feature-Driven Rate Control for Online Transcoding}\label{sec:feature_rate_control}

A driving motivation for my work is to enable the development of faster video analysis applications and content-based rate adaptation. To explore the utility of \adder{} in an end-to-end system, I adapted the FAST feature detection algorithm for the asynchronous paradigm \cite{fast_features}.  The FAST feature detector examines pixels that lie along a circle around a given candidate pixel. For a given streak size $n$ (e.g., 9 pixels), I say that the candidate pixel is a \textit{feature} if at least $n$ contiguous pixels in the circle exceed the candidate pixel's value plus or minus some pre-determined threshold.

\begin{figure*}
        \centering
        \includegraphics[width=\linewidth]{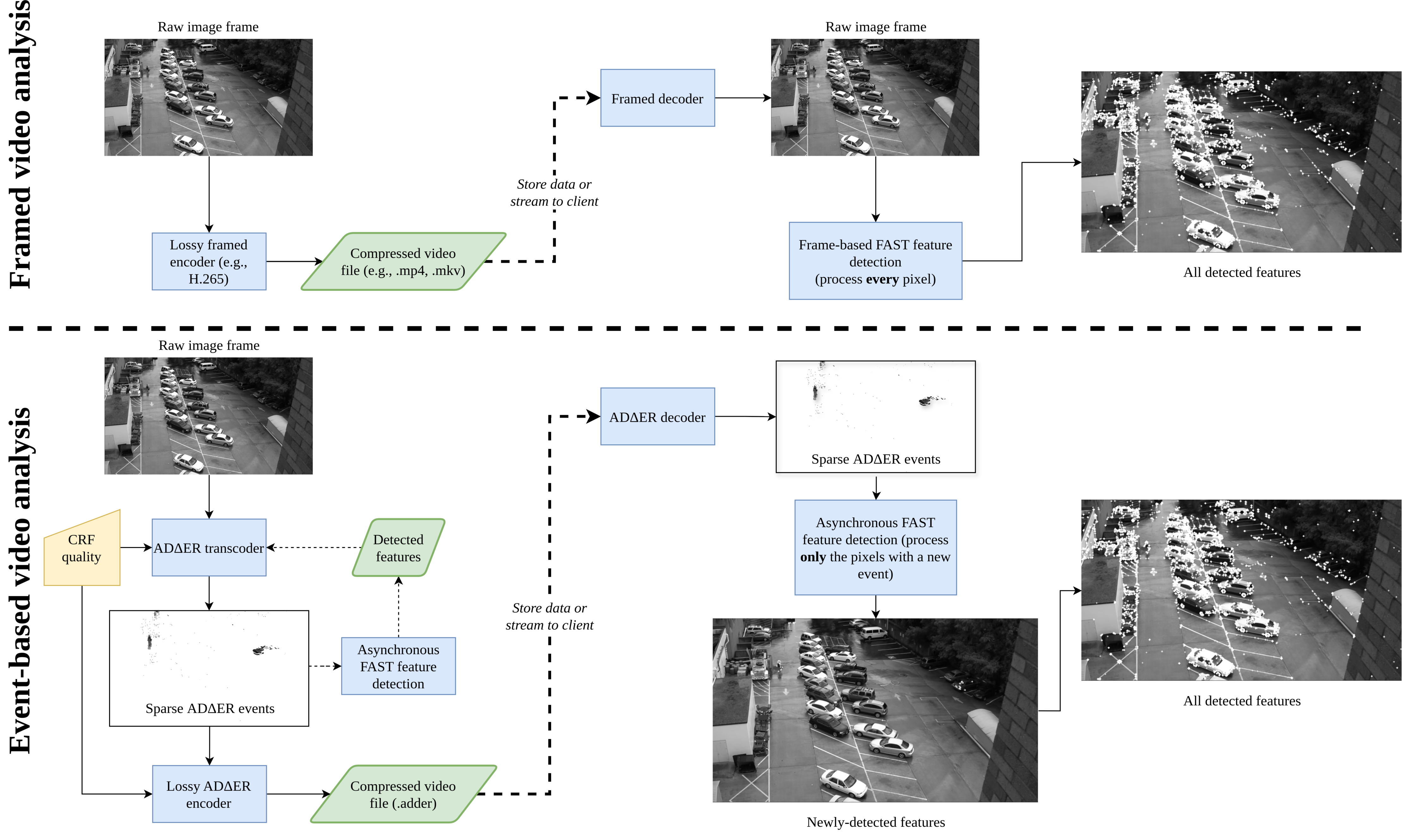}
       \caption[Feature detection pipelines]{Comparison between FAST feature detection in a classical video system (top) and my \adder-based system (bottom). With \adder{}, the decompressed representation is itself sparse, meaning that the application has much less data to process.}
        \label{fig:framed_adder_comparison}
\end{figure*}

In a traditional video analysis pipeline, the input to an application such as the FAST detector is an entire decompressed image frame, as illustrated in \cref{fig:framed_adder_comparison}. Then, the feature detector iterates through every pixel in the image and tests it as a candidate feature. In a video context, we may visit and test many pixels which have not changed since the previous image frame. One could first calculate the pixelwise difference between the current and previous frames, and run feature detection only on the pixels that have changed, but this operation comes with its own computational costs on the decoder end.

With \adder{}, by contrast, \textit{the decompressed representation is already sparse}. My application layer can simply keep a single reconstructed intensity image in memory, and update an individual pixel value for each new event that it receives. When the application ingests a new event, the feature detector may test that \textit{one pixel}, rather than all the pixels in an image.  \cref{fig:framed_numbers} shows that \adder{} can reduce the data burden for downstream surveillance video applications by more than 90\%.

\paragraph{Implementation Details}\label{sec:implementation_fast}

I used the OpenCV \cite{opencv_library} implementation of the FAST feature detector as my reference. The algorithm includes an iterator loop through all pixels in an image, so I adapted only the interior portion to apply the operation to a single pixel with an index argument. I ported the algorithm to Rust for interoperability with the rest of the \adder{} codebase. For this dissertation, I did not explore an asynchronous implementation of non-maximal feature suppression, which is an optional filtering step in the OpenCV implementation. To verify that the choice of programming language by itself does not produce a performance gain, I tested the speed of my Rust algorithm in processing entire image frames, synchronously. I found that my synchronous Rust algorithm is 6-10 times \textit{slower than} its OpenCV counterpart (written in C++), owing to the high optimization level that the OpenCV project has achieved.

\paragraph{Feature-Driven Rate Control}

\begin{figure}
     \centering
     \begin{minipage}{.24\linewidth}
            \begin{subfigure}[t]{\linewidth}
         \centering
         \includegraphics[width=\linewidth]{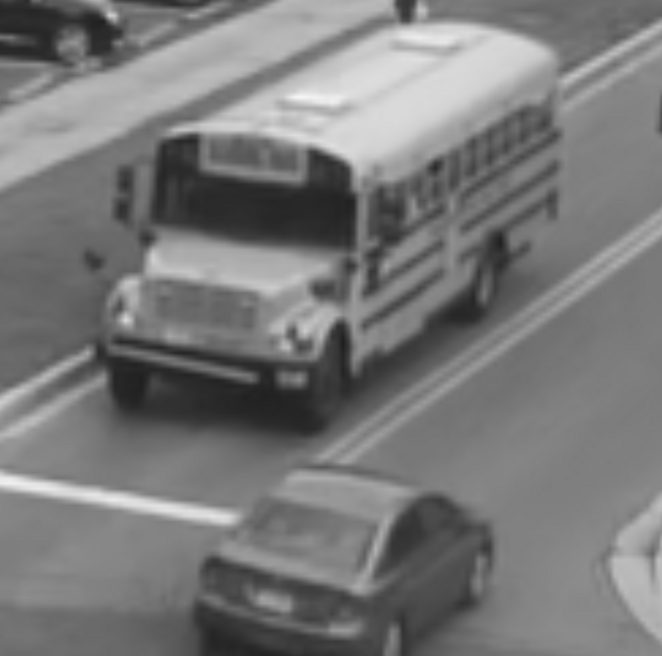}
         \caption{H.265-compressed input}
         \label{fig:source}
         \end{subfigure}
    \end{minipage}
    \begin{minipage}{.74\linewidth}
    \begin{subfigure}[t]{0.32\linewidth}
         \centering
         \includegraphics[width=\linewidth]{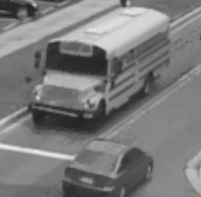}
         \caption{Reconstructed intensities}
         \label{fig:nofeat_int}
     \end{subfigure}
    \begin{subfigure}[t]{0.32\linewidth}
         \centering
         \includegraphics[width=\linewidth]{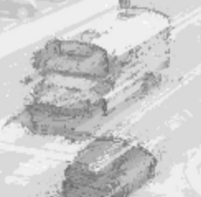}
         \caption{Event \textit{D} values}
         \label{fig:nofeat_d}
     \end{subfigure}
     \begin{subfigure}[t]{0.32\linewidth}
         \centering
         \includegraphics[width=\linewidth]{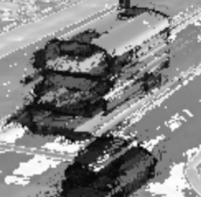}
         \caption{$\Delta t$ between events}
         \label{fig:nofeat_t}
     \end{subfigure} \\
     \begin{subfigure}[t]{0.32\linewidth}
         \centering
         \includegraphics[width=\linewidth]{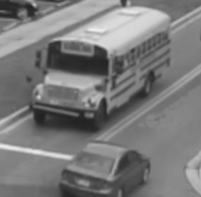}
         \caption{Reconstructed intensities}
         \label{fig:feat_int}
     \end{subfigure}
     \begin{subfigure}[t]{0.32\linewidth}
         \centering
         \includegraphics[width=\linewidth]{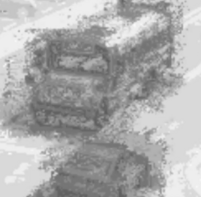}
         \caption{Event \textit{D} values}
         \label{fig:feat_d}
     \end{subfigure}
     \begin{subfigure}[t]{0.32\linewidth}
         \centering
         \includegraphics[width=\linewidth]{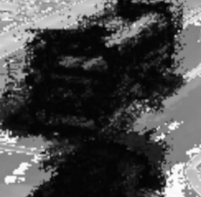}
         \caption{$\Delta t$ between events}
         \label{fig:feat_t}
     \end{subfigure}
    \end{minipage}
     
     \caption[Qualitative effect of feature-driven rate adaptation]{Zoomed-in view of the effect of feature-driven rate adaptation during transcode. (a) shows the input to the transcoder, which was compressed with H.265 at CRF level 23. (b)-(d) show views of the events transcoded under the \texttt{Low} quality setting. (e)-(g) show views of the same transcode setting and feature-driven rate adaptation enabled, as described in \cref{sec:feature_rate_control}. The $D$ and $\Delta t$ images are normalized, such that darker pixels correspond to smaller $D$ and $\Delta t$, respectively.}
     \label{fig:crf_zoom}
\end{figure}

I can achieve significant compression of the raw format by having high contrast thresholds while \textit{transcoding} a video to \adder{}, as I discussed with my CRF parameter in \cref{sec:crf}. Conversely, we may allocate more bandwidth to regions of high salience by \textit{lowering} the contrast thresholds for individual pixels in those regions. As illustrated in \cref{fig:framed_adder_comparison}, I can execute an application such as my asynchronous feature detector not only during video playback, but also during the transcoding process. For the latter case, I devised a straightforward scheme which throttles down the contrast threshold, $M$, of all pixels within a certain distance to a newly detected feature. This distance is determined by the lookup table for the global CRF parameter (\cref{sec:crf}), where higher CRF values correspond to smaller adjustment radii. Since thresholds increase over time, as described in \cref{sec:contrast_thresholds}, these pixels will gradually lower their event rate if their intensities are stable. By this, we may easily choose to prioritize high quality in the spatiotemporal regions known to be of application-level importance. \cref{fig:crf_zoom} shows a \texttt{Low} quality transcode with and without feature-driven rate adaptation enabled.

\paragraph{Dataset and Experiments}\label{sec:dataset_experiments}
I utilized the VIRAT video surveillance dataset \cite{virat}. This dataset contains sequences from stationary cameras recording the movements of people and vehicles in outdoor public spaces. I randomly sampled 132 videos from the dataset and re-encoded them for greater throughput and data diversity with my experiment. I scaled each video to $640\times 360$ resolution, converted it to single-channel grayscale, and encoded the first 480 frames in H.265 with FFmpeg \cite{ffmpeg}. I instructed FFmpeg to use CRF value 23 for moderate loss and have an I-frame interval of 30. 

I fed these H.265-encoded videos as input to the \adder{} transcoder. I transcoded each video at \adder{} CRF \texttt{Lossless}, \texttt{High}, \texttt{Medium}, and \texttt{Low} settings. I ran each transcode level both with and without the feature-driven rate control mechanism described in \cref{sec:feature_rate_control}, for a total of eight experiments on each video. In all cases, I set $\Delta t_{ref} = 255$ ticks and $\Delta t_{adu} = \Delta t_{max} = 7650$ ticks. That is, each compressed \adder{} ADU spans $7650/255=30$ input frames, matching the I-frame interval of the source videos. Notable metrics I collected were feature detection speed, framed reconstruction quality before and after source-modeled arithmetic coding, and compression performance. I ran all experiments on an AMD Ryzen 2700X CPU with 8 cores and 16 threads. The OpenCV implementation of FAST feature detection is single-threaded, so I executed that portion of my system on a single thread for the sake of fair comparison.

\paragraph{Results}

I illustrate various metrics for a single video in \cref{fig:example,fig:example2}. I found that the periodic spikes apparent in \cref{fig:ex_bitrate,fig:ex_mse,fig:ex_feat_speed} are due to the I-frame interval of the H.265-encoded source. Large quality changes occur in the source encoding every 30 input frames, and this leads to a corresponding jump in \adder{} data rate, especially at higher quality levels. \cref{fig:ex_feat_num} illustrates that the feature detector may take up to \dtm{} ticks to detect the first instance of features located in stable pixel regions. In \cref{fig:ex_feat_speed,fig:ex_feat_num}, I compare the performance of running the frame-based OpenCV feature detector on a reconstruction of the raw \adder{} stream at \texttt{Low} quality. For visualization purposes, I do not show the OpenCV results at other quality levels, but the results were virtually identical. The key point is that the frame-based application speed is nearly constant, and does not adapt to the underlying video content, whereas the \adder{} implementation speed can vary widely depending on the number of events. 

\begin{figure}
        \centering
        \includegraphics[width=\linewidth]{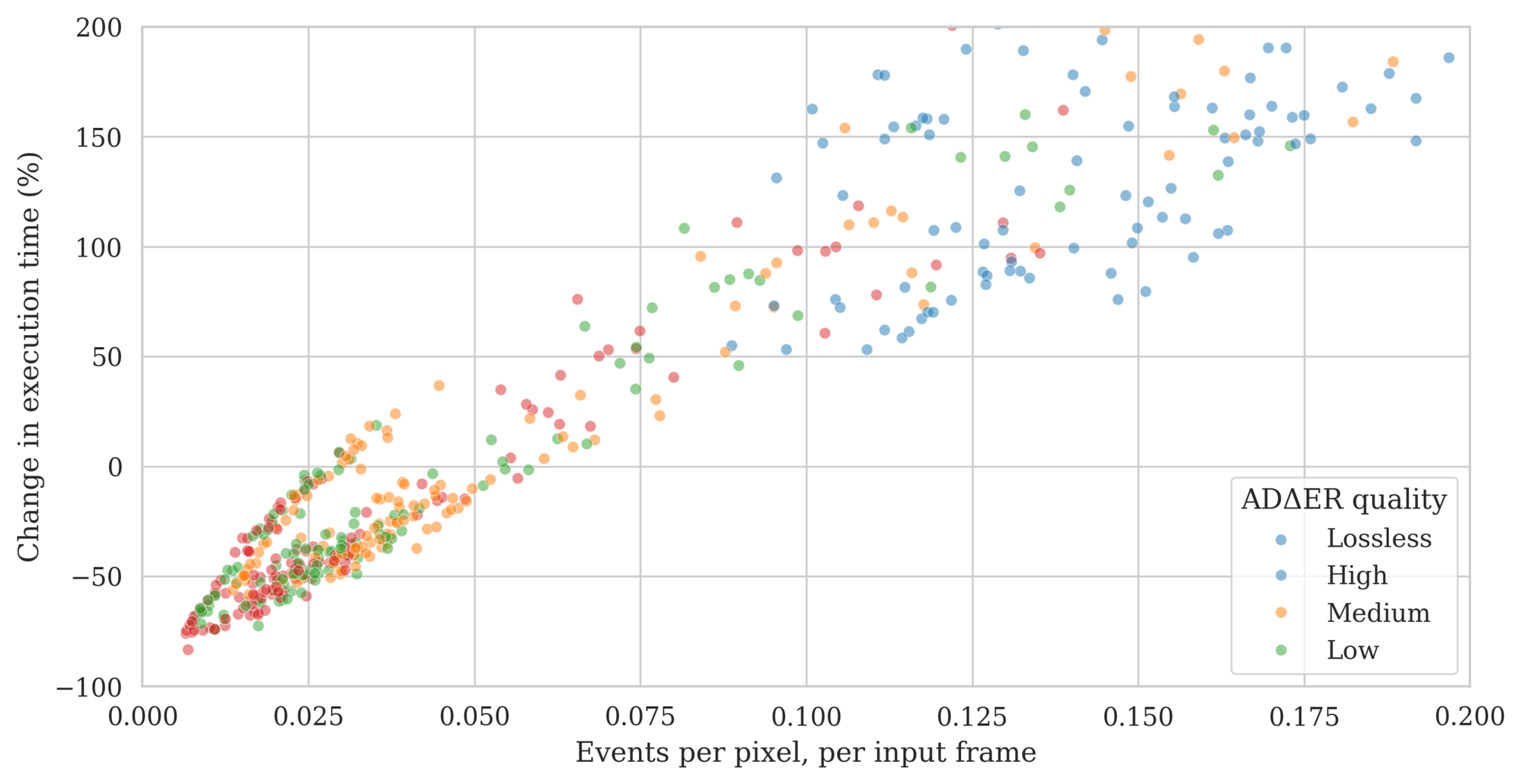}
       \caption[Even rate vs. feature detection speed]{The effect of event rate on FAST feature detection speed, compared to the frame-based OpenCV  implementation. If there is less than 1 \adder{} event for every 40 pixels, the asynchronous FAST detector is faster.}
        \label{fig:all_speed_vs_event_rate}
\end{figure}

Through qualitative examination, I found that the \adder{} compression performance is highly dependent on the amount of motion in a video. I divide the videos into four motion categories, based on the difference between the H.265 bitrate and the compressed bitrate of the \texttt{Low} \adder{} quality transcode without feature detection. These motion categories are \texttt{Low} (lower bitrate than H.265), \texttt{Medium} (within $1\times$-$2\times$ the H.265 bitrate), \texttt{High} (within $2\times$-$3\times$ the H.265 bitrate), and \texttt{Very High} ($3\times$ the H.265 bitrate or greater). Of the 132 videos in my dataset, this division placed 7 videos in the \texttt{Low} motion class, 67 in \texttt{Medium} motion, 25 in \texttt{High} motion, and 33 in \texttt{Very High} motion. The \texttt{Very High} motion videos tended to show moving people or vehicles close to the camera, or high wind activity causing the camera and foliage to move substantially.

\begin{table*}
    \centering
    \begin{tabular}{cc|llll|l}
    & & \multicolumn{5}{c}{Median reconstructed PSNR (dB)}  \\
         & \diagbox[width=6em]{Quality}{Motion}& \texttt{Low} & \texttt{Medium} & \texttt{High} & \texttt{Very High} &  All \\
        \hline
         \parbox[t]{2mm}{\multirow{3}{*}{\rotatebox[origin=c]{90}{Normal}}}
         & \texttt{High}&  40.2& 40.9& 40.7& 39.2& 40.5\\
         & \texttt{Medium}& 38.8& 39.4& 39.5& 35.4& 38.8\\
         & \texttt{Low}&  37.8& 37.9& 38.4& 32.2& 37.3\\
         \hline
         \parbox[t]{2mm}{\multirow{3}{*}{\rotatebox[origin=c]{90}{FAST}}} 
         & \texttt{High}  & 44.0 (+3.8)& 44.7 (+3.8)& 44.3 (+3.6)& 45.2 (+6.0)& 44.6 (+4.1)\\
         & \texttt{Medium}& 42.3 (+3.5)& 42.3 (+2.9)& 41.6 (+2.1)& 41.4 (+6.0)& 42.0 (+1.2)\\
         & \texttt{Low}   & 41.4 (+3.6)& 41.3 (+3.4)& 40.7 (+2.3)& 37.9 (+5.7)& 40.9 (+3.6)\\
    \end{tabular}
    \caption[PSNR \adder{} vs. H.265]{Median PSNR of the compressed \adder{} videos compared to the input H.265 video. I reconstructed framed videos from the \adder{} representations to evaluate the PSNR. The columns indicate the motion category, while the rows indicate the \adder{} transcoder quality under both the standard method and with rate adaptation based on FAST feature detection. I indicate in parentheses that the use of my feature-driven adaptation mechanism described in \cref{sec:feature_rate_control} increases the quality.}
    \label{tab:recon_psnr}
\end{table*}

For the majority of videos tested at the \texttt{Low} quality setting, the \adder{} compression performance is less than $2\times$ that of the H.265-encoded source video, while maintaining a high PSNR value. This result is significant since my compression scheme is naive compared to the advanced techniques of modern frame-based codecs. Specifically, my current scheme does not employ variable block sizes, motion compensation, or frequency transforms. Thus, I expose the inefficiency of frame-based methods for compressing video with high temporal redundancy, and I expect to handily surpass these codecs with future development of my compression scheme. While the feature-driven rate adaptation improves the overall PSNR, as shown in \cref{tab:recon_psnr}, I note that the quality improvement is by design centered around regions with moving features. I visualize this in \cref{fig:crf_zoom}, showing the effect of feature detection on reconstruction quality. When features are detected on the moving vehicles, the transcoder makes nearby pixels more sensitive, so that \cref{fig:feat_int} avoids the temporal smoothing artifacts present in \cref{fig:nofeat_int}. However, we see that the pixels far from the moving bus (where new features were detected) maintain a lower quality due to temporal averaging, and artifacts in those regions are still visible in \cref{fig:feat_int}.

\begin{figure*}
     \centering
     \begin{subfigure}[t]{0.49\textwidth}
         \centering
         \includegraphics[width=\textwidth]{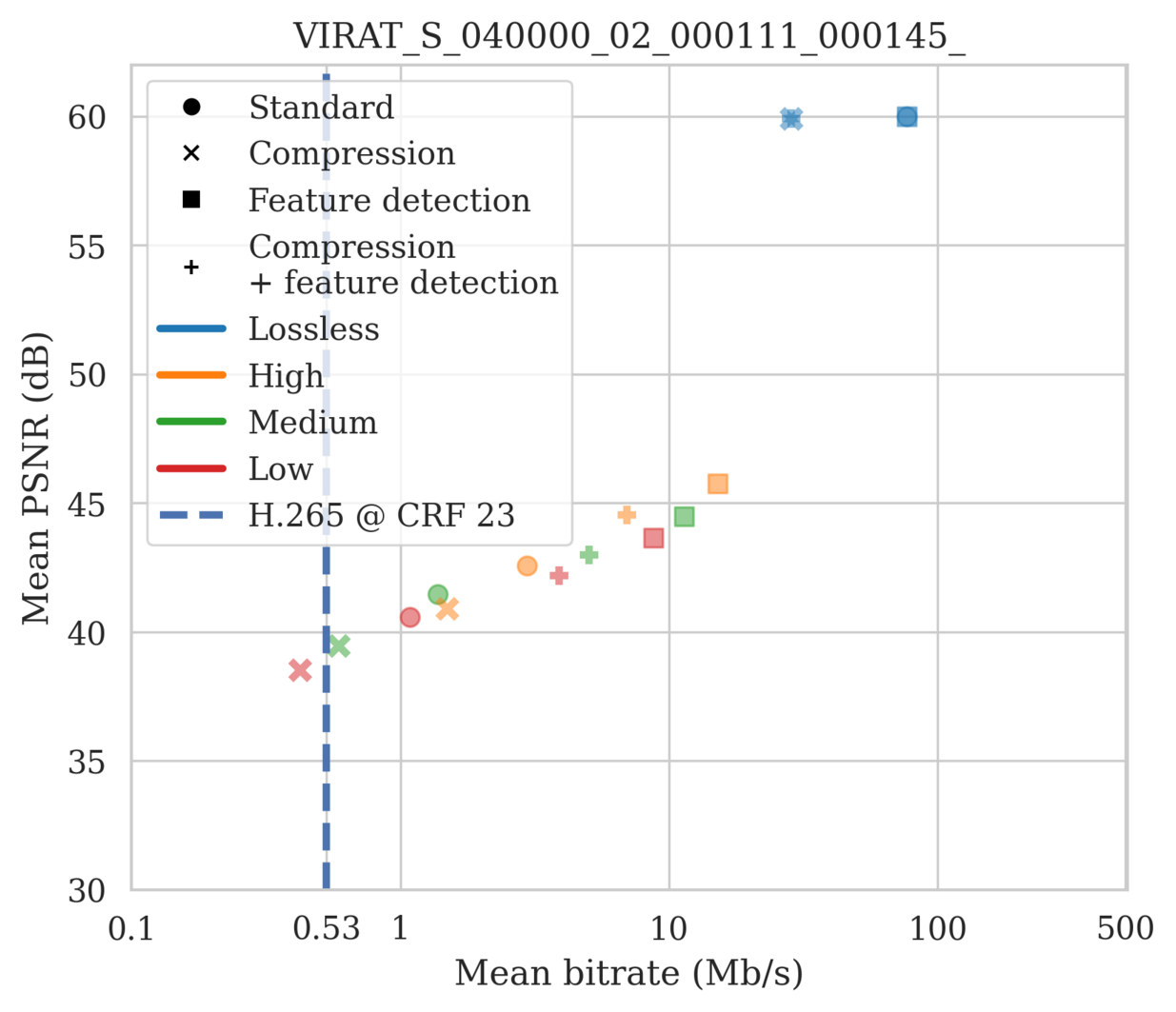}
         \caption{\texttt{Low} motion}
         \label{fig:bitrates1}
     \end{subfigure}
     \begin{subfigure}[t]{0.49\textwidth}
         \centering
         \includegraphics[width=\textwidth]{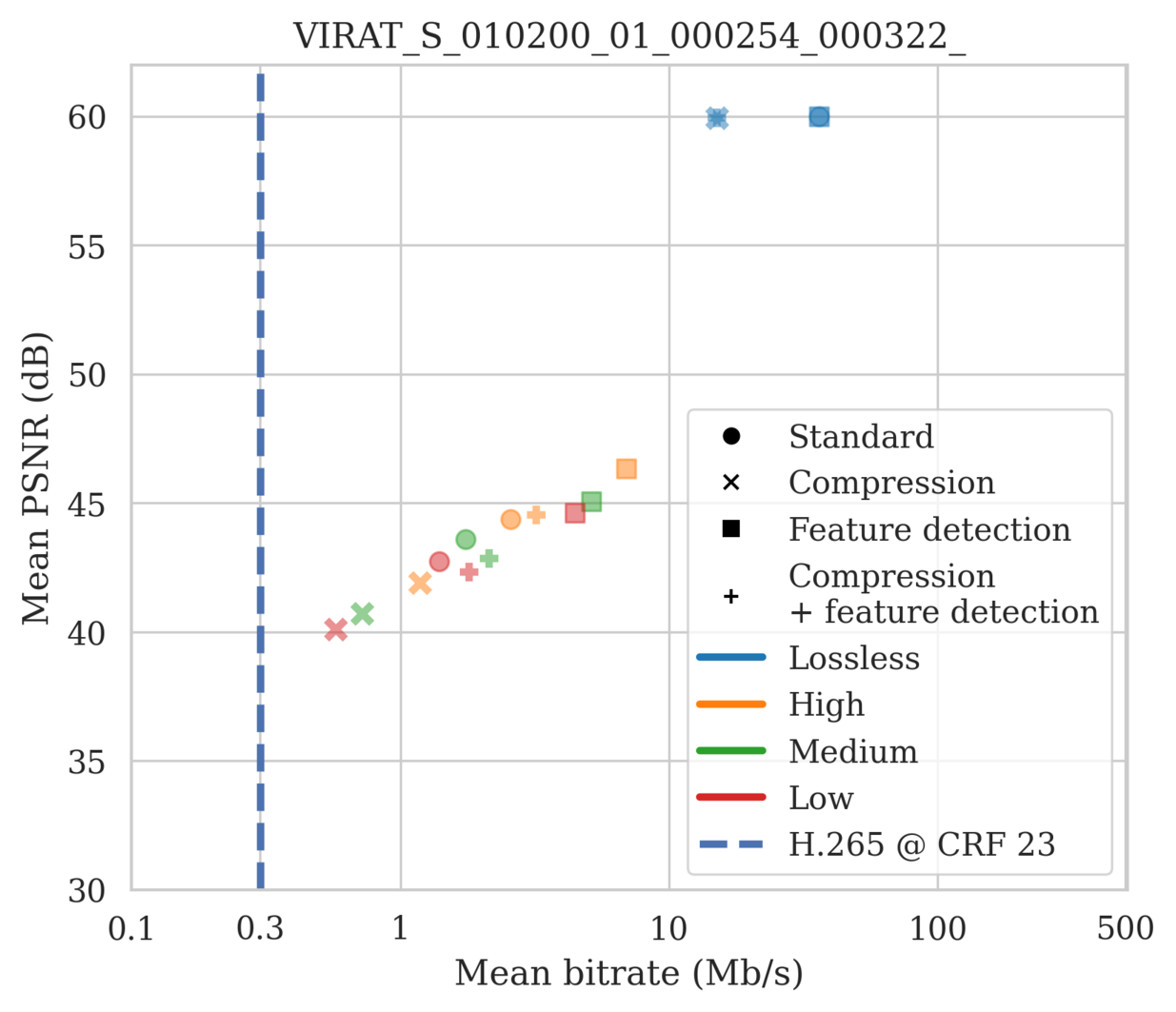}
         \caption{\texttt{Medium} motion}
         \label{fig:bitrates2}
     \end{subfigure}
     \begin{subfigure}[t]{0.49\textwidth}
         \centering
         \includegraphics[width=\textwidth]{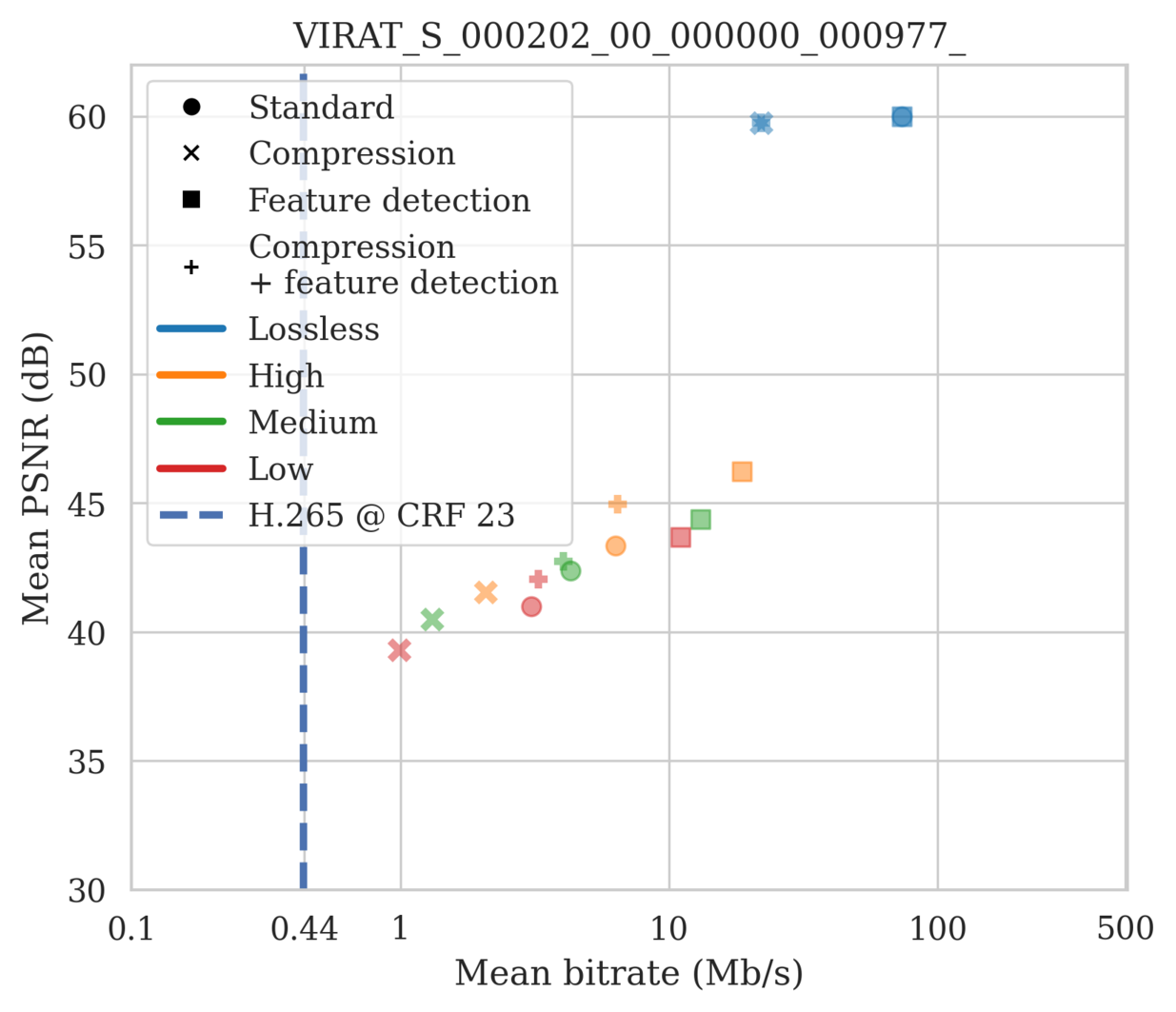}
         \caption{\texttt{High} motion}
         \label{fig:bitrates3}
     \end{subfigure}
     \begin{subfigure}[t]{0.49\textwidth}
         \centering
         \includegraphics[width=\textwidth]{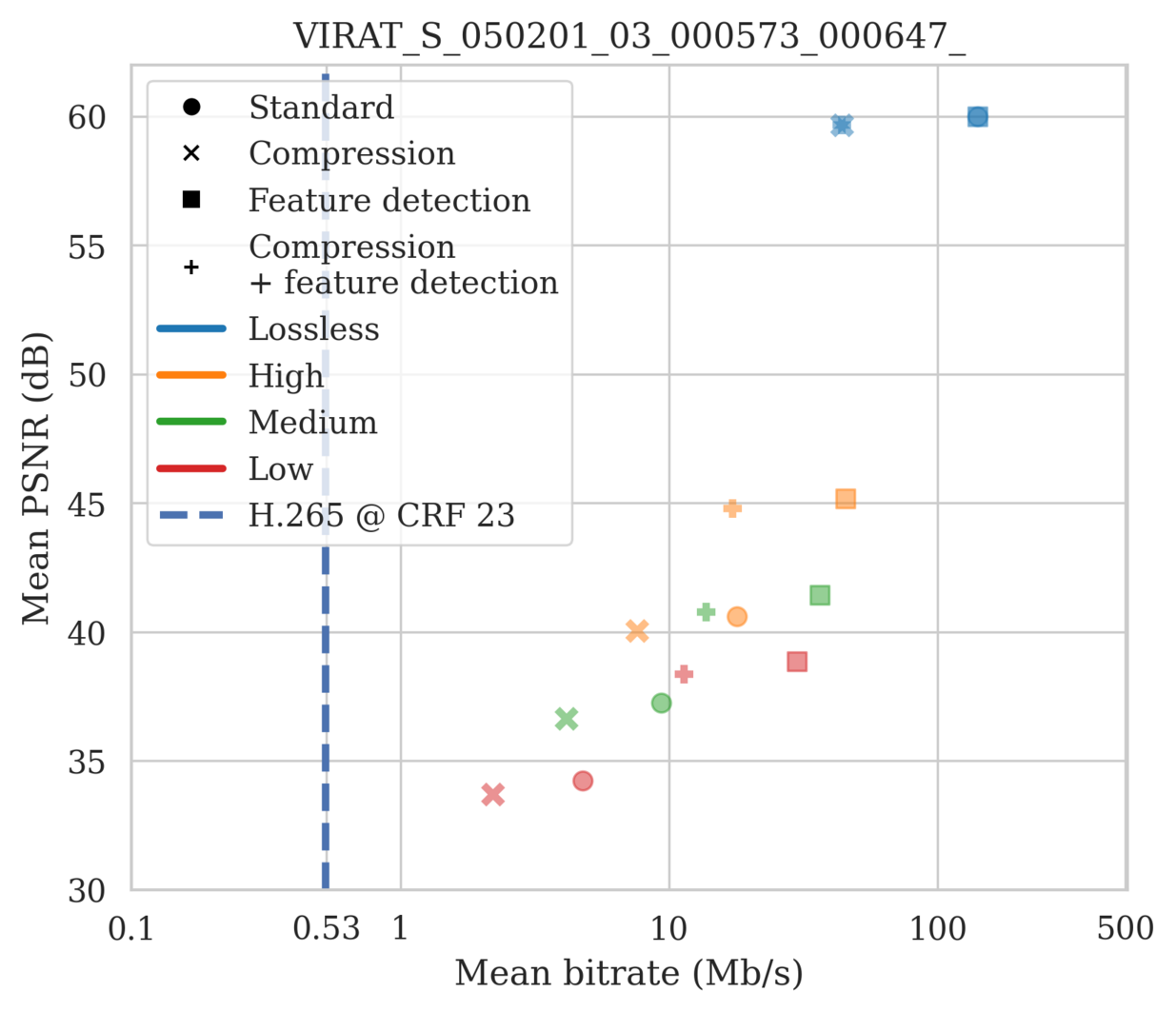}
         \caption{\texttt{Very High} motion}
         \label{fig:bitrates4}
     \end{subfigure}
     \caption[Rate-distortion curves]{Representative rate-distortion curves showing the effect of our transcoder quality settings and feature-driven rate adaptation. At the \texttt{Low} quality setting, our compressed representation approaches or surpasses the bitrate of the H.265-encoded source video, while maintaining a high PSNR value. Note that the H.265 data point for each video expresses only the bitrate, since its PSNR (with reference to itself) is undefined. I show one video per plot for readability, since the other videos in each motion category follow similar patterns.}
     \label{fig:all_bitrates}
\end{figure*}

\cref{fig:all_bitrates} plots the rate-distortion curves for a representative video in each of our motion classes
. I distinguish the four \adder{} quality levels by color. I mark the bitrate of the H.265-encoded source video with a dashed line. The PSNR for each marker is calculated in reference to this source video. I limit the maximal PSNR to 60 dB for visualization, since the \texttt{Lossless} PSNR was effectively infinite. The circular markers show the performance of transcoding to \adder{} without feature-driven rate adaption or source-modeled compression, whereas the x-shaped markers denote the performance of applying compression without feature-driven rate adaption. We see that as the \adder{} quality decreases, the bitrate and PSNR uniformly decrease as well. At any lossy quality level, we can see that applying my compression scheme yields a substantial drop in bitrate and a small reduction in PSNR. In fact, I found that my source-modeled compression scheme achieves ~2.5:1 compression ratios at all quality levels and a reduction in PSNR of just 1.2-1.6 dB. Furthermore, this trend holds true for all four motion classes. The square markers denote the results with feature-driven rate adaption enabled, and the plus-sign markers denote feature-driven rate adaption and source-modeled compression. Examining these two sets of data points, we see the same trends for compression performance. However, the overall data rates with feature detection are higher than the standard transcode results at the same \adder{} quality level, as we increase the pixel sensitivities (and thus the event rate) near detected features. Furthermore, the \texttt{Lossless} quality exhibits extremely high data rates. Enabling feature-driven rate adaptation does not increase the bitrate at the \texttt{Lossless} level, since all pixels are already at their maximum sensitivity.

Notably, we see that the \texttt{Very High} motion videos have a greater reduction in PSNR as the \adder{} quality decreases, as shown in \cref{fig:bitrates4}. However, this motion category also shows a greater increase in PSNR when enabling feature-driven rate adaption. This is evident in the relationship between the circular and square markers with the same color in \cref{fig:bitrates4}, and a higher median increase in PSNR compared to the other motion classes in \cref{tab:recon_psnr}.

 The time to encode an ADU as described in \cref{sec:dataset_experiments} at 360p resolution and \texttt{High} \adder{} quality on a single CPU thread averages 216 ms, while the decode speed averages 123 ms. Thus, excluding the cost of transcoding the input frames to \adder, we can compress up to 138 input frames per second in real time. By contrast, prior compression work on a precursor to \adder{} was only computed offline due to extremely slow performance \cite{FreemanLossyEvent,freeman_mmsys23}.

\begin{table}
    \centering
    \begin{tabular}{l|rrrr|r}
    & \multicolumn{5}{c}{Median change in feature detection time (\%)}  \\
         \diagbox[width=6em]{Quality}{Motion} & \texttt{Low} & \texttt{Medium} & \texttt{High} & \texttt{Very High} &  All \\
        \hline
         \texttt{Lossless}&  150.8& 125.3& 190.3& 444.5&  163.3\\
         \texttt{High}    &  -41.0& -35.1& -13.2& 99.4&  -16.6\\
         \texttt{Medium}  &  -60.4& -58.7& -30.9& 81.5&  -33.3\\
         \texttt{Low}     &  \textbf{-67.5}& -56.7& -38.4& 53.0&  -43.7\\
    \end{tabular}
    \caption[Change in FAST feature detection speed]{Median change in FAST feature detection time between the frame-based OpenCV implementation and my asynchronous implementation. The columns indicate the motion category, while the rows indicate the \adder{} transcoder quality. Results less than 0 indicate that my method performed faster than OpenCV.}
    \label{tab:feature_speed}
\end{table}

\begin{table}
    \centering
    \begin{tabular}{p{0.1cm}l|rrrr|r}
    & & \multicolumn{5}{p{5.5cm}}{\centering Median change in bitrate,\\ H.265 to \adder{} (\%)}  \\
         & \diagbox[width=6em]{Quality}{Motion}& \texttt{Low} & \texttt{Medium} & \texttt{High} & \texttt{Very High} &  All \\
        \hline
         \parbox[t]{2mm}{\multirow{3}{*}{\rotatebox[origin=c]{90}{Normal}}}
         & \texttt{Lossless}  &  5330.3 & 4717.2 & 5916.1 & 6491.2 & 5443.6 \\
         & \texttt{High}      &  181.2  & 335.4  & 411.3  & 1210.4 & 391.7 \\
         & \texttt{Medium}    &  15.3  & 130.2   & 213.5  & 635.8 & 171.6 \\
         & \texttt{Low}       &  \textbf{-9.9} & 51.9   & 292.4   & 214.3 & 84.6 \\
         \hline
         \parbox[t]{2mm}{\multirow{3}{*}{\rotatebox[origin=c]{90}{FAST}}} 
         & \texttt{Lossless}  &  5330.3 & 4717.2 & 5916.1 & 6491.2 & 5443.6 \\
         & \texttt{High}      &  2354.4 & 1279.9 & 1423.3 & 2964.2 & 1384.7 \\
         & \texttt{Medium}    &  768.6  & 848.6  & 929.8  & 2412.2 & 943.2 \\
         & \texttt{Low}       &  601.9  & 646.3  & 702.6  & 2005.3 & 739.0\\
    \end{tabular}
    \caption[Change in compressed video size]{Median change in compressed video size between the H.265 video source and our transcoded \adder{} representation. The columns indicate the motion category, while the rows indicate the \adder{} transcoder quality under both the standard method and with rate adaptation based on FAST feature detection. Results less than 0 indicate that my compressed \adder{} representation has a lower bitrate than the H.265 source.}
    \label{tab:bitrates_to_h265}
\end{table}

Furthermore, I show the relationship between the raw \adder{} event rate and the speed of asynchronous feature detection in \cref{fig:all_speed_vs_event_rate}. We see that if the decoded event rate is less than 1 event for every 40 pixels in the image plane, event-based feature detection on the sparse events executes faster than frame-based feature detection on all the pixel intensities. With surveillance video sources, if I allow a slight amount of temporal loss with a non-zero $M$, my transcoded representation \textit{easily} achieves this sparse event rate. I detail these results at our various quality levels and motion classes in \cref{tab:feature_speed}. Across my 132 videos, the median asynchronous feature detection speed is faster than the frame-based method for all the quality levels except \texttt{Lossless} (which produces a very high data rate). At the \texttt{Low} quality, we see an overall \textbf{43.7\% speed improvement} over the frame-based OpenCV implementation. We see the best results for the \texttt{Low} motion class, where at \texttt{Low} quality my median feature detection speed is \textbf{nearly two-thirds faster} than that of OpenCV. As I noted in \cref{sec:implementation_fast}, any speed improvement here is due to the efficiency of the sparse representation, \textit{not} my particular FAST implementation.

Meanwhile, the \texttt{Very High} motion class produces substantially more \adder{} events (\cref{fig:bitrates4}), and has slower feature detection performance than OpenCV. As noted above, the camera placement and wind-induced camera motion of these videos is atypical within the VIRAT dataset, and I would not expect such results in commercial surveillance camera deployments. Even still, I expect that incorporating motion compensation in my source-modeled encoder will greatly improve the compression performance of such videos. Since my results show that the event-based application speed depends on the \textit{decompressed} data rate, however, a robust \adder{} motion compensation scheme for high-motion video should not merely improve the prediction accuracy of the encoder; rather, it should reduce the raw event rate itself.

Finally, \cref{tab:bitrates_to_h265} shows the median percentage change in bitrate from the H.265 source to our compressed \adder{} representations. We see that my FAST-driven rate adaptation greatly increases the overall bitrate. Since the mechanism concentrates the higher event rate near salient regions, however, this underscores the claim that a more robust event prediction mechanism will help our encoder performance. On the other hand, we see that in scenes with low motion, transcoded at low quality, we outperform H.265 (up to 9.9\%), with only a minor drop in PNSR (\cref{tab:recon_psnr}).

Despite my naive compression scheme, these results are extremely promising: in scenes with high temporal redundancy, we can achieve higher compression ratios than standard video codecs \textit{and} faster speed than standard applications. My decompressed bitrate is concentrated near the beginning of a video, but can drop to near-zero during periods of little motion (\cref{fig:example}).

\subsubsection{Feature Clustering for DVS Object Detection}

\adder{} shows strong performance on applications for framed video sources and native \adder{}-style sensors. For DVS sources, however, I have so far shown only a speed advantage for fusing DVS events and intensity frames. As there is substantial work on bespoke vision applications for standalone DVS sensors, I seek to demonstrate the utility of \adder{} as an intermediate representation for DVS video.

Towards this end, I leverage the Hierarchical Neural Memory Network (HMNet), a state-of-the-art network for DVS-based semantic segmentation, object detection, and depth estimation \cite{hmnet}. The network has multiple latent memories for encoding features at different time scales to improve the computational speed over related methods \cite{hmnet}. For my investigation, I use the variant of HMNet with the object detection task head. This variant was trained on the GEN1 automotive dataset from Prophesee \cite{gen1_dataset}. It contains 39 hours of DVS event video recorded with the Prophesee GEN1 sensor, with 1 Hz bounding box annotations of automobiles and pedestrians \cite{gen1_dataset}. The dataset is given in the \texttt{.dat} format, an uncompressed binary format which is compatible with my transcoder.

    A key strength of \adder{} is the ability to adaptively reduce temporal redundancy in video. With this in mind, I present \adder{} here as an intelligent \textit{temporal filtering} mechanism. With my DVS transcoder, I casted the DVS source videos into \adder{} at the \texttt{Lossless}, \texttt{High}, \texttt{Medium}, and \texttt{Low} quality levels as described in \cref{sec:crf}. Then, I converted these \adder{} videos \textit{back} to DVS for compatibility with the original application.
    
    I used a randomly-sampled 20-video subset from the Test portion of the GEN1 dataset \cite{gen1_dataset} for my evaluation. I encoded the resulting DVS video in the \texttt{.dat} format for compatibility with the HMNet pipeline. I compared the \adder{} performance to a naive filtering mechanism by randomly removing events from the original input videos to match the reconstructed DVS file size. 
    
    \cref{tab:hmnet_results} details the results of this experiment, showing that \adder{}-based temporal filtering actually yields lower performance than random filtering of DVS events. Enabling my FAST feature adaptation at the \texttt{Low} quality level yields marginally better performance than the naive filtering scheme. However, the random filtering scheme may variously produce stronger performance, as with the result at \texttt{High} quality, depending on the actual event data lost. Additionally, this positive result occurs with a median bitrate reduction of only one third. Meanwhile, we see that we can randomly filter out nearly \textit{half} of the DVS events and still maintain high object detection performance. This result suggests that the HMNet model is robust to high data loss if it is evenly distributed in both space and time. \adder{} without feature-driven adaptation, however, filters events only temporally, regardless of spatial saliency.

    \begin{table}
    \centering
    \begin{tabular}{p{2.0cm}|rr|r}
         &  \multicolumn{2}{c}{\centering Mean Average Precision (mAP)} & \\
        \adder{} transcode quality & \multicolumn{1}{p{2.0cm}}{\raggedleft \adder{} \\  $\rightarrow$ DVS} & \multicolumn{1}{p{2.2cm}|}{\raggedleft Random \\ event \\ filtering}  & \multicolumn{1}{p{2.4cm}}{\raggedleft Median change \\ in bitrate (\%)} \\
         \hline \texttt{Lossless} &  38.8 (-0.8) & 39.3 (-0.3) & -15.56\% \\
                \texttt{High}     &  37.4 (-2.2) & 38.4 (-1.2) & -45.44\% \\
                \texttt{Medium}   &  33.2 (-6.4) & 36.1 (-3.5) & -66.34\% \\
                \texttt{Low}      &  20.4 (-19.2) & 27.2 (-12.4) & -85.05\% \\
                \texttt{Low} + FAST     &  38.9 (-0.7) & 37.9 (-1.7) & -33.9\% \\
    \end{tabular}
    \caption[Effect of \adder{} transcoding on DVS object detection]{Effect of \adder{} transcoding on DVS object detection. The GEN1 \cite{gen1_dataset} videos were transcoded to \adder{} at the four quality levels as described in \cref{sec:crf}, then transcoded back to DVS. The ground truth dataset has a mAP of 39.6.}
    \label{tab:hmnet_results}
\end{table}

    With these results in mind, I devised a scheme to evaluate the efficacy of aggressive spatial filtering with \adder{}, in an application-driven context. Since the GEN1 dataset shows forward-facing car dashcam footage, one can easily imagine an autonomous vehicle scenario where impending crashes must be detected and avoided. Then suppose that the object detection model is limited in the number of DVS events it can process at a time while preserving low latency and fast vehicle response, or assume that we have limited communication bandwidth from the imaging source to the application. Therefore, we must substantially reduce the rate of DVS events provided to the application, but we want to preserve application performance. While prior work leveraged APS frames to identify objects and filter events outside the object regions \cite{banerjee2021joint}, this method required multimodal data and a strictly online link between the application receiver and the server. In this case, we have only DVS event data and assume that the application cannot update the state of the imaging server. In this context, we want to quickly identify objects close to the camera (i.e., large objects) and allocate more bandwidth towards these regions.

    \begin{figure*}
        \centering
        \includegraphics[width=\linewidth]{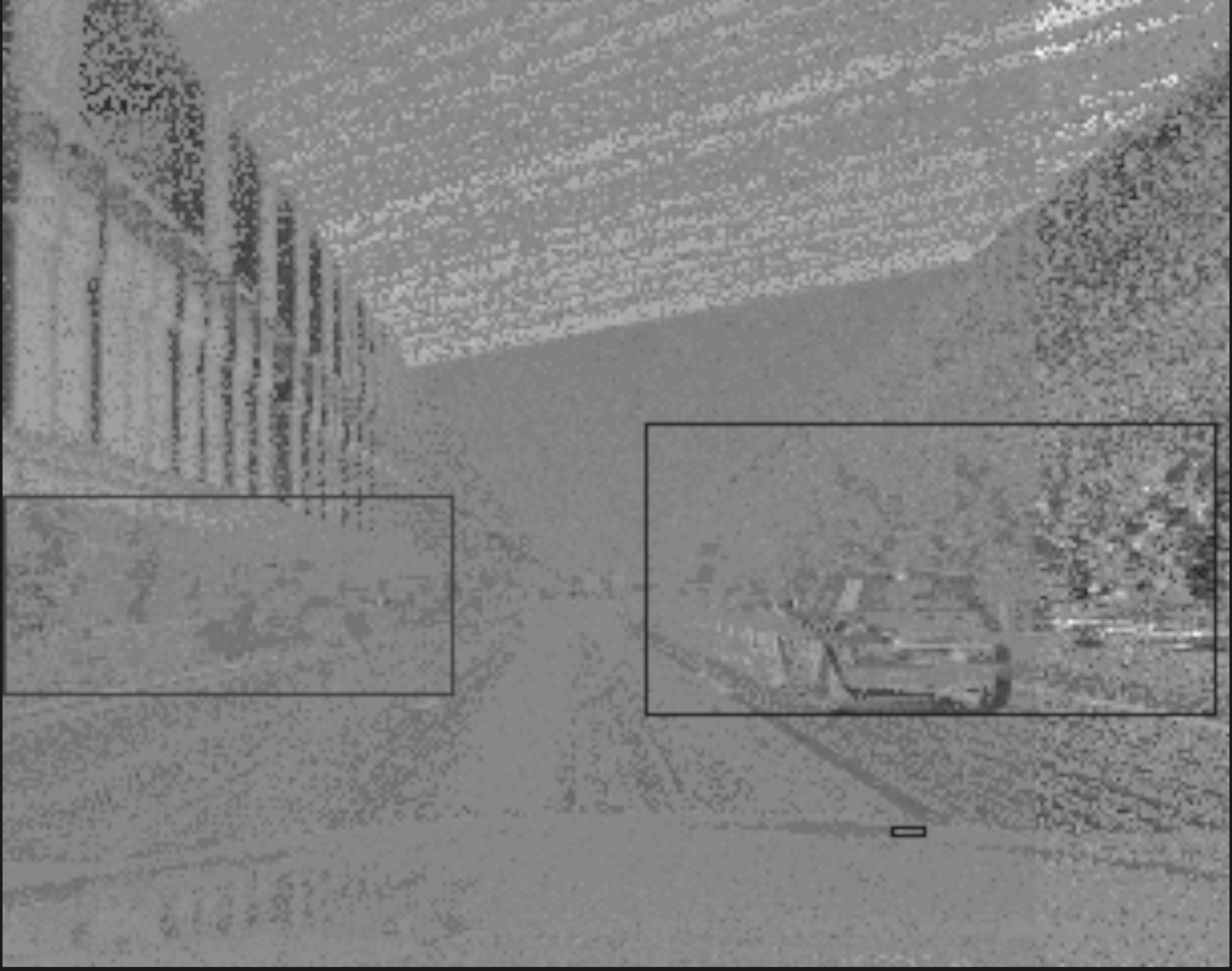}
       \caption[Bounding boxes from feature clusters]{Bounding boxes from feature clusters on the GEN1 \cite{gen1_dataset} dataset. The features are detected based on the transcoded \adder{} intensities, visualized in grayscale. The detected regions contain a vehicle in the adjacent lane and several vehicles traveling in the opposite direction. Note that the artifacts on the right of the image are due to noise characteristics of the camera used during dataset capture, not due to the \adder{} processing.}
        \label{fig:feature_cluster}
\end{figure*}

    I implemented the Density-Based Spatial Clustering of Applications with Noise (DBSCAN) \cite{dbscan} algorithm to cluster the FAST features at regular intervals (30 times per second) detected during \adder{} transcode. These feature clusters then serve as prototypical regions for moving object detection, without the use of a learning-based method. I tuned the feature detection settings to avoid noisy regions and small clusters, instead prioritizing large clusters with several features. Within the bounding boxes of these clusters (\cref{fig:feature_cluster}), I then randomly discard one half of the input DVS events, since \cref{tab:hmnet_results} shows that HMNet still performs well with 50\% data loss. Outside these regions, I discard \textit{all} the input DVS events, such that the only data preserved is centered on spatial regions deemed salient during \adder{} transcode. I compared the performance of the resulting videos to the random filtering scheme described above.

\begin{figure*}
        \centering
        \includegraphics[width=\linewidth]{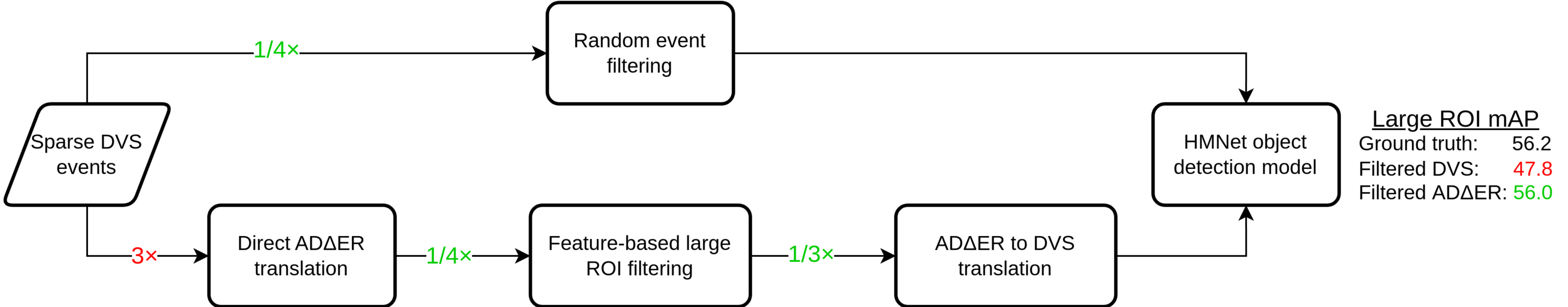}
       \caption[DVS filtering for object detection diagram]{Comparison of the methods and data rate reduction for DVS object detection. The data scaling numbers on the arrows are approximate, for readability; the median bitrate reduction across the 20-video dataset was 76.8\%. With this system, \adder{} maintains much higher mean average precision for large object detection, compared to random event filtering.}
        \label{fig:dvs_numbers}
\end{figure*}

    \cref{fig:dvs_numbers} shows the design and primary results of this secondary experiment. The filtered videos show a median bitrate reduction of $76.8\%$ compared to the ground truth input videos. At this low bitrate, the \adder{} scheme shows an mAP reduction of only 0.2 percentage points compared to the ground truth, whereas the random filtering method shows a loss of 8.4 percentage points. This prioritization of large spatiotemporal regions, however, comes at the cost of accuracy in small- and medium-sized objects, where the random filtering mechanism performs 3.7 and 10.8 percentage points higher, respectively. In the context of our autonomous driving scenario, however, this can be a reasonable tradeoff. Large objects are likely to be those of the greatest semantic interest for obstacle detection and avoidance. We can accommodate different application needs by adjusting the parameters of the feature detection and clustering algorithms. Regardless, the source-agnostic \adder{} system can help to intelligently compress DVS data, which is already quite sparse, for downstream applications.

\section{Future Work}

Although my compression scheme shows promise for end-to-end systems, there is much room for improvement. Central to the question of \adder{} compression is the current lack of an event-based quality metric. I explored traditional metrics for visual quality and application performance, but these operate only on a coarse level. If I instead had a metric which captures the difference between an input event and a lossy-compressed event, I could make more informed decisions during the compression stage. Currently, my metric is only the \textit{intensity} of the event ($\frac{2^D}{\Delta t}$). I do not penalize event loss based on the change in event \textit{timing}. Additionally, an \adder{} quality metric should capture the relative information of two event trains (sequences of events), including the outright \textit{removal} of events. That is, the metric should be resilient to differences in the number of events, since I have shown that subsequent events may be merged to average out variations in intensity. 

Additionally, I will work to improve the sophistication of my source-modeled compression with variable cube sizes, motion compensation, and better prediction schemes. A bespoke \adder{} quality metric would greatly aid motion compensation, but it must be fast enough to operate in real time (unlike my early work adapting a slow DVS metric \cite{Freeman2021mmsys}). I may further investigate the adaptation and use of frequency transforms for sparse spatiotemporal data. An improved compression scheme may enable a layered rate system for adaptive streaming, which I would seek to implement with the DASH \cite{mpeg_dash} or QUIC \cite{quic} protocols.

On the application side, I note that any convolution kernel or iterative image processing algorithm can easily be adapted to operate asynchronously on sparse \adder{} events. In the future, I will explore convolutional algorithms such as edge detection, sharpening, and blur filters, and develop spiking neural networks for fully event-based applications. While my experiments on YOLO and HMNet demonstrate certain modest advantages for \adder, it is clear that a bespoke object detection model will best be able to exploit the unique sparsity of \adder{} events. To this end, I will develop machine learning models which ingest \adder{} data directly. Spiking neural networks (SNNs) are a natural avenue for such an effort, where I may leverage the Norse library \cite{norse2021}.

\section{Big Picture: Robust Video Archival for Post Facto Applications}

Many aspects of traditional video systems have been built around the notion of a human viewer. Lossy compression algorithms are based on perceptual quality metrics. Bitrate adaptation algorithms are similarly designed around human  quality of experience (QoE) metrics. These ubiquitous mechanisms inherently assume that the target application is \textit{human viewership}. Increasingly, however, cameras are deployed at large scales to continuously record and analyze video with little to no human monitoring. Technology companies use such deployments to monitor their data center security, logistics companies track the flow of packages, stores detect shoplifting, transportation departments analyze traffic patterns, and intelligence agencies track suspected threats to national security.

\subsection{Video Surveillance Weaknesses}\label{sec:surveil_weaknesses}

These large-scale surveillance systems pose a number of unique issues. First, a business or government may find long-term storage costs prohibitively expensive. Although stationary cameras enable strong compression performance, 24/7 recording from hundreds or thousands of cameras can quickly saturate the operator's storage or networking budget. In critical systems, indiscriminately discarding old video data can pose a risk of safety, security, or legal liability to the operator.

Second, while traditional video codecs are highly efficient at compressing stationary surveillance video, the analysis applications are largely decoupled from the compression pipeline. Although the compressed representation may indicate that a certain region of a video is not changing over a long period of time, the compressed structure is not amenable to most applications. On the other hand, the \textit{decompressed} representation ingested by the application is a series of standalone images with a uniform sample rate for every pixel. Therefore, vision applications may spend significant time and computational resources processing pixel values that the encoder determined to be of low salience. Meanwhile, any improvement to the computational speed of a real-time video analysis pipeline can make a difference in human safety in emergency situations.

At the same time, we often cannot design video systems strictly around computer vision applications.  That is, we still need to preserve a visually coherent signal for human supervision or intervention. Consider again a large-scale surveillance camera deployment for a government agency. Suppose the agency wants to minimize network usage by intelligently compressing the video data at the edge, before analyzing the data from a centralized server. The automated analysis could include facial recognition, activity recognition, threat detection, or any number of tasks. There may be some combination of these tasks, and the tasks may change over time.

In this scenario, therefore, we require that the application layer is entirely separate from the acquisition and compression layers of the system. We assume that videos may be stored for some period of time, and may be processed by an unknown set of applications. Additionally, we must preserve acceptable quality for human visual analysis. This scenario is one where \adder{} can truly excel. 

\subsection{\texorpdfstring{\adder}{ADDER} for Framed Camera Surveillance}

\begin{figure*}
     \centering
     \begin{subfigure}[t]{0.49\textwidth}
         \centering
         \includegraphics[width=\textwidth]{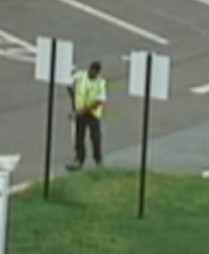}
         \caption{\adder}
         \label{fig:fullres_adder}
     \end{subfigure}
     \begin{subfigure}[t]{0.49\textwidth}
         \centering
         \includegraphics[width=\textwidth]{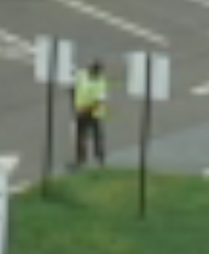}
         \caption{Framed}
         \label{fig:scaled_source}
     \end{subfigure}
     \caption[\adder{} vs. framed quality at equal bitrates]{This \adder-transcoded surveillance video (left) has a low bitrate in its decompressed representation. If we lower the resolution of the framed video (right) so that its decompressed bitrate is equivalent, the visual quality decreases dramatically.}
     \label{fig:scaled_comparison}
\end{figure*}

As shown in \cref{sec:feature_rate_control}, traditional frame-based methods unavoidably have redundant data at the application layer, except in extremely specialized systems with compressed-domain applications. This redundant data slows the application-level performance. One may improve compression performance by aggressively lowering the quality level for non-salient regions in a framed codec. Additionally, in the framed paradigm, quality loss does not directly correlate to application speed. To improve the speed at the application level (in the raw domain), one must either lower the frame rate or decrease the video resolution (\cref{fig:scaled_comparison}) at the compressor. These strategies have a negative impact on both human perceptual quality \textit{and} application accuracy.

\adder{} presents a unique alternative. It nonuniformly encodes intensity samples based on their rate of change---the raw bitrate adjusts dynamically to the motion content. By increasing the temporal sparsity of the video representation, we can highly compress temporally stable regions without degrading the image quality. \cref{fig:scaled_comparison} compares a sample from an \adder-transcoded surveillance video and a framed video scaled such that its raw bitrate is equivalent. We see that with \adder, we can maintain the high source resolution and preserve visual quality. 

\begin{figure*}
        \centering
        \includegraphics[width=\linewidth]{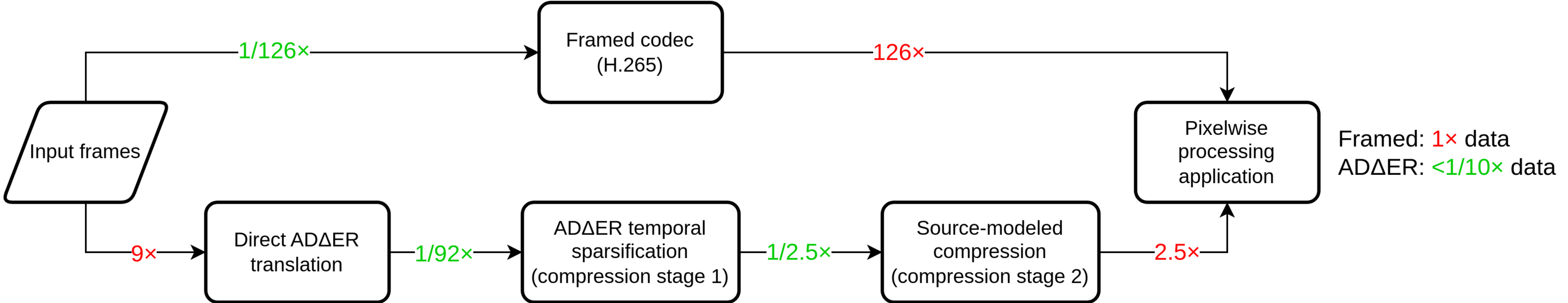}
       \caption[Framed surveillance data reduction]{Comparison of the methods and data rate reduction for framed surveillance video. The data scaling numbers on the arrows are approximate, for readability. Since \adder{} reduces temporal redundancy in the raw representation, it can dramatically reduce the data burden for downstream applications by approximately 10 times..}
        \label{fig:framed_numbers}
\end{figure*}

In a frame-based system, one alternative to lowering the resolution is to discard some image frames during compression. This then has the adverse effect of lowering the temporal resolution (frame rate), which can make both human and computer vision applications less accurate. The question of \textit{which} image frames to discard is also nontrivial. If two consecutive frames are identical, we can simply discard one. If there is some change between them, however, then we must define a threshold for the amount of change necessary to preserve both frames. In practice, this threshold must be application-driven, requiring a tight integration between the compression and application layers of a system. Given that applications may vary, this interdependence violates the separation requirements of our surveillance system scenario (\cref{sec:surveil_weaknesses}).

Despite incurring a major data rate multiplication for lossless transcoding (\cref{fig:framed_numbers}), \adder{} can greatly outperform the framed representation data rate when many pixels are temporally stable. In video surveillance, this broad stability is the typical case. Then, any applications which operate on a pixelwise basis will experience a speed improvement by processing less data, as demonstrated with FAST feature detection in \cref{sec:feature_rate_control}.

\subsection{\texorpdfstring{\adder}{ADDER} for DAVIS Camera Surveillance}

Typical surveillance camera deployments record at a constant 30 or 60 frames per second. When capturing high-speed subjects such as vehicles and weaponry, or in challenging lighting conditions, the low frame rate or slow shutter speed of framed cameras may yield blurry or poorly-exposed videos. While a high-speed framed camera can capture faster motion, their cost, power consumption, and high bandwidth make them impractical for continuous recording (\cref{sec:high_speed_cameras}). A DVS camera can alleviate these issues with its high temporal resolution and log temporal contrast sensing. However, DVS alone does not capture the visual texture in static regions necessary for human visual analysis. A DAVIS camera combines DVS event capture with low-rate APS image frame capture to cover both sensing modalities. Thus, let us explore the implications of a DAVIS camera deployment for video surveillance, using the same scenario outlined above (\cref{sec:surveil_weaknesses}).

\begin{figure*}
     \centering
     \begin{subfigure}[t]{0.46\textwidth}
         \centering
         \includegraphics[width=\textwidth]{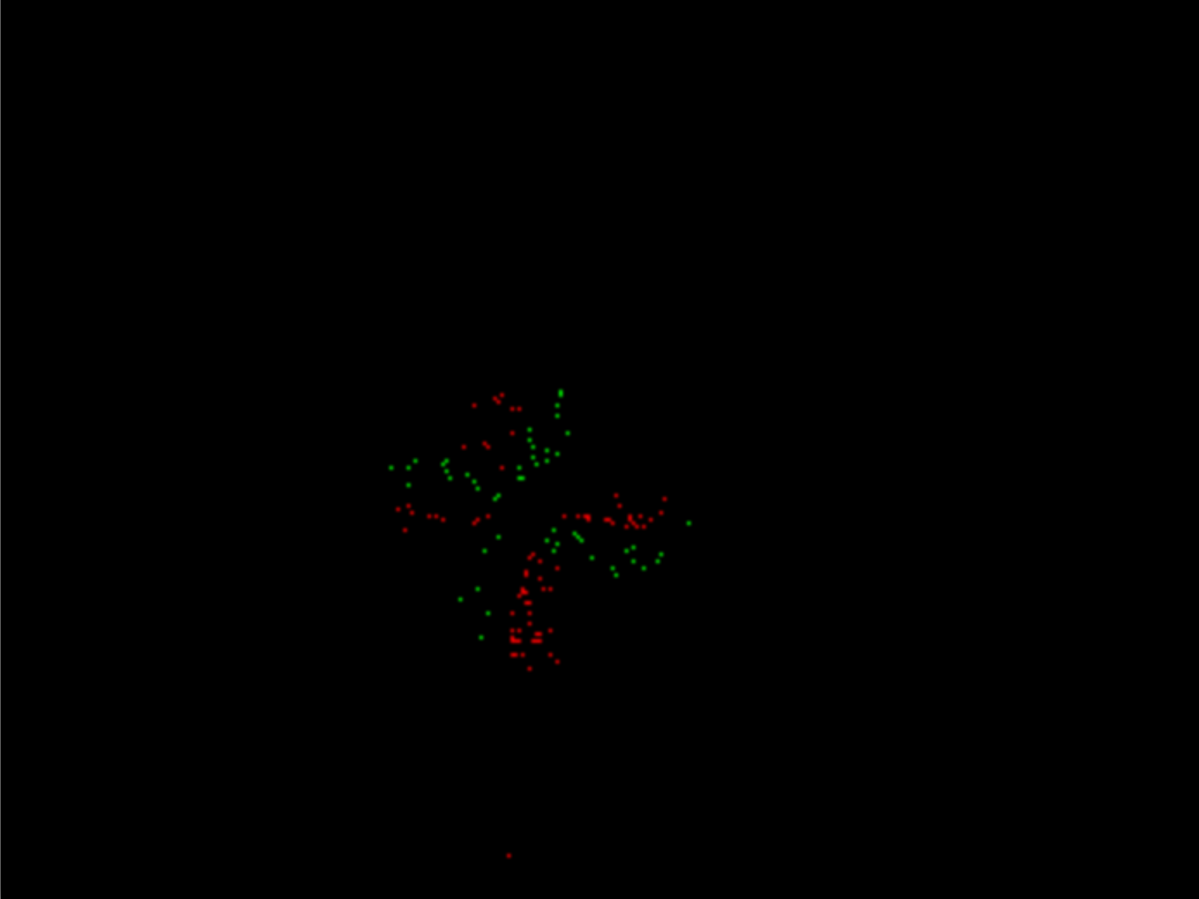}
         \caption{DVS events occurring over 0.01-second interval}
         \label{fig:fan_dvs}
     \end{subfigure}
     \hfill
     \begin{subfigure}[t]{0.46\textwidth}
         \centering
         \includegraphics[width=\textwidth]{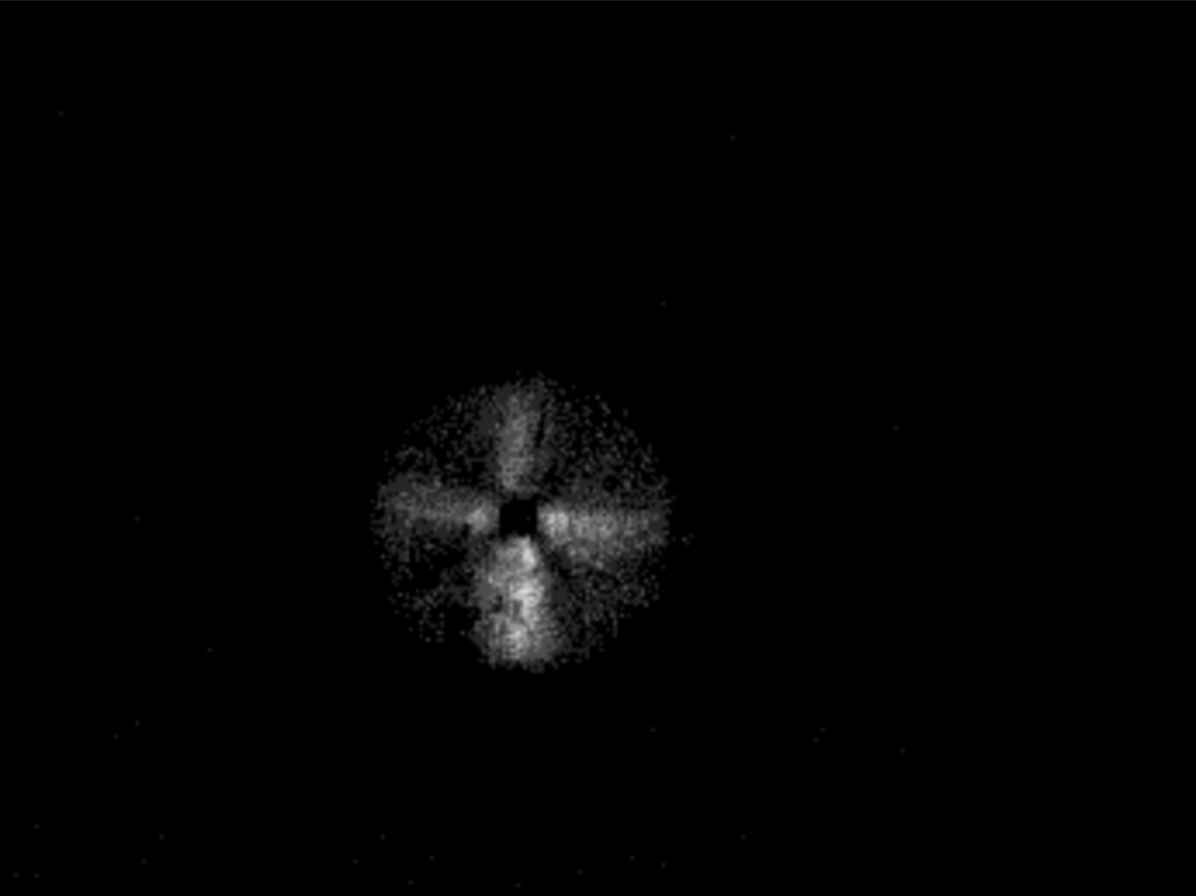}
         \caption{Accumulated DVS event frame with exponential decay}
         \label{fig:fan_dvs_accum}
     \end{subfigure}
     \begin{subfigure}[t]{0.49\textwidth}
         \centering
         \includegraphics[width=\textwidth]{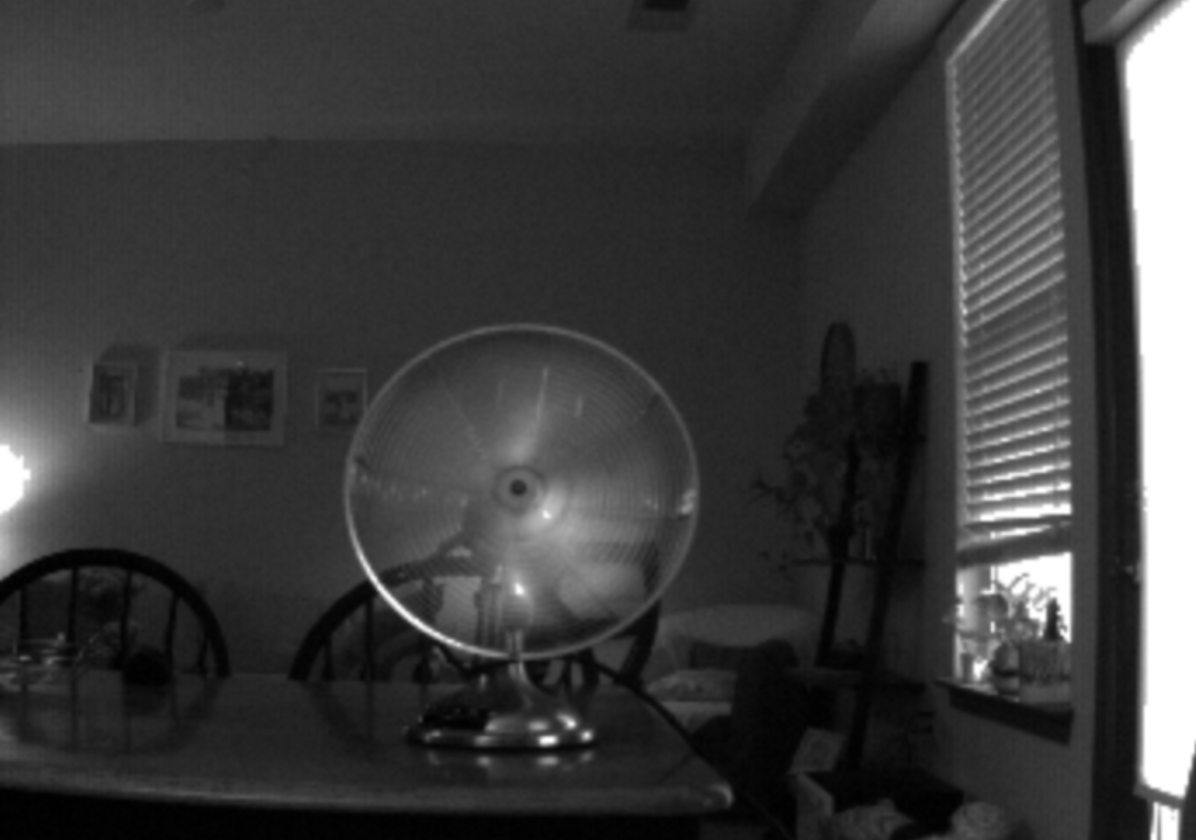}
         \caption{Blurry APS source image}
         \label{fig:fan_aps}
     \end{subfigure}
     \hfill
     \begin{subfigure}[t]{0.46\textwidth}
         \centering
         \includegraphics[width=\textwidth]{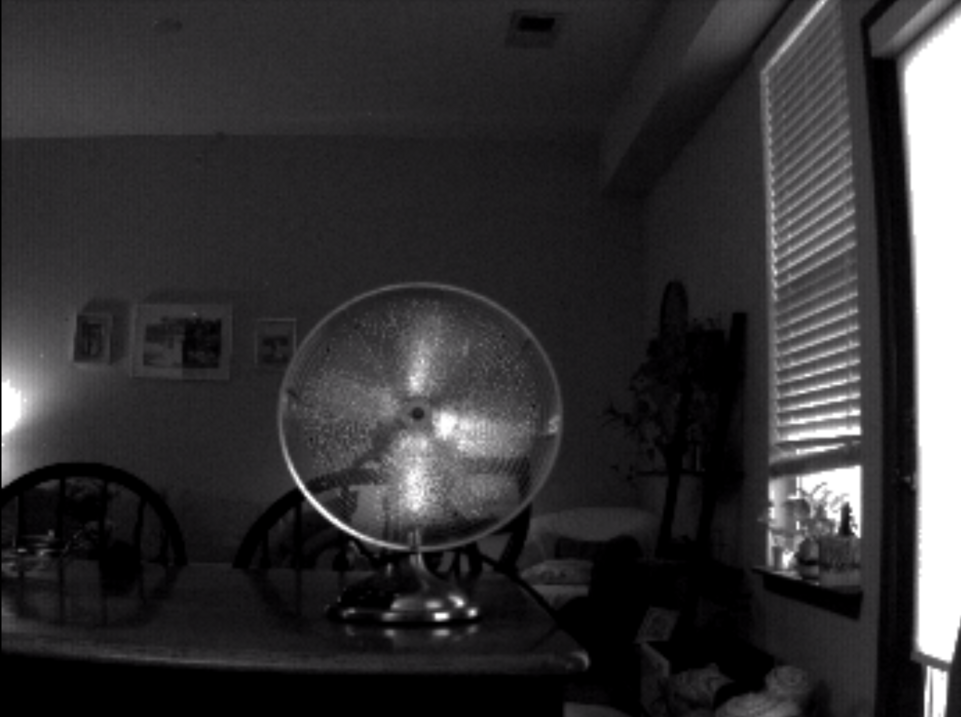}
         \caption{\adder{} framed reconstruction from mode (ii) transcode}
         \label{fig:fan_adder}
     \end{subfigure}
     \caption[Qualitative exhibition of \adder{} for fusing DVS and APS data]{Qualitative exhibition of \adder{} for fusing DVS and APS data. Images (a)-(c) were obtained in the iniVation DV software. (d) is from a 2000 FPS framed reconstruction of the \adder-transcoded video. Here, we can clearly see the four moving fan blades along with the rest of the image texture.}
     \label{fig:fan_comparison}
\end{figure*}

As discussed in \cref{sec:event_video_systems}, typical DAVIS applications either reconstruct a framed image sequence from the two sensing modalities or fuse the information implicitly within a neural network. Under the traditional paradigm, the former approach is necessary to support human visual analysis. As shown in \cref{sec:event_video_adder}, high-speed framed reconstruction from DAVIS data is extremely slow, even with a highly performant implementation.

In contrast, we can fuse DAVIS video in real time by transcoding to \adder{} events under mode (ii), as described in \cref{sec:davis_transcoder_details}. Since the fundamental representation remains asynchronous, we can further realize a performance advantage from downstream event-based applications. A human analyst may interact with the video playback at arbitrarily slow speeds. The video player simply increases the update frequency of the intensity sample frame, which is updated asynchronously as new \adder{} events are digested. \cref{fig:fan_comparison} offers a qualitative comparison of computationally fast visualization schemes for DAVIS data. Only with \adder{} can we observe the high-speed motion detail with natural texture amenable to human vision.

\begin{figure*}
        \centering
        \includegraphics[width=\linewidth]{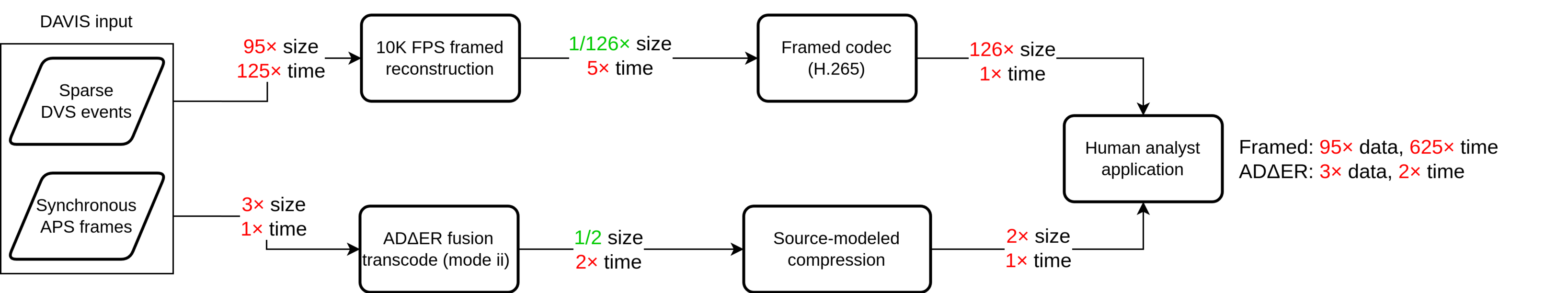}
       \caption[DAVIS fusion data reduction]{Comparison of the methods and data reduction for fusing multimodal DVS and APS data for archival storage and human visual analysis. With settings which preserve the high temporal rate of the DVS information, \adder{} offers a dramatic improvement in data rate and processing time compared to the frame-based method, by factors of 32 times and 312 times, respectively.}
        \label{fig:davis_numbers}
\end{figure*}

\cref{fig:davis_numbers} compares the raw data rate and speed of high-rate framed reconstruction and \adder{} transcoding of DAVIS surveillance data. Although the current source-modeled compression scheme adds some latency, the \adder{} method is two orders of magnitude faster than the frame-based method, even at modest frame rates. Furthermore, with \adder{} we can properly separate the concerns of the acquisition and representation layers from those of the application layer, since the application can decide \textit{post facto} at what temporal rate to sample the event data.

\subsection{Sparse Surveillance Computing on the Edge}

The ultimate realization of an \adder{} surveillance system would place the transcoding processes on the cameras themselves. As further research may unlock higher compression ratios for \adder{} compared to existing frame-based codecs and lossless DVS compression, \adder{} may decrease the network bandwidth necessary to transmit surveillance video to an archival server. For optimal performance, this edge computation must run on dedicated transcoder hardware, akin to the H.264 encoder chips commonly found in traditional cameras.

\begin{figure*}
     \centering
     \begin{subfigure}[c]{0.46\textwidth}
         \centering
         \includegraphics[width=\textwidth]{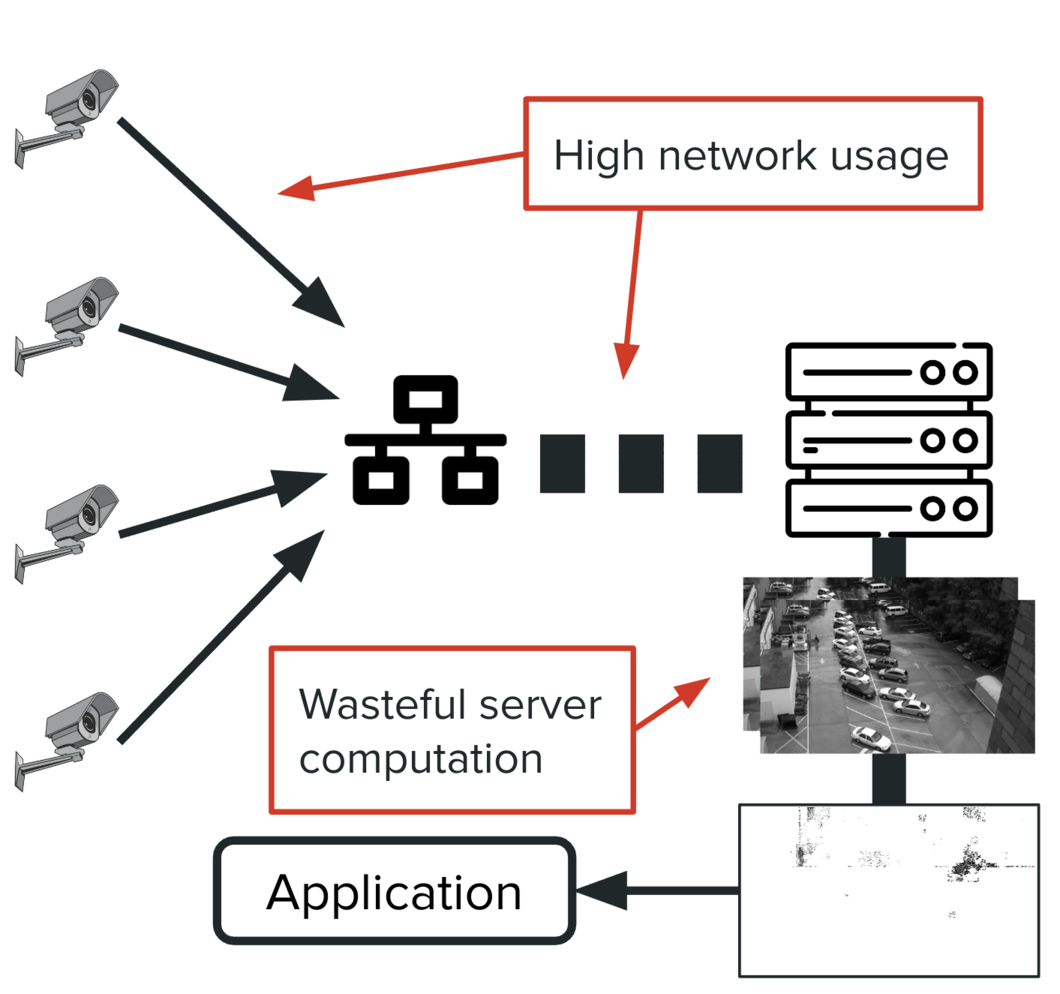}
         \caption{Traditional surveillance system with framed cameras}
         \label{fig:edge_current}
     \end{subfigure}
     \hfill
     \begin{subfigure}[c]{0.46\textwidth}
         \centering
         \includegraphics[width=\textwidth]{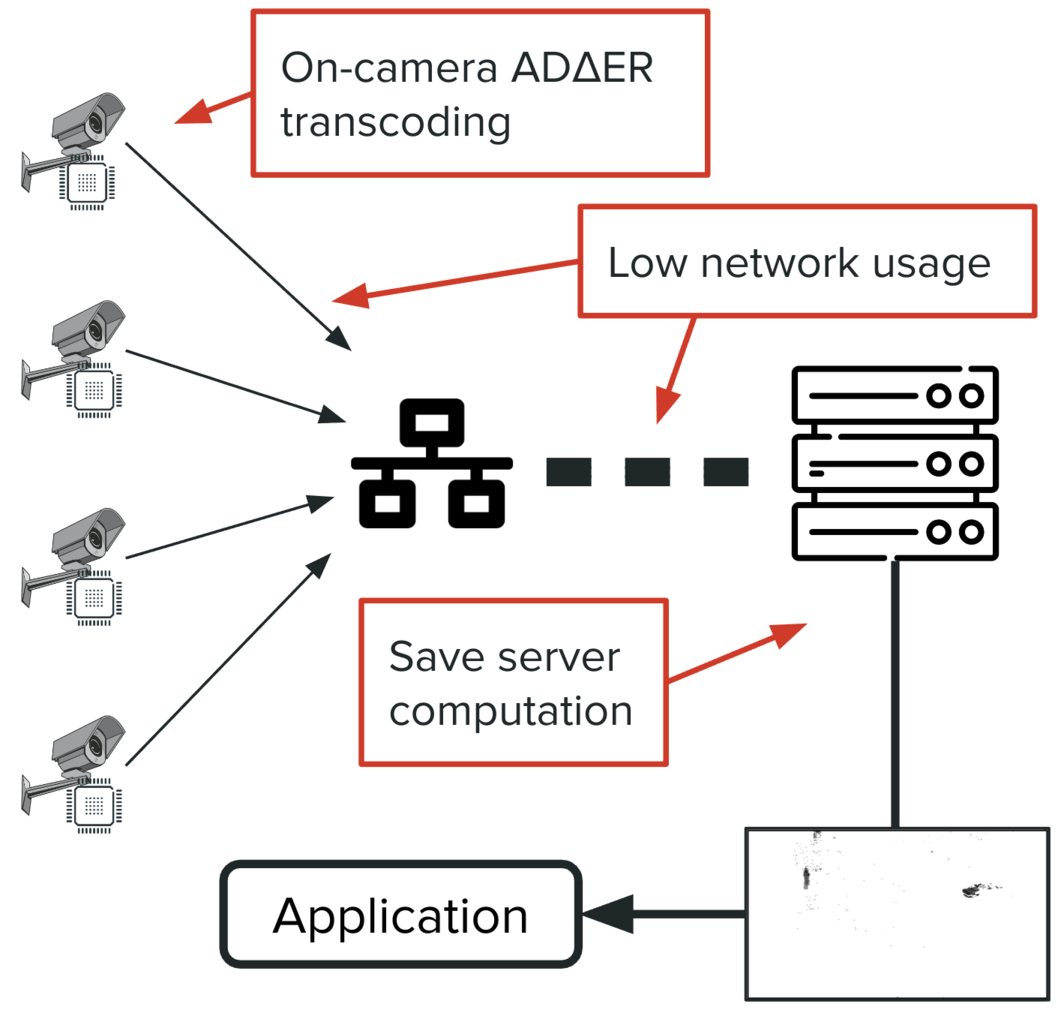}
         \caption{Proposed surveillance system with \adder{} transcoding at the edge}
         \label{fig:edge_adder}
     \end{subfigure}
     \caption[Surveillance system comparison]{Surveillance system comparison. With further development of \adder{} compression and hardware transcoding at the edge, we may unlock lower bandwidth requirements and activity-driven applications.}
     \label{fig:edge_comparison}
\end{figure*}

With edge transcoding, a centralized server can benefit from activity-driven applications. Since the server will receive new information from a camera \textit{only} when that camera records new intensity information, an application need only process the new spatiotemporal regions of activity. This sparse computation contrasts with the classical method for identifying motion, where the application continually thresholds the difference matrix of consecutive frames. \cref{fig:edge_comparison} compares an existing surveillance system to the proposed \adder{} edge system. Since the \adder{} system makes the server and application processing camera agnostic, one could also add DAVIS or future Aevon cameras to a surveillance deployment. \adder's separation of the acquisition, representation, and application layers simplifies multimodal sensor deployment.


\section{Conclusion}
This chapter proposes a number of extensions to the \adder{} video framework to enable simple quality control, application-driven rate adaptation of the raw representation, and robust source-modeled arithmetic coding. Overall, we find a unique set of trade-offs between video quality, application speed, and application accuracy in this asynchronous paradigm. Since vision applications typically operate on decoded intensity representations, the \adder{} representation makes possible application acceleration if the decoded data rate is sufficiently lower than that of a frame-based system. I can achieve lower \adder{} rates by reducing the transcoder quality, but such a change can potentially harm the accuracy of downstream applications. I show that my FAST feature detection can be incorporated into the compression loop itself, to ensure that high quality is preserved in the regions likely to be of highest salience for other downstream applications, while enabling faster application speed. In large-scale video surveillance system deployments, even a modest reduction in application computation time can translate to hundreds of thousands of dollars in savings per year. As native event-based intensity sensors such as Aeveon \cite{aeveon} enter the fold, this work shows a robust system to intelligently reduce high event rates and adapt vision algorithms to the asynchronous paradigm.

 \chapter[~~~~~~~~~~~~Open-Source Software Description]{Open-Source Software Description\footnotemark}\label{ch:software_description}\footnotetext{Significant portions of this chapter previously appeared in the proceedings of 2024 ACM Multimedia Systems \cite{freeman_mmsys_2024_osd}.}

\graphicspath {{ch06_software_description/images/}}

\section{Introduction}

My work in this dissertation has required the ground-up development of an entirely new video codec and application interface. I separate the concerns of researchers in the disparate areas of event-based hardware, networking, and vision: rather than developing techniques for a single event camera and file format, researchers' efforts can have forwards compatibility with future camera types. Furthermore, I bring classical video into the asynchronous paradigm, meaning that event-based applications developed for this framework will also be compatible with traditional frame-based video sources. 

From the beginning, I have placed an emphasis on open-source software development, so that my work may be more accessible for others working in the event video space. In this chapter, I describe in detail my software for transcoding to a unified event representation, rate adaptation, compression mechanisms, event-based applications,  visual playback, stream inspection, and a graphical interface. The software is available from a centralized repository at \href{https://github.com/ac-freeman/adder-codec-rs}{https://github.com/ac-freeman/adder-codec-rs}.

\begin{figure}
  \centering
  \includegraphics[width=1.0\textwidth]{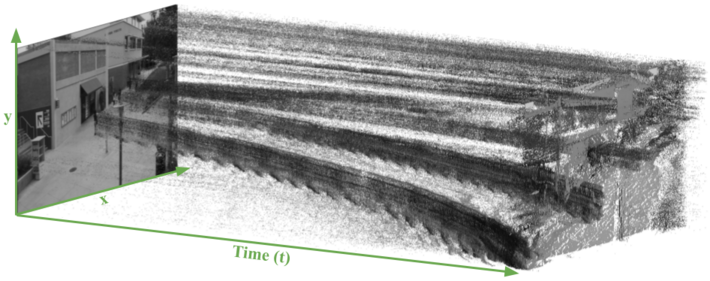}
  \caption[Visualization of an \adder{} video]{Visualization of an \adder{} video. The framed input (left) produces a stream of temporally sparse intensity samples (``events'') which are concentrated on areas of high motion. The burst of events near the beginning (right) ensures that we obtain an initial intensity for every pixel.}
  \label{fig:teaser}
\end{figure}

\section{Software Architecture}

I designed the \adder{} software to be highly modular. \cref{fig:system_diagram_full_repeat} illustrates the various components and their interdependence. The name of each standalone component is conveyed in italics. I wrote the software in the Rust programming language, and the standalone components are available for download from the Rust Package Registry.

\subsection{Common Codec }\label{sec:core}

The core library (\textit{adder-codec-core} in \cref{fig:system_diagram_full_repeat}) handles the encoding and decoding of \adder{} events, irrespective of their generation. I expose an interface whereby an \adder{} transcoder may instantiate an encoder with a set of options and then simply send its raw events for that encoder to handle. The core can write the raw events directly to a file or stream, or (if the user chooses) queue up event sequences to perform source-modeled lossy compression, as described in \cref{sec:adder_compression}. Similarly, one may import the core library to act as a decoder for arbitrary \adder{} streams if building a custom application or video player.

To perform lossy compression, the programmer must import the core library with the ``compression'' feature flag enabled. Currently, this feature requires the nightly release channel for Rust, due to unstable features in the subsequent dependency for arithmetic coding. For this reason, the ``compression'' feature is currently disabled by default.

\begin{figure}
        \centering
        \includegraphics[width=1.0\linewidth]{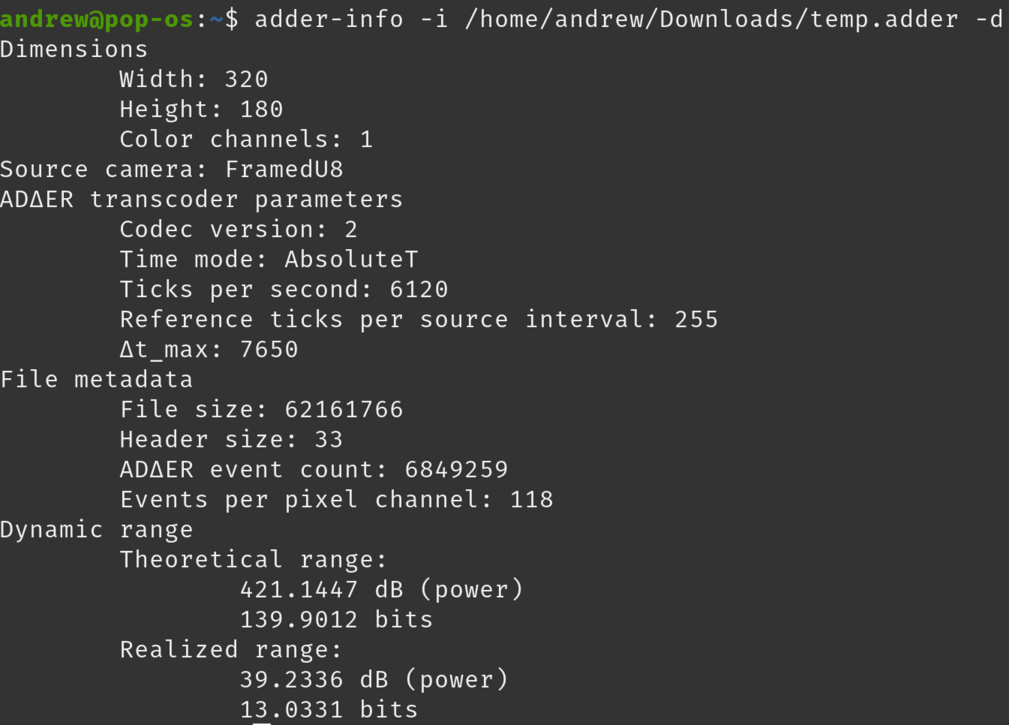}
        \caption[Example output of the \textit{adder-info} utility]{Example output of the \textit{adder-info} utility. The program reports metadata from the file header and calculates the dynamic range of the video.}
        \label{fig:adder_info}
    \end{figure}

\subsection{Metadata Inspection}
The \textit{adder-info} program provides a simple command-line interface to quickly inspect the metadata of an \adder{} file. This program is analogous to the \texttt{ffprobe} utility for framed video \cite{ffmpeg}. It prints information extracted from the file header (encoded by the core library) about the video resolution, time parameters, and video source. Optionally, the program can scan the file to determine the event rate and the dynamic range. In this case, dynamic range refers to the realized precision of the event intensities, given in bits by $\log_2(I_{max}/I_{min})$. Since \adder{} allows stable pixels to average their intensities over time, the precision is often higher than what the source representation allows. Example output is shown in \cref{fig:adder_info}, where the 8-bit framed video source has a reported dynamic range of 13.03 bits in the \adder{} representation. 

\subsection{Middleware}

The programming interfaces for transcoding videos to \adder{}, reconstructing image frames, and running event-based applications are found within the \textit{adder-codec-rs} package.

\subsubsection{Event Generation}

\adder{} currently supports transcoding from frame-based video sources, DVS event sources from camera manufacturers iniVation and Prophesee, and multimodal DVS event \textit{and} framed sources from iniVation DAVIS cameras. The transcoder defines a shared \texttt{Video} interface for generic source video types, including the event encoding mechanism (which calls on \textit{adder-codec-core}, as in \cref{sec:core}), pixel models, and integration functions. A \texttt{Video} has a 3D array (for $x$, $y$, and $c$) of \texttt{EventPixel} structs, which integrate intensity inputs over time to generate \adder{} events. Each \texttt{EventPixel} is independent, determining its optimal $D$ values according to the scheme described in \cite{freeman_mmsys23}.

This scheme uses a linked list to integrate incoming intensities at a range of possible $D$ values. When a node reaches its $2^{D'}$ (for some particular $D'$) intensity threshold, the child of that node is replaced with a new node initialized with decimation $D'$. The parent node stores the generated event in memory and increments its decimation to $D' + 1$ to continue integrating intensities. Then, when the incoming intensity change exceeds the threshold $M$, the \texttt{EventPixel} returns the event stored for each node in the list \cite{freeman_mmsys23}. By design, these events are ordered with monotonically decreasing $D$ and $\Delta t$ values. That is, the first event will have the largest $D$ and implicit $\Delta t$ value, spanning the majority of the integration time.

The multi-node integration process ensures that the full integrated intensity over a long, stable period of time can be precisely represented. However, it can lead to slow performance if recursion is deep, such as when a pixel is very stable and thus has several nodes to integrate. Additionally, the slight variance in intensity precision between a pixel's first event and last event has a negligible effect on reconstruction quality. As such, I introduce a new ``Collapse'' pixel mode as the default integration scheme for \texttt{EventPixel} structs. Under this mode, each pixel integrates \textit{only} a single node, successively incrementing its $D$ when it reaches the integration threshold. When the intensity change exceeds $M$, however, the pixel must account for any time that has elapsed since it generated its candidate event. Therefore, the pixel returns both its candidate event and an ``empty'' event with a reserved $D$ symbol spanning the intervening time. For example, suppose we have an  \texttt{EventPixel} with state $D=9$, $t = 519$, and a running integration of 324 intensity units. The candidate event for the pixel, $\{ D = 8, t = 410\}$ was generated when it reached its last $2^D$ integration, 256. When the incoming intensity changes beyond $M$, this pixel returns the events $\{ D = 8, t = 410\}$ and $\{ D = \texttt{EMPTY}, t = 519\}$. Applications which digest these events will then interpret the latter event as carrying the same average intensity as the first. The collapse mode greatly improves the integration speed, as shown in \cref{tab:perf_framed}.

A particular video source (e.g., framed or DVS) implements the \texttt{Source} trait. Each implementation then defines how to read data a data point from the source representation and convert it into an intensity and timespan. For example, the \texttt{Source} implementation for frame-based video uses an FFmpeg \cite{ffmpeg} backend to decode the next image frame. Then, each 8-bit pixel intensity is integrated in an \texttt{EventPixel} for the same time period (e.g., 255 ticks). In contrast, the multimodal DVS and framed video source must decode packets from a proprietary camera output format, then alternatingly integrate frame-based intensities over a fixed timespan and DVS intensities over variable timespans. 

For event-based video sources, I leverage the \textit{davis-edi-rs} package, which I introduced in \cite{freeman_mmsys23}. This component can ``deblur'' image frames based on their corresponding DVS events, allowing for higher-quality transcodes. Currently, this package depends on OpenCV to process the image frames in log space. One can avoid this dependency by disabling the ``open-cv'' feature flag for the \textit{adder-codec-rs} package, but doing so will remove the ability to transcode from event-based sources. Future work will focus on removing the OpenCV dependency.

\begin{figure}
  \centering
  \includegraphics[width=1.0\textwidth]{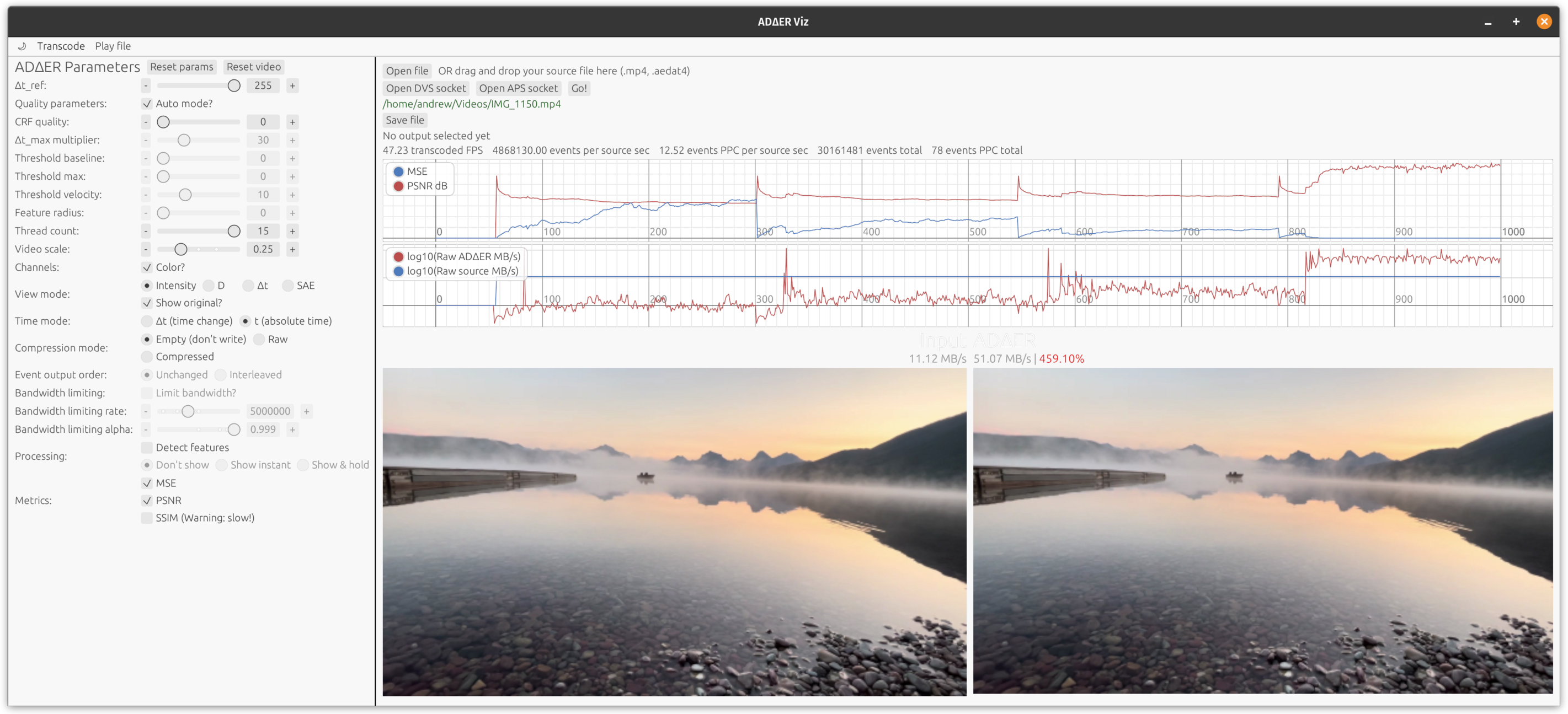}
  \caption[The \textit{adder-viz} transcoder user interface]{The \textit{adder-viz} transcoder user interface. This short video clip was transcoded to \adder{} at successively higher quality levels. As the CRF level decreases, the quality metrics improve (top graph), but the bitrate increases (bottom graph).}
  \label{fig:transcoder}
\end{figure}

\subsubsection{Framed Reconstruction}\label{sec:reconstruction}

While \adder{} is asynchronous, display pipelines and most existing vision applications require framed images. As such, this package also provides a reconstructor to generate an image sequence from arbitrary \adder{} events. The user provides an output frame rate which, based on the ticks per second of the \adder{} stream, determines how many ticks each output frame will span. Then, the reconstructor awaits raw events, scaling their intensities to match the timespan of the output frames. When all the pixels in a frame have been initialized with an intensity, the reconstructor outputs that frame. An example is shown in \cref{fig:player_int}.
\paragraph{\texorpdfstring{$D$}{D}-Value Images}\label{sec:dval}
        Rather than calculating the intensity of the last event for each pixel, I may also sample the $D$ portion of the events alone. $D$ describes both the log intensity captured by an event and an indication of past event stability, since accurate $\Delta t$ predictions will produce a higher $D$ value for a given pixel over time. Thus, a $D$-sampled image is a log intensity image with region smoothing where there are stable pixel histories. These images may find application as supplemental features in machine learning applications, or incorporated in \adder{} post-processing systems. An example is shown in \cref{fig:player_d}.
    
    \paragraph{\texorpdfstring{$\Delta t$}{Delta t} Sampling}\label{sec:dt_sampling}
    If I sample the last event fired by each pixel, I can map the events' $\Delta t$ values onto an image. A small $\Delta t$ value indicates an event fired much sooner than the controller expected, so these images can emphasize the spatial regions of the video with high temporal entropy. An example is shown in \cref{fig:player_dt}.

\subsubsection{Applications}
Finally, this package contains an implementation of the FAST feature detector as described in \cref{sec:feature_rate_control}. I ported and modified the OpenCV FAST detector, which operates on image frames, to instead run on individual pixels. My version receives a pointer to an array which contains the most recent intensity for every pixel, as well as the coordinates of the pixel for which to run the feature test. When the \adder{} event rate is sufficiently low, I found that the event-based version of the algorithm runs upwards of 43\% faster than OpenCV on the VIRAT surveillance dataset.

This application (and future applications) can transparently both during \adder{} video playback and while transcoding to \adder{}. In the latter case, one may use the application results to dynamically adjust the pixel sensitivities, allocating available bandwidth towards the pixels of greatest interest. This option is illustrated with the dashed lines in \cref{fig:system_diagram_full_repeat} and described in greater detail in \cref{sec:feature_rate_control}. Due to the current low-level integration of the transcoder and applications, incorporating or modifying this application-specific sensitivity adjustment requires changes to the transcoder source code. In the future, I will work to make a modular interface for applications and their effect on transcoder behavior, so that one can develop new applications without delving into the transcoder itself.

The key innovation here is that event-based applications can be, for the first time, agnostic to the imaging modality. My FAST feature detector runs identically on frame-based inputs, DVS inputs from iniVation or Prophesee cameras, and multimodal DVS/APS inputs from iniVation DAVIS cameras. It does not require any tuning or modification for different camera sources. Likewise, \adder{} applications can support any future event-based sensors (such as Aeveon \cite{aeveon}), so long as one implements a simple camera driver and \adder{} transcoder module.

\subsection{Graphical Interface}

Recognizing that the command-line interfaces can be slow and esoteric for new users, I created the \textit{adder-viz} application to enable straightforward explorations of the \adder{} framework. 

\begin{figure}
     \centering
     \begin{subfigure}[t]{0.32\textwidth}
         \centering
         \includegraphics[width=\textwidth]{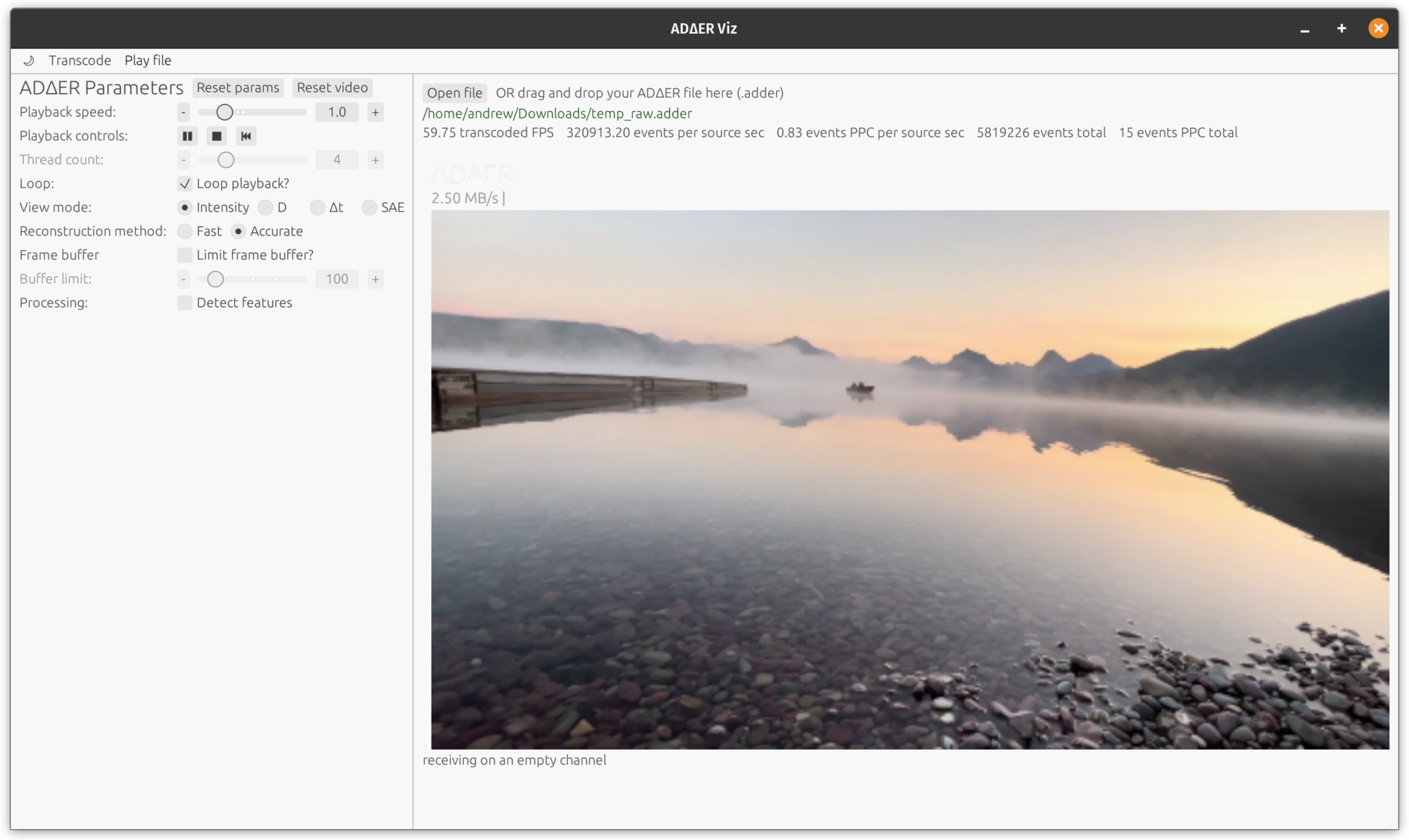}
         \caption{Intensities}
         \label{fig:player_int}
     \end{subfigure}
     \hfill
     \begin{subfigure}[t]{0.32\textwidth}
         \centering
         \includegraphics[width=\textwidth]{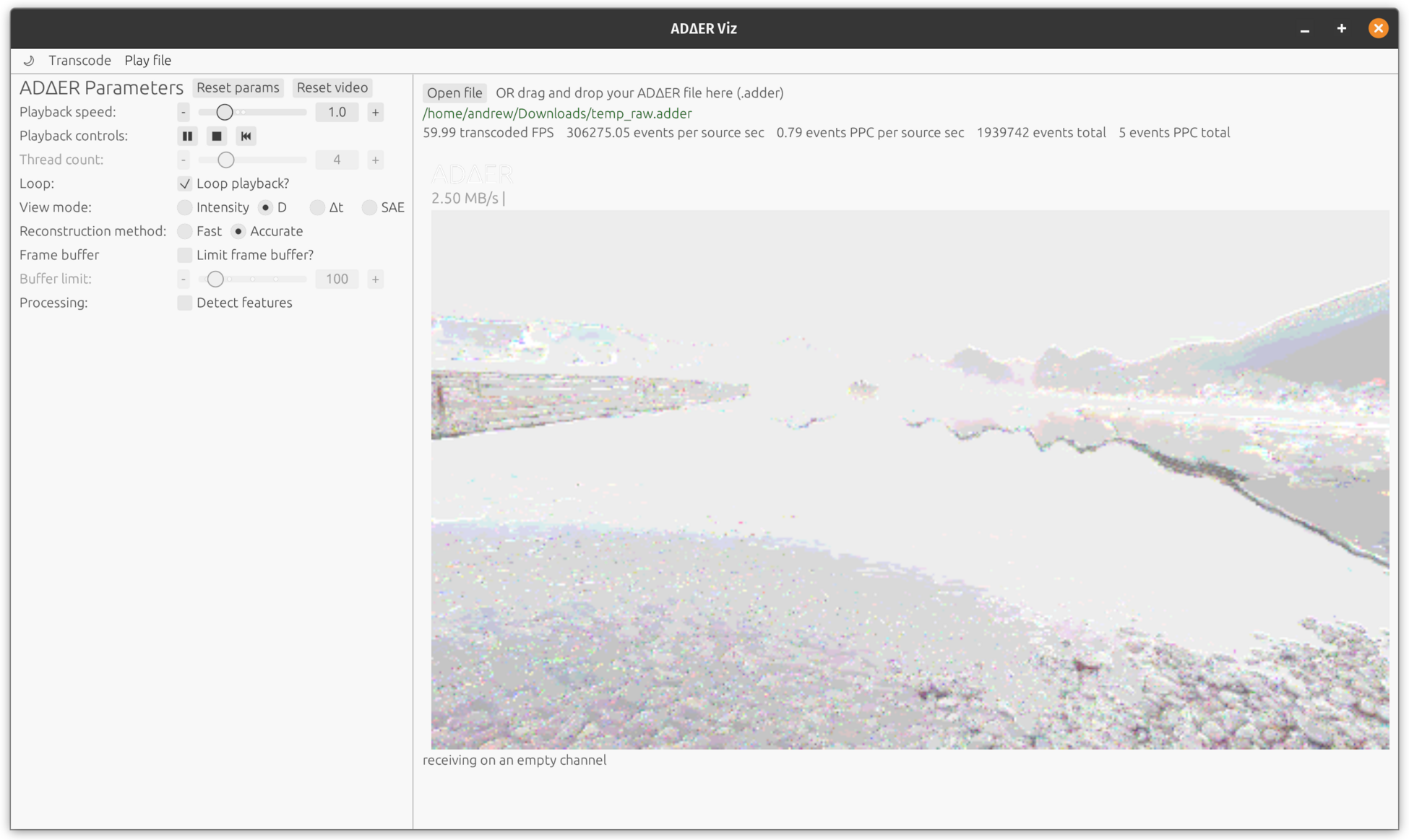}
         \caption{Event $D$ components}
         \label{fig:player_d}
     \end{subfigure}
     \hfill
     \begin{subfigure}[t]{0.32\textwidth}
         \centering
         \includegraphics[width=\textwidth]{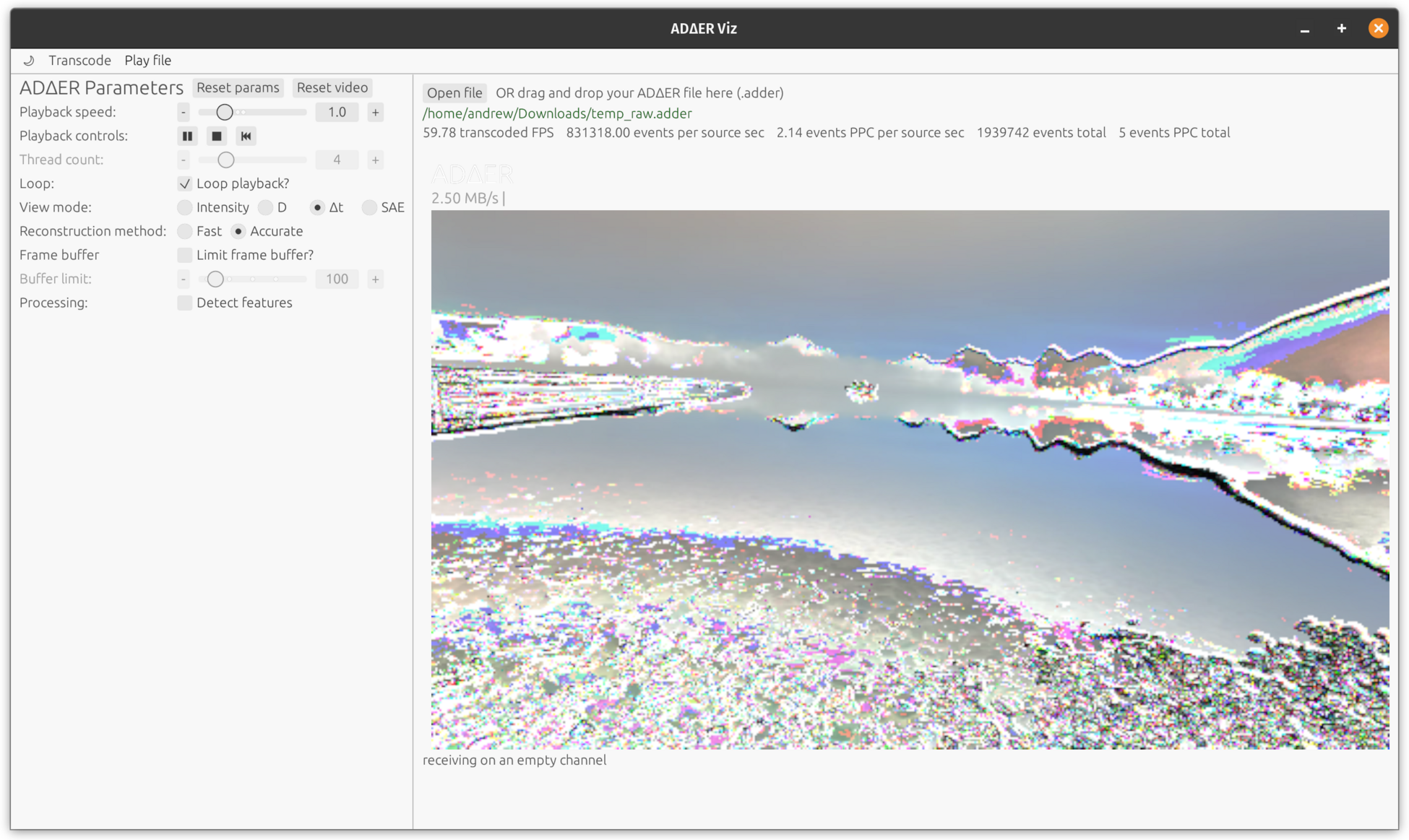}
         \caption{Event $\Delta t$ components}
         \label{fig:player_dt}
     \end{subfigure}
     \caption[The \textit{adder-viz} player interface]{The \textit{adder-viz} player interface for \adder{} video, with different visualization modes shown.}
     \label{fig:players}
\end{figure}

\subsubsection{Transcoder Interface}

The transcoder GUI is shown in \cref{fig:transcoder}. A user can open a framed video file, an AEDAT4 file (from an iniVation-branded DVS camera), or a DAT file (from a Prophesee-branded DVS camera) with a file dialog or by dragging and dropping into the window. Alternatively, a user can open a live connection to a hybrid DVS camera by opening a socket for the DVS events and a socket for the frames. This method requires that the iniVation driver software is running and publishing the data to the respective UNIX sockets. The user can export the \adder{} data by selecting the ``Raw'' or ``Compressed'' output modes and a ``Save file'' destination.

The main panel of the window shows a live view of the transcoded events (the right image in \cref{fig:transcoder}). When the source is a framed video, the application by default displays the input frame on the left. I update the live \adder{} image array with the intensity of a pixel each time it generates a new event. This live update is synchronous with the input, obviating the need for slower framed reconstruction (\cref{sec:reconstruction}).

The left panel shows the various transcoder parameters that a user can adjust. These include settings related to the time representation, pixel sensitivities, resolution, color, and feature-driven rate adaption. Many of these settings can be controlled with a single slider for Constant Rate Factor (CRF) quality, as described in prior work \cref{sec:crf}. Settings specific to DVS/DAVIS camera sources are made available once an appropriate file or socket connection is established \cite{freeman_mmsys23}. These include options related to event-based deblurring of intensity frames from DAVIS \cite{Pan_EDI}. The user can also enable event-based FAST feature detection and visualize the detected features on the live image.

I provide a number of metrics which are visualized above the display views in \cref{fig:transcoder}. The top plot shows frame-based quality metrics, which the user can enable if the source is a framed video. These include mean squared error (MSE), peak signal-to-noise ratio (PSNR), and structural similarity index measure (SSIM). The bottom plot illustrates the decompressed bitrates of the source and the \adder{} representations. The user can reference these plots to see, in real time, the effect of changing \adder{} transcode parameters on quality and bitrate. For example, \cref{fig:transcoder} shows that our transcoded representation has a lower decompressed bitrate than the source video at many lossy quality levels, but a higher bitrate at the lossless quality level.

\subsubsection{Playback Interface}

\cref{fig:players} shows the video playback interface. One can select a file through a drag-and-drop interaction or a file explorer prompt. The user can pause the video, adjust the playback speed, and visualize the $D$ and $\Delta t$ event components. Furthermore, my event-based FAST feature detection application \cite{freeman_mmsys23} is also available during playback. 

The player supports two playback modes: accurate and fast. The accurate mode leverages the framed reconstruction technique describe in \cref{sec:reconstruction}. This method yields the best visual quality, but may introduce high latency. For example, if some pixel is stable for the duration of a video, it will fire only one event near the beginning of the video encoding. The reconstructor  does not have \textit{a priori } knowledge on whether the pixel has additional events in the future, so it must build a queue of frames for the \textit{entire} video before playback begins. To mitigate the high latency this may cause, the user has an option to limit the size of the frame buffer, such that the reconstructor will assume that a pixel intensity has not changed if its last event sufficiently long ago.

In contrast, the fast playback mode simply holds a single image array and updates each pixel intensity when it decodes a new event for that pixel. At a sample rate of the user's choosing, we may derive an intensity frame by calculating
        
        \begin{equation}
            \frac{2^D}{I_{max}} \cdot \frac{\Delta t_{frame}}{\Delta t}
        \end{equation}
        
        for each particular \adder{} event $\{D, \Delta t\}$, where $I_{max}$ is the maximum intensity permitted in the resulting frame, and $\Delta t_{frame}$ is the number of ticks for the resulting frame to span. The parameters $I_{max}$ and $\Delta t_{frame}$ together determine the normalization of the image.

The player then updates the displayed frame once the change in $t$ represented by any event exceeds the frame interval threshold $\Delta t_{frame}$. Thus, the method is beholden to the temporal event order within the file, which is not guaranteed to be perfectly ordered between different pixels. The lower latency allowed by this method, however, makes it suitable for vision applications. An application may sample these images from the \adder{} stream as needed, rather than extracting a video with a fixed sample rate, whereas the accurate playback mode is a better suited visualization for human viewers.

\begin{table}
    \centering
    \begin{tabular}{cc|cc|cc}
        & & \multicolumn{2}{c|}{Raw events} & \multicolumn{2}{c}{Lossy compression} \\
          &
         Resolution & Grayscale & Color & Grayscale & Color \\
        \hline \parbox[t]{2mm}{\multirow{4}{*}{\rotatebox[origin=c]{90}{Normal}}} 
        & 480×270   & 209.0 & 109.7  & 153.9   & 71.1 \\
        & 960×540   & 71.6  & 33.8  &  43.2  & 17.1 \\
        & 1440×810  & 32.9  & 15.7  & 19.6   & 7.6 \\
        & 1920×1080 & 22.7  & 9.5  &  11.5  & 4.1 \\
    \hline \parbox[t]{2mm}{\multirow{4}{*}{\rotatebox[origin=c]{90}{Collapse}}} 
        & 480×270   & 262.3 & 161.6 & 176.6 & 88.9 \\
        & 960×540   & 91.8  & 48.4  & 51.3 & 20.5 \\
        & 1440×810  & 43.2  & 22.2  & 23.0 & 9.0 \\
        & 1920×1080 & 31.1  & 13.8  & 13.5 & 4.7 \\
        
    \end{tabular}
    \caption[Framed transcode FPS]{Frames per second which a representative framed video can be transcoded to \adder{}. Resolution, color depth, and lossy compression are varied. A CRF value of 3 (the default) was used for these experiments.}
    \label{tab:perf_framed}
\end{table}

\section{Performance and Future Work}

I gathered general speed measurements on a machine with a Ryzen 5800x CPU with 8 cores and 16 threads. As shown in \cref{tab:perf_framed}, the \adder{} transcoder achieves fast performance for low-resolution framed inputs, but slows substantially at high definition. At 540p resolution and higher, the time required to transcoding a color video is about double that of the grayscale version.

I use a memory-safe parallelization scheme for matrix integrations (transcoding image frames) and framed reconstruction, whereby pixel arrays are divided into groups of spatial rows. Currently, however, most other processes are serial. According to performance profiling results, roughly $75\%$ of transcoder execution time is spent on pixel integrations. Each software pixel is a struct that is dynamically sized according to the length of its event queue. The transcoder logic will likely benefit from a GPU implementation, though this may require a static maximum queue length. This may prove most advantageous on GPUs with direct memory access, so that high-rate event sequences do not have to move through the CPU cache during encoding or decoding.

Currently, the lossy compression scheme is single-threaded, and it carries a computational overhead over raw event encoding (\cref{tab:perf_framed}). Within the \textit{adder-viz} GUI, this manifests as a notable pause each time an application data unit (ADU) of events undergoes lossy compression. This latency may be mitigated if compression were delegated to its own background thread and if a second ADU were constructed while one is being compressed. Furthermore, compression speed can likely be improved by dividing the events into horizontal spatial regions for parallel processing.

While this tool suite provides an end-to-end system for event-based video, there is ample room to extend it with additional features and performance improvements. I will also work to create a generic application interface and simplify the application integration process for new researchers.  Finally, I will work to incorporate \adder{} events as inputs for novel spiking neural network vision applications, which can take advantage of sparse representations.

\section{Conclusion}

This open-source release and user guide for the \adder{} framework provides researchers with straightforward tools to experiment with forward-looking event video. As the sensor community pushes for ever higher resolution, dynamic range, sample rate, and event-based representations, my \adder{} software unlocks novel approaches to rate adaptation, compression, and applications.
 \chapter[~~~~~~~~~~~~Conclusion]{Conclusion}
\label{ch:conclusion}

This dissertation sought to unify the disparate realms of classical video and modern event-based video. Traditional video systems were initially designed to optimize the human viewing experience, with vision applications being of secondary concern. Existing event-based systems have taken the opposite approach, having bespoke processing mechanisms for individual cameras and applications. I argued that these two modes of thinking have lessons to learn from each other. In response, I proposed a universal video representation, the \textbf{A}ddress, \textbf{D}ecimation, $\Delta t$ \textbf{E}vent \textbf{R}epresentation (\textbf{\adder}). As opposed to the contrast-based sensing mechanism of prominent DVS event sensors, this representation conveys \textit{absolute} intensity measurements.

\section{Summary}

Let us revisit the thesis statement from \cref{ch:intro}, repeated below.

\begin{quote}
\textit{
    Arbitrary spatio-temporal video data can be represented as a single asynchronous, compressible data type. Applications can operate on this unified data representation, rather than targeting a number of source-specific representations. This universal representation unlocks advantages in compression, rate adaptation, and application speed and accuracy.
}
\end{quote}

To motivate this thesis, I described a generalized system model with three layers: acquisition, representation, and application. In the acquisition layer (\cref{ch:adder_transcoding}), \adder{} encompasses the data of both frame-based and DVS-based video sources, through the use of my software transcoders. My method for combining framed and DVS data outperformed the state-of-the-art in fusion speed. The transcode process serves as the first stage of lossy compression, by allowing pixels to average out slight variations in intensity samples as a single event. This effort demonstrates that arbitrary video can be represented with an event representation, and that the representation is lossy-compressible.

In the representation layer (\cref{sec:adder_reprsentation_layer}), I introduced a scheme for organizing sparse \adder{} events into a memory-dense data structure. My lossy source-modeled compression scheme can readily achieve 2:1 compression ratios over the input events. On a framed surveillance video dataset, my compressed \adder{} representation approaches the compression performance of H.265, despite having a much less complex encoding scheme.

The sparse raw representation yields improvements to application quality and speed. In the application layer (\cref{sec:adder_application_layer}), I explored a number of computer vision tasks which leverage different interfaces for \adder{} data. On a classical object detection task, I showed up to a 4\% improvement in precision by simply adding the events' $D$ and $\Delta t$ components to a framed image representation. I demonstrated that we can adapt frame-based algorithms, which operate on each pixel of an image in sequence, to the asynchronous video realm. On FAST feature detection, my event-based implementation performed $43.7\%$ faster on \adder{} events than OpenCV on image frames. These speed results demonstrated cost-saving practical advantages to the \adder{} system. \adder{} exposes an entirely unique control mechanism where the application speed may vary with the both the dynamism of the video content and the level of lossy compression.

Finally, I demonstrated that my end-to-end system has a simple interface to drive rate adaptation based on application-level performance. By adjusting pixel sensitivites based on their proximity to detected features or objects, we can allocate bitrate towards the regions of predicted saliency. I demonstrated that with temporal foveation, we can dramatically reduce the event rate of a simulated ASINT sensor. Furthermore, we can cluster FAST features as prototype object boundaries to filter events from a DVS camera source. With this scheme, I demonstrated that we can maintain large-object detection accuracy near the ground truth performance while reducing the bitrate by 76.8\%.

\section{Future Work}

As discussed throughout this dissertation, there is ample room to expand upon my work across all layers of the \adder{} system. 

As new event cameras gain commercial availability, I intend to add support for them in the \adder{} acquisition layer. For all cameras, I will work to support additional binary formats of the various manufacturers, and to implement live data transcoding through the manufacturer-provided camera drivers.

The transcoding and compression software will greatly benefit from hardware acceleration. I will work to find additional opportunity for CPU parallelization through multithreaded execution and SIMD instructions. I will also seek out collaborations with computer engineering experts to aid with the development of a dedicated video engine with an FPGA. With a hardware implementation, I expect to see extreme gains over traditional codecs in compressing sparse video. If I develop a robust \adder{} quality metric, further investigations into motion compensation, transform coding, and adaptive streaming will likely bear more fruit. Hardware encoding will allow us to temporally ``sparsify'' video data on the edge (at the camera source) and explore computational and bandwidth savings for large-scale surveillance systems.

I will simplify the development and integration of applications with a simplified API for requesting \adder{} data from the representation layer. These requests may be either requests for individual pixel streams (e.g., in regions of interest) or requests for all events spanning some time interval.

Furthermore, I will expand the functionality for end-to-end transcoding and applications with dynamic rate control. I will add an application-generic rate controller API which will set the \adder{} pixel sensitivities. An application will be able to make a request to this API to set the sensitivities based on application-level concerns such as application accuracy. Additionally, the representation layer must have a mechanism for adjusting the data rate according to the network capacity. I will incorporate an ABR mechanism to dynamically choose the lossy compression levels based on the available bandwidth. The network-based ABR scheme must work in tandem with the application-level API.

A rich area of potential research that I hope my colleagues will engage in is new applications for intensity event data. As the DVS literature has shown, many applications require entirely new techniques for the event-based paradigm \cite{survey}. While  \adder{} supports more traditional processing techniques, there will likely be a computational benefit to bespoke neuromorphic processing with spiking neural networks.

\section{Impact}

\adder{} provides inroads towards source-agnostic, adaptable event video systems. For high-frame-rate or stationary-camera video from framed sources, it offers the potential for better compression ratios and faster vision applications. For existing event cameras, it offers faster frame-event fusion, lossy compression, and rate adaptation. For any arbitrary video source, \adder{} exposes has a single interface for lossy compression and applications. Therefore, for the first time, event-based applications can be designed around a universal representation and have compatibility with existing and future event-based cameras. At the same time, \adder{} is trivially backwards compatible with classical frame-based applications, through instantaneous frame sampling or framed reconstruction of \adder{} intensities. My open-source software release makes these tools available to the public, enabling further research into the rapidly-growing world of event video.

My hope is that this work will be used to bridge the gap between cutting-edge computer vision research and practical systems. Especially with the advent of \adder-style sensors such as Aeveon, I believe that the research community will start to take more seriously the practical issues of high-rate video data. I would like to see event cameras gain prominence in real-world products, and I believe that an open ecosystem and generic video interface will be necessary. I hope that an \adder-style representation is at the center of the design considerations for future systems.

In this rapidly growing field, I took an unusual but rewarding path. My colleagues by and large have limited their work only to the event cameras currently in existence, often assuming infinite storage and computation availability, and focused on developing novel computer vision applications. While that work has been extremely compelling and useful, I sought to make the processing pipeline amenable to upcoming sensors and apply event-based techniques to readily-available classical video. Even still, I have only scratched the surface of what is possible with a universal event video representation. I am confident that the next decade of video research will reveal many improvements and extensions of these ideas.

% ...

\clearpage
\phantomsection

{\def\chapter*#1{} 
\begin{singlespace}
\addcontentsline{toc}{chapter}{BIBLIOGRAPHY}
\begin{center}
\textbf{BIBLIOGRAPHY}
\end{center}

\bibliographystyle{abbrvnat}
\bibliography{references.bib}

\begin{thebibliography}{142}
\providecommand{\natexlab}[1]{#1}
\providecommand{\url}[1]{\texttt{#1}}
\expandafter\ifx\csname urlstyle\endcsname\relax
  \providecommand{\doi}[1]{doi: #1}\else
  \providecommand{\doi}{doi: \begingroup \urlstyle{rm}\Url}\fi

\bibitem[avi()]{avi}
URL \url{https://web.archive.org/web/20160317062723/http://www.kk.iij4u.or.jp/~kondo/wave/mpidata.txt}.

\bibitem[blo()]{blooming_overview}
Ccd blooming and anti blooming: Can anti blooming sensors help? - andor learning centre.
\newblock URL \url{https://andor.oxinst.com/learning/view/article/ccd-blooming-and-anti-blooming}.

\bibitem[mkv()]{mkv}
URL \url{https://matroska.org/technical/elements.html}.

\bibitem[pha()]{phantom}
\url{https://web.archive.org/web/20231128210921/https://www.phantomhighspeed.com/products/cameras/tseries/t4040}.
\newblock URL \url{https://web.archive.org/web/20231128210921/https://www.phantomhighspeed.com/products/cameras/tseries/t4040}.

\bibitem[Almatrafi et~al.(2020)Almatrafi, Baldwin, Aizawa, and Hirakawa]{dayton_dataset}
M.~Almatrafi, R.~Baldwin, K.~Aizawa, and K.~Hirakawa.
\newblock Distance surface for event-based optical flow.
\newblock \emph{IEEE Transactions on Pattern Analysis and Machine Intelligence}, 42\penalty0 (07):\penalty0 1547--1556, jul 2020.
\newblock ISSN 1939-3539.
\newblock \doi{10.1109/TPAMI.2020.2986748}.

\bibitem[Alper and Alper(2019)]{blooming_and_smear}
G.~Alper and G.~Alper.
\newblock Ccd versus cmos: blooming and smear performance, Dec 2019.
\newblock URL \url{https://www.adimec.com/ccd-versus-cmos-blooming-and-smear-performance/}.

\bibitem[Baldwin et~al.(2020)Baldwin, Almatrafi, Asari, and Hirakawa]{dvsnoise_dataset}
R.~Baldwin, M.~Almatrafi, V.~Asari, and K.~Hirakawa.
\newblock Event probability mask (epm) and event denoising convolutional neural network (edncnn) for neuromorphic cameras.
\newblock In \emph{2020 IEEE/CVF Conference on Computer Vision and Pattern Recognition (CVPR)}, pages 1698--1707, Los Alamitos, CA, USA, jun 2020. IEEE Computer Society.
\newblock \doi{10.1109/CVPR42600.2020.00177}.
\newblock URL \url{https://doi.ieeecomputersociety.org/10.1109/CVPR42600.2020.00177}.

\bibitem[Baldwin et~al.(2022)Baldwin, Liu, Almatrafi, Asari, and Hirakawa]{TORE_volumes}
R.~Baldwin, R.~Liu, M.~M. Almatrafi, V.~K. Asari, and K.~Hirakawa.
\newblock Time-ordered recent event (tore) volumes for event cameras.
\newblock \emph{IEEE Transactions on Pattern Analysis and Machine Intelligence}, pages 1--1, 2022.
\newblock \doi{10.1109/TPAMI.2022.3172212}.

\bibitem[Banerjee et~al.(2021)Banerjee, Wang, Chopp, Cossairt, and Katsaggelos]{quadtree_compression}
S.~Banerjee, Z.~W. Wang, H.~H. Chopp, O.~Cossairt, and A.~K. Katsaggelos.
\newblock Lossy event compression based on image-derived quad trees and poisson disk sampling.
\newblock In \emph{2021 {IEEE} International Conference on Image Processing ({ICIP})}. {IEEE}, sep 2021.
\newblock \doi{10.1109/icip42928.2021.9506546}.
\newblock URL \url{https://doi.org/10.1109\%2Ficip42928.2021.9506546}.

\bibitem[Banerjee et~al.(2022)Banerjee, Chopp, Zhang, Wang, Kang, Cossairt, and Katsaggelos]{banerjee2021joint}
S.~Banerjee, H.~H. Chopp, J.~Zhang, Z.~W. Wang, P.~Kang, O.~Cossairt, and A.~Katsaggelos.
\newblock A joint intensity-neuromorphic event imaging system with bandwidth-limited communication channel.
\newblock \emph{IEEE Transactions on Neural Networks and Learning Systems}, pages 1--15, 2022.
\newblock \doi{10.1109/TNNLS.2022.3214779}.

\bibitem[Barbier et~al.(2021)Barbier, Teuliere, and Triesch]{Barbier_2021_CVPR}
T.~Barbier, C.~Teuliere, and J.~Triesch.
\newblock Spike timing-based unsupervised learning of orientation, disparity, and motion representations in a spiking neural network.
\newblock In \emph{Proceedings of the IEEE/CVF Conference on Computer Vision and Pattern Recognition (CVPR) Workshops}, pages 1377--1386, June 2021.

\bibitem[Benosman et~al.(2014)Benosman, Clercq, Lagorce, Ieng, and Bartolozzi]{eventflow}
R.~Benosman, C.~Clercq, X.~Lagorce, S.-H. Ieng, and C.~Bartolozzi.
\newblock Event-based visual flow.
\newblock \emph{IEEE Transactions on Neural Networks and Learning Systems}, 25\penalty0 (2):\penalty0 407--417, 2014.
\newblock \doi{10.1109/TNNLS.2013.2273537}.

\bibitem[Berner et~al.(2018)Berner, Brändli, and Zannoni]{dvs_rate_patent}
R.~Berner, C.~Brändli, and M.~Zannoni.
\newblock Data rate control for event-based vision sensor, Jul 2018.

\bibitem[Birnbaum et~al.(2019)Birnbaum, Kuleshov, Enam, Koh, and Ermon]{temporal_film}
S.~Birnbaum, V.~Kuleshov, S.~Z. Enam, P.~W. Koh, and S.~Ermon.
\newblock \emph{Temporal FiLM: Capturing Long-Range Sequence Dependencies with Feature-Wise Modulation}.
\newblock Curran Associates Inc., Red Hook, NY, USA, 2019.

\bibitem[Blanc(2001)]{blanc2001ccd}
N.~Blanc.
\newblock Ccd versus cmos-has ccd imaging come to an end.
\newblock In \emph{Photogrammetric Week}, volume~1, pages 131--137, 2001.

\bibitem[Bradski(2000)]{opencv_library}
G.~Bradski.
\newblock {The OpenCV Library}.
\newblock \emph{Dr. Dobb's Journal of Software Tools}, 2000.

\bibitem[Brandli et~al.(2014{\natexlab{a}})Brandli, Berner, Yang, Liu, and Delbruck]{davis_a}
C.~Brandli, R.~Berner, M.~Yang, S.-C. Liu, and T.~Delbruck.
\newblock A 240 × 180 130 db 3 µs latency global shutter spatiotemporal vision sensor.
\newblock \emph{IEEE Journal of Solid-State Circuits}, 49\penalty0 (10):\penalty0 2333--2341, 2014{\natexlab{a}}.
\newblock \doi{10.1109/JSSC.2014.2342715}.

\bibitem[Brandli et~al.(2014{\natexlab{b}})Brandli, Muller, and Delbruck]{DAVIS}
C.~Brandli, L.~Muller, and T.~Delbruck.
\newblock Real-time, high-speed video decompression using a frame- and event-based davis sensor.
\newblock In \emph{2014 IEEE International Symposium on Circuits and Systems (ISCAS)}, pages 686--689, 2014{\natexlab{b}}.
\newblock \doi{10.1109/ISCAS.2014.6865228}.

\bibitem[Brandli et~al.(2014{\natexlab{c}})Brandli, Muller, and Delbruck]{inherent_compression}
C.~Brandli, L.~Muller, and T.~Delbruck.
\newblock Real-time, high-speed video decompression using a frame- and event-based davis sensor.
\newblock In \emph{2014 IEEE International Symposium on Circuits and Systems (ISCAS)}, pages 686--689, 2014{\natexlab{c}}.
\newblock \doi{10.1109/ISCAS.2014.6865228}.

\bibitem[Bross et~al.(2021)Bross, Wang, Ye, Liu, Chen, Sullivan, and Ohm]{vvc}
B.~Bross, Y.-K. Wang, Y.~Ye, S.~Liu, J.~Chen, G.~J. Sullivan, and J.-R. Ohm.
\newblock Overview of the versatile video coding (vvc) standard and its applications.
\newblock \emph{IEEE Transactions on Circuits and Systems for Video Technology}, 31\penalty0 (10):\penalty0 3736--3764, 2021.
\newblock \doi{10.1109/TCSVT.2021.3101953}.

\bibitem[Butler and Baskin()]{Butler_Baskin}
R.~Butler and D.~Baskin.
\newblock Sony announces a9 iii: World’s first full-frame global shutter camera.
\newblock URL \url{https://www.dpreview.com/news/7271416294/sony-announces-a9-iii-world-s-first-full-frame-global-shutter-camera}.

\bibitem[Cannici et~al.(2019)Cannici, Ciccone, Romanoni, and Matteucci]{asyncconv1}
M.~Cannici, M.~Ciccone, A.~Romanoni, and M.~Matteucci.
\newblock Asynchronous convolutional networks for object detection in neuromorphic cameras.
\newblock In \emph{2019 IEEE/CVF Conference on Computer Vision and Pattern Recognition Workshops (CVPRW)}, pages 1656--1665, 2019.
\newblock \doi{10.1109/CVPRW.2019.00209}.

\bibitem[Chen et~al.(2021)Chen, He, Wang, Ren, Lim, and Shrivastava]{chen2021nerv}
H.~Chen, B.~He, H.~Wang, Y.~Ren, S.-N. Lim, and A.~Shrivastava.
\newblock Ne{RV}: Neural representations for videos.
\newblock In A.~Beygelzimer, Y.~Dauphin, P.~Liang, and J.~W. Vaughan, editors, \emph{Advances in Neural Information Processing Systems}, 2021.
\newblock URL \url{https://openreview.net/forum?id=BbikqBWZTGB}.

\bibitem[Chen et~al.(2022)Chen, Li, Zhang, Sun, and Jia]{focalsparse}
Y.~Chen, Y.~Li, X.~Zhang, J.~Sun, and J.~Jia.
\newblock Focal sparse convolutional networks for 3d object detection.
\newblock In \emph{2022 IEEE/CVF Conference on Computer Vision and Pattern Recognition (CVPR)}, pages 5418--5427, Los Alamitos, CA, USA, jun 2022. IEEE Computer Society.
\newblock \doi{10.1109/CVPR52688.2022.00535}.
\newblock URL \url{https://doi.ieeecomputersociety.org/10.1109/CVPR52688.2022.00535}.

\bibitem[Chen et~al.(2013)Chen, Patel, Shekhar, Chellappa, and Phillips]{jointsparse}
Y.-C. Chen, V.~M. Patel, S.~Shekhar, R.~Chellappa, and P.~J. Phillips.
\newblock Video-based face recognition via joint sparse representation.
\newblock In \emph{2013 10th IEEE International Conference and Workshops on Automatic Face and Gesture Recognition (FG)}, pages 1--8, 2013.
\newblock \doi{10.1109/FG.2013.6553787}.

\bibitem[Cho et~al.(2023)Cho, Lee, Je, Kim, Ryu, and No]{10114931}
M.~Cho, H.~Lee, H.~Je, K.~Kim, D.~Ryu, and A.~No.
\newblock Pynet-q×q: An efficient pynet variant for q×q bayer pattern demosaicing in cmos image sensors.
\newblock \emph{IEEE Access}, 11:\penalty0 44895--44910, 2023.
\newblock \doi{10.1109/ACCESS.2023.3272665}.

\bibitem[de~Tournemire et~al.(2020)de~Tournemire, Nitti, Perot, Migliore, and Sironi]{gen1_dataset}
P.~de~Tournemire, D.~Nitti, E.~Perot, D.~Migliore, and A.~Sironi.
\newblock A large scale event-based detection dataset for automotive.
\newblock \emph{CoRR}, abs/2001.08499, 2020.
\newblock URL \url{https://arxiv.org/abs/2001.08499}.

\bibitem[Delbruck et~al.(2021)Delbruck, Graca, and Paluch]{dvs_feedback_control}
T.~Delbruck, R.~Graca, and M.~Paluch.
\newblock Feedback control of event cameras.
\newblock In \emph{2021 IEEE/CVF Conference on Computer Vision and Pattern Recognition Workshops (CVPRW)}, pages 1324--1332, Los Alamitos, CA, USA, jun 2021. IEEE Computer Society.
\newblock \doi{10.1109/CVPRW53098.2021.00146}.
\newblock URL \url{https://doi.ieeecomputersociety.org/10.1109/CVPRW53098.2021.00146}.

\bibitem[Dikov et~al.(2017)Dikov, Firouzi, R{\"o}hrbein, Conradt, and Richter]{spiking1}
G.~Dikov, M.~Firouzi, F.~R{\"o}hrbein, J.~Conradt, and C.~Richter.
\newblock Spiking cooperative stereo-matching at 2 ms latency with neuromorphic hardware.
\newblock In M.~Mangan, M.~Cutkosky, A.~Mura, P.~F. Verschure, T.~Prescott, and N.~Lepora, editors, \emph{Biomimetic and Biohybrid Systems}, pages 119--137, Cham, 2017. Springer International Publishing.
\newblock ISBN 978-3-319-63537-8.

\bibitem[Dong et~al.(2017)Dong, Huang, and Tian]{vidar1}
S.~Dong, T.~Huang, and Y.~Tian.
\newblock Spike camera and its coding methods.
\newblock In \emph{2017 Data Compression Conference (DCC)}, pages 437--437, 2017.
\newblock \doi{10.1109/DCC.2017.69}.

\bibitem[dos Santos et~al.(2019)dos Santos, Sebe, and Almeida]{c3d}
S.~F. dos Santos, N.~Sebe, and J.~Almeida.
\newblock Cv-c3d: Action recognition on compressed videos with convolutional 3d networks.
\newblock In \emph{2019 32nd SIBGRAPI Conference on Graphics, Patterns and Images (SIBGRAPI)}, pages 24--30, 2019.
\newblock \doi{10.1109/SIBGRAPI.2019.00012}.

\bibitem[Duwek et~al.(2021)Duwek, Shalumov, and Tsur]{Duwek_2021_CVPR}
H.~C. Duwek, A.~Shalumov, and E.~E. Tsur.
\newblock Image reconstruction from neuromorphic event cameras using laplacian-prediction and poisson integration with spiking and artificial neural networks.
\newblock In \emph{Proceedings of the IEEE/CVF Conference on Computer Vision and Pattern Recognition (CVPR) Workshops}, pages 1333--1341, June 2021.

\bibitem[Eng(2023)]{aeveon}
K.~Eng.
\newblock Kynan eng at cvpr 2023 workshop on event-based vision, 2023.
\newblock URL \url{https://www.youtube.com/watch?v=tv-GqKg4Mak&ab_channel=RPGWorkshops}.

\bibitem[Ester et~al.(1996)Ester, Kriegel, Sander, and Xu]{dbscan}
M.~Ester, H.-P. Kriegel, J.~Sander, and X.~Xu.
\newblock A density-based algorithm for discovering clusters in large spatial databases with noise.
\newblock In \emph{Proceedings of the Second International Conference on Knowledge Discovery and Data Mining}, KDD'96, page 226–231. AAAI Press, 1996.

\bibitem[{Fabien Christin}(2020)]{photographer}
{Fabien Christin}.
\newblock Photographer blender add-on, 2020.
\newblock URL \url{https://gumroad.com/l/cQTgl}.

\bibitem[{FFmpeg Project}(2021)]{ffmpeg}
{FFmpeg Project}.
\newblock Ffmpeg, 2021.
\newblock URL \url{https://ffmpeg.org/}.

\bibitem[Fossum(1997)]{cmos_review}
E.~Fossum.
\newblock Cmos image sensors: electronic camera-on-a-chip.
\newblock \emph{IEEE Transactions on Electron Devices}, 44\penalty0 (10):\penalty0 1689--1698, 1997.
\newblock \doi{10.1109/16.628824}.

\bibitem[Freeman(2024)]{freeman_mmsys_2024_osd}
A.~C. Freeman.
\newblock An open software suite for event-based video.
\newblock In \emph{Proceedings of the 15th ACM Multimedia Systems Conference}, MMSys '24, page 271–277, New York, NY, USA, 2024. Association for Computing Machinery.
\newblock ISBN 9798400704123.
\newblock \doi{10.1145/3625468.3652169}.
\newblock URL \url{https://doi.org/10.1145/3625468.3652169}.

\bibitem[Freeman and Mayer-Patel(2020)]{freeman_emu}
A.~C. Freeman and K.~Mayer-Patel.
\newblock Integrating event camera sensor emulator.
\newblock In \emph{Proceedings of the 28th ACM International Conference on Multimedia}, MM '20, page 4503–4505, New York, NY, USA, 2020. Association for Computing Machinery.
\newblock ISBN 9781450379885.
\newblock \doi{10.1145/3394171.3414394}.
\newblock URL \url{https://doi.org/10.1145/3394171.3414394}.

\bibitem[Freeman and Mayer-Patel(2021)]{FreemanLossyEvent}
A.~C. Freeman and K.~Mayer-Patel.
\newblock Lossy compression for integrating event cameras.
\newblock In \emph{2021 Data Compression Conference (DCC)}, pages 53--62, 2021.
\newblock \doi{10.1109/DCC50243.2021.00013}.

\bibitem[Freeman et~al.(2021)Freeman, Burgess, and Mayer-Patel]{Freeman2021mmsys}
A.~C. Freeman, C.~Burgess, and K.~Mayer-Patel.
\newblock Motion segmentation and tracking for integrating event cameras.
\newblock In \emph{Proceedings of the 12th ACM Multimedia Systems Conference}, MMSys '21, page 1–11, New York, NY, USA, 2021. Association for Computing Machinery.
\newblock ISBN 9781450384346.
\newblock \doi{10.1145/3458305.3463373}.
\newblock URL \url{https://doi.org/10.1145/3458305.3463373}.

\bibitem[Freeman et~al.(2023)Freeman, Singh, and Mayer-Patel]{freeman_mmsys23}
A.~C. Freeman, M.~Singh, and K.~Mayer-Patel.
\newblock An asynchronous intensity representation for framed and event video sources.
\newblock In \emph{Proceedings of the 14th ACM Multimedia Systems Conference}, MMSys '23, page 1–12, New York, NY, USA, 2023. Association for Computing Machinery.
\newblock ISBN 979-8-4007-0148-1/23/06.
\newblock \doi{10.1145/3587819.3590969}.
\newblock URL \url{https://doi.org/10.1145/3587819.3590969}.

\bibitem[Freeman et~al.(2024)Freeman, Mayer-Patel, and Singh]{freeman_mmsys_2024}
A.~C. Freeman, K.~Mayer-Patel, and M.~Singh.
\newblock Accelerated event-based feature detection and compression for surveillance video systems.
\newblock In \emph{Proceedings of the 15th ACM Multimedia Systems Conference}, MMSys '24, page 132–143, New York, NY, USA, 2024. Association for Computing Machinery.
\newblock ISBN 9798400704123.
\newblock \doi{10.1145/3625468.3647618}.
\newblock URL \url{https://doi.org/10.1145/3625468.3647618}.

\bibitem[Fu et~al.(2019)Fu, Li, Dong, Tian, and Huang]{towards_dvs_lossy}
Y.~Fu, J.~Li, S.~Dong, Y.~Tian, and T.~Huang.
\newblock Spike coding: Towards lossy compression for dynamic vision sensor.
\newblock In \emph{2019 Data Compression Conference (DCC)}, pages 572--572, 2019.
\newblock \doi{10.1109/DCC.2019.00084}.

\bibitem[{Gallego} et~al.(2020){Gallego}, {Delbruck}, {Orchard}, {Bartolozzi}, {Taba}, {Censi}, {Leutenegger}, {Davison}, {Conradt}, {Daniilidis}, and {Scaramuzza}]{survey}
G.~{Gallego}, T.~{Delbruck}, G.~M. {Orchard}, C.~{Bartolozzi}, B.~{Taba}, A.~{Censi}, S.~{Leutenegger}, A.~{Davison}, J.~{Conradt}, K.~{Daniilidis}, and D.~{Scaramuzza}.
\newblock Event-based vision: A survey.
\newblock \emph{IEEE Transactions on Pattern Analysis and Machine Intelligence}, pages 1--1, 2020.
\newblock \doi{10.1109/TPAMI.2020.3008413}.

\bibitem[{Galoogahi} et~al.(2017){Galoogahi}, {Fagg}, {Huang}, {Ramanan}, and {Lucey}]{need_for_speed}
H.~K. {Galoogahi}, A.~{Fagg}, C.~{Huang}, D.~{Ramanan}, and S.~{Lucey}.
\newblock Need for speed: A benchmark for higher frame rate object tracking.
\newblock In \emph{2017 IEEE International Conference on Computer Vision (ICCV)}, pages 1134--1143, 2017.
\newblock \doi{10.1109/ICCV.2017.128}.

\bibitem[Gehrig et~al.(2019)Gehrig, Loquercio, Derpanis, and Scaramuzza]{Gehrig_2019_ICCV}
D.~Gehrig, A.~Loquercio, K.~G. Derpanis, and D.~Scaramuzza.
\newblock End-to-end learning of representations for asynchronous event-based data.
\newblock In \emph{Proceedings of the IEEE/CVF International Conference on Computer Vision (ICCV)}, October 2019.

\bibitem[Glover et~al.(2018{\natexlab{a}})Glover, Vasco, and Bartolozzi]{8460541}
A.~Glover, V.~Vasco, and C.~Bartolozzi.
\newblock A controlled-delay event camera framework for on-line robotics.
\newblock In \emph{2018 IEEE International Conference on Robotics and Automation (ICRA)}, pages 2178--2183, 2018{\natexlab{a}}.
\newblock \doi{10.1109/ICRA.2018.8460541}.

\bibitem[Glover et~al.(2018{\natexlab{b}})Glover, Vasco, and Bartolozzi]{glover}
A.~Glover, V.~Vasco, and C.~Bartolozzi.
\newblock A controlled-delay event camera framework for on-line robotics.
\newblock In \emph{2018 IEEE International Conference on Robotics and Automation (ICRA)}, pages 2178--2183, 2018{\natexlab{b}}.
\newblock \doi{10.1109/ICRA.2018.8460541}.

\bibitem[Graham et~al.(2018)Graham, Engelcke, and van~der Maaten]{3DSemanticSegmentationWithSubmanifoldSparseConvNet}
B.~Graham, M.~Engelcke, and L.~van~der Maaten.
\newblock 3d semantic segmentation with submanifold sparse convolutional networks.
\newblock \emph{CVPR}, 2018.

\bibitem[Hamaguchi et~al.(2023)Hamaguchi, Furukawa, Onishi, and Sakurada]{hmnet}
R.~Hamaguchi, Y.~Furukawa, M.~Onishi, and K.~Sakurada.
\newblock Hierarchical neural memory network for low latency event processing.
\newblock In \emph{2023 IEEE/CVF Conference on Computer Vision and Pattern Recognition (CVPR)}, pages 22867--22876, 2023.
\newblock \doi{10.1109/CVPR52729.2023.02190}.

\bibitem[Han et~al.(2021)Han, Li, Mukherjee, Chiang, Grange, Chen, Su, Parker, Deng, Joshi, Chen, Wang, Wilkins, Xu, and Bankoski]{av1}
J.~Han, B.~Li, D.~Mukherjee, C.-H. Chiang, A.~Grange, C.~Chen, H.~Su, S.~Parker, S.~Deng, U.~Joshi, Y.~Chen, Y.~Wang, P.~Wilkins, Y.~Xu, and J.~Bankoski.
\newblock A technical overview of av1.
\newblock \emph{Proceedings of the IEEE}, 109\penalty0 (9):\penalty0 1435--1462, 2021.
\newblock \doi{10.1109/JPROC.2021.3058584}.

\bibitem[Horé and Ziou(2010)]{psnrssim}
A.~Horé and D.~Ziou.
\newblock Image quality metrics: Psnr vs. ssim.
\newblock In \emph{2010 20th International Conference on Pattern Recognition}, pages 2366--2369, 2010.
\newblock \doi{10.1109/ICPR.2010.579}.

\bibitem[Hu et~al.(2022)Hu, Zhao, Ding, Ma, Shi, Xiong, and Huang]{spiking_camera}
L.~Hu, R.~Zhao, Z.~Ding, L.~Ma, B.~Shi, R.~Xiong, and T.~Huang.
\newblock Optical flow estimation for spiking camera.
\newblock In \emph{2022 IEEE/CVF Conference on Computer Vision and Pattern Recognition (CVPR)}, pages 17823--17832, 2022.
\newblock \doi{10.1109/CVPR52688.2022.01732}.

\bibitem[Hu et~al.(2020)Hu, Binas, Neil, Liu, and Delbruck]{ddd20}
Y.~Hu, J.~Binas, D.~Neil, S.-C. Liu, and T.~Delbruck.
\newblock Ddd20 end-to-end event camera driving dataset: Fusing frames and events with deep learning for improved steering prediction.
\newblock In \emph{2020 IEEE 23rd International Conference on Intelligent Transportation Systems (ITSC)}, pages 1--6, 2020.
\newblock \doi{10.1109/ITSC45102.2020.9294515}.

\bibitem[Hu et~al.(2021)Hu, Liu, and Delbruck]{v2e}
Y.~Hu, S.-C. Liu, and T.~Delbruck.
\newblock v2e: From video frames to realistic dvs events.
\newblock In \emph{2021 IEEE/CVF Conference on Computer Vision and Pattern Recognition Workshops (CVPRW)}, pages 1312--1321, 2021.
\newblock \doi{10.1109/CVPRW53098.2021.00144}.

\bibitem[ISO/IEC 14496-14:2020()]{mp4_container}
ISO/IEC 14496-14:2020.
\newblock {Coding of audio-visual objects}.
\newblock Standard, International Organization for Standardization, Geneva, CH, Jan. 2020.

\bibitem[ISO/IEC 23009-1:2022()]{mpeg_dash}
ISO/IEC 23009-1:2022.
\newblock {Dynamic adaptive streaming over HTTP (DASH)}.
\newblock Standard, International Organization for Standardization, Geneva, CH, Aug. 2022.

\bibitem[Jacquemont. et~al.(2019)Jacquemont., Antiga., Vuillaume., Silvestri., Benoit., Lambert., and Maurin.]{visapp19}
M.~Jacquemont., L.~Antiga., T.~Vuillaume., G.~Silvestri., A.~Benoit., P.~Lambert., and G.~Maurin.
\newblock Indexed operations for non-rectangular lattices applied to convolutional neural networks.
\newblock In \emph{Proceedings of the 14th International Joint Conference on Computer Vision, Imaging and Computer Graphics Theory and Applications (VISIGRAPP 2019) - Volume 5: VISAPP}, pages 362--371. INSTICC, SciTePress, 2019.
\newblock ISBN 978-989-758-354-4.
\newblock \doi{10.5220/0007364303620371}.

\bibitem[Jocher et~al.(2022)Jocher, Chaurasia, Stoken, Borovec, NanoCode012, Kwon, Michael, TaoXie, Fang, imyhxy, Lorna, Yifu, Wong, V, Montes, Wang, Fati, Nadar, Laughing, UnglvKitDe, Sonck, tkianai, yxNONG, Skalski, Hogan, Nair, Strobel, and Jain]{yolov5}
G.~Jocher, A.~Chaurasia, A.~Stoken, J.~Borovec, NanoCode012, Y.~Kwon, K.~Michael, TaoXie, J.~Fang, imyhxy, Lorna, Z.~Yifu, C.~Wong, A.~V, D.~Montes, Z.~Wang, C.~Fati, J.~Nadar, Laughing, UnglvKitDe, V.~Sonck, tkianai, yxNONG, P.~Skalski, A.~Hogan, D.~Nair, M.~Strobel, and M.~Jain.
\newblock {ultralytics/yolov5: v7.0 - YOLOv5 SOTA Realtime Instance Segmentation}, Nov. 2022.
\newblock URL \url{https://doi.org/10.5281/zenodo.7347926}.

\bibitem[Joubert et~al.(2021)Joubert, Marcireau, Ralph, Jolley, van Schaik, and Cohen]{10.3389/fnins.2021.702765}
D.~Joubert, A.~Marcireau, N.~Ralph, A.~Jolley, A.~van Schaik, and G.~Cohen.
\newblock Event camera simulator improvements via characterized parameters.
\newblock \emph{Frontiers in Neuroscience}, 15, 2021.
\newblock ISSN 1662-453X.
\newblock \doi{10.3389/fnins.2021.702765}.
\newblock URL \url{https://www.frontiersin.org/articles/10.3389/fnins.2021.702765}.

\bibitem[Kaiser et~al.(2016)Kaiser, Vasquez~Tieck, Hubschneider, Wolf, Weber, Hoff, Friedrich, Wojtasik, Roennau, Kohlhaas, Dillmann, and Zöllner]{7862386}
J.~Kaiser, J.~C. Vasquez~Tieck, C.~Hubschneider, P.~Wolf, M.~Weber, M.~Hoff, A.~Friedrich, K.~Wojtasik, A.~Roennau, R.~Kohlhaas, R.~Dillmann, and J.~M. Zöllner.
\newblock Towards a framework for end-to-end control of a simulated vehicle with spiking neural networks.
\newblock In \emph{2016 IEEE International Conference on Simulation, Modeling, and Programming for Autonomous Robots (SIMPAR)}, pages 127--134, 2016.
\newblock \doi{10.1109/SIMPAR.2016.7862386}.

\bibitem[Kang et~al.(2021)Kang, Li, Zhu, and Tian]{retinomorphic_sensing}
Z.~Kang, J.~Li, L.~Zhu, and Y.~Tian.
\newblock Retinomorphic sensing: A novel paradigm for future multimedia computing.
\newblock In \emph{Proceedings of the 29th ACM International Conference on Multimedia}, MM '21, page 144–152, New York, NY, USA, 2021. Association for Computing Machinery.
\newblock ISBN 9781450386517.
\newblock \doi{10.1145/3474085.3479237}.
\newblock URL \url{https://doi.org/10.1145/3474085.3479237}.

\bibitem[Karpathy et~al.(2014)Karpathy, Toderici, Shetty, Leung, Sukthankar, and Fei-Fei]{large_video_classification}
A.~Karpathy, G.~Toderici, S.~Shetty, T.~Leung, R.~Sukthankar, and L.~Fei-Fei.
\newblock Large-scale video classification with convolutional neural networks.
\newblock In \emph{2014 IEEE Conference on Computer Vision and Pattern Recognition}, pages 1725--1732, 2014.
\newblock \doi{10.1109/CVPR.2014.223}.

\bibitem[Khan et~al.(2021)Khan, Iqbal, and Martini]{khan}
N.~Khan, K.~Iqbal, and M.~G. Martini.
\newblock Time-aggregation-based lossless video encoding for neuromorphic vision sensor data.
\newblock \emph{IEEE Internet of Things Journal}, 8\penalty0 (1):\penalty0 596--609, 2021.
\newblock \doi{10.1109/JIOT.2020.3007866}.

\bibitem[Lagorce et~al.(2017)Lagorce, Orchard, Galluppi, Shi, and Benosman]{time_surface}
X.~Lagorce, G.~Orchard, F.~Galluppi, B.~E. Shi, and R.~B. Benosman.
\newblock Hots: A hierarchy of event-based time-surfaces for pattern recognition.
\newblock \emph{IEEE Transactions on Pattern Analysis and Machine Intelligence}, 39\penalty0 (7):\penalty0 1346--1359, 2017.
\newblock \doi{10.1109/TPAMI.2016.2574707}.

\bibitem[Langdon(1984)]{arithmetic_coding}
G.~G. Langdon.
\newblock An introduction to arithmetic coding.
\newblock \emph{IBM Journal of Research and Development}, 28\penalty0 (2):\penalty0 135--149, 1984.
\newblock \doi{10.1147/rd.282.0135}.

\bibitem[Langley et~al.(2017)Langley, Riddoch, Wilk, Vicente, Krasic, Shi, Zhang, Yang, Kouranov, Swett, Iyengar, Bailey, Dorfman, Roskind, Kulik, Westin, Tenneti, Shade, Hamilton, Vasiliev, and Chang]{quic}
A.~Langley, A.~Riddoch, A.~Wilk, A.~Vicente, C.~B. Krasic, C.~Shi, D.~Zhang, F.~Yang, F.~Kouranov, I.~Swett, J.~Iyengar, J.~Bailey, J.~C. Dorfman, J.~Roskind, J.~Kulik, P.~G. Westin, R.~Tenneti, R.~Shade, R.~Hamilton, V.~Vasiliev, and W.-T. Chang.
\newblock The quic transport protocol: Design and internet-scale deployment.
\newblock 2017.

\bibitem[Lee et~al.(2022)Lee, Cho, Shin, Kim, and Myung]{lee2022vivid}
A.~J. Lee, Y.~Cho, Y.-s. Shin, A.~Kim, and H.~Myung.
\newblock Vivid++: Vision for visibility dataset.
\newblock \emph{IEEE Robotics and Automation Letters}, 7\penalty0 (3):\penalty0 6282--6289, 2022.

\bibitem[Lenzen et~al.(2018)Lenzen, Hedtke, and Christmann]{tone_mapping}
L.~Lenzen, R.~Hedtke, and M.~Christmann.
\newblock How tone mapping influences the bit rate and the bit depth of coded sequences.
\newblock \emph{SMPTE Motion Imaging Journal}, 127\penalty0 (5):\penalty0 38--43, 2018.
\newblock \doi{10.5594/JMI.2018.2810021}.

\bibitem[Li et~al.(2019)Li, Dong, Yu, Tian, and Huang]{8784777}
J.~Li, S.~Dong, Z.~Yu, Y.~Tian, and T.~Huang.
\newblock Event-based vision enhanced: A joint detection framework in autonomous driving.
\newblock In \emph{2019 IEEE International Conference on Multimedia and Expo (ICME)}, pages 1396--1401, 2019.
\newblock \doi{10.1109/ICME.2019.00242}.

\bibitem[Li et~al.(2021)Li, Fu, Dong, Yu, Huang, and Tian]{spike_metric}
J.~Li, Y.~Fu, S.~Dong, Z.~Yu, T.~Huang, and Y.~Tian.
\newblock Asynchronous spatiotemporal spike metric for event cameras.
\newblock \emph{IEEE Transactions on Neural Networks and Learning Systems}, pages 1--12, 2021.
\newblock \doi{10.1109/TNNLS.2021.3061122}.

\bibitem[Li et~al.(2016)Li, Aaron, Katsavounidis, Moorthy, and Manohara]{vmaf}
Z.~Li, A.~Aaron, I.~Katsavounidis, A.~Moorthy, and M.~Manohara.
\newblock Toward a practical perceptual video quality metric.
\newblock \emph{The Netflix Tech Blog}, 6\penalty0 (2), 2016.

\bibitem[{Lichtsteiner} et~al.(2006){Lichtsteiner}, {Posch}, and {Delbruck}]{dvs}
P.~{Lichtsteiner}, C.~{Posch}, and T.~{Delbruck}.
\newblock A 128 x 128 120db 30mw asynchronous vision sensor that responds to relative intensity change.
\newblock In \emph{2006 IEEE International Solid State Circuits Conference - Digest of Technical Papers}, pages 2060--2069, 2006.

\bibitem[MacKay(2002)]{information_theory_textbook}
D.~J.~C. MacKay.
\newblock \emph{Information Theory, Inference \& Learning Algorithms}.
\newblock Cambridge University Press, USA, 2002.
\newblock ISBN 0521642981.

\bibitem[Maqueda et~al.(2018)Maqueda, Loquercio, Gallego, Garc{\i}a, and Scaramuzza]{eventcar}
A.~I. Maqueda, A.~Loquercio, G.~Gallego, N.~Garc{\i}a, and D.~Scaramuzza.
\newblock Event-based vision meets deep learning on steering prediction for self-driving cars.
\newblock In \emph{Proceedings of the IEEE Conference on Computer Vision and Pattern Recognition}, pages 5419--5427, 2018.

\bibitem[Mendelovich(2023)]{phantom_article}
Y.~Mendelovich.
\newblock Meet the new phantom t4040 high-speed camera: 9,350 fps at 2.5k of resolution, Mar 2023.
\newblock URL \url{https://web.archive.org/web/20230317100431/https://ymcinema.com/2023/03/06/meet-the-new-phantom-t4040-high-speed-camera-9350-fps-at-2-5k-of-resolution/}.

\bibitem[Messikommer et~al.(2020)Messikommer, Gehrig, Loquercio, and Scaramuzza]{Messikommer20eccv}
N.~Messikommer, D.~Gehrig, A.~Loquercio, and D.~Scaramuzza.
\newblock Event-based asynchronous sparse convolutional networks.
\newblock In \emph{European Conference on Computer Vision. (ECCV)}, 2020.
\newblock URL \url{http://rpg.ifi.uzh.ch/docs/ECCV20_Messikommer.pdf}.

\bibitem[Mittal et~al.(2013)Mittal, Soundararajan, and Bovik]{niqe}
A.~Mittal, R.~Soundararajan, and A.~C. Bovik.
\newblock Making a “completely blind” image quality analyzer.
\newblock \emph{IEEE Signal Processing Letters}, 20\penalty0 (3):\penalty0 209--212, 2013.
\newblock \doi{10.1109/LSP.2012.2227726}.

\bibitem[{Moeys} et~al.(2018){Moeys}, {Corradi}, {Li}, {Bamford}, {Longinotti}, {Voigt}, {Berry}, {Taverni}, {Helmchen}, and {Delbruck}]{sensitive_color_davis}
D.~P. {Moeys}, F.~{Corradi}, C.~{Li}, S.~A. {Bamford}, L.~{Longinotti}, F.~F. {Voigt}, S.~{Berry}, G.~{Taverni}, F.~{Helmchen}, and T.~{Delbruck}.
\newblock A sensitive dynamic and active pixel vision sensor for color or neural imaging applications.
\newblock \emph{IEEE Transactions on Biomedical Circuits and Systems}, 12\penalty0 (1):\penalty0 123--136, Feb 2018.
\newblock ISSN 1940-9990.
\newblock \doi{10.1109/TBCAS.2017.2759783}.

\bibitem[Moorthy and Bovik(2010)]{biqi}
A.~K. Moorthy and A.~C. Bovik.
\newblock A two-step framework for constructing blind image quality indices.
\newblock \emph{IEEE Signal Processing Letters}, 17\penalty0 (5):\penalty0 513--516, 2010.
\newblock \doi{10.1109/LSP.2010.2043888}.

\bibitem[Morrison(1992)]{h261}
G.~Morrison.
\newblock Video coding standards for multimedia: Jpeg, h.261, mpeg.
\newblock In \emph{IEE Colloquium on Technology Support of Multimedia}, pages 2/1--2/4, 1992.

\bibitem[Mueggler et~al.(2017)Mueggler, Rebecq, Gallego, Delbruck, and Scaramuzza]{davis_dataset}
E.~Mueggler, H.~Rebecq, G.~Gallego, T.~Delbruck, and D.~Scaramuzza.
\newblock The event-camera dataset and simulator: Event-based data for pose estimation, visual odometry, and slam.
\newblock \emph{The International Journal of Robotics Research}, 36\penalty0 (2):\penalty0 142--149, 2017.
\newblock \doi{10.1177/0278364917691115}.
\newblock URL \url{https://doi.org/10.1177/0278364917691115}.

\bibitem[Muybridge(c1878)]{horse_in_motion}
E.~Muybridge.
\newblock The horse in motion. "sallie gardner," owned by leland stanford; running at a 1:40 gait over the palo alto track, 19th june 1878 / muybridge., c1878.
\newblock URL \url{https://www.loc.gov/item/97502309/}.
\newblock [Online; accessed February 6, 2024].

\bibitem[Oh et~al.(2011)Oh, Hoogs, Perera, Cuntoor, Chen, Lee, Mukherjee, Aggarwal, Lee, Davis, Swears, Wang, Ji, Reddy, Shah, Vondrick, Pirsiavash, Ramanan, Yuen, Torralba, Song, Fong, Roy-Chowdhury, and Desai]{virat}
S.~Oh, A.~Hoogs, A.~Perera, N.~Cuntoor, C.-C. Chen, J.~T. Lee, S.~Mukherjee, J.~K. Aggarwal, H.~Lee, L.~Davis, E.~Swears, X.~Wang, Q.~Ji, K.~Reddy, M.~Shah, C.~Vondrick, H.~Pirsiavash, D.~Ramanan, J.~Yuen, A.~Torralba, B.~Song, A.~Fong, A.~Roy-Chowdhury, and M.~Desai.
\newblock A large-scale benchmark dataset for event recognition in surveillance video.
\newblock In \emph{CVPR 2011}, pages 3153--3160, 2011.
\newblock \doi{10.1109/CVPR.2011.5995586}.

\bibitem[Omori et~al.(2017)Omori, Onishi, Iwasaki, and Shimizu]{high_fps_hevc}
Y.~Omori, T.~Onishi, H.~Iwasaki, and A.~Shimizu.
\newblock A 120 fps high frame rate real-time hevc video encoder with parallel configuration scalable to 4k.
\newblock In \emph{2017 IEEE Symposium in Low-Power and High-Speed Chips (COOL CHIPS)}, pages 1--3, 2017.
\newblock \doi{10.1109/CoolChips.2017.7946382}.

\bibitem[{OpenCV}(2020)]{opencv}
{OpenCV}.
\newblock Opencv, 2020.
\newblock URL \url{https://opencv.org/}.

\bibitem[Pan et~al.(2019)Pan, Scheerlinck, Yu, Hartley, Liu, and Dai]{Pan_EDI}
L.~Pan, C.~Scheerlinck, X.~Yu, R.~Hartley, M.~Liu, and Y.~Dai.
\newblock Bringing a blurry frame alive at high frame-rate with an event camera.
\newblock In \emph{Proceedings of the IEEE/CVF Conference on Computer Vision and Pattern Recognition (CVPR)}, June 2019.

\bibitem[Pan et~al.(2020)Pan, Hartley, Scheerlinck, Liu, Yu, and Dai]{high-framerate_recon}
L.~Pan, R.~Hartley, C.~Scheerlinck, M.~Liu, X.~Yu, and Y.~Dai.
\newblock High frame rate video reconstruction based on an event camera.
\newblock \emph{IEEE Transactions on Pattern Analysis and Machine Intelligence}, pages 1--1, 2020.
\newblock \doi{10.1109/TPAMI.2020.3036667}.

\bibitem[Pantos and May(2017)]{apple_hls}
R.~Pantos and W.~May.
\newblock {HTTP Live Streaming}.
\newblock RFC 8216, Aug. 2017.
\newblock URL \url{https://www.rfc-editor.org/info/rfc8216}.

\bibitem[Paredes-Vallés et~al.(2020)Paredes-Vallés, Scheper, and de~Croon]{dvs_snn}
F.~Paredes-Vallés, K.~Y.~W. Scheper, and G.~C. H.~E. de~Croon.
\newblock Unsupervised learning of a hierarchical spiking neural network for optical flow estimation: From events to global motion perception.
\newblock \emph{IEEE Transactions on Pattern Analysis and Machine Intelligence}, 42\penalty0 (8):\penalty0 2051--2064, 2020.
\newblock \doi{10.1109/TPAMI.2019.2903179}.

\bibitem[Parger et~al.(2022)Parger, Tang, Twigg, Keskin, Wang, and Steinberger]{deltacnn}
M.~Parger, C.~Tang, C.~D. Twigg, C.~Keskin, R.~Wang, and M.~Steinberger.
\newblock Deltacnn: End-to-end cnn inference of sparse frame differences in videos.
\newblock In \emph{2022 IEEE/CVF Conference on Computer Vision and Pattern Recognition (CVPR)}, pages 12487--12496, 2022.
\newblock \doi{10.1109/CVPR52688.2022.01217}.

\bibitem[Pehle and Pedersen(2021)]{norse2021}
C.~Pehle and J.~E. Pedersen.
\newblock {Norse - A deep learning library for spiking neural networks}, Jan. 2021.
\newblock URL \url{https://doi.org/10.5281/zenodo.4422025}.
\newblock Documentation: https://norse.ai/docs/.

\bibitem[Pennebaker et~al.(1988)Pennebaker, Mitchell, Langdon, and Arps]{cabac}
W.~B. Pennebaker, J.~L. Mitchell, G.~G. Langdon, and R.~B. Arps.
\newblock An overview of the basic principles of the q-coder adaptive binary arithmetic coder.
\newblock \emph{IBM Journal of Research and Development}, 32\penalty0 (6):\penalty0 717--726, 1988.
\newblock \doi{10.1147/rd.326.0717}.

\bibitem[Piechaczek et~al.(2022)Piechaczek, Schrey, Ligges, Hosticka, and Kokozinski]{antiblooming}
D.~S. Piechaczek, O.~Schrey, M.~Ligges, B.~Hosticka, and R.~Kokozinski.
\newblock Anti-blooming clocking for time-delay integration {CCDs}.
\newblock \emph{Sensors (Basel)}, 22\penalty0 (19):\penalty0 7520, Oct. 2022.

\bibitem[Posch et~al.(2010)Posch, Matolin, and Wohlgenannt]{atis}
C.~Posch, D.~Matolin, and R.~Wohlgenannt.
\newblock A qvga 143db dynamic range asynchronous address-event pwm dynamic image sensor with lossless pixel-level video compression.
\newblock In \emph{2010 IEEE International Solid-State Circuits Conference - (ISSCC)}, pages 400--401, 2010.
\newblock \doi{10.1109/ISSCC.2010.5433973}.

\bibitem[Posch et~al.(2014)Posch, Serrano-Gotarredona, Linares-Barranco, and Delbruck]{atis2}
C.~Posch, T.~Serrano-Gotarredona, B.~Linares-Barranco, and T.~Delbruck.
\newblock Retinomorphic event-based vision sensors: Bioinspired cameras with spiking output.
\newblock \emph{Proceedings of the IEEE}, 102\penalty0 (10):\penalty0 1470--1484, 2014.
\newblock \doi{10.1109/JPROC.2014.2346153}.

\bibitem[Qblocks()]{Qblocks}
Qblocks.
\newblock Nvidia tesla a100 gpu benchmarks.
\newblock URL \url{https://web.archive.org/web/20230303144803/https://www.qblocks.cloud/creators/nvidia-tesla-a100-gpu-benchmarks}.

\bibitem[Rassool(2017)]{vmaf_reproducibility}
R.~Rassool.
\newblock Vmaf reproducibility: Validating a perceptual practical video quality metric.
\newblock In \emph{2017 IEEE International Symposium on Broadband Multimedia Systems and Broadcasting (BMSB)}, pages 1--2, 2017.
\newblock \doi{10.1109/BMSB.2017.7986143}.

\bibitem[Rebecq et~al.(2017)Rebecq, Horstschaefer, and Scaramuzza]{eventframe}
H.~Rebecq, T.~Horstschaefer, and D.~Scaramuzza.
\newblock Real-time visual-inertial odometry for event cameras using keyframe-based nonlinear optimization.
\newblock In \emph{British Machine Vision Conference}, 2017.
\newblock URL \url{https://api.semanticscholar.org/CorpusID:30723444}.

\bibitem[Rebecq et~al.(2018)Rebecq, Gehrig, and Scaramuzza]{pmlr-v87-rebecq18a}
H.~Rebecq, D.~Gehrig, and D.~Scaramuzza.
\newblock Esim: an open event camera simulator.
\newblock In A.~Billard, A.~Dragan, J.~Peters, and J.~Morimoto, editors, \emph{Proceedings of The 2nd Conference on Robot Learning}, volume~87 of \emph{Proceedings of Machine Learning Research}, pages 969--982. PMLR, 29--31 Oct 2018.
\newblock URL \url{https://proceedings.mlr.press/v87/rebecq18a.html}.

\bibitem[Rebecq et~al.(2019{\natexlab{a}})Rebecq, Ranftl, Koltun, and Scaramuzza]{Rebecq19cvpr}
H.~Rebecq, R.~Ranftl, V.~Koltun, and D.~Scaramuzza.
\newblock Events-to-video: Bringing modern computer vision to event cameras.
\newblock \emph{{IEEE} Conf. Comput. Vis. Pattern Recog. (CVPR)}, 2019{\natexlab{a}}.

\bibitem[Rebecq et~al.(2019{\natexlab{b}})Rebecq, Ranftl, Koltun, and Scaramuzza]{Rebecq19pami}
H.~Rebecq, R.~Ranftl, V.~Koltun, and D.~Scaramuzza.
\newblock High speed and high dynamic range video with an event camera.
\newblock \emph{{IEEE} Trans. Pattern Anal. Mach. Intell. (T-PAMI)}, 2019{\natexlab{b}}.
\newblock URL \url{http://rpg.ifi.uzh.ch/docs/TPAMI19_Rebecq.pdf}.

\bibitem[{Rebecq} et~al.(2019){Rebecq}, {Ranftl}, {Koltun}, and {Scaramuzza}]{event_cv}
H.~{Rebecq}, R.~{Ranftl}, V.~{Koltun}, and D.~{Scaramuzza}.
\newblock Events-to-video: Bringing modern computer vision to event cameras.
\newblock In \emph{2019 IEEE/CVF Conference on Computer Vision and Pattern Recognition (CVPR)}, pages 3852--3861, 2019.
\newblock \doi{10.1109/CVPR.2019.00398}.

\bibitem[Rebecq et~al.(2019)Rebecq, Ranftl, Koltun, and Scaramuzza]{rebecq2019high}
H.~Rebecq, R.~Ranftl, V.~Koltun, and D.~Scaramuzza.
\newblock High speed and high dynamic range video with an event camera, 2019.

\bibitem[Reinbacher et~al.(2016)Reinbacher, Graber, and Pock]{reinbacher}
C.~Reinbacher, G.~Graber, and T.~Pock.
\newblock {Real-Time Intensity-Image Reconstruction for Event Cameras Using Manifold Regularisation}.
\newblock In \emph{2016 British Machine Vision Conference (BMVC)}, 2016.

\bibitem[Richardson(2010)]{richardson}
I.~E. Richardson.
\newblock \emph{The H.264 Advanced Video Compression Standard}.
\newblock Wiley Publishing, 2nd edition, 2010.
\newblock ISBN 0470516925.

\bibitem[Rosten and Drummond(2005)]{fast_features}
E.~Rosten and T.~Drummond.
\newblock Fusing points and lines for high performance tracking.
\newblock In \emph{Tenth IEEE International Conference on Computer Vision (ICCV'05) Volume 1}, volume~2, pages 1508--1515 Vol. 2, 2005.
\newblock \doi{10.1109/ICCV.2005.104}.

\bibitem[Roy and Bhowmik(2020)]{cv_survey}
S.~D. Roy and M.~K. Bhowmik.
\newblock A comprehensive survey on computer vision based approaches for moving object detection.
\newblock In \emph{2020 IEEE Region 10 Symposium (TENSYMP)}, pages 1531--1534, 2020.
\newblock \doi{10.1109/TENSYMP50017.2020.9230869}.

\bibitem[Scheerlinck et~al.(2018)Scheerlinck, Barnes, and Mahony]{scheerlinck2018continuoustime}
C.~Scheerlinck, N.~Barnes, and R.~Mahony.
\newblock Continuous-time intensity estimation using event cameras, 2018.

\bibitem[Scheerlinck et~al.(2019)Scheerlinck, Barnes, and Mahony]{Scheerlinck}
C.~Scheerlinck, N.~Barnes, and R.~Mahony.
\newblock Continuous-time intensity estimation using event cameras.
\newblock In C.~Jawahar, H.~Li, G.~Mori, and K.~Schindler, editors, \emph{Computer Vision -- ACCV 2018}, pages 308--324, Cham, 2019. Springer International Publishing.
\newblock ISBN 978-3-030-20873-8.

\bibitem[Scheerlinck et~al.(2020)Scheerlinck, Rebecq, Gehrig, Barnes, Mahony, and Scaramuzza]{Scheerlinck20wacv}
C.~Scheerlinck, H.~Rebecq, D.~Gehrig, N.~Barnes, R.~Mahony, and D.~Scaramuzza.
\newblock Fast image reconstruction with an event camera.
\newblock In \emph{{IEEE} Winter Conf. Appl. Comput. Vis. {(WACV)}}, pages 156--163, 2020.
\newblock \doi{10.1109/WACV45572.2020.9093366}.

\bibitem[Schiopu and Bilcu(2022{\natexlab{a}})]{Schiopu_1}
I.~Schiopu and R.~C. Bilcu.
\newblock Lossless compression of event camera frames.
\newblock \emph{IEEE Signal Processing Letters}, 29:\penalty0 1779--1783, 2022{\natexlab{a}}.
\newblock \doi{10.1109/LSP.2022.3196599}.

\bibitem[Schiopu and Bilcu(2022{\natexlab{b}})]{Schiopu_2}
I.~Schiopu and R.~C. Bilcu.
\newblock Low-complexity lossless coding of asynchronous event sequences for low-power chip integration.
\newblock \emph{Sensors}, 22\penalty0 (24), 2022{\natexlab{b}}.
\newblock ISSN 1424-8220.
\newblock \doi{10.3390/s222410014}.
\newblock URL \url{https://www.mdpi.com/1424-8220/22/24/10014}.

\bibitem[Schiopu and Bilcu(2023)]{Schiopu_3}
I.~Schiopu and R.~C. Bilcu.
\newblock Entropy coding-based lossless compression of asynchronous event sequences.
\newblock In \emph{2023 IEEE/CVF Conference on Computer Vision and Pattern Recognition Workshops (CVPRW)}, pages 3923--3930, 2023.
\newblock \doi{10.1109/CVPRW59228.2023.00407}.

\bibitem[Seufert et~al.(2015)Seufert, Egger, Slanina, Zinner, Hoßfeld, and Tran-Gia]{qoe}
M.~Seufert, S.~Egger, M.~Slanina, T.~Zinner, T.~Hoßfeld, and P.~Tran-Gia.
\newblock A survey on quality of experience of http adaptive streaming.
\newblock \emph{IEEE Communications Surveys \& Tutorials}, 17\penalty0 (1):\penalty0 469--492, 2015.
\newblock \doi{10.1109/COMST.2014.2360940}.

\bibitem[Si et~al.(2020)Si, Liu, Bi, and Shan]{deep_multiplicative}
W.~Si, C.~Liu, Z.~Bi, and M.~Shan.
\newblock Modeling long-term dependencies from videos using deep multiplicative neural networks.
\newblock \emph{ACM Trans. Multimedia Comput. Commun. Appl.}, 16\penalty0 (2s), jul 2020.
\newblock ISSN 1551-6857.
\newblock \doi{10.1145/3357797}.
\newblock URL \url{https://doi.org/10.1145/3357797}.

\bibitem[{Singh} et~al.(2017){Singh}, {Zhang}, {Vitkus}, {Mayer-Patel}, and {Vicci}]{montek}
M.~{Singh}, P.~{Zhang}, A.~{Vitkus}, K.~{Mayer-Patel}, and L.~{Vicci}.
\newblock A frameless imaging sensor with asynchronous pixels: An architectural evaluation.
\newblock In \emph{2017 23rd IEEE International Symposium on Asynchronous Circuits and Systems (ASYNC)}, pages 110--117, May 2017.
\newblock \doi{10.1109/ASYNC.2017.19}.

\bibitem[Smith et~al.(2017)Smith, Singh, and Mayer-Patel]{Smith2017ASM}
A.~J. Smith, M.~Singh, and K.~Mayer-Patel.
\newblock A system model for frameless asynchronous high dynamic range sensors.
\newblock In \emph{NOSSDAV'17}, 2017.

\bibitem[Song et~al.(2022)Song, Huang, and Bajaj]{song_ecir}
C.~Song, Q.~Huang, and C.~Bajaj.
\newblock E-cir: Event-enhanced continuous intensity recovery.
\newblock In \emph{2022 IEEE/CVF Conference on Computer Vision and Pattern Recognition (CVPR)}, pages 7793--7802, 2022.
\newblock \doi{10.1109/CVPR52688.2022.00765}.

\bibitem[Stoffregen et~al.(2019)Stoffregen, Gallego, Drummond, Kleeman, and Scaramuzza]{compensation}
T.~Stoffregen, G.~Gallego, T.~Drummond, L.~Kleeman, and D.~Scaramuzza.
\newblock Event-based motion segmentation by motion compensation.
\newblock \emph{CoRR}, abs/1904.01293, 2019.
\newblock URL \url{http://arxiv.org/abs/1904.01293}.

\bibitem[Sullivan et~al.(2012)Sullivan, Ohm, Han, and Wiegand]{h265}
G.~J. Sullivan, J.-R. Ohm, W.-J. Han, and T.~Wiegand.
\newblock Overview of the high efficiency video coding (hevc) standard.
\newblock \emph{IEEE Transactions on Circuits and Systems for Video Technology}, 22\penalty0 (12):\penalty0 1649--1668, 2012.
\newblock \doi{10.1109/TCSVT.2012.2221191}.

\bibitem[Talluri et~al.(1997)Talluri, Oehler, Barmon, Courtney, Das, and Liao]{obvc}
R.~Talluri, K.~Oehler, T.~Barmon, J.~Courtney, A.~Das, and J.~Liao.
\newblock A robust, scalable, object-based video compression technique for very low bit-rate coding.
\newblock \emph{IEEE Transactions on Circuits and Systems for Video Technology}, 7\penalty0 (1):\penalty0 221--233, 1997.
\newblock \doi{10.1109/76.554433}.

\bibitem[{Taverni} et~al.(2018){Taverni}, {Paul Moeys}, {Li}, {Cavaco}, {Motsnyi}, {San Segundo Bello}, and {Delbruck}]{new_color_davis}
G.~{Taverni}, D.~{Paul Moeys}, C.~{Li}, C.~{Cavaco}, V.~{Motsnyi}, D.~{San Segundo Bello}, and T.~{Delbruck}.
\newblock Front and back illuminated dynamic and active pixel vision sensors comparison.
\newblock \emph{IEEE Transactions on Circuits and Systems II: Express Briefs}, 65\penalty0 (5):\penalty0 677--681, May 2018.
\newblock ISSN 1558-3791.
\newblock \doi{10.1109/TCSII.2018.2824899}.

\bibitem[Vavra()]{triage_article}
S.~Vavra.
\newblock The nsa is experimenting with machine learning concepts its workforce will trust.
\newblock URL \url{https://web.archive.org/web/20231004221857/https://cyberscoop.com/nsa-machine-learning-workforce/}.

\bibitem[Viola and Jones(2001)]{viola}
P.~Viola and M.~Jones.
\newblock Robust real-time face detection.
\newblock In \emph{Proceedings Eighth IEEE International Conference on Computer Vision. ICCV 2001}, volume~2, pages 747--747, 2001.
\newblock \doi{10.1109/ICCV.2001.937709}.

\bibitem[Wang et~al.(2020)Wang, He, Yu, Xia, and Yang]{wang2020event}
B.~Wang, J.~He, L.~Yu, G.-S. Xia, and W.~Yang.
\newblock Event enhanced high-quality image recovery.
\newblock In \emph{European Conference on Computer Vision}. Springer, 2020.

\bibitem[Wang et~al.(2021{\natexlab{a}})Wang, Huo, and Zhang]{dynamic_range}
H.~Wang, Y.~Huo, and H.~Zhang.
\newblock Research on dynamic range analysis and improvement of imaging equipment.
\newblock In \emph{2021 Workshop on Algorithm and Big Data}, WABD 2021, page 40–44, New York, NY, USA, 2021{\natexlab{a}}. Association for Computing Machinery.
\newblock ISBN 9781450389945.
\newblock \doi{10.1145/3456389.3456392}.
\newblock URL \url{https://doi.org/10.1145/3456389.3456392}.

\bibitem[Wang et~al.(2019)Wang, Inguva, and Adsumilli]{yt-ugc-dataset}
Y.~Wang, S.~Inguva, and B.~Adsumilli.
\newblock Youtube ugc dataset for video compression research.
\newblock In \emph{2019 IEEE 21st International Workshop on Multimedia Signal Processing (MMSP)}, 2019.

\bibitem[Wang et~al.(2003)Wang, Simoncelli, and Bovik]{ssim}
Z.~Wang, E.~Simoncelli, and A.~Bovik.
\newblock Multiscale structural similarity for image quality assessment.
\newblock In \emph{The Thrity-Seventh Asilomar Conference on Signals, Systems \& Computers, 2003}, volume~2, pages 1398--1402 Vol.2, 2003.
\newblock \doi{10.1109/ACSSC.2003.1292216}.

\bibitem[Wang et~al.(2021{\natexlab{b}})Wang, Ng, Scheerlinck, and Mahony]{async_kalman_filter}
Z.~Wang, Y.~Ng, C.~Scheerlinck, and R.~Mahony.
\newblock An asynchronous kalman filter for hybrid event cameras.
\newblock In \emph{Proceedings of the IEEE/CVF International Conference on Computer Vision (ICCV)}, pages 448--457, October 2021{\natexlab{b}}.

\bibitem[Wiles et~al.(2023)Wiles, Carreira, Barr, Zisserman, and Malinowski]{compressed_vision}
O.~Wiles, J.~Carreira, I.~Barr, A.~Zisserman, and M.~Malinowski.
\newblock Compressed vision for efficient video understanding.
\newblock In L.~Wang, J.~Gall, T.-J. Chin, I.~Sato, and R.~Chellappa, editors, \emph{Computer Vision -- ACCV 2022}, pages 679--695, Cham, 2023. Springer Nature Switzerland.
\newblock ISBN 978-3-031-26293-7.

\bibitem[Wu et~al.(2018)Wu, Zaheer, Hu, Manmatha, Smola, and Krahenbuhl]{compressed_recognition}
C.~Wu, M.~Zaheer, H.~Hu, R.~Manmatha, A.~J. Smola, and P.~Krahenbuhl.
\newblock Compressed video action recognition.
\newblock In \emph{2018 IEEE/CVF Conference on Computer Vision and Pattern Recognition (CVPR)}, pages 6026--6035, Los Alamitos, CA, USA, jun 2018. IEEE Computer Society.
\newblock \doi{10.1109/CVPR.2018.00631}.
\newblock URL \url{https://doi.ieeecomputersociety.org/10.1109/CVPR.2018.00631}.

\bibitem[{Wu} et~al.(2013){Wu}, {Lim}, and {Yang}]{object_tracking}
Y.~{Wu}, J.~{Lim}, and M.~{Yang}.
\newblock Online object tracking: A benchmark.
\newblock In \emph{2013 IEEE Conference on Computer Vision and Pattern Recognition}, pages 2411--2418, 2013.
\newblock \doi{10.1109/CVPR.2013.312}.

\bibitem[Xu et~al.(2021)Xu, Xu, Gao, Lin, and Nie]{vidar2}
J.~Xu, L.~Xu, Z.~Gao, P.~Lin, and K.~Nie.
\newblock A denoising method based on pulse interval compensation for high-speed spike-based image sensor.
\newblock \emph{IEEE Transactions on Circuits and Systems for Video Technology}, 31\penalty0 (8):\penalty0 2966--2980, 2021.
\newblock \doi{10.1109/TCSVT.2020.3034649}.

\bibitem[Xu et~al.(2020)Xu, Qin, Sun, Wang, Chen, and Ren]{learning_freq}
K.~Xu, M.~Qin, F.~Sun, Y.~Wang, Y.~Chen, and F.~Ren.
\newblock Learning in the frequency domain.
\newblock In \emph{2020 IEEE/CVF Conference on Computer Vision and Pattern Recognition (CVPR)}, pages 1737--1746, Los Alamitos, CA, USA, jun 2020. IEEE Computer Society.
\newblock \doi{10.1109/CVPR42600.2020.00181}.
\newblock URL \url{https://doi.ieeecomputersociety.org/10.1109/CVPR42600.2020.00181}.

\bibitem[Yi et~al.(2022)Yi, Wang, Kwong, and Kuo]{task_driven_compression}
X.~Yi, H.~Wang, S.~Kwong, and C.-C.~J. Kuo.
\newblock Task-driven video compression for humans and machines: Framework design and optimization.
\newblock \emph{IEEE Transactions on Multimedia}, pages 1--12, 2022.
\newblock \doi{10.1109/TMM.2022.3233245}.

\bibitem[Yu et~al.(2020)Yu, Chen, Wang, Xian, Chen, Liu, Madhavan, and Darrell]{bdd100k}
F.~Yu, H.~Chen, X.~Wang, W.~Xian, Y.~Chen, F.~Liu, V.~Madhavan, and T.~Darrell.
\newblock Bdd100k: A diverse driving dataset for heterogeneous multitask learning.
\newblock In \emph{2020 IEEE/CVF Conference on Computer Vision and Pattern Recognition (CVPR)}, pages 2633--2642, 2020.
\newblock \doi{10.1109/CVPR42600.2020.00271}.

\bibitem[Yu et~al.(2017)Yu, Jhang, Wei, Tseng, Wen, Liu, Lin, Chen, Chiou, and Lee]{7991046}
Y.-C. Yu, J.-W. Jhang, X.~Wei, H.-W. Tseng, Y.~Wen, Z.~Liu, T.-L. Lin, S.-L. Chen, Y.-S. Chiou, and H.-Y. Lee.
\newblock Chroma upsampling for ycbcr 420 videos.
\newblock In \emph{2017 IEEE International Conference on Consumer Electronics - Taiwan (ICCE-TW)}, pages 163--164, 2017.
\newblock \doi{10.1109/ICCE-China.2017.7991046}.

\bibitem[Zhu et~al.(2018)Zhu, Yuan, Chaney, and Daniilidis]{EV-FlowNet}
A.~Zhu, L.~Yuan, K.~Chaney, and K.~Daniilidis.
\newblock Ev-flownet: Self-supervised optical flow estimation for event-based cameras.
\newblock In \emph{Proceedings of Robotics: Science and Systems}, Pittsburgh, Pennsylvania, June 2018.
\newblock \doi{10.15607/RSS.2018.XIV.062}.

\bibitem[Zhu et~al.(2021{\natexlab{a}})Zhu, Liu, Liu, Luo, Ye, Xu, Huang, Jiao, Xu, Zhang, and Gu]{h266}
B.~Zhu, S.~Liu, Y.~Liu, Y.~Luo, J.~Ye, H.~Xu, Y.~Huang, H.~Jiao, X.~Xu, X.~Zhang, and C.~Gu.
\newblock A real-time h.266/vvc software decoder.
\newblock In \emph{2021 IEEE International Conference on Multimedia and Expo (ICME)}, pages 1--6, 2021{\natexlab{a}}.
\newblock \doi{10.1109/ICME51207.2021.9428470}.

\bibitem[Zhu et~al.(2021{\natexlab{b}})Zhu, Li, Wang, Huang, and Tian]{neuspikenet}
L.~Zhu, J.~Li, X.~Wang, T.~Huang, and Y.~Tian.
\newblock Neuspike-net: High speed video reconstruction via bio-inspired neuromorphic cameras.
\newblock In \emph{2021 IEEE/CVF International Conference on Computer Vision (ICCV)}, pages 2380--2389, 2021{\natexlab{b}}.
\newblock \doi{10.1109/ICCV48922.2021.00240}.

\end{thebibliography}
\end{singlespace}
}

\newpage

\end{document}